\documentclass[a4paper,12pt,openright, useAMS, usenatbib]{book}
\pdfoutput=1
\usepackage{customisations}

\usepackage{booktabs} % booktabs provides professional formatting commands for tables
\usepackage{amsmath} % amsmath provides extra maths symbols
\usepackage{textcomp} % textcomp provides extra text symbols (like a degrees celsius symbol)
\usepackage{graphics}
\usepackage{amssymb} 
\usepackage{amsmath}
\usepackage{placeins}
\usepackage{physics}
\usepackage{floats}
\usepackage{comment}
\usepackage{adjustbox}
\usepackage{blindtext}
\usepackage{csquotes}
\usepackage{afterpage}
\usepackage{booktabs}
\usepackage{multirow}
\usepackage{siunitx}

\usepackage{graphicx}
\usepackage{natbib}
\usepackage{xspace}
\usepackage{cancel}
\usepackage{url}
\usepackage{breqn}
\usepackage{hyperref}
\hypersetup{
    colorlinks=true,
    linkcolor=red,
    filecolor=blue,      
    urlcolor=blue,
    bookmarks=false,
    citecolor=blue
}

\def\[{\begin{equation}}
\def\]{\end{equation}}
\def\gsim{\mathrel{\lower0.6ex\hbox{$\buildrel {\textstyle >}\over {\scriptstyle \sim}$}}}
\def\lsim{\mathrel{\lower0.6ex\hbox{$\buildrel {\textstyle <}\over {\scriptstyle \sim}$}}}
\def\deg{^\circ}

\def\citejap#1{\citeauthor{#1}\ \citeyear{#1}}

\newcommand{\los}{$\boldsymbol{\hat{\eta}}$}
\newcommand{\mpcoh}{\,\ensuremath{h^{-1}}\textrm{Mpc}}
\newcommand{\gpcoh}{\,\ensuremath{h^{-1}}$\textrm{Gpc}$}
\newcommand{\hompc}{\,\ensuremath{h}\textrm{Mpc}^{-1}}
  
\newcommand{\bnabla}{\boldsymbol \nabla}
\newcommand{\diag}{\rm{diag}}
\newcommand{\aj}{AJ}
\newcommand{\aap}{A\&A}
\newcommand{\apj}{ApJ}
\newcommand{\apjl}{ApJL}
\newcommand{\apjs}{ApJS}
\newcommand{\apss}{Ap \& SS}
\newcommand{\araa}{ARA \& A}
\newcommand{\mnras}{MNRAS}
\newcommand{\prd}{Phys.~Rev.~D}

\newcommand{\nat}{Nature}
\newcommand{\pasp}{Publications of the ASP}
\newcommand{\physrep}{Physics Reports}
\newcommand{\pasa}{Publications of the Astron. Soc. of Australia}
\newcommand{\jcap}{Journal of Cosmology and Astroparticle Physics}

\makeatletter
\g@addto@macro\bfseries{\boldmath}
\makeatother

\title{Geometric and growth rate tests of General Relativity with recovered linear cosmological perturbations}
\author{Michael J. Wilson}
\date{June 2016} % of submission

\graphicspath{{./basics_images/}{./RSD_images/}{./clustering_images/}{./VIPERS_images/}{./MaskedRSD_images/}{./Clipping_images/}{./VIPERSRSD_images/}{./thesis-frontmatter/}}

\begin{document}
% Thesis front matter - title page, abstract, acknowledgements, declaration and table of contents
% See customisations.sty to modify the title page or declaration
\singlespacing
\maketitlepage
\frontmatter
\eighteenptleading
\parindent2.5em

\chapter{Abstract}
The expansion of the universe is currently accelerating, as first inferred by \cite{EfstathiouDarkEnergy}, \cite{OstrikerSteinhardt} and directly determined by \cite{RiessNobel} and \cite{PerlmutterNobel}.  Current constraints are consistent with a time independent equation-of-state of $w=-1$, which is to be expected when a constant vacuum energy density dominates.  But the Quantum Field Theory prediction for the magnitude of this vacuum energy is very much larger than that inferred \citep{WeinbergCosmoCnstProb, KoksmaProkopec}.  It is entirely possible that the cause of the expansion has an alternative explanation, with both the inclusion of a quantum scalar field and modified gravity theories able to reproduce an expansion history close to, but potentially deviating from, that of a cosmological constant and cold dark matter, $\Lambda$CDM; see \cite{WeinbergDE} for a review.

In this work I investigate the consistency of the VIMOS Public Extragalactic Redshift Survey (VIPERS) v7 census of the galaxy distribution at $z=0.8$ with the expansion history and linear growth rate predicted by General Relativity (GR) when a \cite{Planck} fiducial cosmology is assumed.  To do so, I measure the optimally weighted redshift-space power spectrum \citep{FKP}, which is anisotropic due to the coherent infall of galaxies towards overdensities and outflow from voids \citep{Kaiser}.  The magnitude of this anisotropy can distinguish between modified theories of gravity as the convergence (divergence) rate of the velocity field depends on the effective strength of gravity on cosmological scales \citep{Guzzo}.  This motivates measuring the linear growth rate rather than the background expansion, which is indistinguishable for a number of modified gravity theories.  In Chapter \ref{chap:VIPERS_RSD} I place constraints of 
\begin{align}
&f \sigma_8(0.76) = 0.44 \pm 0.04, \nonumber \\
&f \sigma_8(1.05) = 0.28 \pm 0.08, 
\end{align}
with the completed VIPERS v7 survey; the combination remains consistent with General Relativity at 95\% confidence.  The dependence of the errors on the assumed priors will be investigated in future work.

Further anisotropy is introduced by the Alcock-Paczy\'nski effect \citep{AP} -- a distortion of the observed power spectrum due to the assumption of a fiducial cosmology differing from the true one.  These two sources of anisotropy may be separated based on their distinct scale and angular dependence with sufficiently precise measurements.  Doing so degrades the constraints: 
\begin{align}
&f \sigma_8(0.76) = \ \ 0.31 \pm 0.10, \nonumber \\
&f \sigma_8(1.05) = -0.04 \pm 0.26,
\end{align}
but allows for the background expansion ($F_{AP} \equiv (1+z) D_A H/c$) to be simultaneously constrained.  Galaxy redshift surveys may then directly compare both the background expansion and linear growth rate to the GR predictions \citep{Ruiz, Linder2016}.  I find the VIPERS v7 joint-posterior on $(f \sigma_8, F_{AP})$ shows no compelling deviation from the GR expectation although the sizeable errors reduce the significance of this conclusion.

In Chapter \ref{chap:VIPERS} I describe and outline corrections for the VIPERS spectroscopic selection, which enable these constraints to be made.  The VIPERS selection strategy is (projected) density dependent and may potentially bias measures of galaxy clustering.  Throughout this work I present numerous tests of possible systematic biases, which are performed with the aid of realistic VIPERS mock catalogues.  These also allow for accurate statistical error estimates to be made -- by incorporating the sample variance due to both the finite volume and finite number density.    

Chapter \ref{chap:maskedRSD} details the development and testing of a new, rapid approach for the forward modelling of the power spectrum multipole moments obtained  from a survey with an involved angular mask; this work has been accepted by the Monthly Notices of the Royal Astronomical Society as \href{http://arxiv.org/abs/1511.07799}{Wilson, Peacock, Taylor and de la Torre (2015)}.  An investigation of the necessary corrections for the VIPERS PDR-1 angular mask is recorded.  This includes an original derivation for the integral constraint correction for a smoothed, joint-field estimate of $\bar n(z)$ and a description of how the mask should be accounted for in light of the Alcock-Paczy\'nski effect.

Chapter \ref{chap:Clipping} investigates the inclusion of a simple local overdensity transform: `clipping' \citep{Fergus} prior to the redshift-space distortions (RSD) analysis.  This tackles the root cause of non-linearity and potentially extends the validity of perturbation theory.  Moreover, this marked clustering statistic potentially amplifies signatures of modified gravity \citep{LombriserClipping, White2016} and, as a density-weighted two-point statistic, includes information not available to the power spectrum.  

I show that a linear real-space power spectrum with a Kaiser factor and a Lorentzian damping yields a significant bias without clipping, but that this may be removed with a sufficiently strict transform; similar behaviour is observed for the VIPERS v7 dataset.  Estimates of $f \sigma_8$ for different thresholds are highly correlated due to the overlapping volume, but the bias for insufficient clipping can be calibrated and the correlation obtained using mock catalogues.  A maximum likelihood value for the combined constraint of a number of thresholds is shown to achieve a $\simeq 16\%$ decrease in statistical error relative to the most precise single-threshold estimate.  The results are encouraging to date but represent a work in progress; the final analysis will be submitted to Astronomy \& Astrophysics as Wilson~et~al.~(2016).  

In addition to this, an original extension of the prediction for a clipped Gaussian field \citep{Fergus} to a clipped lognormal field is presented.  The results of tests of this model with a real-space cube populated according to the halo occupation distribution model \citep{Zheng} are also provided. 
\begin{chapter}{Lay summary}
Cosmology asks many of the grandest questions in science, questioning how the universe in which we find ourselves came to be.  Its remit extends to time and length scales that are beyond comprehension and yet, with concerted effort, a vast range of observations are now well described in a (mostly) consistent framework and often with principles derived from everyday experience.  

While in Edinburgh, my own small part in this work has focused on a subfield of Cosmology named large-scale structure, which seeks to explain the spatial distribution of galaxies and harness this information in order to test the laws of gravity.  To this end, I've had the pleasure of working with the \href{www.vipers.inaf.it/}{VIPERS} galaxy redshift survey, which provides a census of the positions and spectral properties of approximately 90,000 galaxies as the universe was when (at most) half its current age.  

In our current understanding, the galaxy distribution is the product of initial density perturbations set by small-scale, quantum physics at least $13.4$ billion years ago, which have since grown due to gravitational collapse and eventually formed stars and galaxies.  In the interim, evidence suggests the young universe was extraordinarily hot, dense and bright -- in fact, this light can be observed directly (at longer wavelengths) as the cosmic microwave background \citep{Planck}.  That an equally dense universe is not seen today is explained by a dilution of matter with time and is further evidenced by the observed recession of galaxies in our local neighbourhood.         

While this current understanding consistently models a range of observations, cosmologists have had to reluctantly accept various additions to physics for which there is no evidence for in the laboratory -- the matter making up bicycles and books is seemingly only $4\%$ of the total.  This is a sorry state of ignorance and the current focus of cosmology is on learning more about the remainder.

The observed trajectories of stars in galaxies and of galaxies in clusters suggest that a large percentage of this mass is present and simply does not emit light; it is therefore invisible to conventional optical astronomy.  This hypothetical cold dark matter particle(s) is believed to make up $30 \%$ of the total mass.  At most 0.5\% can be contributed by neutrinos, which are a prime candidate as they have been detected, are known to possess mass and do not interact electromagnetically.  Perhaps the most likely candidate is a similar, weakly interacting particle, which is simply more elusive.    

The rate of galaxy recession is observed to be increasing with time, in direct conflict with the expected decrease due to the gravitational pull of baryonic and cold dark matter.  This suggests that the remaining $\simeq 70$\% of the energy content must be very different from normal baryonic matter.  This is to simplify matters, as decades of quantum mechanics research suggests a likely candidate: the energy density of `empty' space; this is a misnomer as truly empty space cannot be physically realised for any length of time.  But the quantum prediction is very much different from the value required by cosmological observations -- when General Relativity is assumed.  This is one instance, in a recurring theme, of a failure to combine our understanding of gravity with small-scale quantum physics.  The problem is so large, one `solution' is to speculate that this energy can't possibly contribute and look for another source of the acceleration.  Numerous alternatives have been postulated, including the introduction of yet more previously undetected particles or modifying the theory of gravity on cosmological scales. 

To proceed it seems legitimate to ensure that gravity is understood as well as we believe.  This is especially the case on cosmological length scales, which have only recently become accessible.  General Relativity has been extensively tested in the solar system in the past, with remarkable success.  If cosmological tests can be performed in a manner that distinguishes between the predictions of alternative theories then so much the better; as the majority of alternatives are flexible enough to completely agree on the expected rate of galaxy recession, alternative tests gain prominence.  Later chapters detail my work in ensuring the rate of gravitational collapse, as evidenced by the VIPERS galaxy distribution, is consistent with the predictions of General Relativity.   

To do so, I exploit the effect of redshift-space distortions \citep{Kaiser}.  As the attraction of gravity typically causes mass to coalesce, there is a large-scale infall of galaxies on the outskirts of a large cluster towards the cluster centre; this is shown in Fig. \ref{fig:CosmicFlows}.  This additional velocity causes the observed light from a given galaxy to be redder -- due to the everyday Doppler effect (\citejap{Doppler},  \citejap{Feynman1}), when the galaxy lies between us and the cluster centre.  By looking for a systematic change in the observed colour of galaxies on the outskirts of a cluster the rate of infall may be inferred.  As this rate is predicted by General Relativity (once the total mass in the universe has been deduced by other means) the observed rate may be used to distinguish between modified theories of gravity \citep{Guzzo}.
\end{chapter}

\singlespacing
% Uncomment this line if you need to declare published work which forms part of the thesis
\declarationpublications{}
\makedeclaration

\chapter{Acknowledgements}
\noindent
\normalsize
This thesis would not have been possible without the aid of many people.  Principally, my parents, Dorothy and Alfie Wilson, whose love, interest and sacrifice has been an everlasting support on which I could always depend.  I could never thank you enough and I hope this work does justice to your efforts.       

I would like to thank my supervisor, John Peacock, for being the guiding hand through what has occasionally seemed like choppy waters. I especially appreciate your dedication, insight and ability to set the highest of standards to aspire to, all of which helped improved this thesis significantly.  I owe a debt of gratitude to Sylvain de la Torre for his help -- especially in my first year; without his obvious hard work for VIPERS, in particular with regard to the mocks on which this analysis is founded, the later chapters would look very different.  My thanks go to Andy Taylor and Catherine Heymans for their always open door.  I'd like to thank Fergus Simpson for his originality, technical support and friendship.  My appreciation extends to the entire VIPERS team, in particular to Gigi Guzzo, Ben Granett, Julien Bel, Stefano Rota and Andrea Pezzotta;  The VIPERS conferences have always been enjoyable thanks to your approachability and willingness to offer helpful advice.  My thanks go to Shaun Cole and Andy Lawrence for their enthusiastic and knowledgeable questioning during my viva. 

Often the Observatory feels like a small place, but looking back it's surprising to think of the number of people who have passed through.  Although too many to name specifically, Ami Choi deserves special praise for being my longest friend here and for her support this year.  My appreciation goes to Marco Lam for his friendship and continued patience as a flatmate.  My time in Edinburgh has been all the better since inheriting an addiction to cycling from Becca, Chris, Sandy and David; relieving stress at five-a-side football; brewing beer with the expert guidance of Edouard, Esther \& Jorge and dominating (for a spell) the Dagda pub quiz -- thanks to Alex Hall and David.  To the only source of stress I anticipate in the next few weeks, my chess nemeses: Fergus, Ami and Alex Hall, I'll be working on my end game.  Finally, my thoughts go to my long-suffering office mates, Shegy, Maria and Alex Amon in particular, who have had to endure my often foul mood in the past few months.  Your company and continued support have made this all immeasurably easier.      

\begin{center}
In memory of my mother.
\end{center}

\cleardoublepage
\phantomsection
\addcontentsline{toc}{chapter}{\contentsname}
\setcounter{tocdepth}{2}
\tableofcontents

\cleardoublepage
\phantomsection
\addcontentsline{toc}{chapter}{\listfigurename}
\listoffigures

\cleardoublepage
\phantomsection
\addcontentsline{toc}{chapter}{\listtablename}
\listoftables

% Include main matter here
\mainmatter
\eighteenptleading
%\section*{Frequently used symbols}
%Frequently used symbols and their place or source of definition.
%
%\begin{center}
%\begin{tabular}{|p{1.5cm} p{5.cm} p{2.5cm} p{3.cm}|}
%\hline
%\hline
%\multicolumn{3}{|c|}{Frequently used symbols} \\
%\hline
%Symbol & Definition & Value & Page or source \\
%\hline
%$t$ & Cosmological time & & \\
%$\Phi$ & Gravitational potential & & \\
%$R(t)$ & Scale factor &  & x \\
%$G$ & Gravitational constant & & \\
%$H(t)$ & Hubble parameter & & \\ 
%$z$ & Redshift & & \\
%$\rho_c(t)$ & Critical density & & \\ 
%$w$ & Equation of state & & \\
%$k$ & Comoving wavenumber & & \\ 
%$\delta$ & Fractional overdensity & & \\
%$\mathbf{u}$ & Comoving peculiar velocity & &\\
%$\mathbf{r}$ & Comoving position & & \\
%$(r, \theta, \phi)$ & Spherical coordinates & & \\ 
%\hline
%\end{tabular}
%\end{center}
%
\begin{chapter}{Principles of cosmology and gravitation}
\label{chap:Basics}
This chapter provides a terse introduction to cosmology and gravitation, in order to better understand the method and results of later chapters.  It borrows heavily from \cite{CP} and \cite{CarrollBook}, to which any appreciation of moments of clarity or insight should be directed.  

\section{The cosmological principle and Hubble's law}
The majority of observations to date corroborate Einstein's Cosmological principle \citep{Milne1935}: the universe is statistically homogeneous (the same at all locations) and statistically isotropic (independent of direction) on large scales; although recent results have brought statistical isotropy into question by providing some evidence for hemispherical anomalies \citep{Planck}.  These laws are only statistical as there can be no plausible reason for physical properties, such as the density or velocity, to have a specific value at a given location. Rather, only the relative probability of these observables (for an ensemble of possible realisations) has significance.  This statistical nature is a problem, as there is only one universe (horizon) to be observed; an escape is manufactured by assuming the universe to be ergodic --  ``the property that the probability of any state can be estimated from a single sufficiently extensive realisation, independently of initial conditions'' \citep{OED}.  

The Cosmological Principle severely restricts the possibilities for our local velocity field; $\grad \cdot \mathbf{v}$ must be a homogeneous constant, which leaves only $\mathbf v = H(\theta, \phi, t) \mathbf x$ as a possibility.  Here $H(\theta, \phi, t)$ is independent of radial position but otherwise without restriction.  By also requiring isotropy, the velocity field must obey Hubble's law:   
\[
\mathbf{v} = H_0 \mathbf{x},
\]
to first order in $x$ (as velocities are ill-defined at large distances).  Despite appearing to suggest we are at a preferred centre, this cannot be the case as homogeneity was a crucial assumption in deriving this law; rather, any two sufficiently close galaxies will perceive the other to obey Hubble's law with themselves at the centre.       

The first evidence this was obeyed by our own (local) universe was obtained by \cite{VSlipher}.  Slipher observed the spectral lines of spiral galaxies (nebulae) to be redshifted with respect to their rest frame wavelengths, as would be measured in the laboratory.  This requires a recession velocity proportional to distance \citep{HubbleLaw} when interpreted as a Doppler shift \citep{Doppler}. 

Considering further afield, there is good reason to expect this law to be an approximation that breaks down for galaxies at sufficiently large distances.  This expectation is based on the foundations of General Relativity \citep{EinsteinGR} -- the equivalence principle and general covariance.  A brief outline of these principles is given in the following primer as they form the foundation of what is to come. 

\section{A primer on Einstein gravity}
As there is no detectable difference between the inertial and passive gravitational masses \citep{Will1993}: $\cancel m_I \mathbf{a} = \cancel m_g \mathbf{g}$, the response to a given gravitational field is independent of `mass'.  As a result, in a small frame that freely falls with an acceleration $\boldsymbol g$ the effects of gravity are undetectable; gravity then assumes the role of a fictitious force to be found in limited (non-inertial) frames.  The laws of physics in the absence of gravity are described by Special Relativity \citep{SR}, which are a complete prescription for physics in a small region about a freely falling observer.  But over a sufficiently large volume $\mathbf g$ will vary and gravity will then be detectable by tidal forces sourced by gradients, $\partial_{i} \mathbf{g}_j$.  These ideas are encapsulated by (three) equivalence principles, which are described in greater detail in \S \ref{sec:MG}. 

To implement General Covariance is to write physical laws in a manner that explicitly applies equally to any freely falling observer.  When written in a generally covariant fashion, the laws of Special Relativity then plausibly remain valid for observers who do not fall freely and therefore remain valid in a gravitational field, by the (Einstein) equivalence principle.  This approach does not necessarily produce unique physical laws -- any (tensorial) term which vanishes in the inertial frame can be added without violating these principles; often the simplest possible law is assumed, in the absence of evidence for further complexity.  The result of this approach for a particle which is free in the inertial frame is the Geodesic equation: 
\[
\frac{d^2 x^{\mu}}{d \tau^2} + \Gamma^{\mu}_{\alpha \beta} \frac{dx^{\alpha}}{d \tau} \frac{dx^{\beta}}{d \tau} = 0,
\]
here $x^{\mu}$ is the $\mu$ component of the observed four-position, $(ct, \mathbf{x})^T$, $\Gamma^{\mu}_{\alpha \beta}$ is the Christoffel connection and $\tau$ is the proper time.  In the absence of gravity, $\Gamma^{\mu}_{\alpha \beta} = 0$ and the particle simply travels in a straight line, with $t \propto \tau$ and $\mathbf x \propto t$; the effect of gravity is then encapsulated in non-zero $\Gamma^{\mu}_{\alpha \beta}$.  

For a given gravitational field, every mass will follow an identical trajectory between two points (in the absence of other forces) -- this property is what separates gravity from other forces, e.g. there is no common trajectory for particles of differing charge in an electric field.  In particular, there are no bodies immune to gravity.  The most economic approach is then to describe this preferred class of trajectories through spacetime directly, rather than as an apparent gravitational force experienced by a given particle.  This idea can be achieved by ascribing a geometry to the space in which preferred trajectories (by an action principle) are not simply straight lines.  That this is the correct approach cannot be proved directly, but its consequences can be derived and compared with experiment.  There are an arbitrarily large number of possible gravitational fields, which must be reflected by a similar number of possible geometries.  This is achieved by encompassing those that have curvature and are therefore non-Euclidean.  The outcome of this approach specifies the effective gravitational potentials, 
\[
\Gamma^{\alpha}_{\lambda \mu} = \frac{1}{2} g^{\alpha \nu} \left(\frac{\partial g_{\mu \nu}}{\partial x^\lambda} + \frac{\partial g_{\lambda \nu}}{\partial x^\mu} - \frac{\partial g_{\mu \lambda}}{\partial x^\nu} \right).
\]
in terms of the metric, $g_{\mu \nu}$.  The metric (tensor) provides the square of the physical distance,
\[
ds^2 = dx_{\mu} dx^{\mu} \equiv g_{\mu \nu} dx^{\mu} dx^{\nu},
\]
between two infinitesimally separated points on a differentiable (locally flat) manifold, e.g. a smooth 2D surface or 3D space that is mapped by coordinates $x^{\mu}$.  As the norm of a four-vector, perceived by a given observer to have components $dx^{\mu}$, the value of $ds^2$ should be independent of the coordinate choice and agreed upon by all observers.  

These relations show how, starting from a non-Euclidean surface or space described by $g_{\mu \nu}$, the trajectories of particles are determined by the geodesic equation.  This has an important physical interpretation; for a given spacetime, the trajectories that solve the geodesic equation (geodesics) are simply those that make the total distance travelled, $s'= \int ds$, between the end points stationary.  Geodesics then simply achieve the closest thing available to the straight line trajectories available in flat (Minkowski) space; this is illustrated on the cover of \cite{CarrollBook} for example.   

Sacrificing Euclidean geometry -- what is a Euclidean plane in the absence of gravity will be distorted according to a given $\mathbf g(\mathbf x)$, has important consequences.  Not least for the four-velocity, as vectors cannot be transported across a curved surface uniquely.  Thus the relative four-velocity of a distant galaxy is not unique and conclusions should not be based on a particular choice; it is this property that accounts for the breakdown of Hubble's law on large scales.  This problem is of no importance for local galaxies as a curved surface (of the type applicable to GR) is flat in a small enough neighbourhood.  It is therefore perfectly valid to interpret Hubble's law as resulting from the recession of galaxies locally.  

So far the Newtonian relation: $\mathbf f = m_I \mathbf a$ has been reconsidered when determining the affect of gravity on astrophysical scales.  It remains to establish how the geometry, $g_{\mu \nu}$, is specifically determined by a given matter distribution.  This is given by Einstein's field equation:        
\[
G^{\mu \nu} \equiv R^{\mu \nu} - \frac{1}{2} g^{\mu \nu} R = \frac{-8 \pi G}{c^4} T^{\mu \nu} - \Lambda g^{\mu \nu} \\ 
\label{eqn:EinsteinField}
\]
which, despite its apparent complexity, is much the same in structure as Poisson's equation.  Here $\Lambda$ is the cosmological constant, which, as we shall see, is identical to a contribution from a perfect fluid with the a $p = - \rho c^2$ equation-of-state.  The gravitational potentials that determine trajectories are present via the Ricci tensor, $R_{\alpha \beta} = R^{\nu}_{\alpha \beta \nu}$, where the Riemann tensor is given by 
\[
{R^{\mu}}_{\alpha \beta \gamma} = \Gamma^{\mu}_{\alpha \gamma, \beta} - \Gamma^{\mu}_{\alpha \beta, \gamma} + \Gamma^{\mu}_{\sigma \beta} \Gamma^{\sigma}_{\gamma \alpha} - \Gamma^{\mu}_{\sigma \gamma} \Gamma^{\sigma}_{\beta \alpha}.
\]
The curvature scalar, $R$, is obtained by the remaining contraction of the Ricci tensor.  In this expression, $\Gamma^{\mu}_{\alpha \gamma, \beta} \equiv \partial_\beta \Gamma^{\mu}_{\alpha \gamma}$.  Eqn. (1.6) is simply a generally covariant (observer or coordinate independent) expression of the curvature of the geometry at a given point based on the physical change of a vector when (parallel) transported around a small closed loop \citep{CarrollBook}; see Fig. 1.2 of \cite{CP} for an illustration. 

The distortions from Euclidean geometry are sourced by the matter distribution via the energy-momentum tensor, $T^{\mu \nu}$; a given component of which specifies the momentum flux, $P^{\mu} = m U^{\mu}$, across a surface at constant $x^{\nu}$.  The most pertinent example in cosmology is a perfect fluid -- when an extended number of particles are characterised as a continuum by macroscopic quantities such as density and pressure.  In a frame travelling at the mean particle velocity, $U(X)$, neighbouring fluid elements in a perfect fluid experience no shear and an isotropic pressure.  In this case, $T^{\hat \mu  \hat \nu} = \text{diag}( \rho c^2, p, p, p)$, where $\rho c^2$ is the rest-frame energy density, $p$ is the pressure and $\hat x$ denotes the coordinates of a local inertial frame.  A manifestly covariant expression that reduces to this rest-frame result is 
\[
T^{\mu \nu} = (\rho + \frac{p}{c^2}) \ U^{\mu} U^{\nu} - p g^{\mu \nu},
\]
which therefore gives the components required in an arbitrary frame. 

The Newtonian limit of the field equation should result in Poisson's equation.  Assuming the deviations of $g_{\mu \nu}$ from flat space are small and time-independent fields gives   
\[
\nabla^2 \Phi = \frac{4 \pi G}{c^2} (\rho c^2 + 3p),
\]
where $\Phi = c^2 g_{00}/2$. This shows that relativistic fluids have an increased gravitational influence in GR, with an effective source term given by the `active mass density': $\rho c^2 + 3p$.

\section{The Robertson-Walker metric}
Having established the approach of Einstein gravity, an evident question is: how is a universe satisfying the Cosmological Principle to be described?  The answer is the Friedmann-Robertson-Walker metric:     
\[
ds^2 = -c^2 dt^2 + R^2(t) [ dr^2 + S_k^2(r) \ d\psi^2], 
\]
for $d \psi^2 = d \theta^2 + \sin^2(\theta) d\phi^2$ and a cosmological scale factor $R(t)$, which has dimensions of length.  The dimensionless variable $r$ is interpreted as a radial coordinate -- often denoted $\chi$ in other notations, and therefore for $r \ll 1 $ this metric should describe the Hubble expansion.  As the radial distance to be travelled to a galaxy at coordinate $r$ is $s(t) = R(t)r$, it follows that $\dot s = \dot R r$ and therefore Hubble's law is recovered locally with $(\dot s/s) = (\dot R/R) \equiv H(t)$.  A similar metric with $R(t, \theta, \phi)$ would then correspond to anisotropic expansion, which would violate the Cosmological Principle.  

A comoving observer -- one at constant $(r, \theta, \phi)$, thus recedes from the origin locally, carrying a clock that ticks at an interval $dt$.  The ticking rate, obtained from $g_{00}$, is independent of $\mathbf r$ and common to all such observers due to homogeneity, which also allows clocks to be synchronised when a physical property, such as the density, reaches a prespecified value.  There are no cross terms, of the type $g_{0i} dt dx^{i}$, as the metric is assumed to possess the time invariance symmetry of Newtonian physics, i.e. invariant under $t \mapsto -t$; to see this is true requires the Friedmann equation for $R(t)$, which is given in \S \ref{dynamics}.   

In the local limit, $S_k(r) \mapsto r$ and the spatial part of the metric (at a given time) is 3D Euclidean space written in spherical coordinates.  Non-Euclidean deviations should be isotropic, irrespective of $r$, which is explicitly satisfied when the metric coefficients have no dependence on $(\theta, \phi)$.  While given that the radial coordinate may always be redefined such that $dr' = \sqrt{g_{11}(r)} dr$, it follows that $g_{11} = 1$ may always be assumed.  It remains to justify the possibilities for the function $S_k(r)$; due to isotropy, only the radial dependence of the possible geometries for the 2D surface: $\mathbf{r} = (r, \pi/2, \phi)$ need be considered. Two familiar examples of 2D surfaces are the (infinite) Euclidean plane and the surface of a sphere; both possess constant spatial curvature -- every point is then equivalent, as required by homogeneity.  Given that any smooth curved surface looks locally flat, in the absence of cosmological tests, what evidence is there that the spatial geometry is flat on $\simeq 100 \mpcoh$ scales?  Thus determining the spatial geometry is a principal goal of cosmology.  A final, less familiar, possibility is a 2D surface of constant negative curvature -- a hyperbolic geometry.  These cases are specified by the curvature constant, with $k=\{0, 1, -1 \}$ respectively. The functional form of $S_k(r)$ is  
\[
S_k(r) =
            \begin{cases}
            \sinh(r)   & \text{$(k=-1)$} \\
            r          & \text{$(k=0)$} \\
            \sin(r)    & \text{$(k=1)$},
            \end{cases}
\]
in each case; similarly $C_k(r) = \sqrt{1 - kS_k^2(r)}$.  These are of no surprise for the flat and spherical cases ad $(r, \psi) \mapsto (\theta, \phi)$ recovers the usual notation for the metric on the surface of a sphere and emphasises $R(t)$ plays the role of the radius (with units of distance).  The (Lagrangian) coordinate $r$ is simply a label for a fundamental observer in the Hubble flow; conventionally, when specifying a given fundamental observer, $R_0r$ is quoted (with units of $\text{Mpc}$) rather than $r$ itself.  Of course this is just the physical distance to be travelled to the position of that observer today and termed the comoving distance.           

More formal arguments for deriving or justifying the Friedmann-Robertson-Walker metric may be found, e.g. \S 3.1 of \cite{CP} or Chapter 8 of \cite{CarrollBook}. But being the simplest, these are perhaps the most compelling.   

\subsection{Expansion dynamics}
\label{dynamics}
The observed expansion results simply from an initial condition in classical cosmology.  The subsequent evolution of the scale factor, $R(t)$, is then subject to the equation of motion or Friedmann equation:     
\[
\dot R^2 - \frac{8 \pi G}{3} \rho R^2 = - k c^2.
\]
This may be obtained from Einstein's field equation, eqn. (\ref{eqn:EinsteinField}), by assuming a perfect fluid source.  Remarkably, this is simply the Newtonian energy equation for an expanding sphere of mass; an outer shell with velocity $\dot Rr$ typically decelerates due to the gravitational attraction of the body.  This Newtonian argument assumes no attraction from outer shells -- Newton's shell theorem \citep{Feynman1}, which should fail for large radii, where curvature becomes important.  It does not as a consequence of Birkhoff's theorem, first derived by Jebsen \citep{NotBirkhoff}, which states that: ``any spherically symmetric solution of the vacuum equations is both static and asymptotically Minkowski space''.  Consequently, the metric must be the Schwarzchild solution \citep{SchwarzschildMetric}, for which the shell theorem remains applicable \citep{CarrollBook}.  The total `energy', which determines if $R(t)r$ will expand to infinity or recollapse, is fixed by the curvature constant; this dependence of the rate of the Hubble expansion on the spatial curvature is one of the greatest surprises of GR; however, more fundamentally both are determined by the matter distribution.  

In addition to Friedmann's equation, the Raychaudhuri or acceleration equation:  
\[
\frac{\ddot R}{R} = -\frac{4 \pi G}{3 c^2} (\rho c^2 + 3p),
\]
may be obtained from the independent components of the field equation.  Alternatively, a time derivative of Friedmann's equation and an appeal to adiabatic expansion, $dU = d(\rho c^2 R^3) = -pd(R^{3})$ for $dS=0$, suffices.  This equation illustrates the importance of the active mass density, $(\rho c^2 + 3p)$, which will double the influence of radiation due to the $p = \rho c^2/3$ equation-of-state. 

To solve the expansion history for a given component, an approximation is made in which $\rho$ is decomposed into various components specified by an equation-of-state, $w$, defined by 
\[
p = w \rho c^2.
\]
Those key to cosmology are: pressureless dust ($w=0$), which represents particles that are stationary in the fluid rest-frame; radiation ($w=1/3$), representing a relativistic gas, and a vacuum term with equation-of-state: $w=-1$.  This surprising component is further discussed in \S \ref{sec:lambda}.  The ratio of a component of the density to the density in a flat universe with the same expansion rate is conventionally quoted; the latter is termed the critical density and is given by 
\[
\rho_c = \frac{3H^2}{8 \pi G},
\]
for $H = (\dot R/R)$.  The time-dependent ratio or density parameter is then  
\[
\Omega \equiv \frac{\rho}{\rho_c} = \frac{8 \pi G \rho}{3 H^2}.  
\]
Universes with a density greater (lesser) than critical are therefore spatially closed (open), with a spatial curvature of $k=1$ and $k=-1$ respectively.  This convention may be extended to the cosmological constant and curvature by defining    
\[
\Omega_v \equiv \frac{8 \pi G \rho_v}{3H^2} = \frac{\Lambda c^2}{3 H^2}, \qquad
\Omega_k \equiv -\frac{k c^2}{H^2 R^2}.
\] 
The Friedmann equation is
\[
\Omega_m(t) + \Omega_r(t) + \Omega_v(t) + \Omega_k(t) = 1.
\]
in this parametrisation.  The present day value of the scale factor, or curvature length, is given by
\[
R_0 = \frac{c}{H_0} \left [ \frac{(\Omega_0 -1)}{k}\right ]^{-1/2}.
\]

\section{Physical separation from observables}
\subsection{Redshift}
The Hubble expansion is observed locally by the Doppler effect, which induces a frequency ratio:   
\[
\frac{\nu_{\rm{emit}}}{\nu_{\rm{obs}}} \simeq \left ( 1 + \frac{v}{c} \right ) \equiv (1 + z),
\]
due to a recession velocity $v$; this relation defines the redshift, $z$.  This law is correct for small radial distances, where $S_k(r) \simeq r$ -- e.g. the distance to M31 at $R_0 r \simeq 0.7 h^{-1} \rm{Mpc}$, but fails on scales approaching the curvature length, $r \simeq 1$ or $R_0$ in comoving distance.  For this regime, consider a local inertial frame for which Special Relativity states the invariant interval is 
\[
ds^2 = -c^2 dt^2  + (dx^2 + dy^2 + dz^2) \equiv \eta_{\mu \nu} dx^{\mu} dx^{\nu},
\]
in which case $ds=0$ for a photon travelling at the speed of light, $c$.  As the norm of a four-vector, $dx^{\mu}$, this is an invariant agreed upon by all observers including those in non-inertial frames.  Therefore, assuming $ds^2 =0$ ing the FRW metric, a photon receding from the origin passes subsequently more distant comoving observers with passing cosmological time (providing $R(t)$ remains shallower than $t$): 
\[
r(t) = \int_0^t \frac{cdt'}{R(t')}.   
\]
By considering two such photons, with emission times separated by $dt_{\text{emit}}$, received by a comoving observer fixed in the expansion at constant $r$, at times separated by an interval $dt_{\text{rec}}$, it follows that events on distant galaxies must be time dilated by 
\[
\frac{dt_{\text{rec}}}{dt_{\text{em}}} = \frac{R(t_{\text{rec}})}{R(t_{\text{em}})}.
\]
Due to the Cosmological Principle, this is the same as observing the light emitted by a distant galaxy; the observed frequency, $\nu \propto (1/dt)$, will then be redshifted by    
\[
\frac{\nu_{\rm{emit}}}{\nu_{\rm{obs}}} = \frac{R(t_{\rm{obs}})}{R(t_{\rm{emit}})} \equiv 1 + z \equiv \frac{1}{a}.
\]
As two sufficiently close comoving observers will observe the other to obey Hubble's law, this may be thought of as simply the accumulation of repeated Doppler shifts along a chain of fundamental observers stretching to the distant galaxy; this is analogous to approximating a curved 2D surface by a number of sufficiently small Euclidean planes.

A redshift of the observed spectral line for a local galaxy is therefore correctly interpreted as a recession velocity, but for a more distant galaxy the redshift represents the ratio of the scale factor at emission and at observation.  In the following section I outline how the observed redshift and angular separation may be used to determine the comoving distances to sources, once a given density composition is assumed.      

\subsection{Distances}
For any massive particles subject to an expansion initiated at an early time, the (proper) rest-mass energy density will dilute with the scale factor, $\rho_m c^2 \propto R^{-3}$.  This is similar for photons, but photon wavelengths also stretch with the expansion and therefore $\rho_r c^2 \propto R^{-4}$; photons emitted at late times will inherit the expansion velocity of the source.  Given that the cosmological constant is equivalent to a perfect fluid with $w=-1$ and constant energy density, $\rho_v = \Lambda c^2/(8 \pi G)$, the Friedmann equation predicts the time evolution of $H(t)$ to be        
\[
H(a) \equiv \left( \frac{\dot R}{R} \right)= H_0 [\Omega_v + \Omega_m a^{-3} + \Omega_r a^{-4} + (1 - \Omega) a^{-2}]^{1/2},
\]
for $a\equiv1/(1+z)$.  Given that $Rdr = cdt = cdR/ \dot R = cdR/(RH)$, it follows that the comoving distance to a galaxy observed at redshift $z$ is therefore 
\[
R_0 r(z) = \int_0^z \frac{c \ dz'}{H(z')}.
\]
This highlights the importance of the Hubble constant, $H(a) \propto H_0$, in determining the comoving distance to a galaxy with an observed redshift.  Our nescience of $H_0 (\equiv 100 h \ \rm{km s^{-1} Mpc^{-1}})$ is conventionally incorporated into the units of distance; $R_0 r$ is then quoted in units of $\mpcoh$ rather than assuming a given $H_0$ to quote $R_0 r$ in Mpc.      
 
Similarly, as $cdt=cdR/(RH) = -Rdz/(R_0H)$, the age of the universe when a source at redshift $z$ emitted the light we see today is  
\[
t_{\rm{age}}(z) = \int_z^{\infty} \frac{dz'}{(1+z') H(z')}.
\]
Lower limits to this integral are provided by dating the oldest stellar systems (see \citejap{WMAP3} and references therein).  Within the `big bang' framework, the current best constraint on the minimum age of the universe is \citep{Planck}
\[
t_{\text{age}}(0) = 13.799 \pm 0.021 \text{Gyr};
\]
see \S \ref{sec:inflation} for an explanation of this minimum caveat.  

The physical volume of a shell at coordinate $r$ is obtainable by inspection for the orthogonal FRW metric; it is
\[
dV = 4 \pi \left ( R S_k(r) \right )^2 R dr. 
\]
When evaluated $R=R_0$, this gives the comoving volume element -- the physical volume spanned by comoving observers today that bounded a physical volume $dV$ at an earlier time and have separated with the expansion in the interim.  This is a consequence of the physical distance between two sources separated by an angle $d \psi$ radians, which, from the FRW metric, is given by
\[
dl_\perp = \frac{R_0}{(1+z)} S_k(r) d\psi \equiv D_A \ d\psi;
\]
this defines the angular diameter distance, $D_A$, as that which makes the mapping resemble Euclidean arc length.

Of equal importance in observational cosmology is the relation between the intrinsic luminosity of a source at emission and the flux density observed today:
\[
S(\nu_{\rm{obs}}) d \nu_{\rm{obs}} = \frac{L(\nu) d \nu}{4 \pi R_0^2 S_k^2(r) (1+z)^2} \equiv \frac{L(\nu) d\nu}{4 \pi D^2_L(z)}. 
\]
Due to the cosmological redshift, the flux observed at frequency $\nu_{\rm{obs}}$ was emitted at $\nu = (1+z) \nu_{\rm{obs}}$; the observed bandwidth therefore corresponds to a emission bandwidth of $d \nu/(1+z)$.  This is countered by the $(1+z)^2$ in the denominator, which accounts for the cosmological redshift of photon energies and the time delay of the emission and arrival rates.  The remaining factor is simply the physical area of a sphere centred on the source, from which we are separated by the physical radius $R_0 S_k(r)$ -- this appears comoving simply because the comoving distance is the physical distance today.  The final equivalency makes the mapping look Euclidean and defines the luminosity distance:
\[
D_L(z) = (1+z) R_0 S_k (r).
\]
For a flat $\Lambda$CDM universe with negligible radiation density today $D_L(z)$ is approximately
\[
D_L(z) \simeq \frac{c}{H_0} \left [ z + \frac{z^2}{2} \left(1 - q_0 + \mathcal O(z^3) \right) \right ],
\]
where $q_0$ is the present-day value of the deceleration parameter,
\[
q(z) \equiv - \frac{\ddot R R}{\dot R^2} = \frac{\Omega_m(z)}{2} + \Omega_r(z) - \Omega_v(z).
\]
Measurements of the observed flux density of local objects, $z \ll 1$, for which the intrinsic luminosity can be inferred -- e.g. with the period-luminosity relation of Cepheids \citep{Riess16}, then constrain $H_0^{-1}$.  If a second `standard candle' -- a population of sources thought to possess a common intrinsic luminosity, can be found that overlaps in redshift with the Cepheids then the luminosity of the candles may be calibrated.  Assuming this second population extends to $z \simeq 1$, the curvature of the $D_L(z)$ relation may be traced and $q_0$ measured; the archetypal example of the latter population are type-Ia supernovae (\citejap{RiessNobel}, \citejap{PerlmutterNobel}).  These are `standardisable' candles in practice -- the timescale over which the flux density decays is used to remove the dependence of the intrinsic luminosity on progenitor mass.  Measurements on $D_L(z)$ at low-$z$ may then be bootstrapped to high-$z$ with multiple populations.  This is referred to as a `distance ladder'.  Constraints on $(\Omega_m, \Omega_v)$ from the $D_L(z)$ relation traced by the JLA compilation of type-Ia supernovae are shown in Fig. \ref{fig:OmOv_Ia}.
\begin{figure}
\centering
\includegraphics[width=0.9\textwidth]{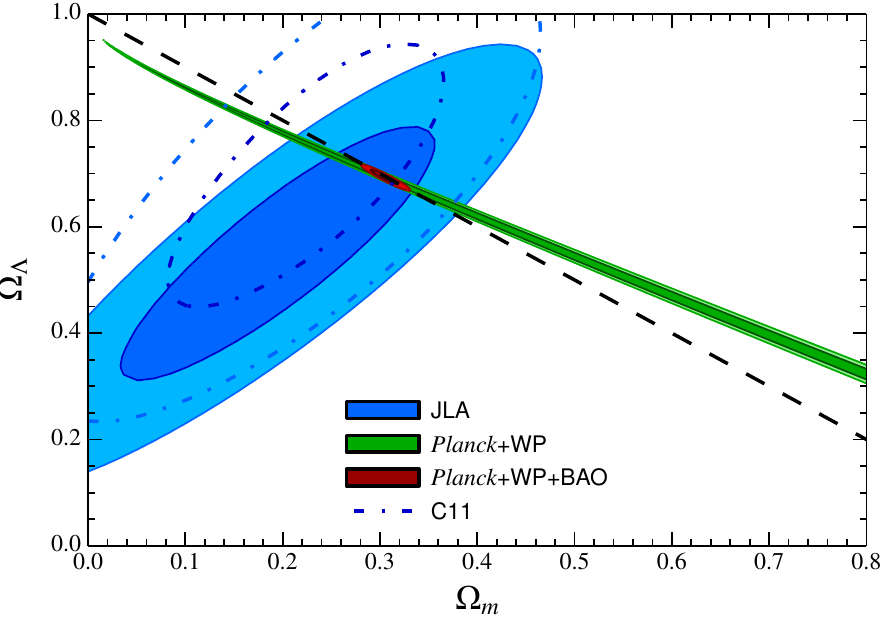}
\caption[JLA supernovae constraints on the expansion history.]{$(\Omega_m, \Omega_v)$ posterior from the JLA compilation \citep{JLA} of 740 type-Ia supernovae extending to $z =1$.  The plotted contours show $68 \%$ and $95 \%$ confidence levels following marginalisation over spatial curvature, $k$.  Dotted constraints show a previous analysis \citep{Conley2011}, together with various combinations of \cite{Planck2013}, cosmic microwave background polarisation \citep{WMAP9} and baryon acoustic oscillations analyses \citep{Beutler2011,Padmanabhan2012,Anderson2012}.  The prominent degeneracy -- ellipticity of the contours, is a consequence of the likelihood being sensitive to $q_0$ only (at second order in $z$).  Reproduced from \cite{JLA}.}
\label{fig:OmOv_Ia}
\end{figure}

Constraints on $D_L(z)$ with type-Ia supernovae clearly show the expansion to be accelerating, although the seminal analyses of \cite{RiessNobel} and \cite{PerlmutterNobel} required the assumption of flatness.  Together with theoretical motivation provided by the theory of inflation, additional constraints, e.g. the cosmic microwave background anisotropies \citep[CMB,][]{Planck} and the baryon acoustic oscillations peak \citep[BAO,][]{BassettHlozek}, suggest the curvature is indeed negligible and the universe accelerating.  Constraining and providing an explanation for this accelerating expansion history is perhaps the principal goal of cosmology currently; models for the `dark energy' component responsible are discussed in \S \ref{sec:lambda}.        

Typically, magnitudes will be quoted rather than the observed flux density, $S(\nu_{\rm{obs}})d\nu_{\rm{obs}}$.  The apparent magnitude of a galaxy is defined as
\[
m(\nu_{\rm{obs}}) \equiv -2.5 \log_{10} \left (\frac{S(\nu_{\rm{obs}})}{F(\nu_{\rm{obs}})} \right ) = -2.5 \log_{10} \left( \frac{L(\nu) d\nu}{4 \pi D^2_L F(\nu_{\rm{obs}}) d \nu_{\rm{obs}}}\right ),
\]
which is relative to the flux density of a calibration source, $F (\nu)d \nu$.  By further defining the absolute magnitude, $M$, as the apparent magnitude when the same source is a distance of $10$pc away, $z \simeq 0$, and assuming a power-law intrinsic luminosity, $L(\nu) \propto  \left ( [1+z] \nu_{\text{obs}} \right)^{- \alpha}$, it follows that 
\[
m(\nu_{\rm{obs}}) - M(\nu_{\rm{obs}}) =  5 \log_{10} \left( \frac{D_L}{10 \text{pc}} \right )  + 2.5(\alpha -1) \log_{10} (1+z).
\]
The second term, the k-correction, accounts for the redshifting of light in cosmology -- a $(1+z)^{\alpha}$ term as the source at $10$pc emits a luminosity $L(\nu_{\rm{obs}})$ rather then $L(\nu)$ and a further $(1+z)$ factor due to the stretching of bandwidth, $d \nu = (1+z) d \nu_{\text{obs}}$.  The convention on distance units, $[D_L]$ are $\mpcoh$, may be extended to apparent magnitudes.  Quoting $m(\nu_{\text{obs}}) + 5\log_{10}(h)$ does not require the assumption of a given $H_0$; a summary of distances in a FRW universe is shown in Fig. \ref{fig:fig8Hamilton98}.  
\begin{figure}
\centering
\includegraphics[width=\textwidth]{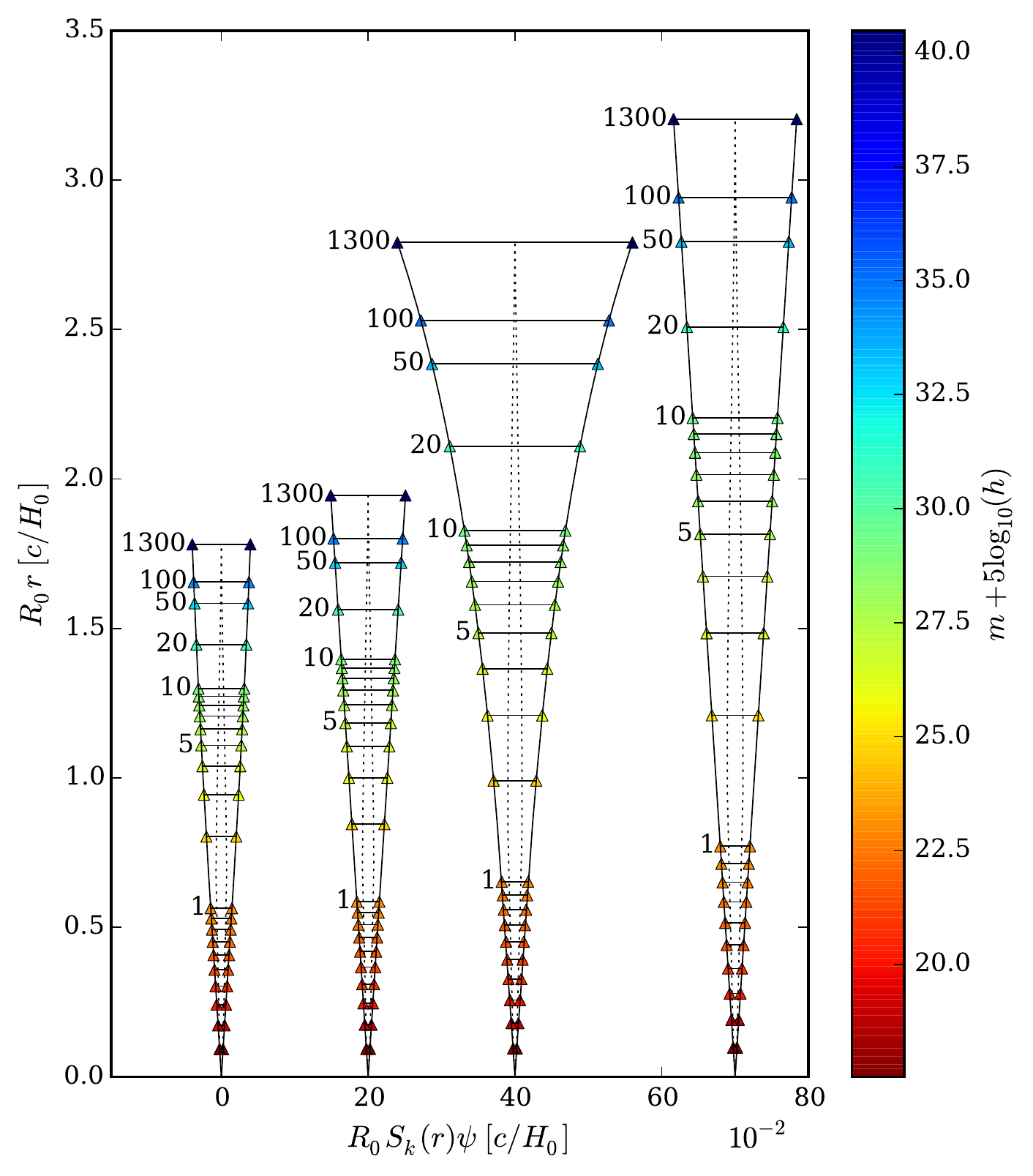}
\caption[Cosmological distances in an Freidmann-Robertson Walker universe.]{Distances in an FRW universe: shown are the physical distance today (comoving distance) to the midpoint of a galaxy pair at redshift $z$ and the corresponding transverse separation, for an angular separation of $3 \deg$.  From left to right corresponds to $(\Omega_m, \Omega_v) = (1.3, 0.0), (1.0, 0.0), (0.3, 0.0)$ and $(0.3, 0.7)$ respectively; this includes closed, flat and open cosmologies.  Dotted lines show the physical transverse separation at a given redshift, $RS_k(r) \psi$, illustrating how $D_A(z)$ peaks at a larger redshift when moving from left to right.  This may be interpreted as gravitational lensing from the intervening (homogeneous) matter distribution in flat cases or as sensitivity to the spatial curvature at high redshift.  The apparent magnitudes quoted are for a source with $M=-20.0$ and a spectral slope of $\alpha=1$.  This figure is adapted from Fig. 8 of \cite{Hamilton98}.}
\label{fig:fig8Hamilton98}
\end{figure}

\section{Dark matter}
The current standard model of cosmology proffers a weakly interacting, cold dark matter particle (CDM) as a solution to the following observations: 

\textit{Galaxy rotation curves} -- the rotation curves of spiral galaxies are observed to flatten at large radii \citep{Rubin}.  Within the cold dark matter (CDM) model, this is explained by an embedding of the baryonic disk in a dark matter halo with density profile: $\rho(r) \propto r^{-2}$ at large radii.  This is also a requirement for the stability of such systems \citep{OstrikerPeebles}.  See page 371 of \cite{CP}, \cite{DM_review} or \S 2.6.5 of \cite{ModGravReview} for further details.

\textit{Galaxy clusters} -- as first inferred by \cite{Zwicky}, the velocity dispersion of galaxies in clusters suggests a much greater mass than that present in the luminous component.  In Zwicky's original analysis this was limited to the stellar component but there is also a hot baryonic intracluster medium (ICM) visible in the X-ray spectrum.  The ratio of these components is inferred to be 85\% CDM, 14\% ICM and 1\% stars in the largest virialised systems.  This estimate is corroborated by additional observations of the ICM X-ray emission and projected measures of the mass, such as strong and weak gravitational lensing.         

\textit{Galaxy power spectrum} -- the galaxy power spectrum would show much stronger oscillatory features in the absence of dark matter (akin to those in the CMB) from the affect of radiation pressure on the photon-baryon plasma at early times -- see \S \ref{sec:BAO}.  The suppressed amplitude of these oscillations in the observed power spectrum, e.g Fig. 2 of \cite{2dF_SDSS_Pk}, suggests the matter power spectrum is dominated by a pressureless CDM component.

Given the detection of neutrino oscillations \citep{NeutrinoOscillation}, at least two of the three neutrino types are known to be massive and therefore a candidate for non-baryonic DM.  The lower limit on the total mass is $\simeq 60-100$ meV.  At early times, when $k_B T \gg m_{\nu}$, the neutrinos behave relativistically; this `free streaming' prevents the growth of perturbations to the neutrino (DM) density on scales smaller than the horizon.  This effect, together with the early-time integrated Sachs-Wolfe \citep{Sachs-Wolfe} and CMB lensing, constrains the total neutrino mass to be: $\sum m_{\nu} < 0.23$ eV \citep{Planck}, which corresponds to $\Omega_{\nu} h^2 < 2.5 \times 10^{-3}$.  Hence neutrinos are not massive enough to solely satisfy the $(\Omega_m, \Omega_b) \simeq (0.3, 0.05)$ constraints required by observations (when $\Lambda$CDM is assumed).  However, a measurement of the individual masses by cosmological experiments would be a historic achievement in itself.              

Additional weakly interacting candidates that do not couple electromagnetically must therefore be invoked.  To ensure small-scale structure is not suppressed to an extent incompatible with observations, like the massive neutrino, the candidate must be sufficiently massive that thermal velocities are effectively zero.  The allowed mass range of such relics are shown in Fig. 12.5 of \cite{CP} and are further constrained by direct detection experiments \citep[][ and references therein]{CDMS}.  

While CDM achieves many great successes in fitting a wide variety of cosmological observations, several important discrepancies with observations remain.  In particular, numerical simulations of structure formation in a $\Lambda$CDM cosmology predict a universal NFW density profile \citep{NFW} for the dark matter halo, which diverges for $r \ll 1 \mpcoh$.  In contrast, observations favour $\rho \mapsto$ cnst. at small radii \citep{CoreCusp}.  The influence of baryonic feedback is a plausible solution to this problem, which may also solve the satellite problem -- around 500 satellites are predicted to orbit in the halo with the mass of the Milky Way but only 30 such dwarf galaxies are observed \citep{Dwarves}.    

\section{Dark energy and the cosmological constant}
\label{sec:lambda}
As alluded to above, the cosmological constant term, $\Lambda g^{\mu \nu}$, is equivalent to the energy-momentum tensor of a perfect fluid with a $w=-1$ equation-of-state.  There is no classical physical system that possesses a negative pressure but the same is not true of Quantum Mechanical (QM) systems.  The lowest energy state in QM is the vacuum, $\ket{0}$, which, as suggested by Heisenberg's uncertainty principle: $\Delta E \Delta t = (\hbar/2)$, is one in which particle-antiparticle pairs may come into existence for a short time; $\expval{\hat H}{0}$ may then be non-zero.  This is the case for the quantum harmonic oscillator (QHO), for which $E_n = \hbar \omega(2n+1)/2$.  Forces generated by gradients in the zero-point energy have been measured -- by the Casimir effect \citep{CasimirMeasurement}, and a zero-point energy of $\hbar \omega/2$ in the lattice has been confirmed by diffraction experiments \citep{Ziman}.  Prior to GR however, the absolute energy had no bearing on the physical evolution of systems, as both Special Relativity and Newtonian gravity are sensitive to only changes in potential energy, $\grad \rho c^2$. 

In contrast, Einstein's field equation shows geodesics to be determined by $\rho c^2$.  The vacuum contribution to the energy density must then have an associated energy-momentum tensor, $T^{\mu \nu}_{\text{v}}$.  Assuming this representation is invariant under Lorentz transforms in locally inertial frames, i.e. all such observers agree $\ket{0}$ possesses the same energy density and pressure, requires an isotropic (2,0) tensor.  The only (non-zero) candidate is the Minkowski metric, which results in $T^{\hat \mu \hat \nu}_v =  - \rho_v c^2 \eta^{\hat \mu \hat \nu}$.  By appealing to General Covariance, this is 
\[
T^{\mu \nu}_v =  - \rho_v c^2 g^{\mu \nu},
\]
in an arbitrary frame.  Therefore the vacuum contribution enters the field equation in an identical manner to the cosmological constant, possessing a $w=-1$ equation-of-state and constant energy density, $\rho_v c^2$.  

In quantum field theory, a field $\hat \phi(\mathbf x)$ is expanded in harmonic modes and (second) quantisation is then applied to each.  This results in a number of frequency states, each of which has an associated zero-point energy (analogous to a very large number of QHOs).  A naive QFT estimate gives 
\[
\rho_v c^2 = 2 \int_0^{\Lambda} \frac{\hbar \omega}{2} \ \frac{d^3 k}{(2 \pi)^3},  
\]
when two polarisation states are assumed, as is appropriate for radiation.  Assuming QFT may be trusted to the Planck scale, $\Lambda = M_{\text{Pl}} \equiv \sqrt{\hbar c/(8 \pi G)} \simeq 10^{18}$ GeV, at which point quantum gravity is expected to become apparent, yields 
\[
\rho_v = \left ( 10^{18} \text{GeV} \right )^4
\]
in natural units. But this is both an ambitious and fundamentally wrong prediction. Firstly, QFT may fail for $\Lambda \ll M_{\rm{Pl}}$ -- it has only been confirmed up to TeV energies by experiments at the Large Hadron collider.  Secondly, this is the contribution of a single scalar field; the total from the $\simeq 100$ particles of the standard model and their interaction energies will be much greater again.  Finally, applying this non-covariant cut-off results in the wrong equation-of-state -- that of radiation, $w = (1/3)$, rather than the $w=-1$ required for dark energy.  A covariant calculation predicts $\rho_v c^2 \simeq M^4 \ln|\Lambda/M|$, where $M$ is the particle rest mass.  Assuming the most massive elementary particle is the top quark, which has a mass of $\simeq 200$ GeV, the predicted vacuum density is $10^9$ (GeV)$^4$ \citep{KoksmaProkopec}.  The theoretical prediction is therefore much greater than that observed: $\left ( 10^{-12} \text{GeV} \right )^4$ or
\[
\Omega_v = 0.6911 \pm 0.006,
\]
for a combined posterior derived from CMB (TT, TE, EE), CMB lensing, BAO, supernovae and local $H_0$ measurements \citep{Planck}.  This fundamental problem in our formulation of physics is termed the cosmological constant problem \citep{WeinbergCosmoCnstProb, WeinbergDE}.  

An alternative possibility is that the accelerating expansion is caused by a quantum scalar field, which behaves like a perfect fluid with a time-dependent equation-of-state:
\[
w = \frac{\frac{1}{2} \dot \phi^2 - V(\phi) - \frac{1}{6}(\grad \phi)^2}{\frac{1}{2} \dot \phi^2 + V(\phi) + \frac{1}{2}(\grad \phi)^2}.
\]
The Higgs boson is the first detected scalar \citep{HiggsPeter,HiggsDetection}.  In contrast to vector fields, the importance of scalars stems from their ability to produce an isotropic pressure \citep{KaiserElements}.  Slowly varying gradients, $\partial_\mu \phi \ll 1$, then results in mimicry of the cosmological constant.  

In short, while surprising from a classical perspective, $w \simeq -1$ is not an unusual prediction for quantum systems.  As the energy density of radiation and matter redshifts away with time, while $\rho_v$ is a constant -- the sum of at least a `bare' cosmological constant and vacuum energy contribution, the vacuum will come to dominate the evolution of $R(t)$ at late times.  The Friedmann equation predicts a deceleration: $R(t) \propto t^{1/2}, \ t^{2/3}$ for the radiation and matter dominated epochs respectively, while that in a flat $\Lambda$-dominated universe is accelerating: 
\[
R \propto \exp(Ht), \qquad H = \sqrt{\frac{8 \pi G \rho_v}{3}} = \sqrt{\frac{\Lambda c^2}{3}}.
\]
This vacuum dominated de Sitter geometry is one instance of the violation of Mach's principle in GR -- that locally inertial frames (the metric tensor) should be defined by the large-scale matter distribution.  In contrast, ``in de Sitter space there is no matter at all'' \citep{WeinbergCosmoCnstProb}.   

The detection of an accelerating expansion is therefore less surprising in hindsight; there are certainly likely candidates that yield the necessary criterion for acceleration, $(\rho c^2 + 3p)<0$.  Indeed, if $w(a)$ only just turns negative then the universe becomes progressively more vacuum dominated with time.  As adiabatic expansion gives $\partial_t(\rho c^2 R^3) = -p \partial_t (R^3)$, a time-varying $w(t)$ results in 
\[
\rho_v(a) = \frac{3H_0^2 \Omega_v}{8 \pi G } \exp \left ( \int_a^{1} 3 [1+w(a)] d \ln a \right). 
\]
But the inferred magnitude of $\rho_v c^2 $ is very surprising; it came as no shock that a magnitude very much less than the QFT prediction was found: the effect of the cosmological constant is a strong suppression of the rate of gravitational collapse and hence, if $\rho_v c ^2$ was of the magnitude predicted by QFT, we would neither be in existence nor able to measure it \citep{WeinbergAnthropic}.  The more serious quandary was that the observed magnitude was just right for $\Lambda$ domination to occur today.  This led to the obvious question of: ``why now?''.   

In the absence of extreme fine-tuning between the bare $\Lambda$ and the vacuum contribution, resolving this problem requires the replacement of QFT with a new theory able to predict the observed magnitude or a means of preventing the vacuum energy from contributing to the field equation -- with a symmetry principle or otherwise.  The latter case solves the cosmological constant problem, it then remains to explain the small, non-zero, $\Omega_v$ observed and why matter-vacuum equality should occur today.  Obvious theories to explain this are homogeneous scalar fields \citep{RatraPeebles} with a potential appropriate for $\Lambda$ domination today; but it is not easy to avoid parameter fine-tuning in order to achieve this.  In many cases, these models are equivalent to a modified theory of gravity.  

In the absence of a compelling theory, deviations of the background expansion from that of $w=-1$ are sought; most simply by constraining a two-parameter linear model, $w(a) = w_0 + (1-a) w_a$.  The measurement of these parameters to percent level accuracy is a fundamental goal of future galaxy surveys such as DESI and Euclid.  Current constraints on $(w_0, w_a)$ from geometric measurements, including the angular diameter distance to last scattering \citep{Planck}, are shown in Fig. \ref{fig:w0wa}.   
\begin{figure}
\centering
\includegraphics[width=0.8\textwidth]{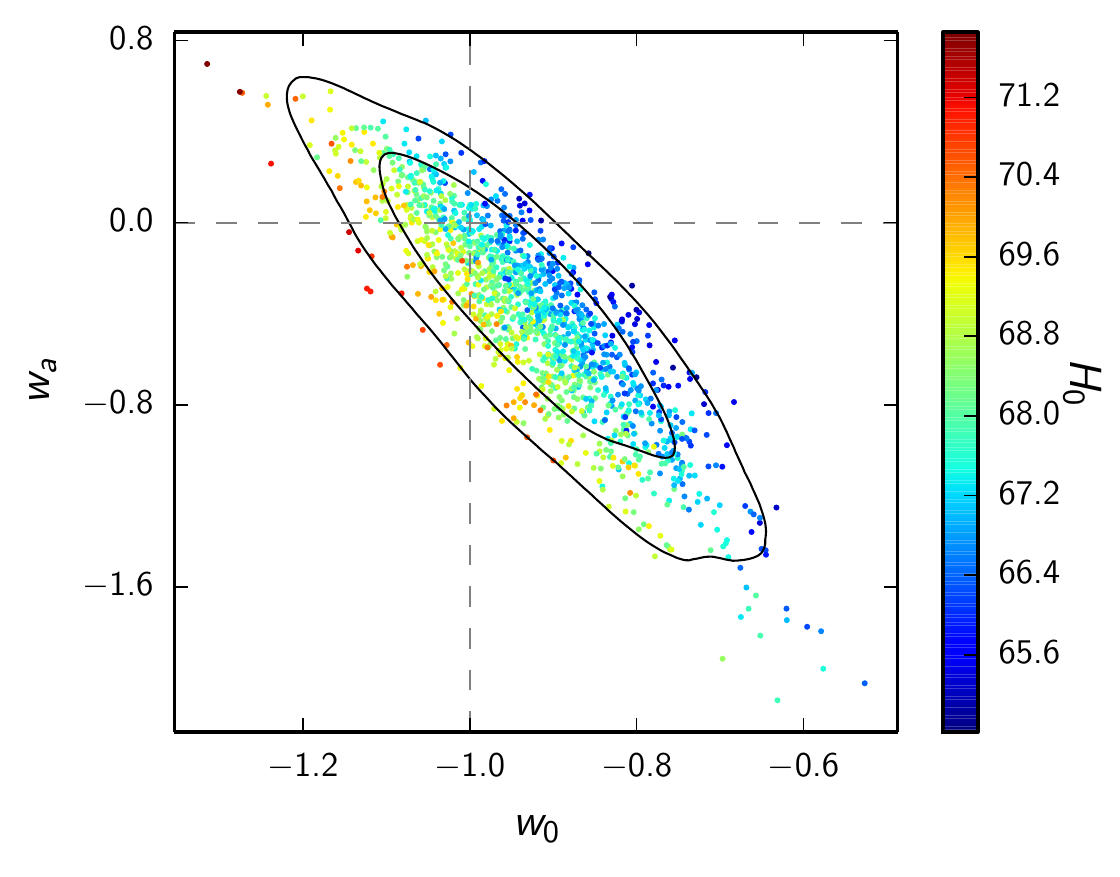}
\caption[Planck constraints on the time-dependent dark energy equation-of-state.]{68\% and 95\% confidence limits on the evolution of the dark energy equation-of-state, $w(a) = w_0 + (1-a) w_a$.  The likelihood is constructed from \cite{Planck} $C_\ell^{\rm{TT}}$ data, which provides a measurement of $D_A$ at the redshift of last scattering, low-$\ell$ polarisation data, BAO measurements of the Alcock-Paczy\'nski effect (see \S \ref{sec:BAO}) and the $D_L(z)$ relation from type-Ia supernovae \citep{JLA}.  No evidence against $\Lambda$ is shown; a fundamental goal of future surveys is to confirm this remains true with percent level precision.  Reproduced from \cite{Planck}.}
\label{fig:w0wa}
\end{figure}

\section{Modified gravity}
\label{sec:MG}
General Relativity is a theory founded on three equivalence principles:  

\textit{The weak equivalence principle} -- all test bodies, defined to have negligible active gravitational mass, follow an identical trajectory in a gravitational field when free of other forces.  With an E{\"o}tv{\"o}s experiment (see \citejap{TorsionBalance} for a review) the difference in acceleration of two such masses has been measured to be \citep{EotWash}
\[
2 \frac{|a_1 - a_2|}{|a_1 + a_2|} = (0.3 \pm 1.8) \times 10^{-13}.
\]
This verifies the weak equivalence principle holds to at least this precision; see \S 2.1.1 of \cite{ModGravReview} for further detail.      

\textit{The Einstein equivalence principle} -- The weak equivalence principle holds and in a freely falling frame (such as that of a test particle) the laws of Special Relativity are recovered locally. 

\textit{The strong equivalence principle} --  Massive bodies, with significant gravitational binding energy, follow the same trajectories as test particles when free of non-gravitational forces.  The relative acceleration of the earth and the moon has been measured with lunar ranging experiments -- by reflecting lasers from panels left on the Moon by the Apollo 11 mission.  The most stringent constraint on deviations from the strong equivalence principle is  
\[
2 \frac{|a_1 - a_2|}{|a_1 + a_2|} = (-1.0 \pm 1.4) \times 10^{-13}.
\]

The geodesic equation is a result of Einstein's equivalence principle, which suggests the gravitational theory should be a metric one.  Einstein's field equation is the simplest possible law that satisfies all of these principles, but it may not be unique. By adding additional generally covariant terms that vanish in a local inertial frame, modified theories may be derived that typically break the strong equivalence principle \citep{JoyceReview}.  This distinguishes modified gravity models from exotic contributions to the energy-momentum tensor -- by particles that have been undetectable in the laboratory to date.       

The observed acceleration of the expansion may then be evidence that the field equation is an approximation, as opposed to evidence for a new quantum scalar field.  This is reminiscent of the situation that led to the development of General Relativity -- posited explanations for the anomalous precession of the perihelion of Mercury were that Newton's theory was correct and a new planet, `Vulcan', would soon be discovered or the theory itself was an incomplete description.  As has been the case many times before in the history of science, Vulcan was falsely discovered numerous times before a better explanation was unearthed \citep{Vulcan}.       

Simple extensions to Einstein's field equation, e.g. adding an additional $f(R)$ term to the Einstein-Hilbert action, are able to reproduce any expansion history; see Fig. 1 of \cite{SongHuSawicki} for example.  As a result, with only measurements of the background expansion, the effect of a quantum scalar field with a given equation-of-state can always be reproduced with an $f(R)$ theory; additional measurements are required to distinguish between the two. 

The constraints placed by solar system tests can be eluded by admitting our selection bias.  The solar system is a particular place in which e.g. the gravitational potential and its derivatives differ greatly from the majority of the horizon volume.  Certain $f(R)$ theories exploit this fact by making predictions indistinguishable from GR in the large (or rapidly changing) curvature limit, e.g. any for which $f(R) \mapsto \text{cnst.}$ for $R \gg 1$ \citep{HuSawicki}; this large curvature limit corresponds to an effective cosmological constant, but the cosmological constant problem remains.  Constraints placed by solar system tests cannot distinguish between GR and such `shielded' modified gravity theories \citep{ModGravReview}.      

Additional observables are required to advance our knowledge.  One possibility is to ensure the rate of gravitational collapse -- directly measured by redshift-space distortions (see Chapter~\ref{chap:RSD}), is consistent with the GR prediction.  A key prediction of General Relativity is a scale-independent growth rate that is determined solely by the expansion history;  this property is typically not true of modified gravity theories.  This has motivated confirming the linear growth rate remains consistent with the GR prediction.  This is a principle goal of future galaxy surveys such as \href{http://sci.esa.int/euclid/}{Euclid} and \href{http://wfirst.gsfc.nasa.gov}{WFIRST}.  Even if this is confirmed, there remains a number of possible models that are degenerate with respect to both the background expansion and growth rate.  Further diagnostics are provided by weak gravitational lensing and may be augmented by constraints on the propagation of gravitational waves in the future \citep{LombriserTaylor}.               

\section{Inflation}
\label{sec:inflation}
\subsection{Deficiencies of classical cosmology}
\label{sec:Deficiencies}
Despite the great successes of the FRW framework in accurately describing large-scale cosmological observations, fundamental questions remain unanswered.  The first is the origin of the expansion itself, which is simply assumed to be an initial condition.  The second is the observed isotropy of the sky at $\simeq 150 \rm{GHz}$ -- this is the Cosmic Microwave Background (CMB): a relic blackbody spectrum from the radiation dominated era, which has redshifted with the expansion:
\[
T(z) = T_0 (1+z); 
\]
the observed temperature today is measured to be $T_0 = 2.718 \pm 0.021$K \citep{Planck}.  As the universe was opaque before the formation of neutral hydrogen at $k_BT(z) \simeq 13.6$eV, due to Thompson scattering between photons and the free protons and electrons, this radiation last scattered at $z^* = 1089.90 \pm 0.23$ \citep{Planck} -- in time, not necessarily radial distance (due to post-reionization scattering).  The comoving radius of a causally connected volume at this redshift, the particle horizon, is
\[
R_0r_p = \frac{c}{H_0} \int_{z^*}^{\infty} \frac{dz'}{\sqrt{\Omega_m ( 1+z )^{3}}} \simeq \frac{2c}{H_0 \sqrt{\Omega_m}}(1+z^*)^{-1/2}; 
\]
this assumes the distance is dominated by the short period before recombination when $\Omega_m(a)$ is significant.  Given $D_A(z^*)$, e.g. from Fig. \ref{fig:fig8Hamilton98}, this causally connected volume may be found to span $1\deg$ on the sky.  The observed isotropy of the microwave sky then suggests a very different mechanism for establishing causal contact in the early universe.   

Thirdly, if the spatial curvature is not exactly zero then $\Omega(a)$ rapidly tends to unity as $a \mapsto 0$:
\begin{align}
\Omega(a) &= \frac{8 \pi G \rho(a)}{3H^2(a)} = \frac{\Omega_v + \Omega_m a^{-3} + \Omega_r a^{-4}}{\Omega_v + \Omega_m a^{-3} + \Omega_r a^{-4} + (1 - \Omega)a^{-2}} \nonumber \\ &\simeq 1 + \frac{(\Omega-1)}{\Omega_r} a^2 \qquad \text{for }a \ll 1.
\label{eqn:oma}
\end{align}
We measure $\Omega \simeq 1$ today but if $|\Omega -1|$ is not exactly zero then at an early time it is arbitrarily small as $|1 - \Omega(a)| \propto a^2$.  This arbitrarily small but non-zero deviation represents a fine-tuning deserving of an explanation.  The maximum fine-tuning occurs when the smallest plausible $a$ is assumed; a plausible prediction of the maximum fine tuning in classical cosmology is $(a_{\rm{Pl}})^2 \sim 10^{-64}$, as anything smaller than the Planck scale will be in the regime of quantum gravity.

Finally, successful theories of galaxy formation currently require initial density fluctuations to originate further collapse.  But what is the origin of the initial fluctuations?  When many theories for the cause of the accelerating expansion seem to have limited explanatory or predictive power, it is remarkable that the theory of inflation was able to answer each of these and more.  The following section gives a brief description of the simplest inflationary mechanisms and their central role in modern cosmology.      

\subsection{An early period of vacuum domination}
The horizon problem would be solved if our current was once in causal contact; this requires that the comoving size of the particle horizon,
\[
R_0 r_h(t) = \int_0^t \frac{cdt'}{R(t')} = c \int_0^R \frac{dR'}{R' \sqrt{\rho {R'}^2}};  
\]
was once bigger than the current horizon.  Providing $\rho R^2$ is either finite or tends to zero as $R \mapsto 0$ the integral diverges and the horizon problem is solved.  Alternatively, in a complete quantum description of gravity the contribution to $R_0 r_h(t)$ from $t' < t_{\rm{Pl}} \equiv 10^{-43} s$ could be sufficiently larger and there would be no horizon problem \citep{Baumann}.  Remaining with the more securely founded classical explanation, the Friedmann equation gives $\ddot R \propto n$ for $\rho R^2 \propto R^n$.  As $n \geq 0$ is required to solve the horizon problem: $\ddot R > 0$, in which case an early epoch during which $w < (-1/3)$ will establish the necessary casual contact (from the Raychauduri equation).  
This period will also solve the flatness problem: from eqn. (\ref{eqn:oma}), if $\Omega(R_i) \simeq \mathcal{O}(1)$ initially then at a later time: $\Omega(R) = 1 + \mathcal{O}(f^{-2})$ for $f=(R_i/R)$.  Thus, as $\dot R$ increases without limit, $\Omega(R)$ is driven arbitrarily close to $1$.  A solution to the (maximum) flatness problem then requires $\ln |f| \gsim 60$.  However, if this early inflationary epoch is maintained, a likely consequence is $f \gg 1$ and an inescapably flat universe is a natural prediction.  We observe just a universe, with the best constraint to date given by:
\[
\Omega_k = \left( 8 \pm 40 \right ) \times 10^{-4}
\]

As we have seen, a quantum scalar field can achieve a time-dependent equation of state with $w(a)<(-1/3)$. This is a plausible mechanism for inflation in which the vacuum does not always dominate.  During inflation, flat-space quantum fluctuations in $\phi$ are stretched far beyond a finite, proper event horizon:
\[
R_0 r_{\rm{EH}} = R_0 \int_{t_0}^{\infty} \frac{c dt}{R(t)} = \left ( \frac{c}{H} \right ),
\]
where the final equality assumes a potential dominated de Sitter period.  The fluctuations span a causally disconnected (physical) volume following crossing and therefore become `frozen' as classical fluctuations.  These classical fluctuations provide the initial density perturbations necessary for gravitational collapse, eventually leading to the formation of galaxies and galaxy clusters.  In the simplest models for $V(\phi)$, the predicted density fluctuations are Gaussian and adiabatic with a nearly scale-invariant spectrum \citep{CP}. 

Inflation must end in practice and therefore diverges from the de Sitter behaviour, $w=-1$, towards the final stages -- when $\dot \phi^2 \simeq V(\phi)$.  If inflation begins at the GUT scale $(10^{15}$ GeV), only the final $\ln |f| \simeq 60$ stages are observable as a mode that does not cross the horizon cannot seed a classical fluctuation, while those that do so at early times will likely be stretched far beyond any observable horizon.  More precisely then, a common prediction for the  spectrum of density fluctuations is a small `red tilt' to  $|\delta_{\mathbf{k}}|^2 \propto k^{n_s}$, with $n_s$ slightly less than unity.

In addition to seeding scalar density perturbations, quantum fluctuations in the (linearised) metric create primordial gravitational waves.  The ratio of the squared amplitude of the tensor-to-scalar perturbations is $r \propto \partial_{\phi} \ln |V|$.  A measurement of $r$ may then be used to distinguish between the possible potentials; $r$ may be inferred from the magnitude of the effective Sachs-Wolfe effect added to the CMB fluctuations by primordial gravitational waves.  The current best CMB constraint on $r$ is shown in Fig. \ref{fig:ns_r}.  This field has been given added impetus by the first confirmed observation of gravitational waves \citep{LIGO}, but a direct detection of the primordial signal will require an improvement of $\simeq 10^{6.5}$ in detector (strain) sensitivity \citep{CP}.     
\begin{figure}
\centering
\includegraphics[width=0.7\textwidth]{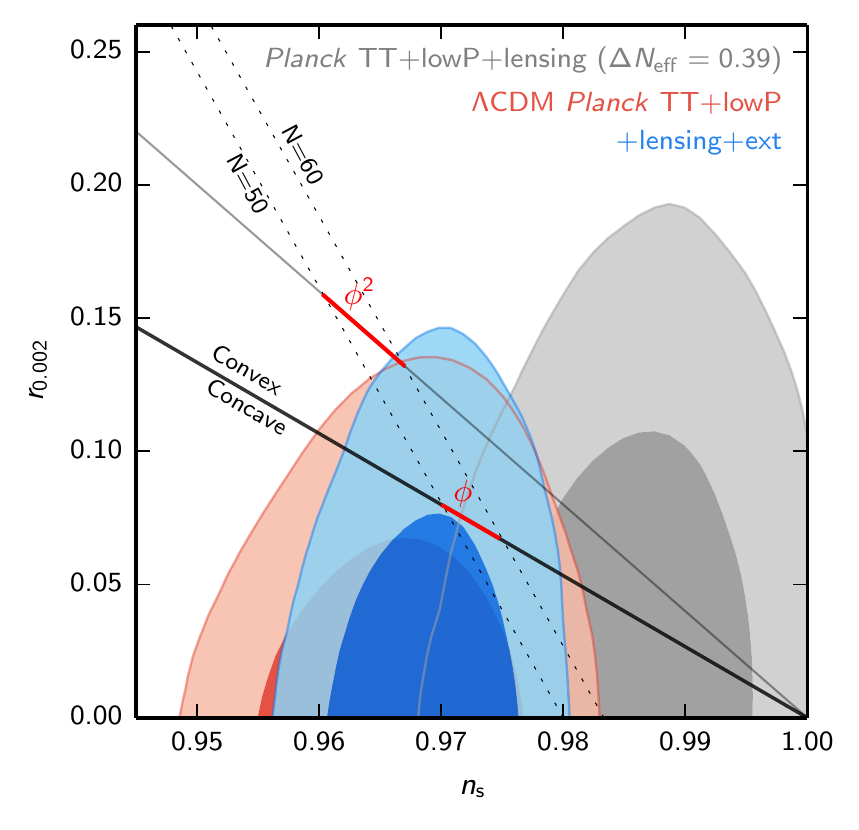}
\caption[Planck posterior on the inflationary parameters.]{\cite{Planck} joint-posterior on $(n_s, r)$.  The   scalar spectral index, $n_s$, is defined by $|\delta_{\mathbf{k}}|^2 \propto k^{n_s}$; this is predicted to be slightly less than unity in the simplest inflationary models and provides a measure of $V(\phi)$ in the closing stages of inflation.  The squared amplitude ratio of the tensor-to-scalar fluctuations, $r$, may be inferred from the effective Sachs-Wolfe effect generated by primordial gravitational waves.  This provides a second independent measure of the potential.  Predictions for the simplest, single scalar field models, e.g. $V(\phi) \propto m \phi^2$, are shown for a number of different e-foldings, $N = \ln |f|$.  The number of e-folds is determined by the reheating process \citep{LiddleLeach}.  These constraints are now sufficiently precise that the possibilities for $V(\phi)$ are restricted; both $\phi^4$ and $\phi^3$ have now been conclusively excluded.}
\label{fig:ns_r}
\end{figure}

Although the classical theory of inflation has solved many of the problems outlined in \S \ref{sec:Deficiencies}, two of the necessary assumptions are of note: to achieve $\ln |f| \simeq 60$ the magnitude of $\phi$ must be $\simeq M_{\rm{Pl}}$ initially.  It is therefore far from clear that quantum gravity has been successfully avoided.  Moreover, it must be assumed that the contribution to $\rho c^2$ from $V(\phi_{\rm{end}})$ is negligible, but the physical motivation for this is unclear.     

Finally, an inflationary universe may not begin with a big bang, $R(0)=0$.  Looking back in time, $R(t)$ exponentially asymptotes to zero but may never reach it.  This period can last for an arbitrarily large number of e-foldings with no observable consequences for any except the last 60 \citep{LiddleLeach}.  Therefore, when invoking an inflationary universe to solve the horizon, flatness and expansion problems (in addition to sourcing initial perturbations), it is impossible to place an upper limit on $t_{\rm{age}}$ with observations.  Claims that cosmology unambiguously determines $t_{\rm{age}} \simeq 13.7$ billion years are therefore somewhat deceptive.     

\section{Structure formation}
\begin{figure}
\centering
\includegraphics[width=\textwidth]{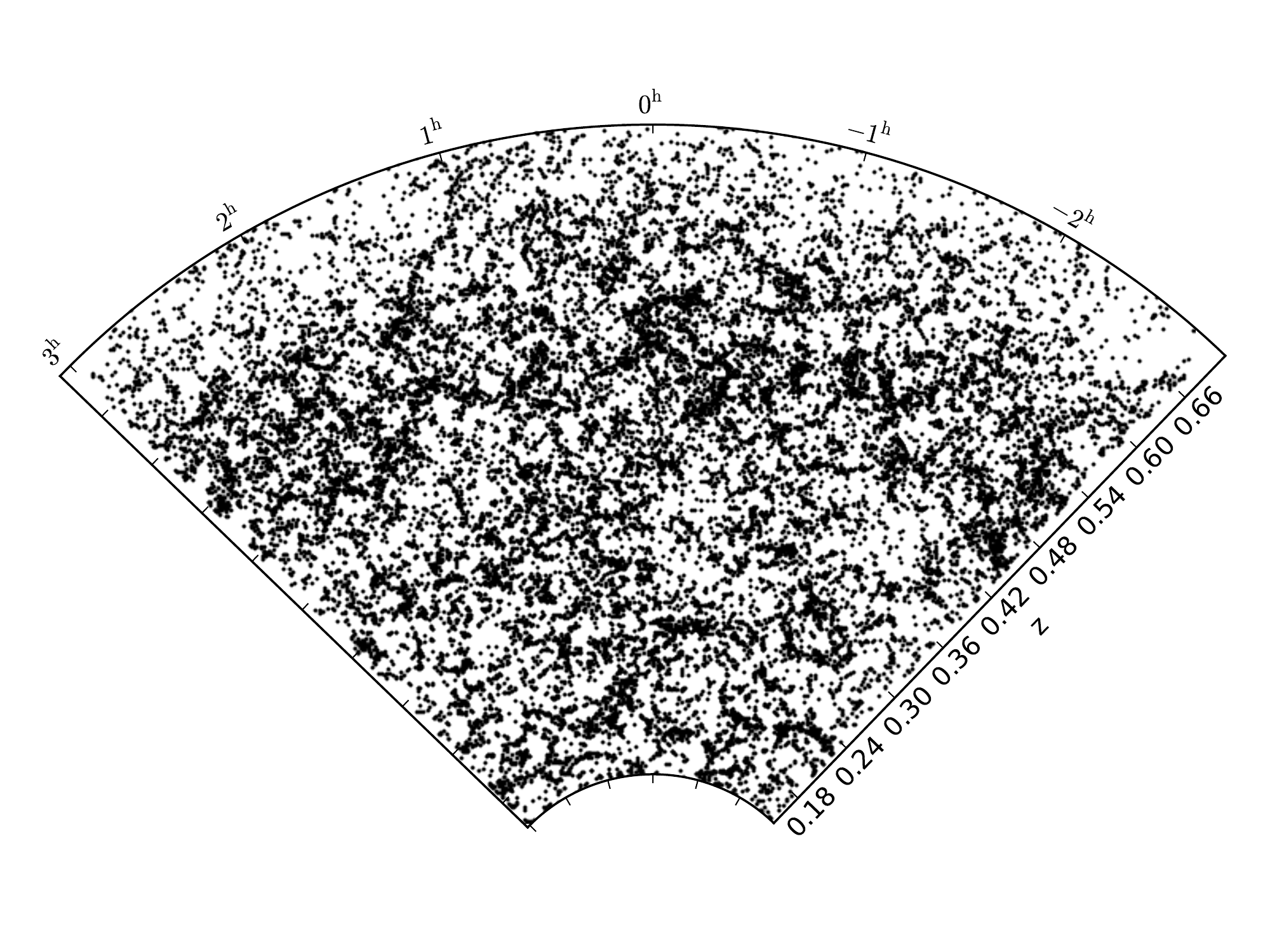}
\caption[A BOSS slice of large-scale structure.]{Observed large-scale structure for a $2\deg$ slice in declination of the BOSS South field, including both the Low-$z$ and CMASS surveys.  Galaxies are thought to linearly trace the underlying matter perturbations, which is dominated by the cold dark matter component.  Clearly visible in this figure are large galaxy clusters, smaller galaxy groups and voids -- large underdense volumes containing very few galaxies.  This galaxy distribution is a relic containing information of both particle physics in the early universe and the gravitational physics of the late universe.}
\label{fig:BOSS_lightcone}
\end{figure}

\subsection{The perturbed metric}
The FRW metric describes a perfectly homogeneous and isotropic universe, which is very different from the one  we observe on almost all spatial scales; see Fig. (\ref{fig:BOSS_lightcone}).  But having determined a mechanism for seeding classical density perturbations (all of which were once superhorizon), perturbations may be included in the metric.  The most general possibility for scalar perturbations is 
\[
ds^2 = - \left( 1 + \frac{2 \Psi}{c^2} \right) c^2 dt^2 + R^2(t) \left( 1 - \frac{2 \Phi}{c^2} \right)(dr^2 + S^2_k(r)d\psi^2), 
\]
in the Newtonian gauge.  Here $(\Psi, \Phi)$ are gauge-invariant scalar perturbations.  These remain invariant under coordinate transforms which relabel the origin, $x^{\mu} \mapsto x^{\mu} + \epsilon^{\mu}(x^{\mu})$; this is not true of scalars generally -- see \S 2 of \cite{MFB} or \cite{Mukhanov} for further detail.  Inserting this metric into Einstein's field equation requires $\Phi = \Psi$, when the energy-momentum tensor is that of a perfect fluid with no anisotropic stress, $T^{i j} = 0$ for $i \neq j$.  Under these circumstances, $\Phi$ satisfies the usual properties of the Newtonian gravitational potential.  

The evolution of linear perturbations on superhorizon scales is obtained by identifying physical solutions to the linearised field equation \citep{LinearGR_perturbations} or by explicitly constructing gauge invariant scalars \citep{BardeenGauge}.  The following discussion presents a rough derivation that obtains the correct result by starting from the Friedmann equation.  This separate universe model follows \S 9.3.6 of \cite{KolbTurner}.  

The effective Friedmann equation for the evolution of a spherical overdensity with radius $R_s(t)$ in an otherwise homogeneous and spatially flat universe with density $\rho_0$ is
\[
H_s^2(t) = \frac{8 \pi G \rho_s}{3} - \frac{k}{R_s^2} \qquad (k=1).  
\]
As the physical laws of GR are manifestly covariant -- explicitly invariant under all spacetime coordinate transforms, there is an inherent ambiguity in defining a perturbation from the background on superhorizon scales.  A gauge choice must be made, in this case, by asking what is the perturbation when the Hubble rate is the same?  The answer is   
\[
\delta_s \equiv \left ( \frac{\rho_s}{\rho_0} \right ) - 1 = \frac{3k}{8 \pi G \rho_0 R_s^2}.
\]
Providing $\delta$ is small, $R_s(t) \simeq R(t)$ and the gauge-invariant solutions are 
\[
\delta \propto \left ( \frac{1}{\rho_0 R^2} \right ) \propto     
    \begin{cases}
            R^2 &        \text{radiation dominated} \\
            R &          \text{matter dominated}.     
    \end{cases}
\]
As the Newtonian relation: $\Phi(t) \propto - \rho_0 \delta R^2$ remains valid on superhorizon scales, the superhorizon perturbations to $\Phi$ are time-independent during the radiation and matter dominated eras.  Modes that enter the horizon during the radiation era are therefore already collapsing with $\partial_t \ln \delta = 2H(a_{\rm{enter}}) > 0$.   

The curvature of spacetime simply manifests itself as Newtonian gravitational physics on subhorizon scales.  If the universe is flat, or comoving separations are much smaller than $R_0$, then spatial curvature may also be neglected.  In this regime, Newtonian gravity in a universe with the Hubble expansion occurring about every point is then a perfectly valid description for structure formation.  This is most simply approached by continuing with the fluid approximation, which is outlined in the following section. 

\subsection{Newtonian structure formation}
Prior to the onset of non-linearity and multi-streaming -- at which point the velocity field is no longer single valued, the equations of motion for the matter component (CDM + baryons) are given by the non-relativistic fluid equations: 
\begin{align}
\bnabla^2 \Phi &= 4 \pi G \rho, \nonumber \\
\frac{D \mathbf{v}}{Dt} &= - \frac{\bnabla p}{\rho} - \bnabla \Phi, \nonumber \\
\frac{D \rho}{Dt} &= - \rho \nabla \cdot \boldsymbol v.
\end{align}
These are the Poisson, continuity and Euler equations respectively.  The continuity equation is a conservation law -- if the density is to decrease in a given volume there must be a flux of matter outwards through the enclosing  boundary.  Note that Poisson's equation is sourced by the density, as opposed to the active mass density, in this non-relativistic limit.  The convective derivative,
\[
\frac{D}{Dt} \equiv \frac{\partial }{\partial t} \biggr |_{\boldsymbol{x}} + \mathbf{v}(\mathbf{x}) \cdot \boldsymbol \nabla  \bigr |_{t},
\]
measures the rate of change from the perspective of an observer moving with the flow, i.e. one travelling at $\mathbf{v}(\mathbf{x})$, and is therefore a result of both traversing a local gradient: $\partial_{\mathbf{x}}|_{t}$, and time evolution: $\partial_t|_{\mathbf{x}}$, as would be experienced by an observer at rest.  The time dependence of the local gradient is higher than linear order and is therefore neglected.  

For initial conditions corresponding to Hubble expansion about every point, convenient comoving co-ordinates and peculiar velocities may be defined to be
\[
\mathbf{r}         = a^{-1} \mathbf{x},  \qquad \qquad 
\mathbf{u}         = a^{-1} \left ( \mathbf{v} - H \mathbf{x} \right ),
\]
in terms of the normalised scale factor, $a \equiv R/R_0$; this is the natural variable for scales on which the spatial curvature is negligible.  Henceforth $\grad$ will denote the gradient with respect to $\mathbf{r}$, $\grad \equiv \grad_{\mathbf r} = a \grad_{\mathbf{x}}$, and by further defining the fractional perturbation to the density field:
\[
(1  +\delta) \equiv \left ( \frac{\rho}{\rho_0} \right ),
\]
the linear theory equations, first order in the (assumed) small perturbations $\delta$ and $\mathbf{u}$, are given by
\begin{align}
\dot{\mathbf{u}} + 2H \mathbf{u} &= -\frac{\grad{\Phi}}{a^2} - \left ( \frac{1}{a^2}\right ) \frac{\grad{p}}{\rho_0}, \nonumber \\
\dot \delta &= - \grad \cdot \mathbf{u}.
\end{align}
Here $\dot{\mathbf u}$ denotes the convective derivative of an observer comoving with the Hubble flow, i.e. travelling at $\mathbf{v_0} = H\mathbf x$ rather than $\mathbf{v}$, as the complete convective derivative reduces to this limit when acting on perturbed quantities (to linear order).  Both $\Phi$ and $p$ represent the deviations from the corresponding quantity in a homogeneous universe.  By modelling the distribution of matter particles as a perfect fluid with an equation-of-state or sound speed of $c_s^2 \equiv \partial p/\partial \rho |_s \equiv w c^2$, where $s$ is the entropy, the linear growth equation may be obtained: 
\[
\partial_t^2 \tilde \delta(\mathbf k) + 2H \partial_t \tilde \delta(\mathbf k) = \tilde \delta(\mathbf k) \left( \frac{3}{2} H^2 \Omega(a) - \frac{c_s^2 k^2}{a^2} \right).
\label{eqn:lineartheory2ODE}
\]
when this is applied to a multi-component analysis, e.g. photons, baryons and CDM, each species experiences solely its own pressure gradient but the total gravitational acceleration.  Here the Fourier transform of $\delta (\mathbf{x})$ has been defined as \citep{FourierT}
\[
\tilde \delta(\mathbf{k}, a) = \int d^3 r \ \delta(\mathbf{r}, a) \ \rm{e}^{-i \mathbf k \cdot \mathbf r},
\]
for $k = (2\pi/ \lambda)$.  The comoving wavenumber, $\lambda$, is equivalent to a physical wavelength of $a \lambda$ at any given time, i.e. the Fourier basis stretches with the expansion.  A key property of linear theory is that each Fourier mode evolves independently.  

The first modes to enter the horizon do so in the radiation era; radiation pressure prevents the further gravitational collapse of these initial perturbations on subhorizon scales.  The sound speed in the tightly coupled photon-baryon fluid is $c_s = (c/\sqrt{3})$ and therefore the sound horizon is roughly the horizon size at any given time.  Consequently, the radiation density simply provides an unperturbed background and the evolution is determined by 
\[
\partial_t^2 \tilde \delta_m(\mathbf k) + 2H \partial_t \tilde \delta_m(\mathbf k) = 4 \pi G \rho_m \tilde \delta_m(\mathbf{k}),
\label{eqn:meszaros}
\]
for the perturbations to the pressureless CDM component.  This equation has a growing mode solution of $\delta_m(a) = (a/a_{\text{eq}}) + (2/3)$ for $\rho_m(a_{\text{eq}}) = \rho_r(a_{\text{eq}})$.  

The amplitude of modes that enter the horizon during the radiation dominated is therefore effectively frozen until matter-radiation equality.  Those still superhorizon continue to grow as $R^2$ during this time.  This M\'{e}sz\'{a}ros effect \citep{Meszaros} is responsible for a characteristic bend in the linear matter power spectrum, $P(\mathbf{\mathbf k}) = \langle |\tilde \delta (\mathbf{k})|^2 \rangle$, at $k_{\text{bend}}^{-1} \simeq R_0 r_p(z_{\text{eq}}) = 16 (\Omega_m h^{-1}) \mpcoh$.  Larger wavelength modes were superhorizon until at least $a_{\text{eq}}$ and therefore experience no suppression.  The inflationary power law, $P(k) \propto k^{n_s}$, is then preserved for $k< k_{\text{bend}}$.  For smaller wavelength modes that enter the horizon, the suppression ratio is $(\delta_{\rm{sub}}/\delta_{\rm{super}}) = (a_{\rm{eq}}/a_{\rm{enter}})^2$ as entry occurs when $k^{-1} = R_0 r_p(a_{\rm{enter}}) \simeq (c/a_{\rm{enter}})H^{-1}(a_{\rm{enter}})$, for the constant $a^2 H$ of this era. Therefore the linear small-scale power spectrum is $P(k) = (a_{\rm{eq}}/a_{\rm{enter}})^4 k^{n_s} = k^{(n_s - 4)}$, which is almost a $k^{-3}$ powerlaw in the simplest inflationary models.  But this is only half of the story; the density perturbation is frozen when the collisonless component sources its own gravitational collapse but each mode is already collapsing on horizon reentry: $\partial_t \ln \delta = 2H(a_{\rm{enter}}) > 0$.  This will be more significant than the sourced collapse and, when ignoring the latter, we have $\partial_t^2 \tilde \delta_m(\mathbf k) + 2H \partial_t \tilde \delta_m(\mathbf k) =0$.  Hence the growth is logarithmic, $\tilde \delta_m(\mathbf{k}) \propto \ln(t)$, following reentry.               

In the matter dominated era, effectively an Einstein-de Sitter universe with $\Omega(a) \simeq \Omega_m \simeq 1$, $a \propto t^{2/3}$ and $H = (2/3)t^{-1}$, the pressureless CDM component evolves as $\delta(t) = D(t) \delta(t_0)$ with the growing and decaying modes given by
\begin{align}
D_+ = \left (t/t_0 \right )^{2/3}, \qquad 
D_- = \left (t/t_0 \right )^{-1},
\end{align}
respectively.  As this subhorizon growing mode solution, which determines the late time behaviour, is identical to the superhorizon solution, the overall shape of the power spectrum is determined by the radiation era in linear theory.  Following this period, the linear behaviour is simply for the amplitude to grow according to $D_+ \propto t^{2/3}$ on both subhorizon and superhorizon scales.

The late-time behaviour in our own universe is seemingly dominated by the vacuum, $\rho_v c^2$.  In this case, a formal solution to eqn. (\ref{eqn:lineartheory2ODE}) for a $w=-1$ equation-of-state is given by (\citejap{Heath}, \citejap{LambdaCarroll})   
\[
\delta_m(a) = \frac{5}{2} H_0^2 \Omega_m H(a) \int_0^a \frac{da'}{H^3(a')}.
\]
This has an exact solution for flat universes, $\Omega_m(a) + \Omega_v(a) = 1$, in terms of elliptical integrals \citep{EisensteinLambdaGrowth}.  The qualitative conclusion is that the accelerating expansion suppresses the growth rate of structure.  

\subsection{Baryonic acoustic oscillations}
\label{sec:BAO}
This linear theory treatment has focused on the pressureless dark matter component so far, but prior to decoupling the baryons are tightly coupled to the photons due to Thomson scattering.  The radiation pressure then prevents the baryons from further collapsing on scales below the Jeans length, $\lambda_J \propto c_s \rho^{-1/2}$.  For $L = \partial_t^2 + 2H \partial_t$, repeated application of eqn. (\ref{eqn:lineartheory2ODE}) results in 
\[
L \tilde \delta_b = L \tilde \delta_c - \frac{c_s^2 k^2 \tilde \delta_b}{a^2},
\]
which has a solution of
\[
\left ( 
    \frac{\tilde \delta_b}{\tilde \delta_m} \right ) 
= \frac{6}{k^2 c_s^2 \eta^2} \left ( 1- \frac{\sin(k c_s \eta)}{k c_s \eta} \right) 
\]
for the evolution of modes smaller than the sound horizon up to decoupling (when a constant $c_s$ is assumed).  This decaying oscillation reflects the competing effects of pressure and gravity on the initial adiabatic perturbation.  The pressure wave initially expels the baryons outwards, which dissipates the perturbation until the radiation pressure, $p \propto \rho c^2$, dispels and recollapse occurs, again raising the pressure.  This process is repeated until decoupling, at which point the photons are freed from the baryons and $c_s$ quickly drops to zero; the pressure is never sufficient to reverse the initial perturbation.  The overall amplitude scaling, $\eta^{-2}$ simply reflects the continued CDM growth, $\delta_c \propto \eta^{2}$ for $\eta \propto t^{1/3}$.  The comoving size of the sound horizon at decoupling is therefore an important `standard ruler' -- an observable of known comoving size that can be established from well established physical processes.  

The evolution in the matter dominated era is again given by eqn. (\ref{eqn:lineartheory2ODE}) following decoupling:
\[
L \begin{pmatrix}
    \tilde \delta_b \\
    \tilde \delta_c
  \end{pmatrix}
  = \frac{4 \pi G \rho}{\Omega_m}
  \begin{pmatrix}
    \Omega_b & \Omega_c\\
    \Omega_b & \Omega_c
  \end{pmatrix}
  \begin{pmatrix}
    \tilde \delta_b \\
    \tilde \delta_c
  \end{pmatrix},
\]
which, when diagonalised, results in a dominant growing mode solution with $\rho_b = \rho_c \propto t^{2/3}$.  Therefore, once decoupling occurs, the baryons rapidly fall into the dark matter potential wells, which have continued to deepen even below the Jeans length.  These baryon acoustic oscillations are imprinted as a series of oscillations on the power spectrum and are further summarised in Fig. \ref{fig:BAO}.        
\begin{figure}
\subfloat{\includegraphics[width=.45\linewidth]{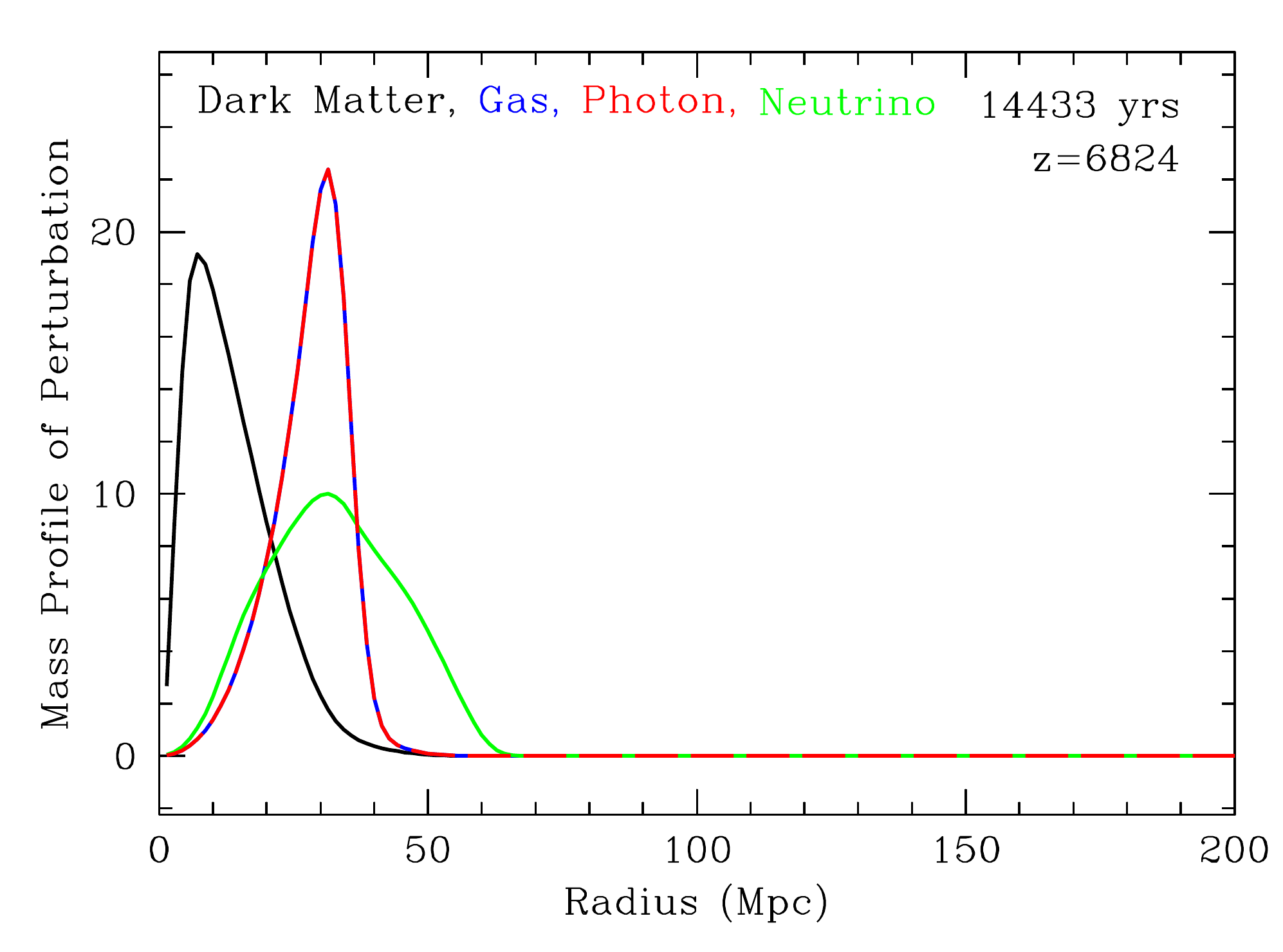}} 
\subfloat{\includegraphics[width=.45\linewidth]{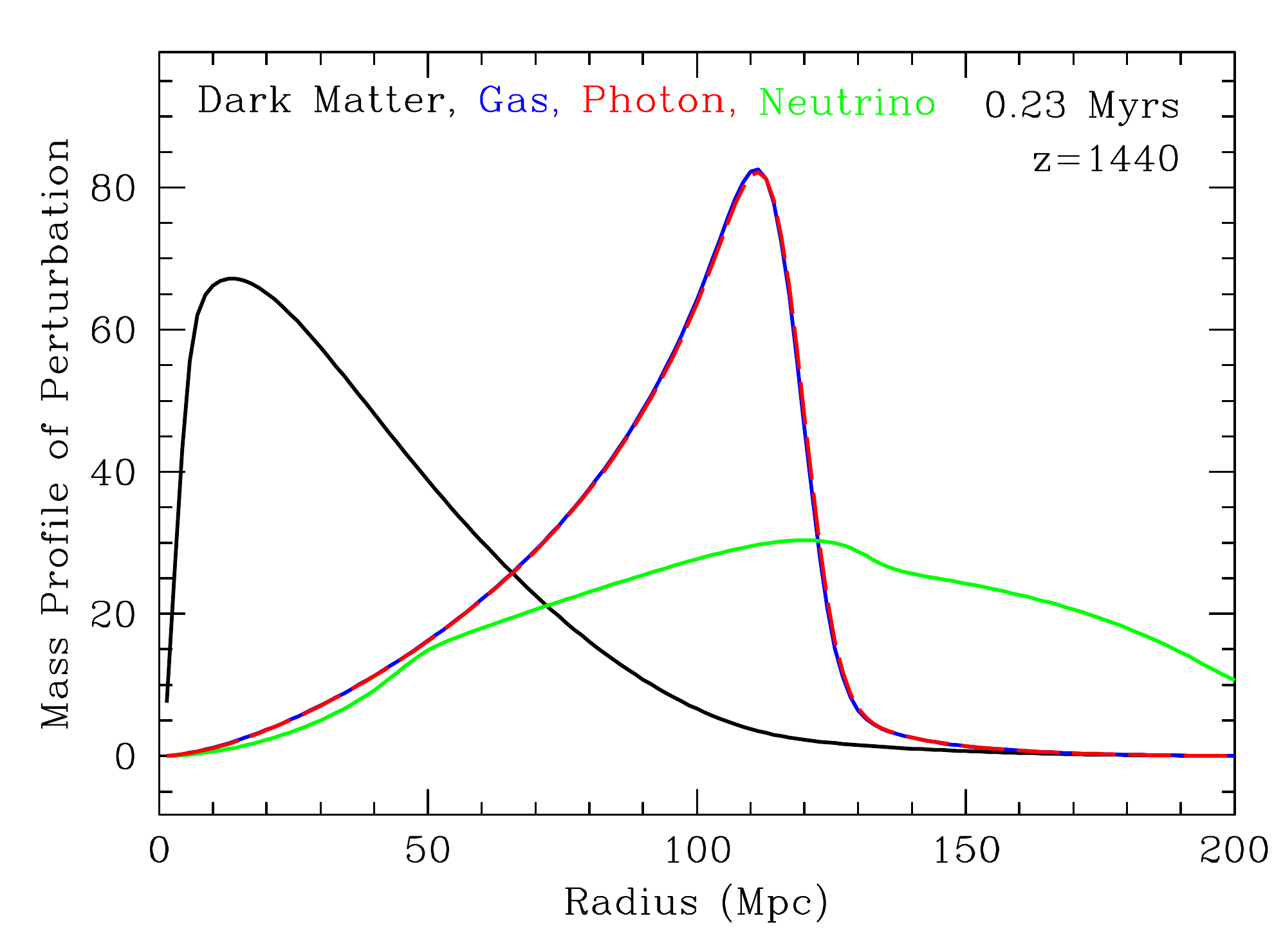}} \\    
\subfloat{\includegraphics[width=.45\linewidth]{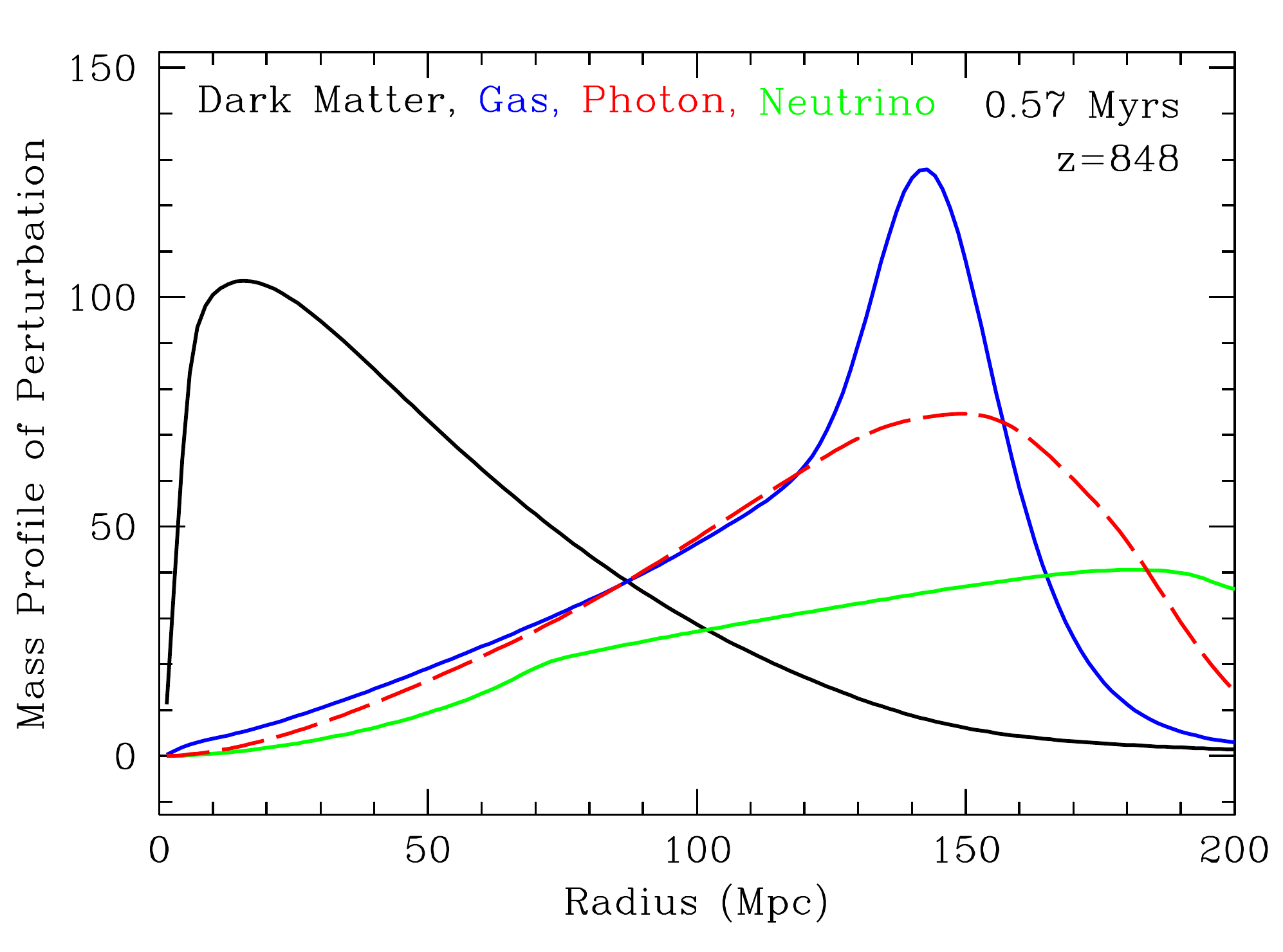}}
\subfloat{\includegraphics[width=.45\linewidth]{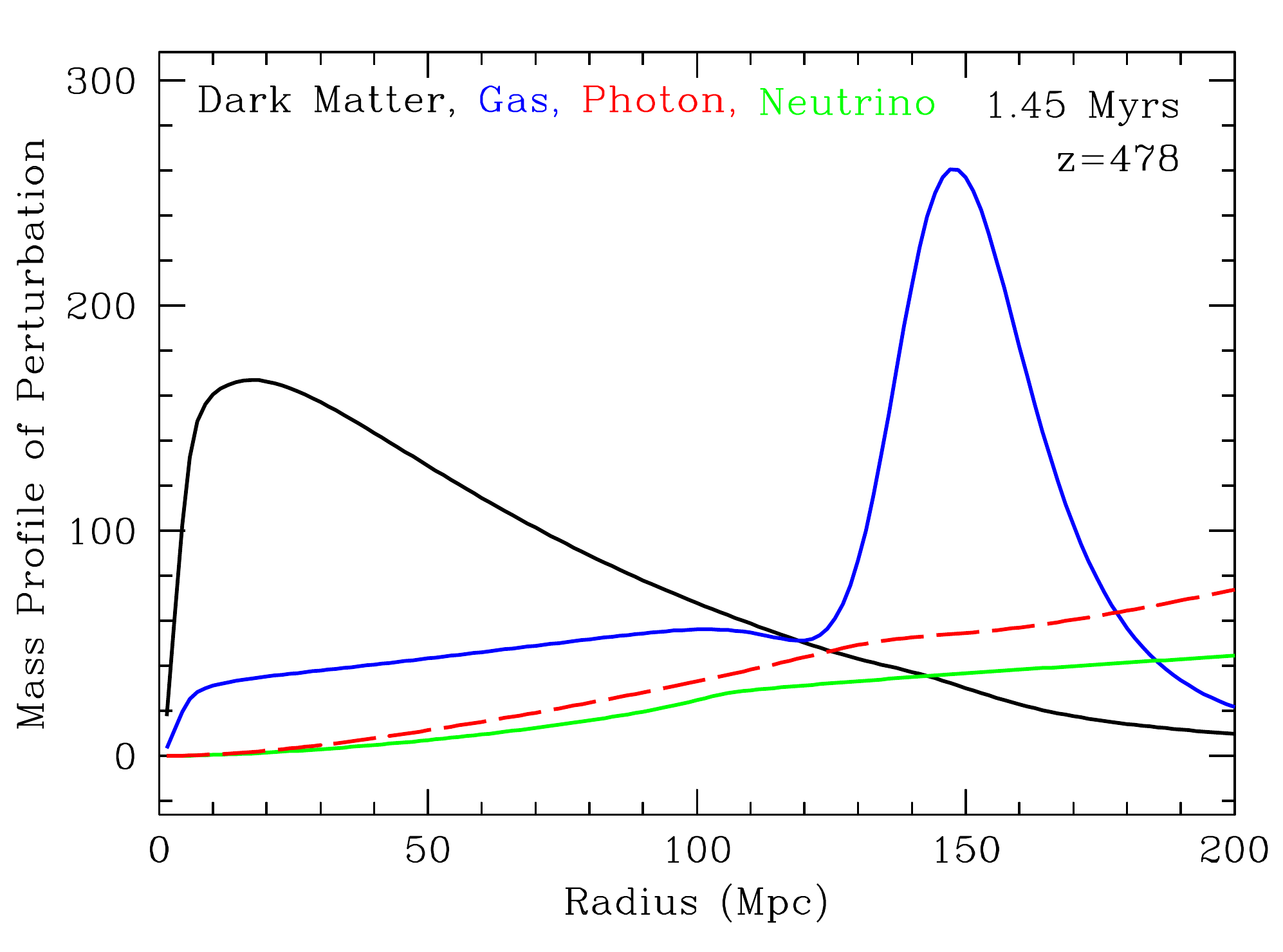}} \\
\subfloat{\includegraphics[width=.45\linewidth]{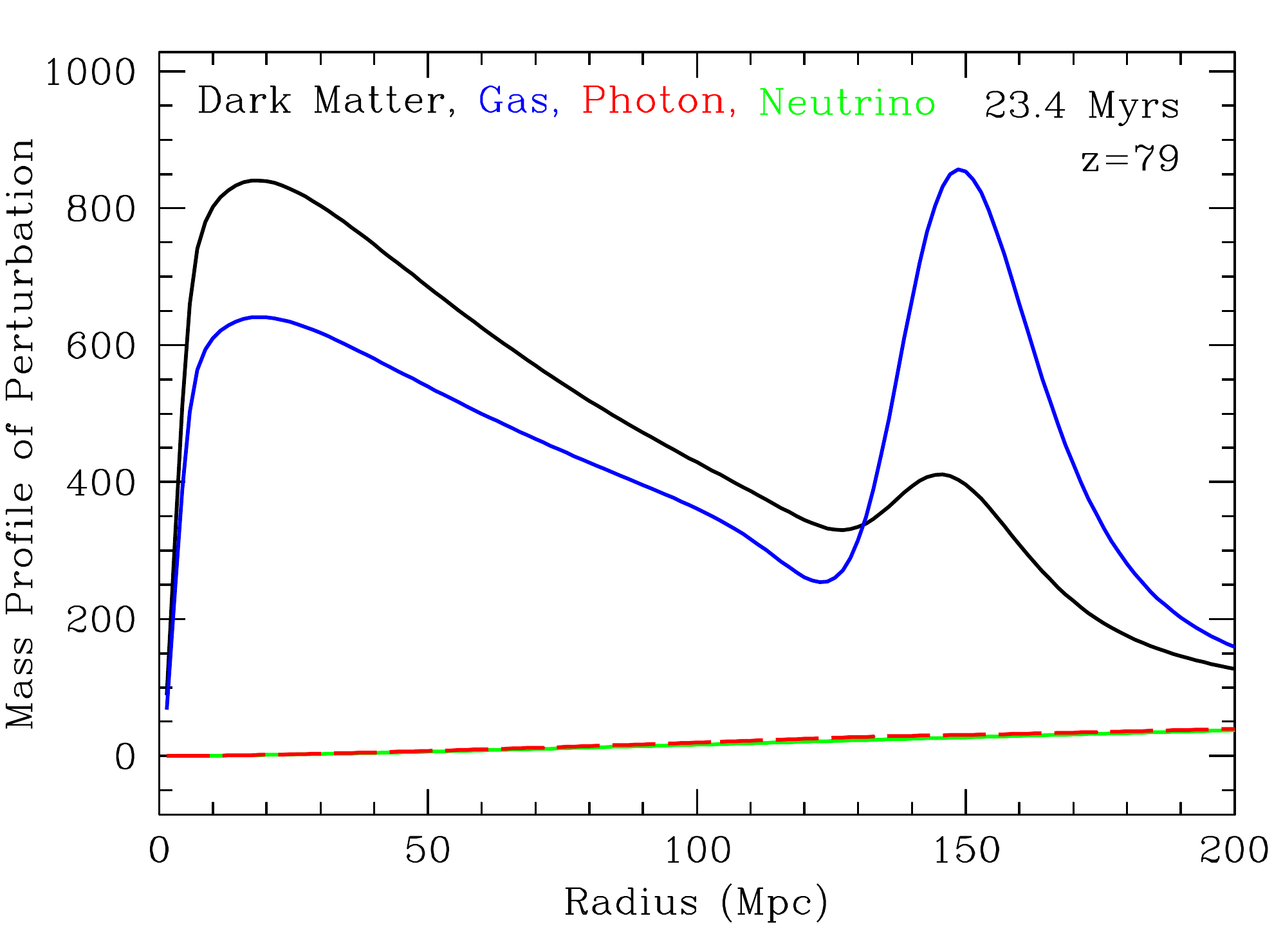}}
\subfloat{\includegraphics[width=.45\linewidth]{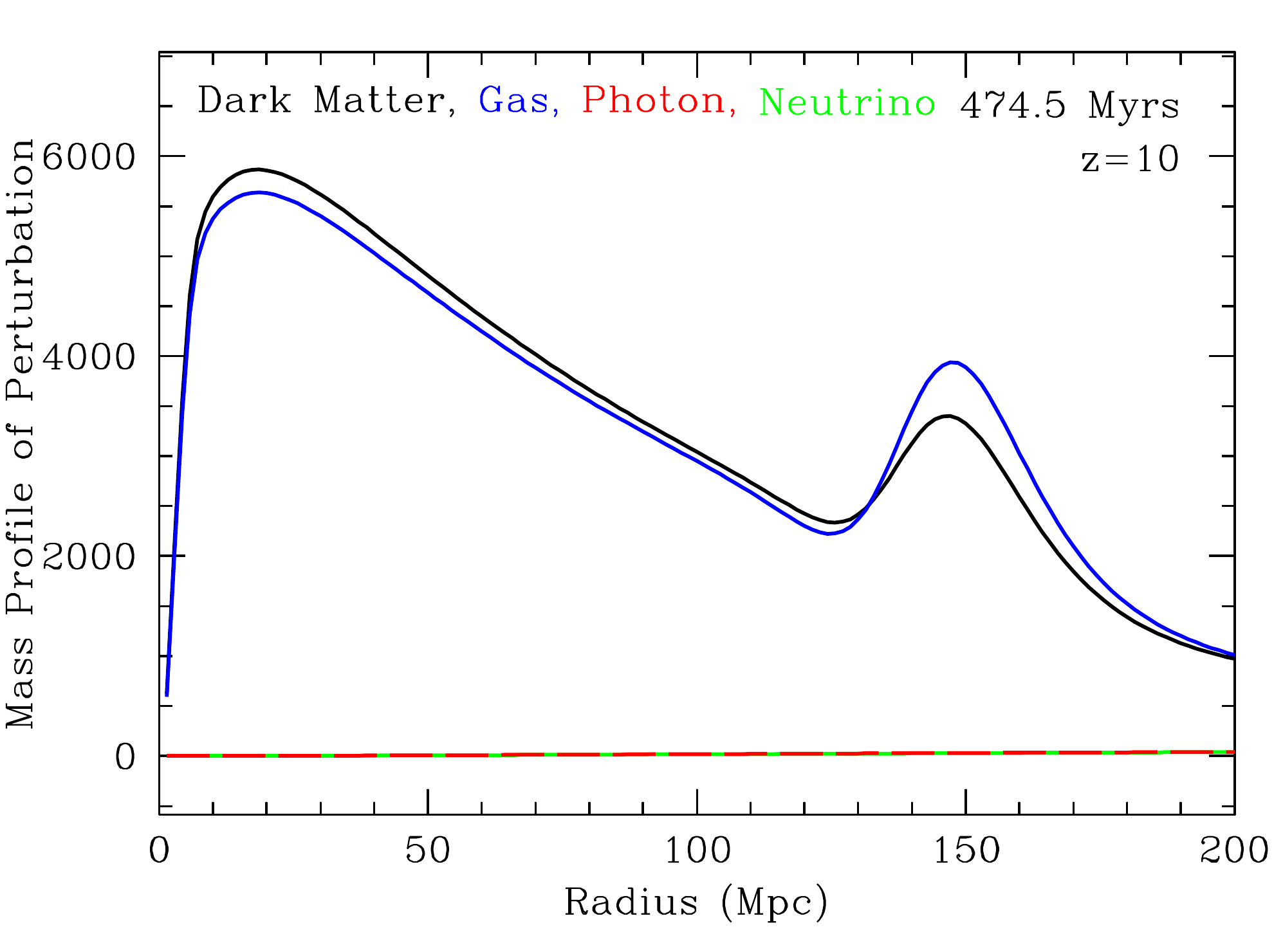}} \\
\caption[An illustration of baryon acoustic oscillations.]{The evolution of the comoving radial mass profile (Green's function; see \citejap{Slepian}) of initially point-like adiabatic perturbations.  Shown are the normalised fractional perturbations which satisfy $\delta_r = \delta_m$ initially.  At early times the photons and baryons are expelled outwards as a sound wave in the tightly coupled (due to Thomson scattering) plasma.  Prior to recombination, this causes a `wake' in the CDM due to the gravitational pull of the radiation -- in reality $\delta_r$ is (4/3)$\times$ larger (as the photon number density is perturbed, $n \propto T^3$, and $\delta_r \propto T^4$) and the force is sourced by the active mass density.  The photons are freed at recombination and $c_s$ falls rapidly.  The CDM is then located near the centre and there is a baryonic shell at a comoving radius of $R_0 c_s t(z^*) \simeq 150$\emph{Mpc}.  Without pressure, further growth is driven by gravity and new matter falls into the potential wells present.  The baryonic fraction of the perturbation is almost the cosmic mean, $\Omega_b = 0.0486$, at late times; this is removed by the choice of normalisation.  Reproduced from \cite{BAORobustness}.}
\label{fig:BAO}
\end{figure}

\subsection{Cosmic microwave background}
Some of the most important geometric tests in cosmology are provided by measurements of the angular diameter distances of the baryon acoustic oscillations (BAO) peak, in particular $D_A(z^*)$.  This may be inferred from the angular size of the imprint of the sound horizon at decoupling on the cosmic microwave background (CMB); this is shown in Fig. \ref{fig:planck_cmb}.  CMB observations, e.g. \cite{Planck}, currently provide many of the most precise cosmological constraints, as has been shown repeatedly throughout this introduction.  

The classical density perturbations created by inflation generates anisotropies in the CMB via a number of physical processes; with the perturbed radiation number density, $n_{\gamma}(T (\boldsymbol{ \hat \eta}))$, effectively still that of a blackbody.  Firstly, the simplest models of inflation predict adiabatic perturbations and therefore the radiation is perturbed from the outset, with $\delta_r = (4/3) \delta_m$.  Recombination occurs at a fixed temperature however, so rather than generating a temperature perturbation $\delta T$ from $\delta n_\gamma$ directly, recombination occurs later and therefore photons redshift less between $z^*(T)$ and $z=0$.  

Following decoupling, photons travel outwards from the potential well associated with the density perturbation, which introduces both a gravitational redshift and a time dilation -- time ticks slower in the well and hence the universe is younger and therefore hotter than in a volume at the background density.  The net effect results in an anisotropy of $\delta T/T = \Phi/(3c^{2})$ \citep{Sachs-Wolfe}.  Finally, there is an effect identical to that which leads to linear redshift-space distortions (see Chapter \ref{chap:RSD}): to fuel collapse the velocity field must converge on overdensities and hence there is a temperature difference induced by a Doppler shift coherent with the density field.  The magnitude of this effect is determined by the growth rate at recombination.

The temperature variance resulting from the combination of these effects is shown in Fig. \ref{fig:planck_tt}.  The position of the main peak, $\ell \simeq 220$, corresponds to the angular size of the sound horizon at last scattering; as this is of a known comoving size, once $\Omega_b$ is inferred from the amplitude of the remaining acoustic peaks, $D_A(z^*)$ may be inferred.  Amongst other things, this provides the constraints on $(w_0, w_a)$ shown earlier.    
\begin{figure}
\centering
\includegraphics[trim={0 .5cm 0 0}, width=0.95\textwidth]{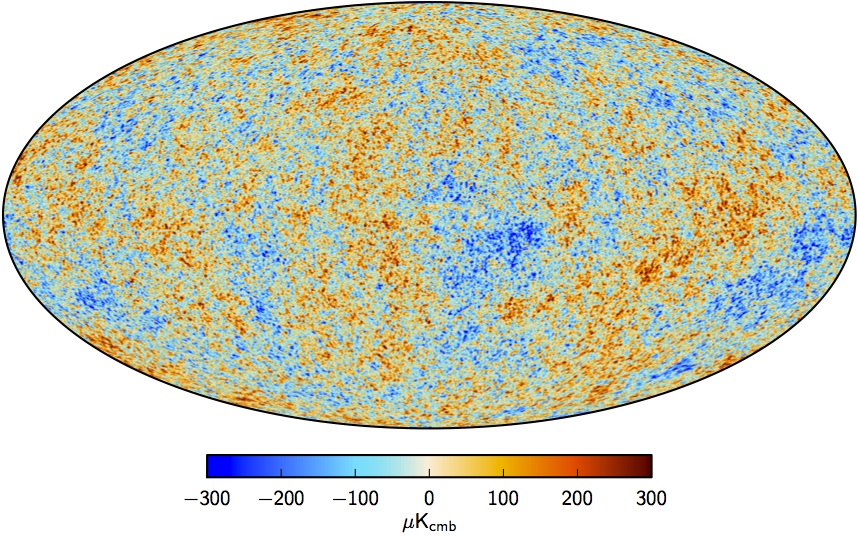}
\caption[The cosmic microwave background observed by Planck.]{A working assumption of cosmology is that the universe is isotropic on large scales, but on smaller scales CMB temperature anisotropies of magnitude $10^{-5}$ are observed.  In our current understanding, these temperature perturbations result from density perturbations seeded by small-scale quantum fluctuations that were made classical by an inflationary period.  The anisotropies are typically patches of one degree, which corresponds to the comoving size of the sound horizon at decoupling.}
\label{fig:planck_cmb}
%\end{figure}
%\begin{figure}
\centering
\includegraphics[trim={0 1.5cm 0 0}, width=0.95\textwidth]{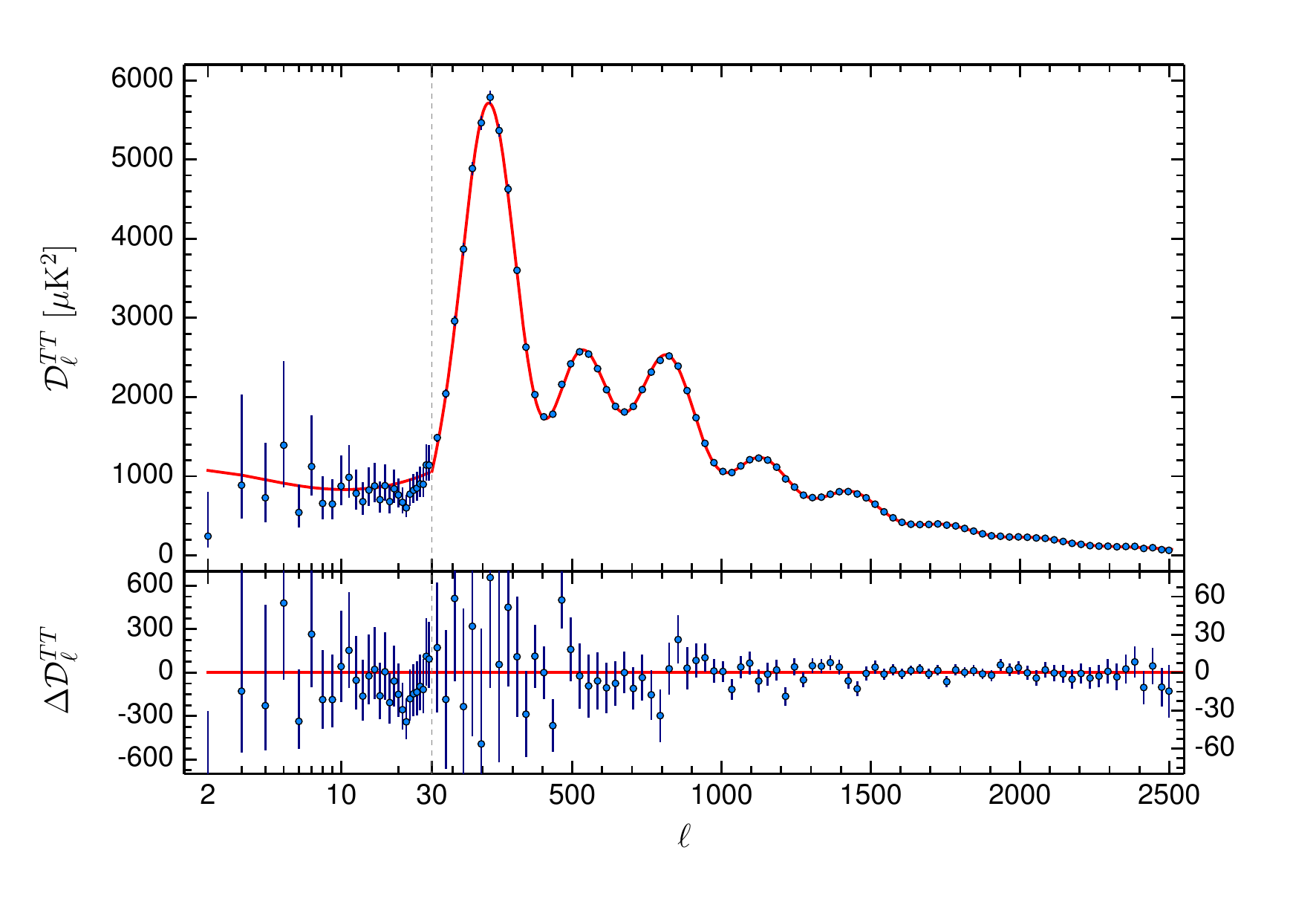}
\caption[Planck constraints on the angular scale-dependent temperature variance.]{The temperature variance, $\mathcal{D}_{\ell} \equiv d\langle (\delta T/T)^2 \rangle/d \ln \ell = \ell(\ell+1)C_{\ell}/(2\pi)$, as a function of angular scale, $\theta \simeq \ell^{-1}$.  The position of the main peak, $\ell \simeq 220$, corresponds to the angular size of the sound horizon at last scattering.  As this is a known comoving size (once $\Omega_b$ is inferred from the remaining acoustic peaks, this constrains $D_A(z^*)$.  Amongst other things, this provides the constraints on $(w_0, w_a)$ shown earlier.}
\label{fig:planck_tt}
\end{figure}

Given its central importance in modern cosmology, it is worth noting that had reionization occurred earlier the optical depth to last scattering would be much greater than unity and much of the small-scale CMB anisotropy would have been strongly suppressed.  It would therefore be impossible to determine the physics of the early universe and the cosmological parameters so successfully.

\section{Thesis outline}
This brief introduction has outlined the successes and remaining problems of contemporary cosmology.  In particular, those associated to the directly observed but unexplained current acceleration of the expansion \citep{WeinbergDE}.  A variety of dark energy and modified gravity models have been posited to explain this observation, with the strong equivalence principle being respected by the former but not the latter \citep{JoyceReview}, but these often make indistinguishable predictions for the expansion history \citep{LinderCahn}.  Further diagnostics are then a requirement for distinguishing between the possible models.  An opportunity is presented by measurements of the linear growth rate of density fluctuations, which is directly measured by redshift-space distortions analyses and predicted to discriminate between modified gravity models \citep{Guzzo, ModGravReview}.

The remainder of this thesis presents a practical algorithm for quantifying the anisotropy of the VIPERS v7 redshift-space power spectrum.  The scale and angular dependence of this anisotropy is sensitive to both the linear growth rate and the expansion history; this allows for simultaneous constraints to be placed on the growth and expansion history, which is a stringent test of modified gravity \citep{Ruiz, Linder2016}.  Consistency tests of the observed anisotropy with the fiducial cosmology assumed are presented in Chapter \ref{chap:VIPERS_RSD}.  To do so requires a number of observational systematics to be modelled or corrected for, including the survey geometry and selection.  These are discussed in detail in Chapter \ref{chap:VIPERS} and Chapter \ref{chap:maskedRSD} respectively.

This work builds upon the VIPERS PDR-1 analysis of \cite{sylvainClustering} by both analysing the larger v7 release and performing a Fourier space analysis; as the power spectrum has been shown to deliver most precise constraints than the correlation function \citep{Alam, Beutler_2016A}.  Chapter \ref{chap:Clipping} improves upon this conventional approach by including a simple local overdensity transform: `clipping' \citep{Fergus} prior to the RSD analysis.  This tackles the root cause of non-linearity and potentially extends the validity of perturbation theory.  Moreover, this marked clustering statistic have been shown to potentially amplify signatures of modified gravity \citep{LombriserClipping, White2016}.  As a higher order statistic, it also includes information that is not available with the power spectrum.  
\end{chapter}
%\newpage
%\begin{figure}
%\centering
%\includegraphics[scale=0.3]{CosmicFlows.eps}
%\caption[The cosmic flows velocity field]{Cosmic flows: the measured velocity field in our local neighbourhood.  The position of galaxies are shown as white spheres, streamlines trace the inferred velocity field.  The infall of galaxies onto the Great Attractor introduces a Doppler effect, redshifting the spectra for those on the near side and blueshifting those on the far.  This serves to compress the density profile along the line-of-sight, when radial positions are inferred from observed redshifts.  The predominant effect of which is to amplify the clustering in an anisotropic manner.  Reproduced from \citep{CosmicFlows}.}
%\label{fig:CosmicFlows}
%\end{figure}
%\clearpage
\begin{chapter}{Redshift-space distortions}
\label{chap:RSD}
\section{Linear theory in the distant observer approximation}
\begin{figure}
\centering
\includegraphics[scale=0.3]{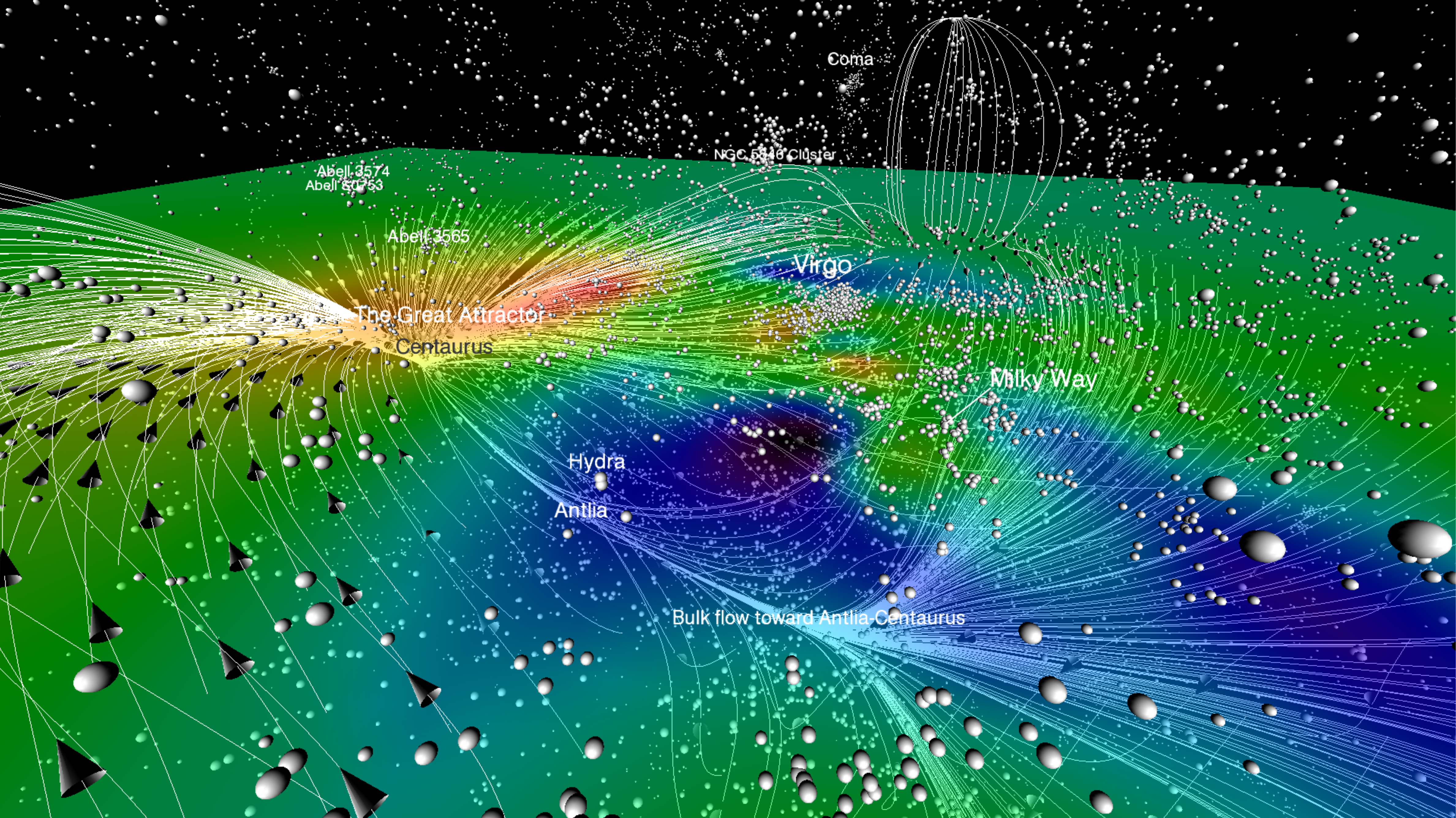}
\caption[The Cosmic Flows velocity field]{Cosmic flows: the inferred velocity field in our local neighbourhood.  Shown are the alaxy positions (white spheres) together with streamlines that trace the inferred velocity field.  The infall onto the Great Attractor introduces an added Doppler effect, which redshifts the spectra for galaxies on the near side and blueshifts those on the far side.  This compresses the density profile along $\boldsymbol{\hat \eta}$ when radial positions are inferred from observed redshifts according to the FRW framework.  This amplifies the large-scale clustering in an anisotropic manner.  Reproduced from \cite{CosmicFlows}.}
\label{fig:CosmicFlows}
\end{figure}
The observed redshift would be an unbiased estimate of the radial comoving position in a perfectly FRW universe.  But large-scale structure is undergoing gravitational collapse, which generates velocities deviating from the global Hubble expansion. The radial component of this peculiar velocity superimposes a Doppler shift on top of the cosmological redshift:
\[
(1+ z_{\rm{obs}}) \simeq (1 + z_{\rm{cos}}) \left (1 + \frac{v_{\rm{pec}}}{c} \right ) + \mathcal{O} \left( \frac{v_{\rm{pec}}^2}{c^2} \right ).
\]
Therefore the density field is distorted in the line-of-sight direction, $\boldsymbol{\hat{\eta}}$, when inferred from observed redshifts.  This effect is illustrated in Fig. \ref{fig:CosmicFlows}.  The starkest characteristic of the redshift-space density field are the `fingers-of-God' \citep[FOG,][]{Jackson}, which are a consequence of the line-of-sight stretching of galaxy groups and clusters due to the virialised motion in deep potential wells.  Although very useful as a means to measure the typical rms velocity in clusters \citep{DavisPeebles}, this greatly complicates the use of redshift-space distortions (RSD) as a test of modified gravity on cosmological scales.
% redshift-space as adjective. 
% redshift space as noun. 
  
Within linear theory dynamics each Fourier mode, $\tilde \delta(\mathbf{k})$, evolves independently.  Consider the observed overdensity for such a mode when observed in redshift space.  The planar symmetry requires the associated peculiar velocity field to be directed along $\mathbf{k}$.  Given the key conclusions of linear theory dynamics:
 \begin{eqnarray}
 \nabla \cdot \mathbf{u} &=& - \dot \delta, \nonumber \\ 
 \delta_+(\mathbf{x}, a) &=& D_+(a) \ \delta(\mathbf{x}, 1),
 \end{eqnarray}
it follows that in Fourier space:  
\[
\tilde{\mathbf{u}}(\mathbf{k}) = ifH \frac{\mathbf{k}}{k^2} \tilde{ \ \delta}(\mathbf{k}).
\]
 Here the logarithmic growth rate is denoted by  
 \[
 f = \frac{d \ln D_+}{d \ln a}
 \]
and $\mathbf{u}$ is the comoving peculiar velocity.  The distance-redshift relation is $R_0 dr  = \left ( c/H \right) dz$ and by further defining the line-of-sight peculiar velocity in units of distance, 
\[
\bar U= \left( \frac{\mathbf{u}(\mathbf{r}) - \mathbf{u}(\mathbf{0})}{H(z)} \right) \cdot \mathbf{\hat{r}},
\]
the mapping from real space to redshift space is found to be
\[
\mathbf{s} = \mathbf{r} \left(
1 + \frac{ \bar U}{r} \right).
\]
The affect of this mapping on the volume element, $dV=r^2 dr d \Omega$, is
\[
d^3 s = d^3 r \left(
1 + \frac{\bar U}{r} \right)^2 \left( 1 + \frac{d \bar U}{dr}\right).
\]
The first factor is a consequence of the change in area at fixed solid angle and the second is due to the radial compression, $ds/dr$.  I assume a volume limited sample for simplicity and therefore terms including derivatives of the selection function vanish.  The number of galaxies is conserved: $\bar n(1 + \delta_s) d^3 s = \bar n (1 + \delta_r) d^3 r$, which gives
\[
\delta_s = \delta_r -2 \left( \frac{\bar U}{r}\right) -\frac{d \bar U}{dr},
\]
to linear order in the perturbations $(\delta_r,\bar U, d \bar U/dr)$.  If $\delta$ is a plane wave with amplitude $\Delta$, it follows from above that $\bar U \simeq (f/k) \Delta$ and $d \bar U / dr \simeq k \bar U \simeq f \Delta $. Thus, providing $r_{\rm{max}} \gg 1/k$, the second term can be safely neglected. For a given angle to the line-of-sight, the cosine of which is $\mu = \mathbf{\hat k} \cdot \mathbf{\hat {r}}$, the velocity field corresponding to a plane wave disturbance, $\delta_r = \tilde \delta_{\mathbf k} \cos(\mathbf{k} \cdot \mathbf{r})$, is
\[
\bar U = - \frac{f \mu}{k} \delta_{\mathbf k} \sin(\mathbf{k} \cdot \mathbf{r}).
\]
Hence $d \bar U/dr = - \mu^2 f \delta_r$ such that $\delta_s = \delta_r (1 + f\mu^2)$.  Therefore the clustering is simply amplified in an anisotropic manner.  

Thus far the derivation has assumed that the mass can be observed directly, but this is not the case for the dominant CDM component.  We observe light emitting tracers instead, which are biased towards the deep potential wells that allow baryons to cool efficiently \citep{WhiteRees}.  On sufficiently large scales linear bias is valid, $\delta_g = b \delta_m$ \citep{KaiserClusters}, which amplifies the intrinsic clustering but not that induced dynamically. Consequently, $\delta_s = \delta_m(b + f \mu^2)$ and the Kaiser model of RSD \citep{Kaiser} for linearly biased tracers follows as a result: 
\[
P_s(\mathbf{k}) = (1 + \beta \mu^2)^2 P_g(\mathbf{k}),
\]
for $\beta = f/b$.  If a reliable estimate of $P_g(k)$ can be made -- by measuring the RSD-free angular clustering for instance, then the Kaiser amplification is obtained simply by measuring the spherically averaged redshift-space power spectrum.  Alternatively $f$ may be deduced from the anisotropy; the most direct means to achieve this is with the quadrupole-to-monopole ratio \citep{ColeFourierOmega}.  When the polar axis is taken to be $\boldsymbol{\hat{\eta}}$, the azimuthal symmetry and $\mu^2$ dependence of the redshift-space power spectrum allows for a Legendre expansion of $P_s(k)$:
\[
\frac{P_s(k)}{P_g(k)} = \left( 1 + \frac{2}{3} \beta + \frac{1}{5} \beta^2 \right ) + \left (\frac{4}{3} \beta + \frac{4}{7} \beta^2 \right) L_2(\mu) + \frac{8}{35} \beta^2 L_4(\mu). 
\]
Here $L_\ell$ is a Legendre polynomial of order $\ell$; these form a complete basis for $-1 \leq \mu \leq 1$.  The non-zero moments in the Kaiser model are given by the monopole, quadrupole and hexadecapole; these are $L_0=1$, $L_2 = \left ( 3 \mu^2 -1 \right )/2$ and $L_4=(35\mu^4 -30 \mu^2 +3)/8$ respectively.  The quadrupole-to-monopole ratio is therefore: 
\[
\frac{P_2(k)}{P_0(k)} = \frac{\frac{4}{3} \beta + \frac{4}{7} \beta^2}{1 + \frac{2}{3} \beta + \frac{1}{5} \beta^2}.
\]
The $\beta$ value to be inferred for a given quadrupole-to-monopole ratio is shown in Fig. \ref{fig:monoquadratio_beta}.
\begin{figure}
\centering
\includegraphics[width=\textwidth]{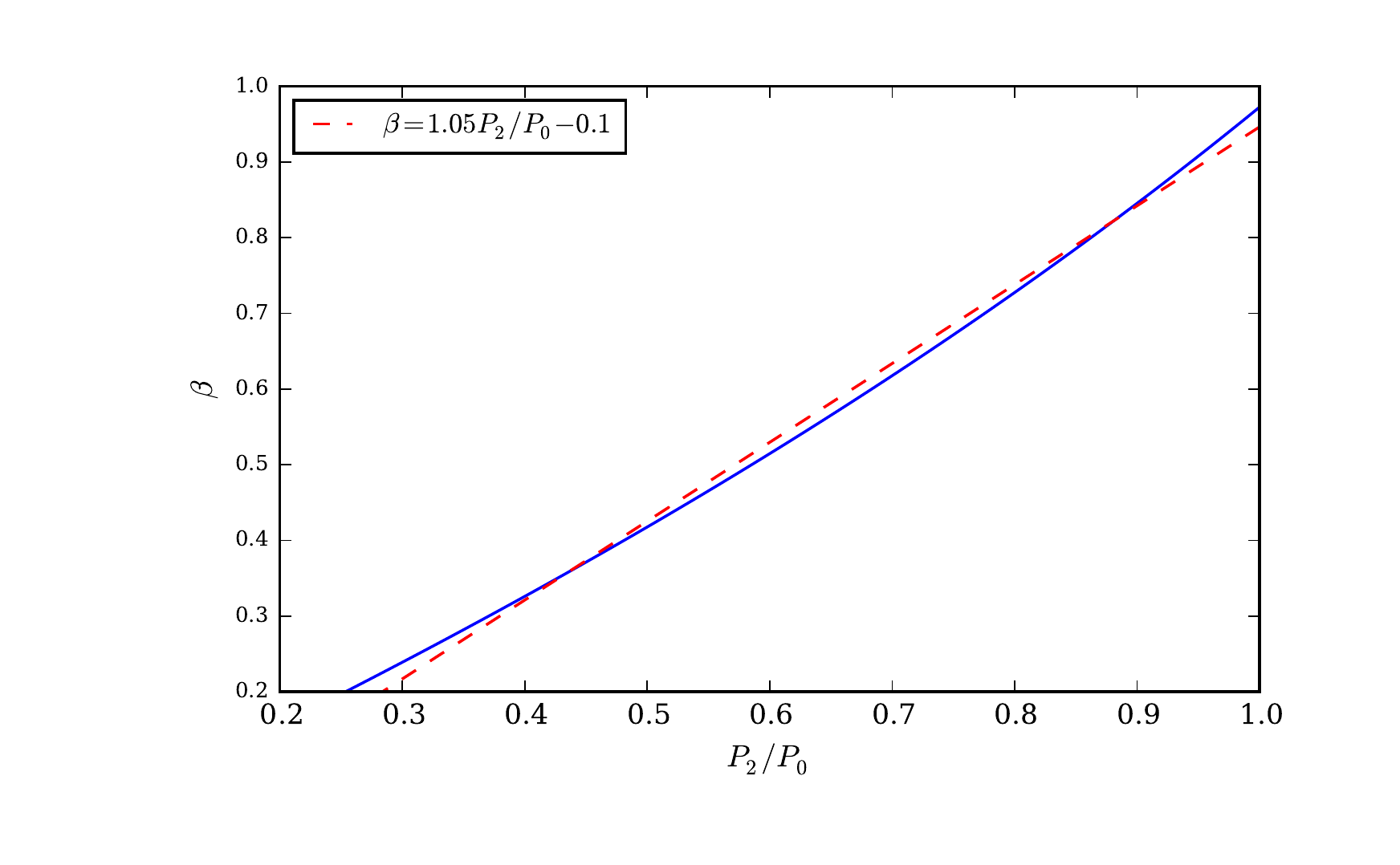}
\caption[Estimating $\beta$ from the quadrupole-to-monopole ratio.]{The most direct means of obtaining $\beta$ from the redshift-space power spectrum (in the absence of survey mask systematics) is the quadrupole-to-monopole ratio.  This figure shows the $\beta$ value to be inferred for a given quadrupole-to-monopole ratio (blue).  Over the range of interest $\beta$ is close to linear in $P_2/P_0$, with $\beta = 1.05P_2/P_0 -0.1$ providing a reasonable fit; this is shown in red.}
\label{fig:monoquadratio_beta}
\end{figure}
\section{Growth rate tests of modified gravity and dark energy models}
There are perhaps three significant limitations to the General Theory of Relativity \citep[GR,][]{EinsteinGR}, which has been a very successful theory of gravity to date.  Firstly, GR predicts a gravitational influence from all forms of energy, including that of the vacuum.  If Quantum Field Theory (QFT) is valid up to the Planck scale then a naive estimate of the zero-point energy contribution of the vacuum energy is $\expval{\hat \rho_{\rm{vac}}}{0} \simeq M_{\rm{Pl}}^4 \simeq 10^{76} \text{(GeV)}^4$ \citep{CarrollBook}; but this is not a Lorentz invariant cutoff and is therefore invalid.  Requiring Lorentz invariance negates the zero-point energy of massless particles and yields $\expval{\hat \rho_{\rm{vac}}}{0} \simeq 10^{9} \text{GeV}^4$, when the most massive elementary particle is assumed to be the top quark \citep{KoksmaProkopec}.  In stark contrast, observations of the Cosmic Microwave Background measure $\simeq 10^{-47} (\text{GeV})^4$ when GR is assumed. 

A symmetry principle that forbids the vacuum energy from contributing to the energy-momentum tensor is often invoked to remedy this situation.  Despite the fact that even the next order term -- the gravitational self-energy of the vacuum fluctuations, is also $\simeq 10 \times$ larger than the measurement \citep{WeinbergCosmoCnstProb}.  If the acceleration is not driven by the vacuum energy, another source of (perhaps dynamical) dark energy is required to provide the small, yet non-zero, effective vacuum density first inferred by \cite{EfstathiouDarkEnergy}, \cite{OstrikerSteinhardt} and directly determined by \cite{RiessNobel} and \cite{PerlmutterNobel}.  The search for departures of the expansion history from that of $w=-1$ is critical in this case.  A further question must also be addressed: why is it today that the expansion begins to accelerate?  This property typically requires excessive fine-tuning in dynamical models.  Alternatively, it may be the Friedmann equation that is at fault.  This motivates the addition of further terms to the Einstein-Hilbert action, which gives rise to new field equations and therefore modified (metric) theories of gravity.  

Secondly, GR is a classical theory that cannot be quantised in the manner that has been so successfully applied to electromagnetism.  This is tied to the cosmological constant problem as the Planck mass assumed for the cutoff above is the energy scale at which quantum effects are expected.  Lastly, the reconciliation of the observed light distribution with the gravitational dynamics in galaxies and galaxy clusters requires the introduction of a dark matter component (\citejap{Zwicky}, \citejap{Rubin}); gravitational lensing provides further strong evidence for this addition.  But the validity of GR will be in question as long as there remains no direct detection of a suitable dark matter candidate.  The consideration of modified theories which address any or perhaps all of these concerns is therefore strongly justified.  These modifications are already subject to stringent constraints from solar system tests and therefore typically include `screening' mechanisms -- reverting to GR in the large (or rapidly changing) curvature limit \citep{ModGravReview}.  

I measure the signature of RSD on the VIPERS power spectrum in Chapter \ref{chap:VIPERS_RSD} and test the consistency of this measurement with the GR prediction. The added Doppler shift from the peculiar velocity distorts the observed clustering along $\boldsymbol{\hat{\eta}}$ when $z$ is used as a proxy for comoving distance.  This picks out a preferred direction (in the distant observer approximation) that breaks the statistical isotropy.  As the peculiar velocities are a consequence of gravitational dynamics -- infall onto clusters and outflow from voids, they are sensitive to the strength and therefore theory of gravity on cosmological scales \citep{Guzzo}.  More specifically, as the amplitude of the power spectrum $(b \sigma_8)^2$ may also be measured, RSD place an independent constraint on $f \sigma_8 = \beta \times b \sigma_8$.  Here $\sigma_8^2$ is the variance of the density field after smoothing on a scale of $8 \mpcoh$.  

In Peebles's parametrisation, $f$ is given by \citep{Wang, LinderCahn}: 
\[
f(a) = \Omega_m(a)^{\gamma},
\]
\begin{figure}
\centering
\includegraphics[width=\textwidth]{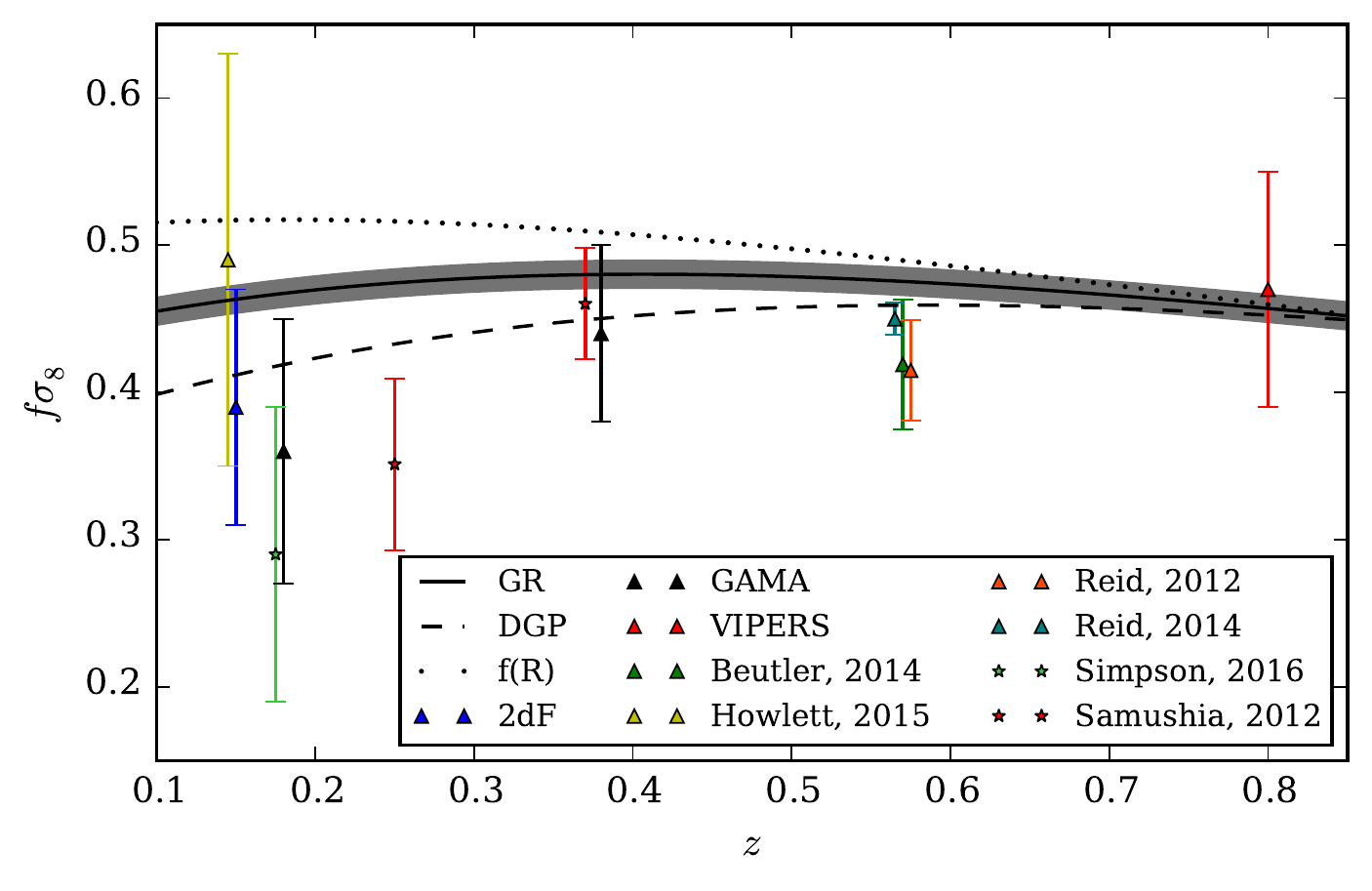}
\caption[Model predictions for $f \sigma_8(z)$ and recent constraints]{Predictions for $f \sigma_8(z)$ in a selection of modified gravity models: GR, $f(R)$ and DGP; these span $0.4 < \gamma < 0.7$ and are shown for a \cite{Planck} fiducial cosmology.  Recent measurements are also shown; here VIPERS denotes the analysis of \cite{sylvainClustering} as opposed to the v7 analysis presented in Chapter \ref{Chap:VIPERS_RSD}.  With the greater precision of future surveys, e.g. eBOSS, DESI, Euclid and WFIRST, it will be possible to either confirm or refute modified gravity models based on the observed $f \sigma_8 (z)$. A rough error on the GR prediction, $d(f \sigma_8) \sim 0.02$, is overplotted; this has been estimated from Fig. 18 of \cite{FastSound}.  This error demonstrates that the expansion will need to be more precisely determined in order to constrain modified gravity models at $z \simeq 0.8$ -- where greater volume is available.}
\label{fig:fsig8_z}
\end{figure}
\noindent
where the exponent $\gamma$ is determined by the gravitational theory.  The predictions of GR, the Dvali-Gabadaze-Poratti model \citep[DGP,][]{DGP} and the \cite{HuSawicki} $f(R)$ model are $0.545$, $0.68$ and $\simeq 0.4$ respectively.  Of course there are many such models, these simply serve as an illustrative selection to which the measurement may be compared.  Together with independent constraints on the background cosmology, e.g. \cite{Planck}, gravitational physics on cosmological scales may be stringently tested with RSD (\citejap{Guzzo}, \citejap{John}, \citejap{sylvainClustering}, \citejap{Beutler}).  The expansion history in many modified theories is indistinguishable from that in GR and therefore these theories cannot be separated by geometric measurements -- the $D_L(z)$ relation of type-Ia supernovae and baryon acoustic oscillations for example.  But they do make distinct predictions for the linear growth rate. As a measure of the differential growth rate, $d\sigma_8/d\ln a$, RSD complement the constraints on the cumulative growth, $\sigma_8(z)$, that may be obtained from tomographic weak gravitational lensing for example.  GR predicts that $f(z)$ is fully determined by the background expansion; this may be seen from the linear growth equation:
\[
\ddot \delta + 2H \dot \delta = 4 \pi G \rho_m \delta,
\]
in which the linear evolution is determined only by measures of the expansion, $H$ and $\rho_m \propto a^{-3}$ in the matter dominated era.  In contrast, many modified gravity theories replace Newton's gravitational constant \citep{Principia} with an effective time and perhaps scale-dependent term, $G_{\rm{eff}}(k, t)$.

\section{Beyond the linear velocity field}
The anisotropic amplification of the power spectrum is a consequence of the large-scale convergence (divergence) of the velocity field at peaks (voids) of the density field.  But the virialised motions of galaxies strongly suppress the small-scale non-linear power.  This effect is commonly modelled by neglecting the coherence of the rms virial velocity with the depth of the potential well and assuming galaxies have simply been scattered along $\boldsymbol{\hat \eta}$ by a characteristic pairwise velocity dispersion, $\sigma_p$, which is commonly quoted in distance units.  For small pair separations, the distribution of pairwise velocities found in N-body simulations is well modelled by a scale-independent exponential distribution \citep{Peebles1976}, which is to be expected on simply physical grounds \citep{Sheth_exp}.  An appropriate model for $P_s(k, \mu)$ is therefore the `dispersion model' with a Lorentzian damping \citep{PeacockDodds}: 
\[
P_s(k, \mu) = \frac{(1 + \beta \mu^2)^2}{1 + \frac{1}{2}k^2 \mu^2 \sigma_p^2} P_g(k).
\]
\begin{figure}
\centering
\includegraphics[width=\textwidth]{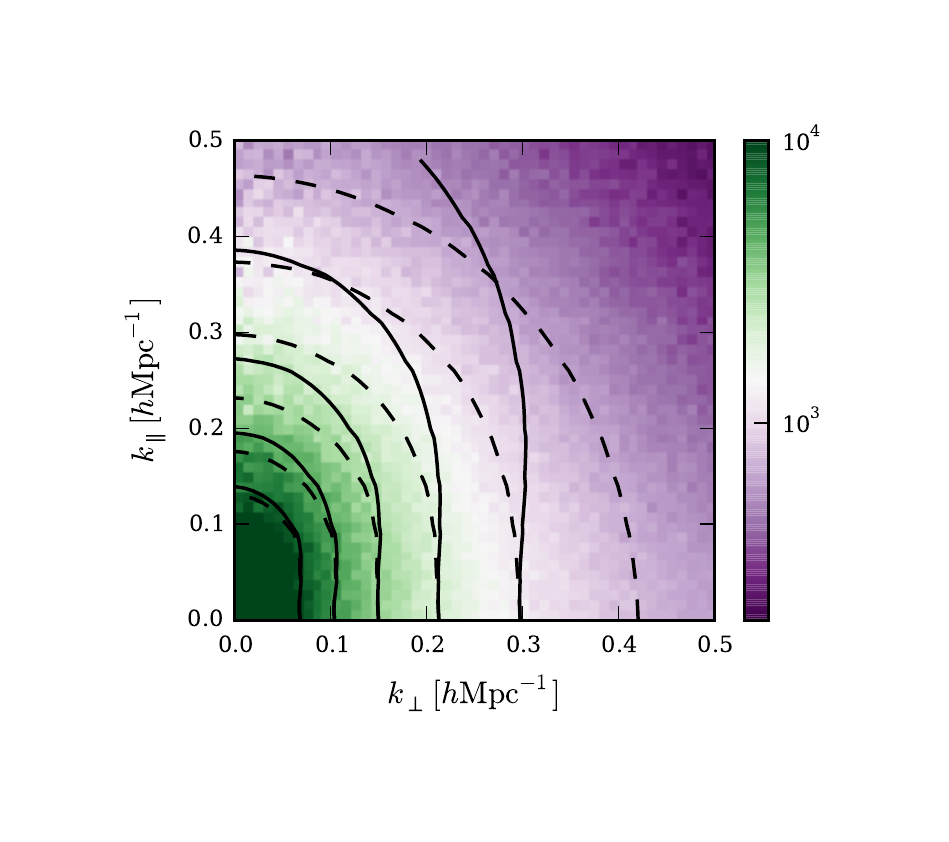}
\caption[An estimate of the 2D redshift-space power spectrum.]{The estimated 2D redshift-space power spectrum for a $(1 \ h^{-1} \emph{Gpc})^3$ cube populated with $M_B=-20.0$ galaxies according to a Halo Occupation Distribution prescription.  Clearly apparent is the Kaiser amplification of the clustering along $\boldsymbol{ \hat{\eta}}$.  Contours for the Kaiser (solid) and Kaiser-Lorentzian model (dashed) are overplotted; the Lorentzian damping washes out the amplification, which results in more circular contours on small scales.  These models have been evaluated for the maximum likelihood parameters, $(\beta=0.542, \sigma_p=3.20 h^{-1} \emph{Mpc})$, which were obtained by fitting $(P_0, P_2)$ to $k_{\rm{max}} = 0.3 h^{-1} \emph{Mpc}$.}
\label{fig:cube_2dpk}
\end{figure}
This choice differs from the other common assumption of a Gaussian damping only at $\mathcal O(k^4 \sigma_p^4)$.  In addition to the non-linear damping term, it is common practice to also include a non-linear model for $P_g$ as RSD have no affect on modes transverse to the line-of-sight.  

In the distant observer approximation, a single quadrant of an azimuthal slice of the power spectrum contains all of the independent information.  This independent quadrant is fully specified by the power spectrum multipole moments:
\[
P_{\ell}(k) = \frac{(2 \ell+1)}{2} \int_{-1}^{1} d \mu \ P(k, \mu) \ L_{\ell}(\mu), 
\]
which are non-zero for even $\ell$ only.  But, in the era of precision cosmology, these symmetries are broken to some degree by both gravitational lensing \citep{Bonvin} and gravitational redshift \citep{Wojtak}.  The correlation function is the Fourier transform of the power spectrum, which, when the power spectrum possesses the symmetries of the RSD, yields a Hankel transform relation between the two:
\begin{align}
\label{Hankelpair}
\xi_\ell(s) &= \frac{i^\ell}{2 \pi^2} \int_0^\infty k^2 dk P_\ell(k) \ j_\ell(ks), \\
P_{\ell}(k) &= 4 \pi (-i)^\ell \int_0^{\infty} \Delta^2 d\Delta \ \xi_{\ell}(\Delta) \, j_{\ell}(k\Delta).
\end{align}
Here $j_\ell(ks)$ is a spherical Bessel function (of the first kind).  The lowest order multipole moments in the Kaiser-Lorentzian model are
\begin{align}
\label{eqn:KL_multipoles}
\frac{P_0(k)}{P_g(k)} &= M_0(\kappa) + 2 \beta M_2(\kappa) + \beta^2 M_4(\kappa), \nonumber \\ 
\frac{P_2(k)}{P_g(k)} &= \frac{5}{2} \left ( -M_0 + (3 - 2 \beta) M_2  + (-\beta^2 + 6\beta) M_4 + 3 \beta^2 M_6 \right), \nonumber \\
\frac{P_4(k)}{P_g(k)} &= \frac{9}{8} \left( 35 \beta^2 M_8 + 10 \beta (7 -3 \beta) M_6 + (35 - 60 \beta + 3 \beta^2 ) M_4 + 6(\beta -5 )M_2 + 3M_0  \right),
\end{align}
for $\kappa = k \sigma_p$ and
\[
M_n \equiv \int_0^1 \frac{\mu^n d \mu}{1 + \frac{k^2 \sigma_p^2 \mu^2}{2}}.
\]
The lowest order integrals are given by 
\begin{align}
&M_0(\kappa) = \frac{\sqrt{2}}{\kappa} \arctan(\kappa /  \sqrt{2}),&  \nonumber \\
&M_2(\kappa) = \frac{2}{\kappa^3} \left( \kappa - \sqrt{2} \arctan(\kappa /  \sqrt{2}) \right),& \nonumber \\
&M_4(\kappa) = \frac{2}{\kappa^5} \left( -2 \kappa + \frac{\kappa^3}{3} + 2\sqrt{2} \arctan(\kappa /  \sqrt{2}) \right),& \nonumber \\
&M_6(\kappa) = \frac{2}{\kappa^7} \left( 4 \kappa - \frac{2}{3} \kappa^3 + \frac{\kappa^5}{5} -4 \sqrt{2} \arctan(\kappa /  \sqrt{2}) \right),&  \nonumber \\ 
&M_8(\kappa) = \frac{2}{\kappa^8} \left ( -8 + \frac{4}{3 \kappa^2} - \frac{2}{5} \kappa^4 + \frac{\kappa^6}{7} + \frac{8 \sqrt{2}}{\kappa} \arctan(\kappa/ \sqrt{2}) \right);&
\raisetag{-.25em}
\end{align}
these higher order expressions are original to my knowledge.

As the motivation of clipping -- described in Chapter \ref{chap:Clipping}, is to extend the scale to which linear theory is valid I do not consider more developed models (\citejap{Taruya}, \citejap{Okumura} and references therein), but restrict the analysis to a relatively simple approach.  However, the dispersion model neglects various physical properties of RSD.  Questions can be raised of both the Kaiser factor and the applied damping, while the ratio represents an unphysical model for the line-of-sight pairwise velocity pdf, $F(\nu)$ (\citejap{Scoccimarro}, S04).  Moreover, it was shown in S04 that the redshift-space power spectrum can always be written as a streaming model \citep{PeeblesBook}, if the distant observer approximation is assumed.  This is somewhat surprising for the non-linear power spectrum as early derivations of the streaming model relied on a Jacobian, $d^3s/d^3r$, which cannot be appealed to when multi-streaming develops. The Kaiser model corresponds to a streaming model of
\[
1 + \xi_s(s_\perp, s_\parallel)  = \int_{-\infty}^{\infty} dr_\parallel \ \left[1 + \xi(r) \right ] \ F_V(s_\parallel - r_\parallel - \frac{r_\parallel}{r} v_{12}(r)),
\]
%\thisfloatpagestyle{empty}
\begin{figure}
%\thisfloatpagestyle{empty}
%\begin{figure}
\centering
\includegraphics[width=0.85\textwidth]{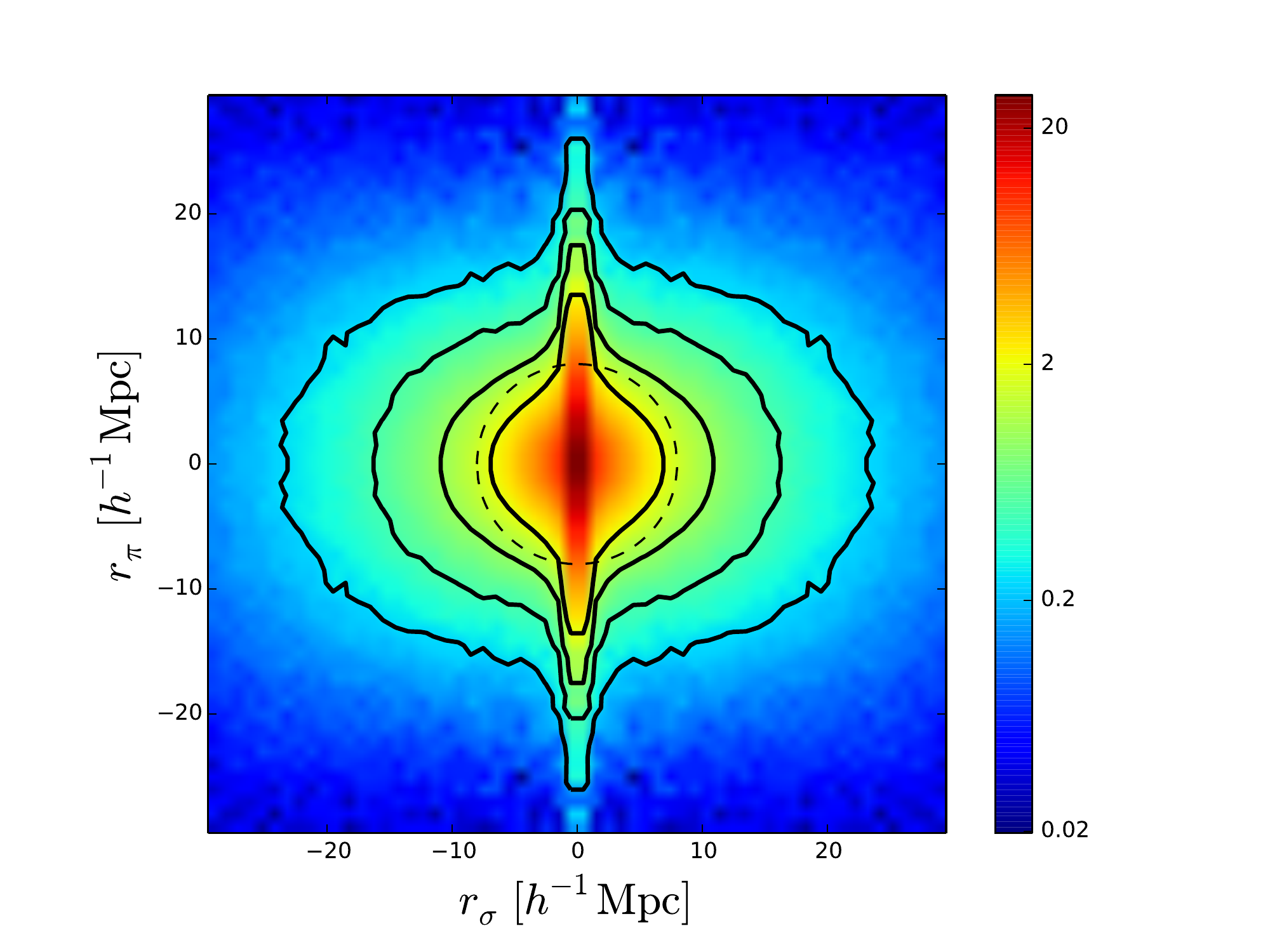}
\caption[The SDSS-III measurement of the 2D $\xi(\mathbf{r})$.]{The CMASS 2D correlation function, $\xi_s(s_\perp, s_\parallel)$.  The large-scale RSD effect is a flattening of the contours for $\mu \simeq 1$ in configuration space; this is in contrast to the stretching in Fourier space.  Galaxies are effectively scattered along \los on small scales, which produces the pronounced streak for $r_\sigma<2 h^{-1}$\emph{Mpc}.  Reproduced from \cite{Reid2.5}.}
\label{fig:Reid_xi}
%\end{figure}
\centering
\includegraphics[width=0.6\textwidth]{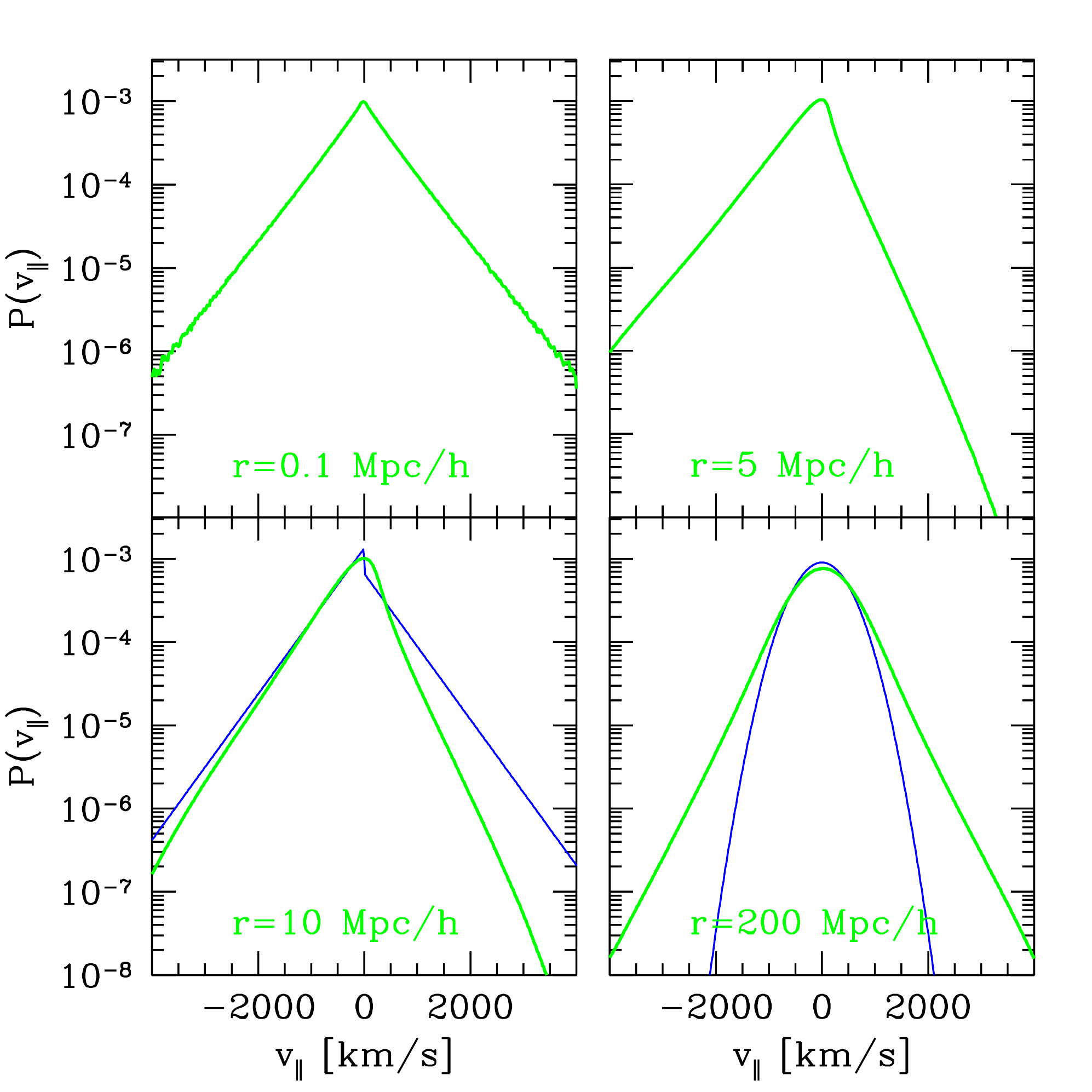}
\caption[The line-of-sight pairwise velocity pdf.]{The line-of-sight pairwise velocity pdf for a N-body simulation (green) and the dispersion model (blue).  The linear prediction of a Gaussian pdf is incorrect, even beyond the BAO scale -- the density and velocity fields are Gaussian but the RSD mapping generates the prominent exponential tails present.  Empirically combining the Kaiser factor and damping yields the discontinuous pdf shown for $r=10 h^{-1}$\emph{Mpc}.  Reproduced from \cite{Scoccimarro}.}
%\floatpagestyle{empty}
\label{fig:Scocc}
\end{figure}
%\pagenumbering{gobble}
%\addtocounter{page}{-1}
%\pagenumbering{arabic}
\noindent
with $F_V$ given by a Gaussian distribution; the first and second moments of which are the scale-dependent mean infall, $v_{12}(r) = \langle (\mathbf{v}' - \mathbf{v})(1+ \delta)(1+ \delta') \rangle$, and anisotropic velocity dispersion, $\langle v_i(\mathbf x) v_j(\mathbf x') \rangle = A \delta_{ij} + B_{ij}$, predicted by linear theory (\citejap{Fisher1995}, F95); a SDSS-III CMASS measurement of $\xi_s(s_\perp, s_\parallel)$ is shown in Fig. \ref{fig:Reid_xi}.  This form for the streaming model is known to be approximate.  Even in the case of Gaussian overdensity and velocity fields satisfying linear theory, the non-linear redshift-space mapping generates a non-Gaussian $F(\nu)$ with exponential tails (see eqn. (20) of F95 or S04); this is shown in Fig. \ref{fig:Scocc}.  However, it is unclear that the exact case is more accurate for the biased tracers actually observed by redshift surveys \citep{ReidWhite}.
\begin{figure}
\centering
\includegraphics[width=0.6\textwidth]{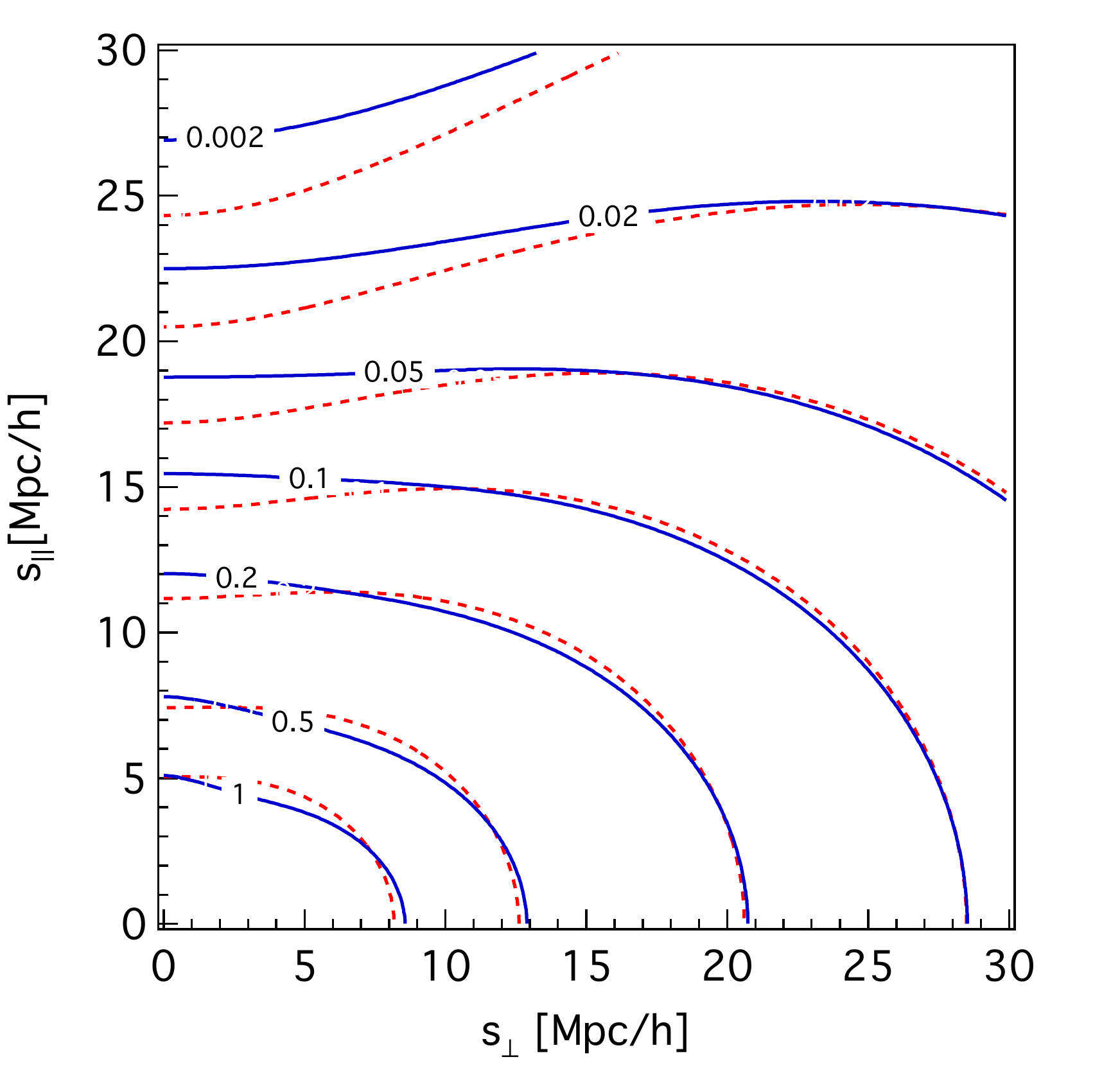}
\caption[Alterations in $\xi_s(s_\parallel, s_\perp)$ due to the large-scale non-Gaussian limit of the line-of-sight pairwise velocity pdf.]{Contours of $\xi_s(s_\parallel, s_\perp)$ for the large-scale limit of Gaussian density and velocity fields (solid blue) -- corresponding to a non-Gaussian $F_V$, and the Kaiser limit, for which $F_V$ is Gaussian (red).  Despite systematic changes in the large-scale $\xi_s(\mathbf{s})$, the multipole moments are obtained by a line integral at constant $s$ and are therefore insensitive to regions where the deviation is large (as the amplitude of $\xi_s(s)$ in this region is small).  Reproduced from \cite{Scoccimarro}.}
\label{fig:Scocc04_fig4.eps}
%\end{figure}
%\begin{figure}
\centering
\includegraphics[width=0.6\textwidth]{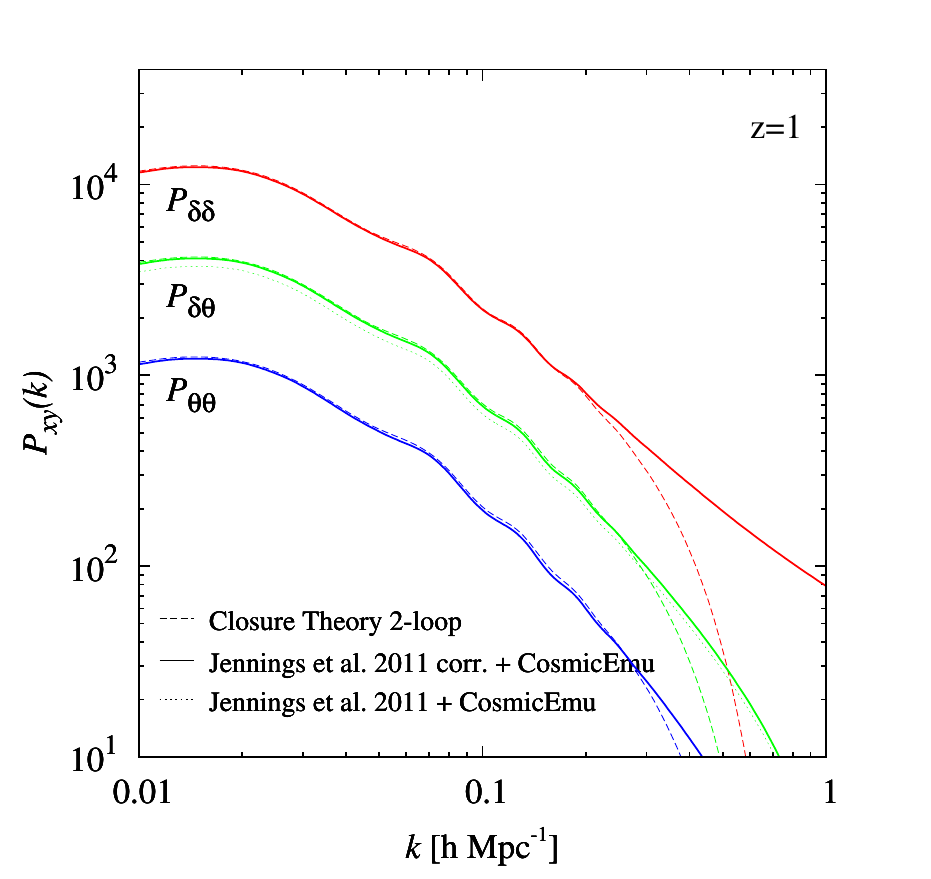}
\caption[Expected deviations from the linear symmetry: $P_{\delta \delta} = P_{\delta \theta} = P_{\theta \theta}$.]{Predictions for $P_{\delta \delta}, P_{\delta \theta}$ and $P_{\theta \theta}$ from an emulator \citep{Lawrence2010} (solid), from fitting functions \citep{Jennings2011Jan} and by closure theory \citep{Taruya2009}. $P_{\delta \theta}$ and $P_{\theta \theta}$ have been scaled by a factor of 3 and 10 respectively, which disguises the expected deviation from the $P_{\delta \delta} = P_{\delta \theta} = P_{\theta \theta}$ relation assumed by the Kaiser factor on even the largest scales surveyed, $k> 0.05$.  Reproduced from \cite{sylvainModels}.}
\label{fig:sdlt_Pdd_Pdt_Ptt}
\end{figure} 

Predictions from perturbation theory (validated with N-body simulations by \citejap{CarlsonWhitePadmanabhan}) have shown the pairwise velocity field to be acutely sensitive to quasilinear corrections on surprisingly large scales, $k \gsim 0.03 \hompc$.  This is because the relative velocity of a pair is sensitive only to wavelengths smaller than the separation and the damping is mild in the linear regime, $\tilde{\mathbf{u}} \propto (\tilde{\delta}/ k)$.  Therefore the transition to non-linearity is more rapid for densities than velocities.  This is compounded by the pair weighting, which amplify the highly overdense volumes in which linearity fails first.  On quasilinear scales, velocities are strongly affected by tidal fields (more so than densities) and as a result grow more slowly than linear theory predicts (S04); this breaks the linear theory symmetry: $P_{\delta \delta} = P_{\delta \theta} = P_{\theta \theta}$, for the normalised divergence $\theta \equiv - \nabla \cdot \mathbf u /\left(aHf\right)$ \citep{Jennings2010}.  Replacing the Kaiser factor, $(b + f \mu^2)^2 P_{\delta \delta}  \mapsto (b^2 P_{\delta \delta} + 2fb \mu^2 P_{\delta \theta} + f^2 \mu^4 P_{\theta \theta})$, is then a necessary step for improving beyond $5 \%$ accuracy \citep{Jennings2011Jan}, see \cite{sylvainModels} for more on this point.  The expected deviation can be seen in Fig. \ref{fig:sdlt_Pdd_Pdt_Ptt}.
\begin{figure}
\centering
\includegraphics[width=0.7\textwidth]{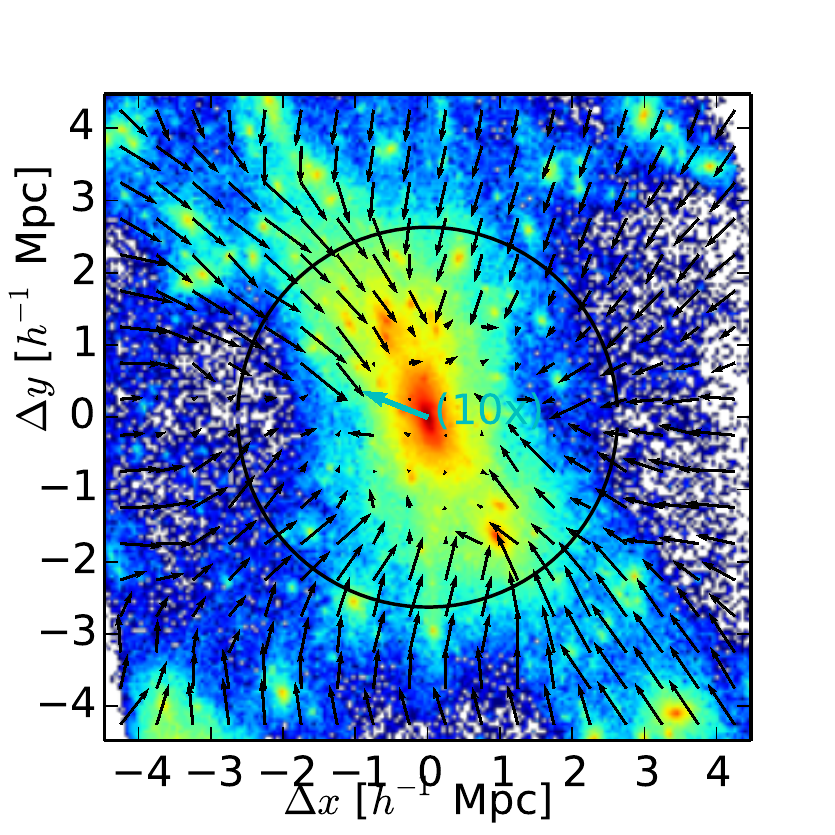}
\caption[The velocity difference between centrals and their host halo.]{Shown in this figure is $\ln(1+\delta)$ (colour) and the matter velocity field, scaled down by a factor of 20 (black arrows).  The black circle denotes $r_{\rm{vir}} = 2.7 h^{-1} \emph{Mpc}$ of the halo.  The velocity at the central galaxy position (the densest part of the halo) is shown in cyan and has been scaled down by a factor of 2.  The large infall from surrounding filaments makes a significant contribution to the halo centre-of-mass (COM) velocity, which is determined by including all matter within $r_{\rm{vir}}$.  The velocity field at the central differs from the halo COM velocity by $0.3 \times$ the rms virial velocity in the halo; the direction of this velocity difference is strongly correlated with the local, major filamentary structure.  As central galaxies reside in haloes at the termini of filaments, this effect will preferentially move pairs along the separation vector and hence systematically distort the $\xi_s(\mathbf s)$ of centrals.  Reproduced from \cite{Reid2.5}}
\label{fig:ReidBias}
\end{figure}

Another cause for concern has been raised by \cite{Reid2.5}, regarding a deviation of the `central' galaxy velocity from the halo centre-of-mass velocity; the latter is follows the linear velocity field \citep{WhiteHOD}.  In this work (R14), the position of the central or brightest galaxy is defined to coincide with the densest part of the halo and the central galaxy is assigned the velocity at this point.  R14 found this velocity had a dispersion with respect to the host halo of $0.3\times$ the rms halo virial velocity; it was also strongly correlated in direction with the local major filamentary structure -- see Fig. \ref{fig:ReidBias}.  As centrals reside in haloes at the termini of filaments, this correlation in direction will preferentially move pairs along the separation vector and hence systematically distort the $\xi_s(\mathbf s)$ of centrals.  However, the effect is small and will be significant only when d$(f \sigma_8) \sim 1.5\% $ is achievable (R14), which is beyond that possible with the VIPERS v7 survey and a realistic choice of $k_{\rm{max}}$.  A further caveat is that these conclusions are drawn from simulations that neglect baryonic effects, which will be large in these dense regions.

Although I restrict the modelling to linear theory when clipping (viz. a Kaiser factor and a linear model for $P_g$), I retain a dispersion term for two reasons; firstly, the measured redshifts have a rms error of $\sigma_z = 4.7(1+z) \times 10^{-4}$ = $141(1+z)\rm{km s^{-1}}$, which introduces a non-zero dispersion that is well modelled by a damping term \citep{PeacockDodds}.  This redshift error is approximately equal to the fingers-of-God dispersion, $(\sim 300 \rm{km s^{-1}})$, at $z \simeq 1$, but clipping will reduce the latter significantly; following this redshift errors will have a greater effect.  Secondly, clipping the redshift-space galaxy distribution is less clean than doing so in real space.  The radially smearing is present prior to the transform and it is likely to be an artefact that remains to some extent.  As such, retaining a damping term with a linear theory limit, $\sigma_p \simeq \sigma_z$, is a conservative choice that allows linear theory to be favoured to the necessary degree by the data.   Ultimately the best justification is to ensure the inferred constraints are unbiased and competitive when analysing realistic VIPERS mocks; this is investigated in Chapter \ref{chap:VIPERS_RSD}.  These simulations remain the benchmark by which models should be judged and I apply the simplest model that has no significant systematic error. 

A further assumption of these models is the validity of the plane parallel approximation -- when the variation of $\boldsymbol \eta$ across the survey is neglected.  Further systematic errors can be introduced when this fails for finite-angle surveys, e.g. BOSS \citep{Beutler}.  In this case, both finite-angle models \citep{Szalay97} and finite-angle estimators of the power spectrum \citep{Yamamoto} are required.  I assume the distant observer approximation throughout as the $24 \  \rm{deg}^2$ surveyed by VIPERS is well within this small angle limit.  In the following section I outline the remaining symmetries and ambiguities of finite-angle RSD and comment on their relevance for future RSD surveys; this discussion is repeated from \cite{maskedRSD}.

\section{Symmetries of finite-angle RSD}
\label{sec:finiteAngle}
I assume the validity of the distant observer approximation in this work, in which the variation of $\boldsymbol{\hat{\eta}}$ across the survey is neglected; the redshift-space $P(\mathbf{k})$ then possesses the symmetries outlined above.  If this assumption is relaxed and finite-angle pairs are included in the analysis then the redshift-space $\xi$ is dependent on the triangular configuration formed by a given pair and the observer \citep{SphericalRSD}.  As statistical isotropy about the observer remains, any such configuration may be rotated into a common plane \citep{Szalay97}, following which the remaining degrees of freedom are vested solely in the triangular shape.  Alternative parametrisations of the resulting configuration are possible and there is a subsequent ambiguity in the definition of the `line-of-sight'.  One possibility is to define $\boldsymbol \eta$ as that bisecting the opening angle.  In this case the triangle is fully defined by $s$, $\mu$, and the opening angle, $\theta$ (see Fig. \ref{fig:YooGeometry}).  
\begin{figure}
\centering
\includegraphics[width=0.65\textwidth]{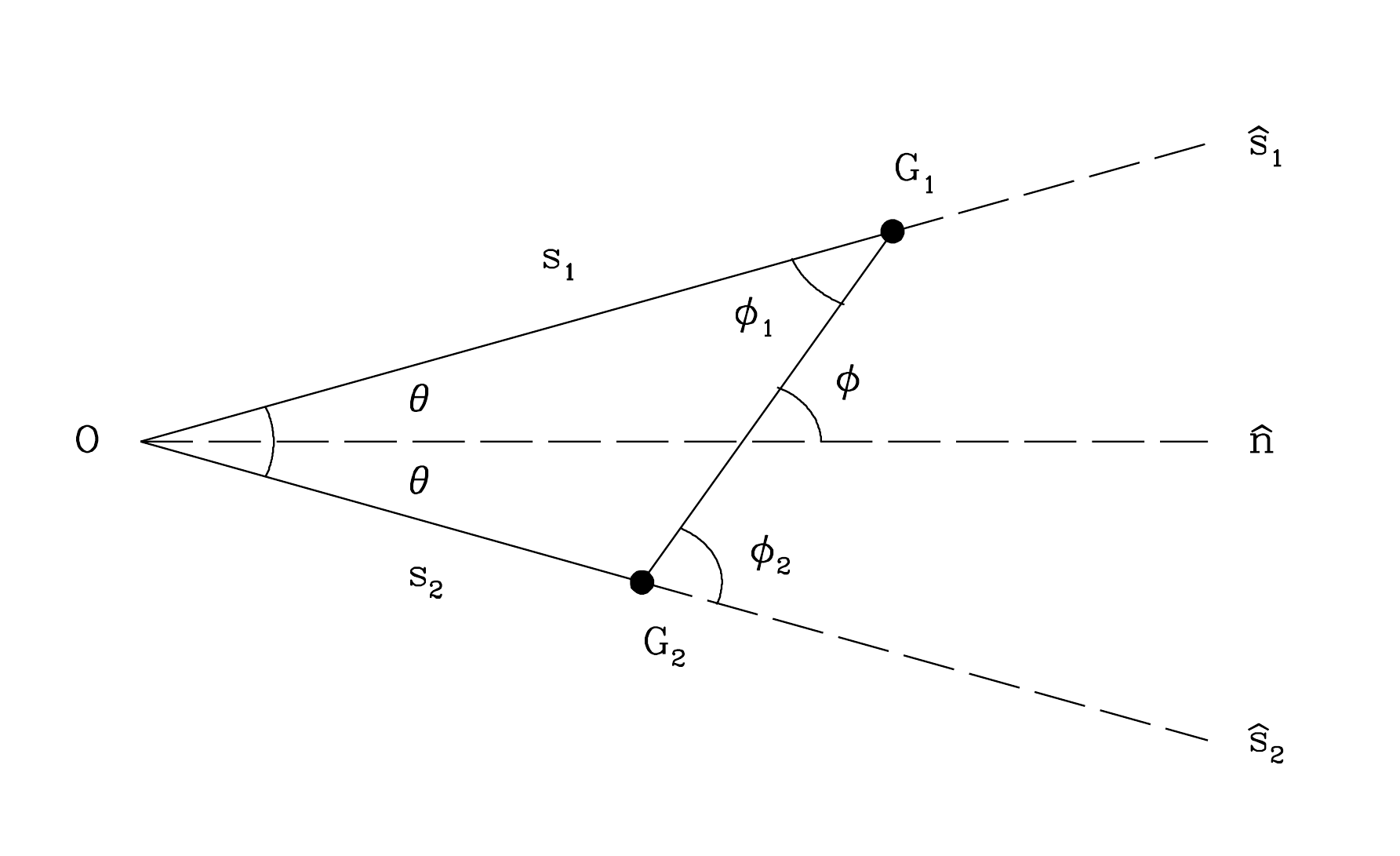}
\caption[Geometry of finite-angle RSD.]{Geometry of finite-angle RSD: the statistical isotropy about the observer allows any galaxy pair, G1 \& G2, to be rotated into a common plane, following which the configuration may be parameterised by the opening angle, $2\theta$, the pair separation and the cosine of the angle to the line-of-sight, $\cos(\phi)$.  In this case, the `line-of-sight' is chosen to be the bisector of the opening angle, $\mathbf{\hat{n}}$.  An additional variable $\theta$ is required to completely specify $\xi_s(\mathbf{s})$ when finite-angle pairs are included.  For a given pair, the opening angle is dependent on the position of the observer and statistical homogeneity is broken as a result.  Reproduced from \cite{Yoo}.}
\label{fig:YooGeometry}
\end{figure}

It is clear that the finite-angle $\xi_s(\mathbf{s})$ does not possess the symmetries I exploit in Chapters \ref{chap:maskedRSD} \& \ref{chap:VIPERS_RSD}, which assume the distant-observer limit, $\theta \mapsto 0 \deg$.  As such the approach described must be applied with some care to large area surveys.  While this is currently a limitation, the median redshift of future surveys will be considerably larger and the modal opening angle of pairs separated by the BAO scale will be $\simeq 4-6 \deg$ as opposed to the $\simeq 20 \deg$ of current surveys, e.g. SDSS (\citejap{Yoo}, Y15).  In fact, in Fig. 7 of Y15 it has been shown that the systematic error introduced by assuming the distant observer approximation (in the modelling) is negligible for both Euclid and DESI, provided the redshift evolution of the density field and bias across the survey is correctly accounted for.  In any case, a practical perspective is to accept that any bias introduced by using an approximate model may be calibrated with simulations and a correction applied to the data analysis.  Finite-angle effects are merely one instance where this approach may be taken.

\section{The Alcock-Paczy\'nski effect}
\label{sec:AP}
The common approach in the reduction of observed redshifts to a $N$-point clustering statistic is the conversion of redshifts and angular positions to comoving coordinates.  From the FRW metric, the comoving distance between a pair separated by $d \theta$ at redshift $z$ is
\[
d \ell_{\perp} = R_0 S_k(r) d\theta = (1 + z) D_A(z) \, d\theta,
\]
while that of a pair separated radially is given by 
\[
d \ell_{\parallel} = \frac{c \ dz}{H(z)}.
\]
For pairs with large separation these are replaced by $C_k(r_{12}) = C_k(r_1) C_k(r_2) + kS_k(r_1)S_k(r_2) \cos(\theta)$; see Chapter \ref{chap:Basics}.  It is therefore more difficult to extract a simple cosmology dependence for large separation pairs.  If quantities estimated in an assumed cosmology, as opposed to the true one, are denoted with primes, this conversion yields
\[
d \ell_{\perp}' = (1 + z) D_A' d\theta = \left ( \frac{D_A'}{D_A} \right ) \ d \ell_{\perp} =  \frac{d \ell_{\perp}}{f_{\perp}},
\]
and 
\[
d \ell_{\parallel}' = \frac{c \ dz}{H'} = \left ( \frac{H}{H'} \right ) \ d \ell_{\parallel} = \frac{d \ell_{\parallel}}{f_{\parallel}};
\]
therefore $f_\parallel$ ($f_\perp$) greater (less) than unity will appear as a squashing of a spherical system.  For instance, only when the fiducial cosmology matches the true one will the comoving sound horizon at recombination (the BAO) appear spherical in real-space comoving coordinates.  More generally, the induced anisotropy has a magnitude:
\[
F  = \frac{f_{\parallel}}{f_{\perp}} = \left( \frac{H'}{H} \right ) \left( \frac{D_A'}{D_A} \right),
\]
which corresponds to a flattening along the line-of-sight when $F >1$.  This Alcock-Paczy\'nski (AP) effect will be approximately degenerate with the Kaiser flattening of redshift-space distortions \citep{Ballinger}.  
\begin{figure}
\centering
\includegraphics[width=0.8\textwidth]{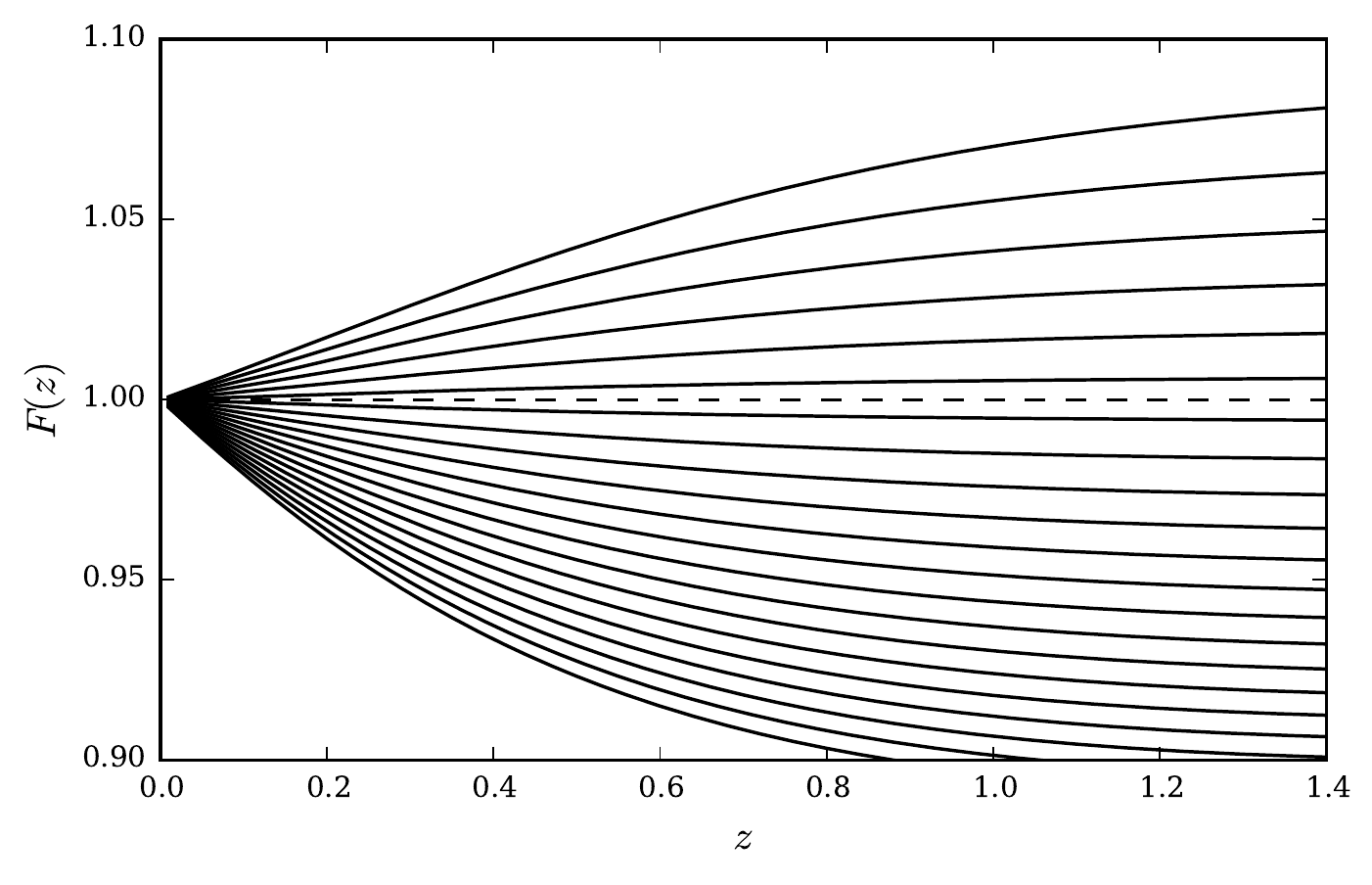}
\caption[Expected magnitude of the geometric anisotropy.]{The fiducial cosmology for this analysis is a flat $\Lambda$\emph{CDM} model with $(\Omega_m, \Omega_v) = (0.69, 0.31)$.  Shown in this figure is the expected geometric anisotropy, $F(z)$, for a flat universe when the true $\Omega_v$ is equal to $0.4$ (bottom), $0.8$ (top) or that in between (spaced by $ \Delta \Omega_v = 0.02$).  Given current constraints, this geometric distortion will be less than $10 \%$ and is expected to be larger at high$-z$.  This confirms the applicability of expressions for $P'_\ell(k')$ truncated at $\mathcal{O}(\epsilon)$, although the validity will be determined by the VIPERS precision to some extent.}
\label{fig:F(z)}
\end{figure}
The expected geometric anisotropy is shown for a range of flat cosmologies in Fig. \ref{fig:F(z)}.  The redshift-space flattening approximately adds to the geometric distortion when $\beta$ is small; the latter is equivalent to 
\[
\beta_F(k) \approx \frac{n}{2}(1-F),
\]
for $n = d\ln P_r/d \ln k$.  The $\mu$ dependence of the combination, which is distinct from either in isolation, may be used to separate $\beta$ from $F$ with sufficient angular resolution. Alternatively, in the realistic case of a power spectrum differing from a power law, which is especially true with FOG, $\beta$ and $F$ may be separated based on their distinct scale dependence.  

There is a corresponding remapping between modes in Fourier space: 
\[
k'_{\perp} = f_{\perp} \, k_{\perp} = \left ( \frac{D_A}{D_A'} \right ) k_{\perp}, 
\qquad \quad
k'_{\parallel} = f_{\parallel} \, k_{\parallel} = \left ( \frac{H'}{H} \right ) k_{\parallel}, 
\] 
hence the measured amplitude in the assumed cosmology is $P'(\mathbf{k}') \propto P(\mathbf{k})$, with a further normalisation due to the affect of the misestimated survey volume on the density of states, $P' \ d^3 k' = P \ d^3k$.  The resulting dilation of scale and change in $\mu$ are \citep{Beutler}
\[
k = \frac{k'}{f_{\perp}} \sqrt{ 1 + {\mu'}^2 \left ( \frac{1}{F^2} - 1 \right ) }, 
\qquad 
\mu = \frac{\mu'}{F} \left [ 1 + {\mu'}^2 \left ( \frac{1}{F^2} - 1 \right ) \right ]^{- \frac{1}{2}}.
\]
As this simple prescription for the forward modelling of the AP effect is available, it is not necessary to remeasure the power spectrum when comoving positions have been redetermined for a given point in the parameter space, which must vary the expansion history.  The power spectrum need be measured only once, in a fiducial cosmology parameterised by $\boldsymbol \theta_F$.  For a given cosmology, $\boldsymbol{\theta}$, postulated to be the truth, the expected model power spectrum is calculated and the AP distortion introduced by assuming $\boldsymbol \theta_F \neq \boldsymbol \theta$ is included by the simple remapping: $ P(\mathbf{k}) \mapsto P'(\mathbf{k'})$.

\cite{PadmanabhanWhite} derive the lowest order $P'_\ell(k', \boldsymbol{\theta_F})$ in terms of an isotropic dilation of scale, parameterised by $\alpha$ -- the geometric mean of the two transverse and one radial scaling factors, $(f_\perp^2 f_\parallel)^{1/3}$, and an anisotropic `warping', $\epsilon$.  In this parametrisation, $F = (1 + \epsilon)^3$ and the mode remapping is given by  
\[
k_{\perp} = \alpha^{-1} (1 + \epsilon) k'_{\perp}, \qquad
k_{\parallel} = \alpha^{-1} (1 + \epsilon)^{-2} k'_{\perp};
\]
\begin{figure}
\centering
\includegraphics[width=\textwidth]{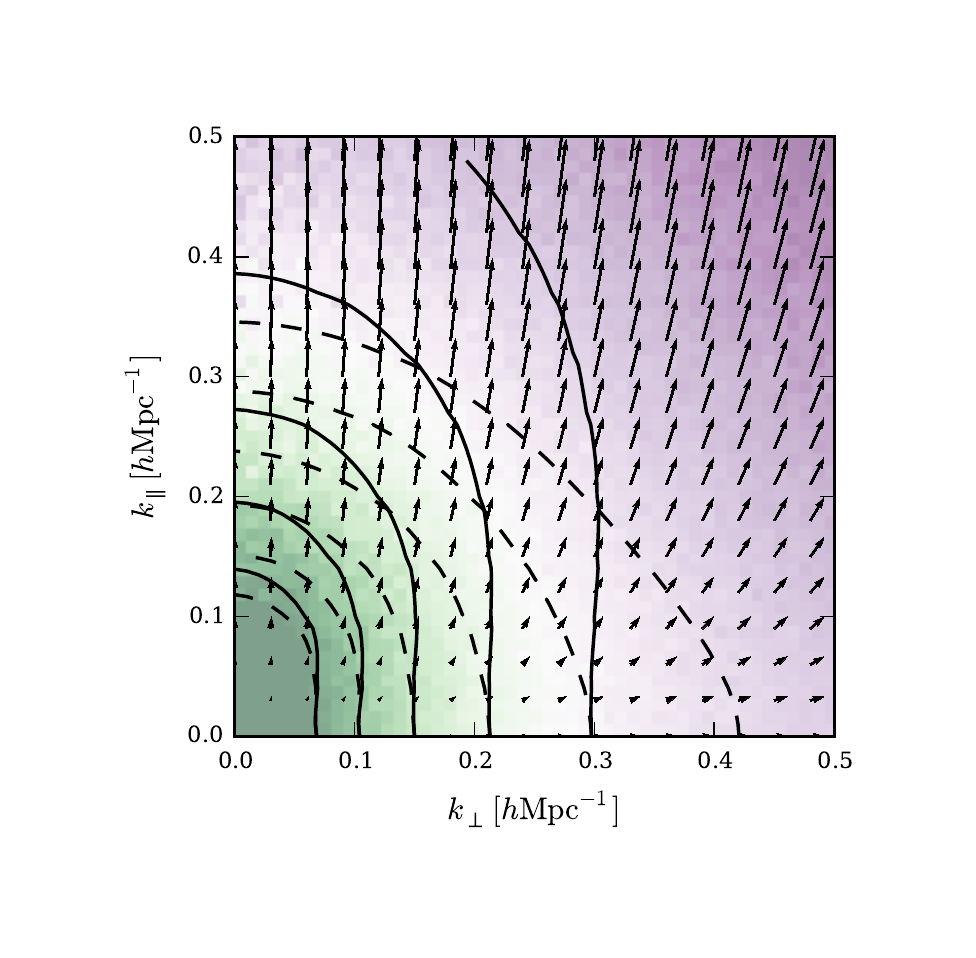}
\caption[The Alcock-Paczy\'nski effect as a mode remapping.]{Required mode remapping for an Alcock-Paczy\'nski distortion of $(\alpha, \epsilon) = (1.1, 0.05)$.  Contours label the Kaiser (solid) and Kaiser-Lorentzian models (dashed) for $P(\mathbf k)$ with $(\beta, \sigma_p/(h^{-1} \emph{Mpc}) ) = (0.54, 6.0)$.  It is clear from this figure that the measurement of $P'(k')$ will yield an estimate of $\beta$ that is biased high if the AP effect is unaccounted for.}
\label{fig:BallingerRemapping}
\end{figure}
this remapping is illustrated by Fig. \ref{fig:BallingerRemapping}.  Consequently,
\[
k = \left ( \frac{k'}{\alpha} \right ) (1+\epsilon)  \left [  1 + {\mu'}^2 \left ( (1+\epsilon)^{-6}  -1 \right ) \right ]^{\frac{1}{2}},
\]
which is 
\[
k = \frac{k'}{\alpha} \left ( 1 - 2 \epsilon L_2(\mu') \right ), 
\qquad \qquad
\mu^2 = {\mu'}^2 - 6 \epsilon \left ( {\mu'}^2 - {\mu'}^4 \right ), 
\]
to linear order in $\epsilon$.  For the power spectrum multipoles in a postulated cosmology, $P_\ell(k, \boldsymbol \theta)$, those measured in the fiducial cosmology, $P'_\ell(k', \boldsymbol \theta_F)$ are therefore given by \citep{Xu} 
\[
P'_\ell (k', \boldsymbol \theta_F) = P_\ell \big | _{\frac{k'}{\alpha}} - 2 \epsilon L_2 (\mu') \frac{dP_\ell}{d \ln k} \Biggr | _{\frac{k'}{\alpha}}.
\]
Using the relation:
\[
\frac{dL_\ell}{d \mu} = \frac{(\ell + 1) \left ( \mu L_\ell - L_{\ell +1}\right )}{(1 - \mu^2)}, 
\]
results in the approximation:
\[
L_\ell(\mu) \approx L_\ell(\mu') - 3 \epsilon (\ell + 1) \left ( {\mu'}^2 L_\ell (\mu') - \mu' L_{\ell+1}(\mu') \right ).
\]
Using this expressions, the lowest-order AP distorted multipole moments are given by 
\begin{flalign}
\label{eqn:AP_multipoles}
 && P'_0(k', \boldsymbol \theta_F) &= P_0 \big |_{\frac{k'}{\alpha}} - \frac{2 \epsilon}{5}   \left (3 P_2 \big |_{\frac{k'}{\alpha}}  + \frac{dP_2}{d\ln k} \Bigg |_{\frac{k'}{\alpha}} \right ), \nonumber && \\
&& P'_2(k', \boldsymbol \theta_F) &= \left ( 1 - \frac{6 \epsilon}{7} \right ) P_2 \bigr |_{\frac{k'}{\alpha}}  -2 \epsilon \frac{dP_0}{d \ln k}  \Bigg |_{\frac{k'}{\alpha}}  - \frac{4 \epsilon}{7} \frac{dP_2}{d \ln k} \Bigg |_{\frac{k'}{\alpha}} 
\nonumber  && \\ 
&& & \quad - \frac{4 \epsilon}{7}   \left ( 5 P_4 \big |_{\frac{k'}{\alpha}}   +  \frac{dP_4}{d \ln k} \Bigg |_{\frac{k'}{\alpha}} \right ), \nonumber && \\
&& P'_4(k', \boldsymbol \theta _F) &= P_4 \big |_{\frac{k'}{\alpha}} -3 \epsilon \left ( -\frac{24}{35} P_2 \bigr |_{\frac{k'}{\alpha}} + \frac{20}{77} P_4 \big |_{\frac{k'}{\alpha}} + \frac{210}{143}P_6 \big |_{\frac{k'}{\alpha}} \right) \nonumber  && \\ 
&& &  \quad -2 \epsilon \left( \frac{18}{35}  \frac{dP_2}{d\ln k} \Bigg |_{\frac{k'}{\alpha}} + \frac{20}{77}  \frac{dP_4}{d\ln k} \Bigg |_{\frac{k'}{\alpha}} + \frac{45}{143} \frac{dP_6}{d\ln k} \Bigg |_{\frac{k'}{\alpha}} \right ). &&
\end{flalign}
These expressions are reproduced from the discussion given by \cite{Xu}.  Note that excluding terms higher than $P_4(k)$ limits the $k_{\rm{max}}$ to which these expressions are valid, which is a greater restriction when $\sigma_p$ is large.  

These expansions allow the multipole moments to be corrected directly, without requiring memory intensive and slow manipulations of the 3D power spectrum -- which would be forward modelled according to eqn. (A8) of \cite{Ballinger}.  The derivatives required above may be calculated analytically in the Kaiser-Lorentzian model.  As an integration by parts gives
\[
\frac{dM_m}{d \ln k} = -(m+1) M_m + \frac{1}{(1 + k^2 \sigma_p^2/2)},
\]
the required derivatives are 
\begin{align}
\frac{dP_0}{d \ln k} &= P_0 \frac{d \ln P_R}{d \ln k} + \left ( -M_0 -6 \beta M_2 -5 \beta^2 M_4 + \frac{(1+\beta)^2}{1 + k^2 \sigma_p^2 /2} \right) P_R(k), \nonumber \\
\frac{dP_2}{d \ln k} &= P_2 \frac{d \ln P_R}{d \ln k} \nonumber \\ & \quad + \frac{5P_R}{2} \left ( M_0 -3(3-2\beta)M_2 -5\beta(6-\beta)M_4 -21 \beta^2 M_6 + \frac{2(1+\beta)^2}{1 + k^2 \sigma^2 /2} \right), \nonumber \\
\frac{dP_4}{d \ln k} &= P_4 \frac{d \ln P_R}{d \ln k} \nonumber \\ 
& \quad + \frac{9P_R}{8} \bigg ( -6M_0  -18(\beta -5) M_2  -5(3 \beta^2 - 60 \beta +35) M_4 \nonumber \\ & \qquad \qquad \qquad -70 \beta (7 -3 \beta) M_6 - 315 \beta^2 M_8 + \frac{8(1+\beta)^2}{1 + k^2 \sigma^2 /2} \bigg),
\end{align}
these expressions are original to my knowledge.  

\begin{figure}
\centering
\includegraphics[scale=1.0]{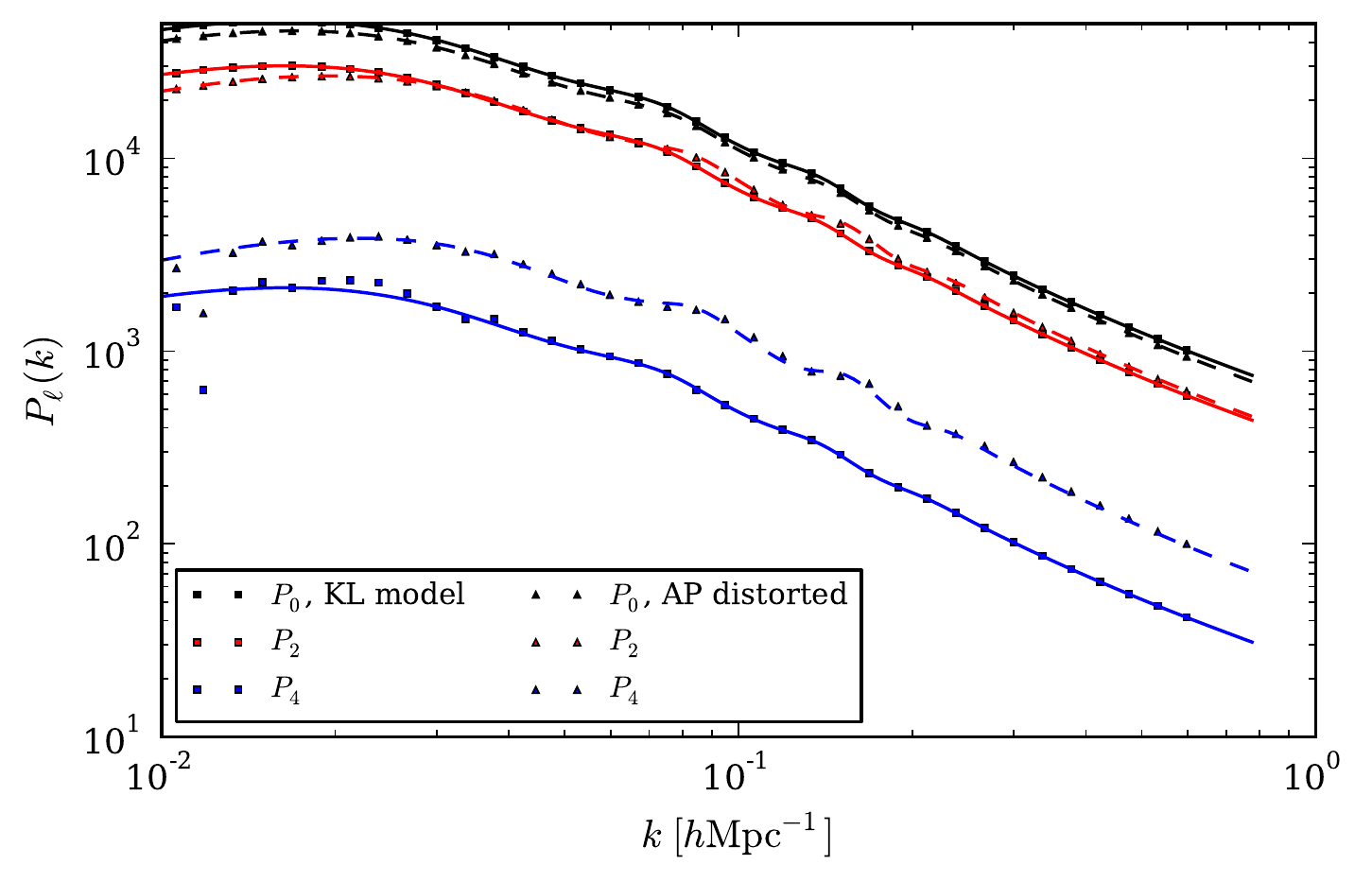}
\caption[A simple consistency test of \cite{PadmanabhanWhite} and  \cite{Ballinger}]{A simple test of the \cite{PadmanabhanWhite} predictions for the AP distorted multipole moments.  Square points show a multipole decomposition of the Kaiser-Lorentzian (KL) power spectrum by linear regression, for the harmonic modes available in a $(800 \hompc)^3$ box; triangles result from replacing this KL model with eqn. (A8) of \cite{Ballinger}; $(\alpha, \epsilon, f\sigma_8, b\sigma_8, \sigma_p)=(1.03, 0.03, 0.50, 1.0, 0.0)$ is assumed.  The expected KL multipoles and \cite{PadmanabhanWhite} predictions are shown by the solid and dashed lines respectively.  A good match to the expectation is shown in both cases.}
\label{fig:BallingerMultipoles}
\end{figure}
In the absence of redshift-space distortions, a flattening due to the AP effect is easily detectable (providing the local spectral index: $n \neq 0$).  But detecting a dilation requires a known scale. Baryon acoustic oscillations introduce just this: the comoving size of the sound horizon at last scattering, $ r_s$ (more correctly, $r_s$ is a shorthand for $R_0 r_s$).  This scale is imprinted as acoustic wiggles on $P(k)$, which are separated by an interval $\Delta k = (\pi/r_s)$ in the true cosmology.  The observed wiggle separation in the radial and transverse directions then constrains the combinations: $\Delta k_\parallel' = (H'/H) (\pi / r_s)$ and $\Delta k_\perp'= (D_A/D'_A) (\pi/r_s)$ respectively. Defining $D_V^3 = cz  D_A^2 /H$, the spherical dilation of the BAO scale allows for the measurement of 
\[
\left ( \frac{D_V }{r_s} \right )^3 = \alpha^3 \left ( \frac{D'_V}{r'_s} \right )^3 = \frac{\alpha^3}{{r'_s}^3}  \bigg ( (1+z)D'_A \bigg )^2 \frac{c z}{H'}. 
\]
Similarly,
\[
F_{AP} \equiv (1+z) D_A \left ( \frac{H}{c} \right ) = (1+z) D'_A \left( \frac{H'}{c} \right) \left( \frac{1}{F} \right ),
\]
may be inferred since $F = (F'_{AP}/F_{AP})$ is observable and $F'_{AP}$ is known.  As this constraint derives from an anisotropy, a known physical scale is not required and hence there is no degeneracy with $r_s$.  Note that while both $F_{AP}(z)$ and $D_V(z)$ may be constrained by redshift surveys, both are measures of the expansion history and hence are subject to the degeneracy between modified gravity models that motivates measurements of the linear growth rate. 

Beyond the BAO wiggles, the broad-band shape of $P(\mathbf{k})$ will have a dependence on $\boldsymbol \theta$ beyond simply $r_s$; for instance, the degree of non-linearity and hence small-scale power will depend strongly on $\Omega_m$.  The best fit dilation parameter is then $\alpha(\boldsymbol \theta)$.    
But in the case of a precise detection of the BAO peak, e.g. as is the case for the BOSS survey, the $\chi^2$ for a given $\alpha(\boldsymbol \theta)$ may be dominated by the tightly constrained wiggles (F. Beutler, private communication); in this case the derived posterior retains the $(D_V/r_s)$ degeneracy shown above.  This will not be the case in this work given the smaller VIPERS volume and lack of a BAO detection to date. 

I follow the common approach of omitting a marginalisation over the $\Lambda$CDM parameters (\citejap{sylvainClustering}, \citejap{BlakeGAMA}, \citejap{Beutler}) in the RSD and clipping analyses presented in Chapters \ref{chap:VIPERS_RSD} and \ref{chap:Clipping}.  This is in order to build a practical algorithm that delivers informative constraints.  It is not possible to account for the AP distortion perfectly in this case, as both $F$ and $P_g(k)$ will vary with the fiducial cosmology.  However, a start is to include $F$ as a free parameter in order to determine if the errors on $f \sigma_8$ are underestimated by neglecting the AP distortion.  One argument for this is that $f\sigma_8$ is constrained by the quadrupole-to-monopole ratio, which is independent of $P_g$ when the effect of the survey mask is negligible.  Moreover, it is good practice to ensure the data prefers the fiducial cosmology by ensuring $\alpha=1$ and $\epsilon=0$ to within the statistical errors. A transparent analysis should also explicitly demonstrate the inherent $(\beta, F)$ degeneracy of redshift-space galaxy clustering.  Such a test serves as an important consistency test of the fiducial cosmology and establishes greater independence from the \cite{Planck} analyses;  the $(f \sigma_8, F_{\rm{AP}})$ posterior of recent surveys are shown in Fig. \ref{fig:VIPERS_ruiz}. 

To neglect a marginalisation over the $\Lambda$CDM parameters in this consistency test requires the VIPERS error bar on $F_{AP}$ to be large enough that it remains representative, despite ignoring the likely variation in the best-fitting $\alpha$ when the fiducial cosmology is varied; i.e. the derived $(\beta, F)$ contours should be consistent when a range of $P_g(k)$ models spanning the $\Omega_m$ values preferred by \cite{Planck} are assumed.
\begin{figure}
\centering
\includegraphics[width=\textwidth]{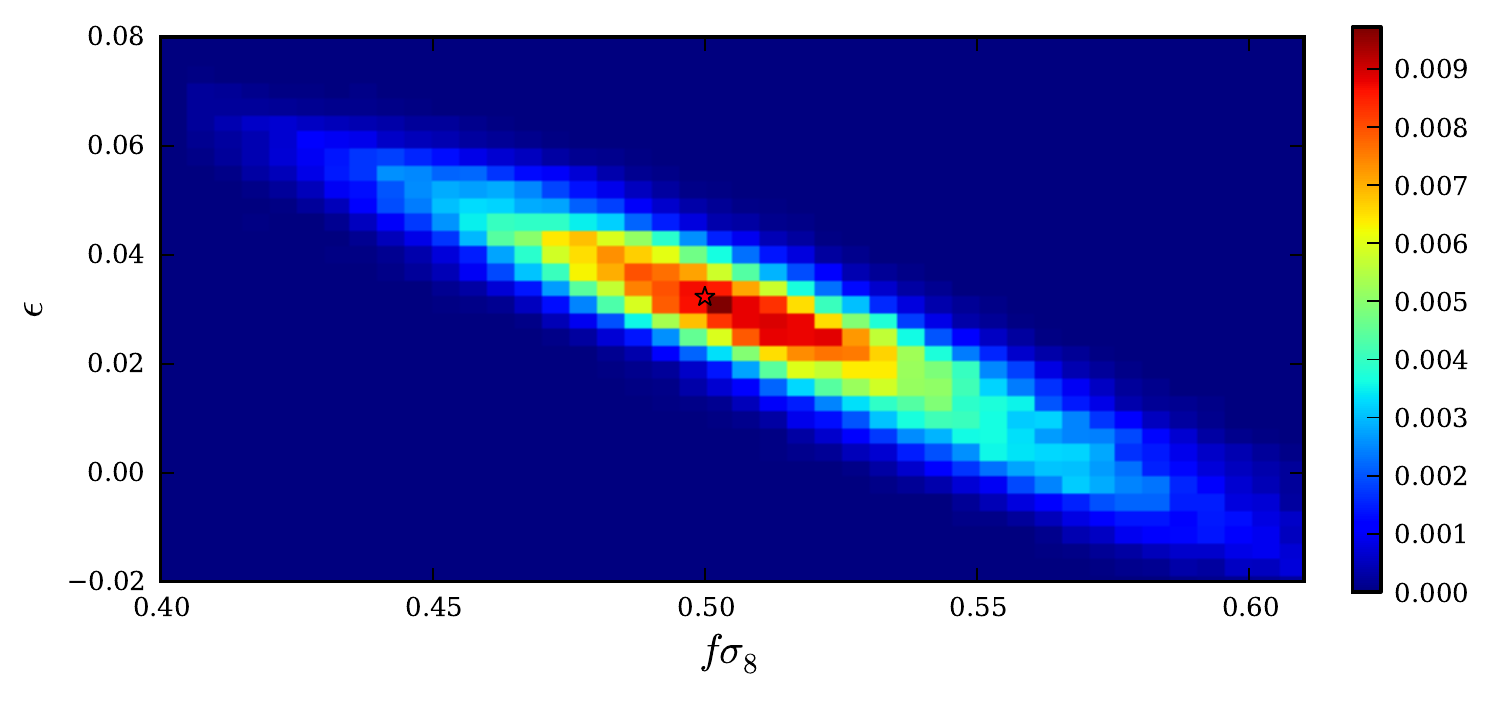}
\caption[A test of the correct recovery of a known input with MCMC.]{A validity test of the pipeline for producing a $(f \sigma_8, \epsilon)$ posterior from a VIPERS measurement of $P_0$ \& $P_2$.  A `dataset' was created, with $P_0$ \& $P_2$ determined by a multipole decomposition of the \cite{Ballinger} prediction; this was shown in Fig. \ref{fig:BallingerMultipoles}.  A diagonal covariance was assumed when computing the likelihood, with non-zero elements set to $\sigma/2$; here $\sigma$ is the expected error of the VIPERS W1 low-$z$ volume.  The posterior was obtained by Markov Chain Monte Carlo (MCMC) and is shown to be unbiased with respect to the input value (red star).  The posterior was displaced by an amount consistent with the degeneracy contour when appropriate VIPERS errors were added, as should be the case.  See \S 4.1 of \cite{HeavensStats} for a description of the Metropolis algorithm (symmetric proposal) used.}
\label{fig:chain_apnonoise}
\end{figure}

\begin{figure}
\centering
\includegraphics[width=0.9\textwidth]{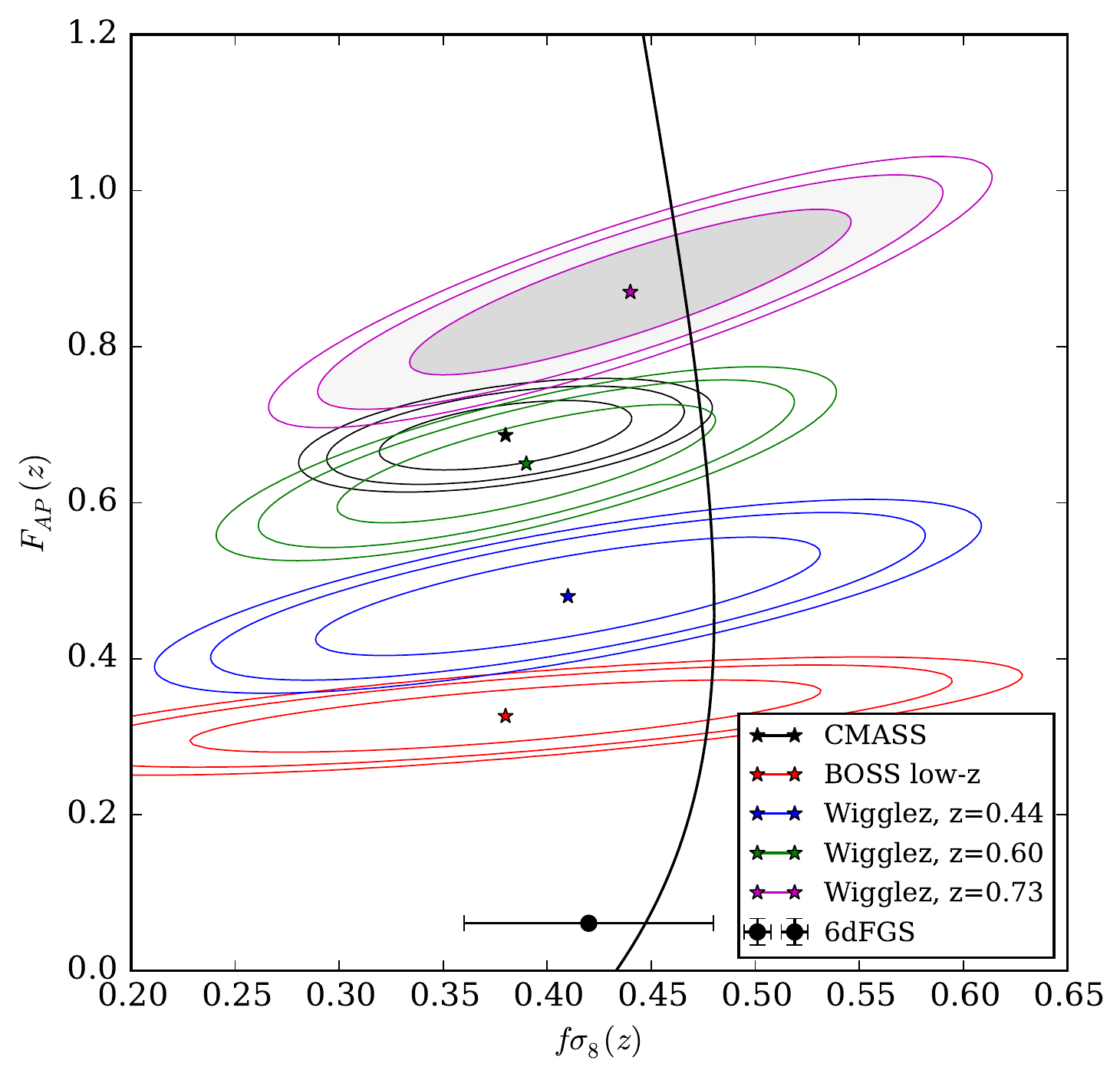}
\caption[Recent tests of the GR predictions for the $(f \sigma_8, F_{\rm{AP}})$ locus when a \cite{Planck} cosmology is assumed]{The GR prediction for $F_{AP}(z) \propto D_A H(z)$ and $f \sigma_8(z)$ is shown as the black locus, when a \cite{Planck} cosmology is assumed; this locus neglects the width resulting from uncertainties in the $\Lambda$CDM parameters.  Plotted contours show the 68\%, 90\% and 95\% confidence limits, $\Delta \chi^2 = (2.3, 4.61, 6.17)$ respectively, for a number of recent surveys; stars mark the peak in each case.  Although $F_{AP}(z)$ is a measure of the expansion history and is therefore degenerate amongst modified gravity models, these may be separated with the linear growth rate, $f \sigma_8(z)$.  In addition to the correlation matrices given in Appendix B, the variances in Table IV and eqn. (B3) of \cite{Ruiz}, a further expression for $\emph{Cov}(F, f \sigma_8) = (1+z) \left( D_A \cdot \emph{Cov}(f \sigma_8, H) + H \cdot \emph{Cov}(f \sigma_8, D_A) \right)/c$ is required for plotting these contours (E. Ruiz, private communication).  I add new VIPERS v7 constraints in Chapter \ref{chap:VIPERS_RSD}.}
\label{fig:VIPERS_ruiz}
\end{figure}

\section{Real-space correlations by deprojection}
 Redshift-space distortions displace galaxies radially from their true position, but their position on the sky is preserved. This results in the corruption of $r_\pi$ for a given pair but $r_\sigma$ is unaffected in the distant observer approximation.  A RSD-free clustering estimate may then be obtained if the $r_\pi$ values are neglected and pairs are simply binned in $r_\sigma$.  This projected correlation function is given by 
 \[
 w(r_\perp) = 2 \int_0^\infty \xi_s(r_\perp, s_\parallel) \ ds_\parallel.
 \]
An upper limit of $s_\parallel \simeq 70 \mpcoh$ is placed on the integral in practice \citep{RossDeproject}, in order to remove the contribution of a noisy $\xi_s$ estimate on large scales.  The large-scale $\xi_s$ is expected to be small due to homogeneity and hence the integral should be unbiased.  As this operation negates RSD, the same result must be obtained when performed on the real-space correlation function:
\[
w(r_\perp) = 2 \int_{r_\perp}^\infty \frac{r  \xi(r)}{\sqrt{r^2 - r^2_\perp}} \ dr,
\]
which results from $r dr = \sqrt{r^2 - r^2_\perp} \ dr_\pi$ at fixed $r_\perp$. Inverting this relation allows for the estimation of $\xi_s(\mathbf{s})$ \citep{Saunders}:
\[
\xi(r) = - \frac{1}{\pi r} \frac{d}{dr} \int_r^{\infty} \frac{r_\perp w(r_\perp)}{\left( r_\perp^2  - r^2 \right)^{1/2}} \ d r_\perp.
\]
\begin{comment}
\begin{figure}
\centering
\includegraphics[scale=0.9]{deprojection.pdf}
\caption[Obtaining the 3D real-space correlation function by deprojection.]{Results of applying a deprojection algorithm to a known angular correlation function, in order to obtain the  3D real-space correlation function.  In principle, this approach would be superior to the ESR weights described in Chapter \ref{chap:VIPERS}, although it is difficult to make an error estimate.  I demonstrate that the weighting approach is sufficient in Chapter \ref{chap:VIPERS_RSD}.}
\label{fig:deprojection}
\end{figure}
\end{comment}
\noindent
A practical implementation is given by eqn. (26) of \cite{Saunders}.
%, the result of which is shown in Fig. \ref{fig:deprojection} for a power-law correlation function: $\xi(r) = \left( r/5.1 \mpcoh \right)^{-1.57}$.  
As the streaming model has been shown to be a fundamental property of the redshift-space power spectrum \citep{Scoccimarro}, the ability to estimate $\xi(r)$ from the survey itself is a huge asset, which allows for the involved modelling of the non-linear real-space clustering and galaxy bias to be avoided.  

There is an additional advantage for VIPERS; the result of the VIPERS target selection algorithm (see Chapter \ref{chap:VIPERS}) would be identical if applied to the redshift-space and real-space galaxy distribution.  Deprojection then determines the affect of the target selection on $\xi(r)$ directly; the RSD modelling can then proceed as in the Kaiser-Lorentzian case.  In principle, this is a more precise approach than applying correcting weights to the spectroscopic galaxies prior to the power spectrum estimate.  However, estimating a robust error on the deprojection estimate is difficult.  Accordingly, I incorporate correcting weights rather than deprojection; this is demonstrated to be sufficient in Chapter \ref{chap:VIPERS_RSD}.  A final, albeit similar, alternative is to determine the real-space power spectrum from the modes transverse to $\boldsymbol{\hat{\eta}}$. Forecasts of this method have been shown to be disappointing; $f$ may be determined to 26\% precision for $k_{\rm{max}} =0.4 \hompc$ with the forthcoming DESI experiment  \citep{Jennings16}.  Although it is more robust, this is less stringent than current constraints. 
\end{chapter}
\begin{chapter}{Galaxy clustering and its estimation}
Chapter \ref{chap:Basics} has outlined the origin and linear evolution of density perturbations at early times.  It is clear from this discussion that the influence of particle physics in the early universe, e.g the baryon fraction $(\Omega_b/\Omega_m)$, is imprinted on the linear matter power spectrum.  However, the research in later chapters focuses on constraining departures from General Relativity with the VIPERS galaxy distribution.  As biased tracers of the local matter distribution, this galaxy distribution differs from the linear theory prediction in many important aspects -- on all but the largest scales.  This chapter focuses on these differences and presents (often toy) models that are used for understanding and simple tests throughout.  More realistic numerical simulations are then discussed, together with the necessary Bayesian statistics for robustly and efficiently, but perhaps not optimally, extracting the desired information.   

\section{Mildly non-linear growth}
\subsection{Breakdown of linear theory}
Linear theory is valid for $\delta~\ll~1$, which is appropriate for describing the density perturbations at early times or on large scales.  But galaxy surveys are apparent magnitude limited, e.g. $m(\nu_{\rm{obs}}) \lsim 20$ for VIPERS, which places a limit of $z~\lsim~1$.  As large-scale structure has been further collapsing between $z~\simeq~1100$ and $z~\simeq~1$, the rms value $\delta$ is typically much greater than unity -- on all but the largest scales.  The volume of current surveys allows for only the measurement of a single decade in wavenumber that is well modelled by linear theory and does so with (relatively) limited statistical significance.  Therefore the first step to placing stringent constraints on cosmological parameters and gravitational theories with the observed galaxy distribution is `modelling' the (mildly) non-linear regime on intermediate scales.  Although this is perhaps best done with numerical simulations, the following subsections outline approximate models for the non-linear density field.  

\subsection{The Zel'dovich approximation}
In Lagrange's approach, each hypothetical fluid element is first labelled by its position at an early time, $\mathbf{q}$, and the subsequent dynamics are then fully described by the trajectory of each:
\[
\mathbf{r} = \mathbf {q} + \boldsymbol{\Psi}(\mathbf{q}, t),  
\]
where comoving positions are used from the outset.  These elements are assumed to move only in straight lines in the Zel'dovich approximation \citep{Zeldovich}.  Requiring the results to match the Eulerian theory prediction for $D_+(a)$ at early times fixes the time dependence as
\[
\boldsymbol{\Psi}(\mathbf{q}, t) = D_+(t) \ \boldsymbol{\Psi}^0(\mathbf{q}).   
\]
Assuming $\mathbf{q}$ is defined sufficiently early, such that the overdensity is effectively homogeneous at that time, the perturbations generated by a given displacement follow from mass conservation, $\rho \ d^3x = \rho_0 \ d^3 q$.  This results in:  
\[
\rho(\mathbf{q}) = \rho_0 \ \Pi_i [1 + D_+(t) \ \boldsymbol \psi^0_{i, i} ]^{-1},
\label{eqn:zeldovich_rho_t}
\]
for $\boldsymbol \psi^0_{i, i} \equiv \partial \boldsymbol \psi^0_i/\partial \mathbf{q}_i$; this is the case when a basis for $\mathbf{q}$ is chosen locally such that the strain tensor, $\boldsymbol \psi_{i, j}$, is diagonal.  To do so requires the overdensity to be in the growing mode, which means that the peculiar velocities derive from potential flow, $\mathbf{u} = - \grad \phi_v$, with no vorticity, $\curl \mathbf{v} = \mathbf{0}$.  Conversely, a linear overdensity field $\delta (\mathbf{x})$ may be generated from a homogeneous distribution by adding displacements of
\[
\tilde{\boldsymbol{\psi}}(\mathbf{k}) = -i D_+(t) \left( \frac{\tilde \delta(t_0)}{k^2} \right ) \mathbf{k}, 
\label{zeldisp}
\]
which allows for the rapid generation of mock galaxy catalogues with a known power spectrum; this applicaction is shown in Fig. \ref{fig:zeldovichMocks}.  

The Zel'dovich approximation has proven to be surprisingly accurate in describing the mildly non-linear regime of structure formation -- certainly more so than Eulerian theory, as the displacement of peaks is included rather than just assumed to be static (as in Eulerian linear theory).  It is clear from eqn. (\ref{eqn:zeldovich_rho_t}) that the density first grows due to collapse along the principal stress axis -- the one with the largest eigenvalue.  Two-dimensional Zel'dovich pancakes rapidly form as a result.  Following this, the evolution resembles the 1D gravitational collapse (neglecting variations in orientation) described exactly by the approximation.  However, being equivalent to first order Lagrangian perturbation theory, the Zel'dovich approximation may be further improved upon, at least prior to shell crossing;  See \cite{White_Zeldovich} and references therein for further detail.
\begin{figure}
\centering
\includegraphics[width=\textwidth]{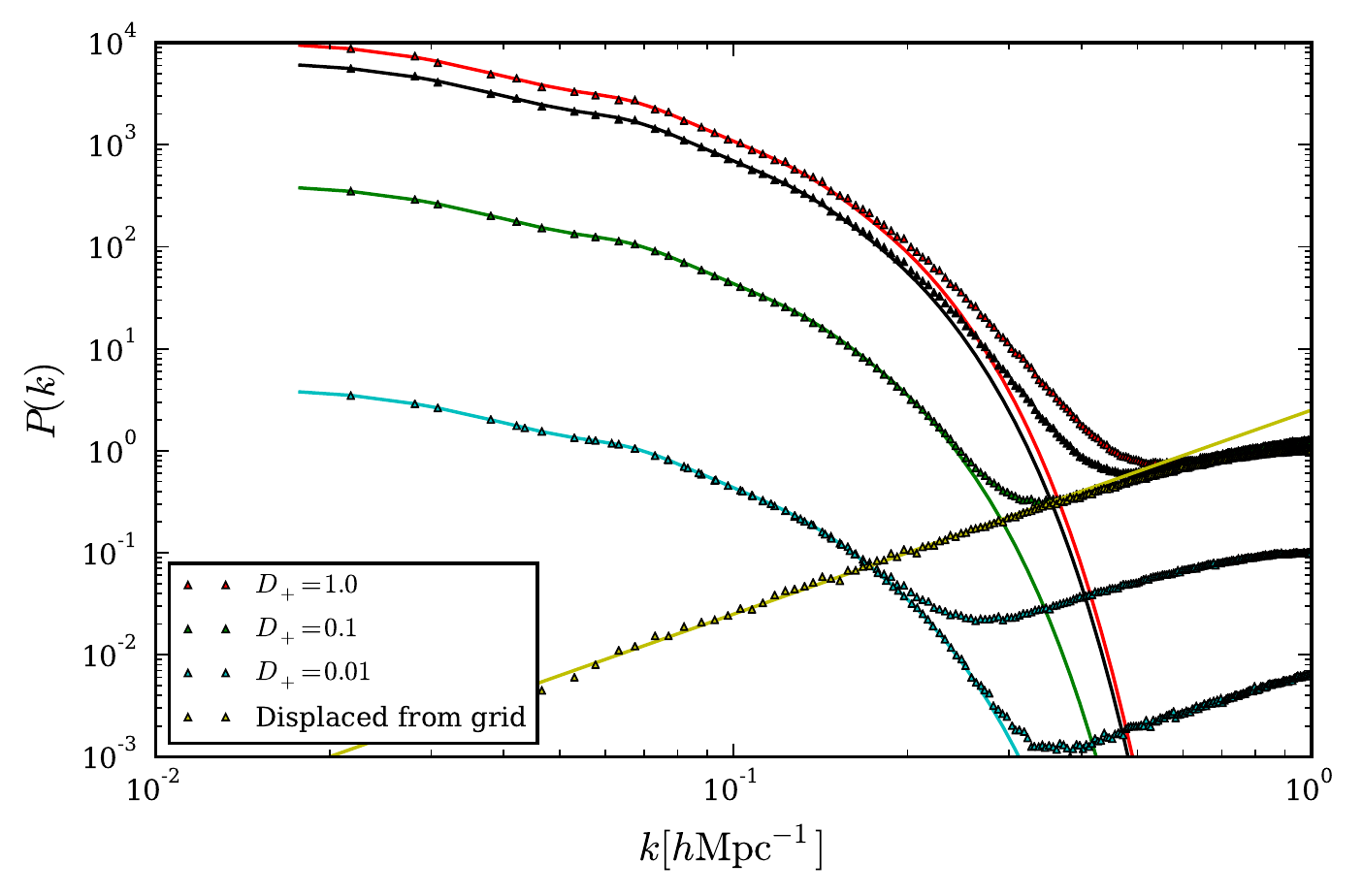}
\caption[Power spectra of Zel'dovich approximation mocks.]{Mock galaxy catalogues generated by the Zel'dovich approximation \citep{Zeldovich}.  A homogeneous (Lagrangian) density field was created by placing one particle randomly per cell, which results in $P(k) \propto k^2$ (yellow); this is shown for both the prediction (solid) and that measured from the realisations (triangles); this discrete contribution may be reduced by increasing $\bar n$.  The particles were then displaced according to eqn. (\ref{zeldisp}) for a range of $D_+(t)$ values.  This scheme may be used to add a two-halo term to halo model catalogues, which is exploited in \S \ref{haloCats}.  The variance of the Lagrangian density approximately adds to that generated by the displacement field, which is an artefact affecting the decade within the cell size.  This may be avoided by initially placing particles on a grid as this has no power on any scale except the cell size and its harmonics.}
\label{fig:zeldovichMocks}
\end{figure}

\subsection{Lognormal model}
\label{sec:lognormal}
As first outlined in \cite{ColesJones}, an approximate model for the mildly non-linear real-space density field is a lognormal distribution.  Explicitly, the continuity equation in comoving coordinates is
\[
\frac{D \rho}{Dt} = - \rho \grad \cdot \mathbf{u},
\]
where $\rho$ is the comoving density.  The primordial velocity field is Gaussian, which implies that $\grad \cdot \mathbf u$ is also; this Gaussianity may be maintained even when $\delta \simeq 1$ as $\mathbf{\tilde u} \propto (\tilde \delta /k)$ in linear theory;  therefore $\mathbf{\tilde u}$ is somewhat protected from the onset of non-linearity on small scales.  In the matter-dominated period: $D_+ \propto t^{2/3}$, which gives $\mathbf{\tilde u} \propto fH \tilde \delta \propto t^{-1/3}$.  By further assuming that in this `weakly' non-linear regime the displacement field is small, $|\mathbf{\Psi}| \ll 1$, such that $\mathbf{r}(\mathbf{q}, t) \simeq \mathbf q$ or equivalently $(\mathbf{u} \cdot \grad)\rho \ll \rho \grad \cdot \mathbf{u}$, the density field may be approximated by 
\[
\rho(\mathbf{r}) \propto \exp \left [ - \frac{\epsilon(\mathbf r)}{2} \left (\frac{t^2}{t_0} \right )^{\frac{1}{3}} \right]; 
\label{eqn:rho_ln}
\]
this is adapted from eqn. (12) of \cite{ColesJones}.  Here $\epsilon = (1/3) \grad \cdot \mathbf u(t_0)$ and the overdensity field is then 
\[
(1 + \delta) = \exp ( \delta_G  - \sigma_G^2/2).
\label{delta_lnnormal}
\]
The increasing mean of eqn. (\ref{eqn:rho_ln}) has been corrected for by rescaling $(1 + \delta)$ such that $\langle 1+\delta \rangle =1$, i.e. imposing large-scale homogeneity but preserving $\delta=-1$ in voids.  Here $\sigma_G^2$ is the variance of the Gaussian field, which is assumed to have zero mean.  Unlike the Gaussian distribution assumed for the primordial $\delta_G$, which assigns a non-zero probability to negative $\rho$ for large $\sigma_G$, this lognormal model is bounded by $\delta \geq -1$ at all times.  It also provides an improved match to the long tail to positive $\delta$ observed in N-body simulations at moderate smoothing lengths.   

\section{Baryonic effects and galaxy bias}
The previous sections have focused on the observation that galaxy surveys sample a local volume, which is typically too evolved to be modelled by linear theory.  In addition to this, only light-emitting galaxies may be observed as opposed to the dominant dark matter component.  Multiple lines of evidence have shown the galaxy distribution to be a biased indicator of perturbations to the mass: 
\[
\delta_g(k) = b(k) \ \delta_m(k);
\]
the simplest of which is the relative clustering of galaxy subpopulations, e.g. when divided by luminosity or colour (\citejap{PeacockDodds}, \citejap{NorbergPeder}, \citejap{MarulliFederico}).  But this linear bias model is a simplification; galaxy biasing may be non-linear and stochastic: $\delta_g = f(\delta_m) + \epsilon$ \citep{DekelLahav} even if it is local.  However, studies have shown the stochastic element to be small \citep{Wild}.  From the perspective of investigating fundamental physics, such as the role of zero-point energies in the present cosmic acceleration, (naively) this is an unwanted complication but forming a coherent picture of galaxy formation is a central goal of cosmology; see \cite{Baugh_bias} and references therein.  If biasing can be sufficiently understood, there is a silver lining in that the relative biasing can be exploited to make a cosmic variance-free RSD measurement of  $f \sigma_8(z)$ (\citejap{McDonaldSeljak}, \citejap{BlakeGAMA}, \citejap{Abramo}).   

Baryons must cool sufficiently by atomic line emission and Bremsstrahlung if stars and galaxies are to form.  As two-body processes, these are most efficient in very dense regions -- i.e. the deep potential wells of the most massive dark matter haloes.  A simple model of this situation is to posit that galaxies form only where the density is above a given threshold, which is analogous to the formation of treelines due to altitude.  But a more realistic picture includes the suppression of galaxy formation by non-linear baryonic feedback from active galactic nuclei in the most massive systems and by supernovae in the loosely bound, small mass haloes.  Due to this preferential residence in haloes of specific masses, the clustering of galaxies are biased with respect to the majority of the dark matter, which comprises the less massive systems.  This is shown quite clearly in Fig. \ref{fig:tinker_biasedclustering}.  The biasing is expected to be scale-independent on scales much greater than those affected by baryonic effects and is commonly modelled as such.
\begin{figure}
\centering
\includegraphics[width=0.75\textwidth]{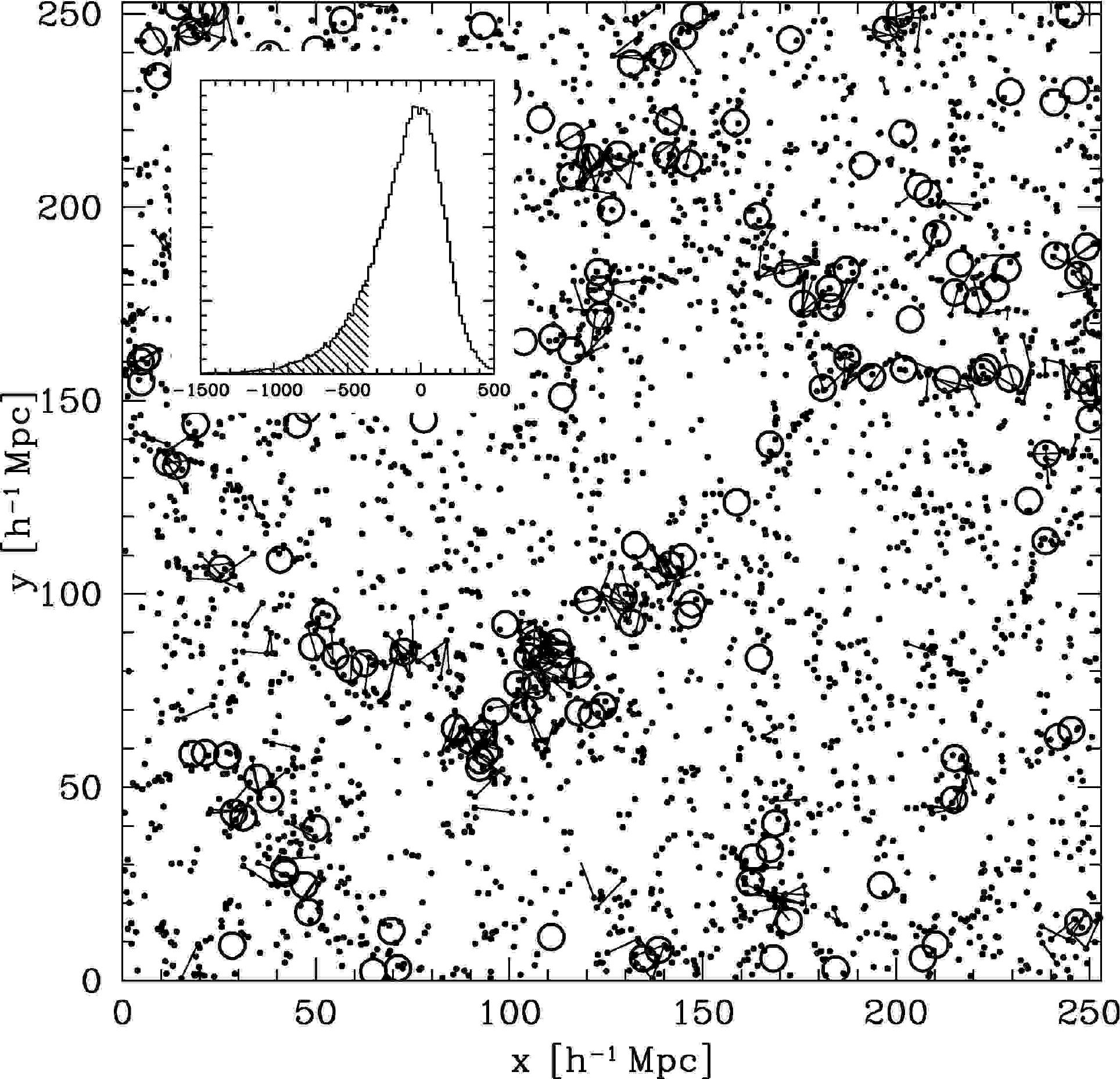}
\caption[The biased clustering of haloes.]{An illustration of the biased clustering of haloes in an N-body simulation.  Shown are haloes of two mass intervals: large open circles show $M > 3 \times 10^{13} \ h^{-1} $\emph{M}$_{\odot}$ and small filled circles represent $M \simeq 2 \times 10^{12} \ h^{-1} $\emph{M}$_{\odot}$.  In a threshold model of galaxy formation, at early times galaxies reside in only the most massive systems, which form at rare peaks of the primordial density field.  Consequently galaxies are formed in compact groups and are hence more biased than the majority of the mass, which also forms the less massive systems.  Reproduced from \cite{TinkerRSD_analytic}.}
\label{fig:tinker_biasedclustering}
\end{figure}

\section{Mock galaxy catalogues}
\label{sec:mocks}
I analyse mock catalogues that reproduce the expected clustering of the VIPERS selected galaxies in order to determine the significance of the VIPERS v7 measurement and ensure there is no systematic bias.  Measurements of the variation amongst the mocks, which includes the sample variance due to the finite volume and Poisson sampling, provide an estimate of the expected statistical error.  I briefly describe the construction of these mocks in this section, which are provided to the VIPERS collaboration by S. de la Torre; see \cite{sylvainModels} for further detail. 

The MultiDark N-body simulation \citep{bigMD} provides the dark matter density field on which the mock galaxy catalogues are founded.  Under the assumption of a flat $\Lambda$CDM  cosmology with 
\[
\begin{pmatrix}
\label{eqn:fiducial}
&\Omega_m \
&\Omega_v \ 
&\Omega_b \\ 
&h        \ 
&n_s      \ 
&\sigma_8 \ 
\end{pmatrix}
=
\begin{pmatrix}
&0.31 \
&0.69 \ 
&0.048 \\ 
&0.673 \
&0.96 \ 
&0.82
\end{pmatrix},
\]
which is consistent with the first release of the Planck mission \citep{Planck2013}, $3840^3$ particles are evolved within a $(2.5 \gpcoh)^3$ periodic volume.  This represents a brute force approach to the calculation of the fully non-linear density field -- by solving the equations of motion for an initially expanding Newtonian universe numerically (as described in Chapter \ref{chap:Basics}).  Multiple `snapshots' taken during this evolution are combined to build a light-cone.  This is an approximation to the redshift evolution of the density field within the VIPERS volume;  see \cite{sylvainModels}, \cite{Manera} and references therein.

Each light-cone is populated with galaxies according to the Halo Occupation Distribution (HOD) model \citep{Zheng}.  The number of VIPERS selected galaxies residing in a given halo is assumed to be dependent on halo mass, redshift and absolute magnitude, $\langle N | m, z, M_B \rangle$, which neglects any possible dependence on environment or formation redshift, for example.  This occupancy may be calibrated by a HOD modelling of the projected correlation function in the VIPERS spectroscopic catalogue, $\hat w(r_p | z, M_B)$.  This approach has a number of advantages, as the angular correlation function of the parent photometry is independent of both RSD and the VIPERS selection (see Chapter \ref{chap:VIPERS}).  The occupancy of central galaxies is distinguished from that of satellites \citep{Zheng} when populating haloes; the latter are assumed to trace the Navarro-Frenk-White density profile \citep[NFW,][]{NFW}, which is opposed to being placed at the location of sub-haloes for example.    

Haloes are provided to a limiting mass of $M_0 = 10^{11.5} \ h^{-1} $M$_{\odot}$ by the MultiDark consortium; these haloes having been identified with a `friends-of-friends' algorithm.  As the least luminous galaxies observed by VIPERS are expected to reside in haloes with $M < M_0$, smaller mass haloes are generating according to \cite{lowmasshaloes}; the limiting mass is $10^{10} \ h^{-1} $M$_{\odot}$ after this repopulation.  Having populated haloes with galaxies, the mocks are placed into redshift space by assigning central galaxies the velocity of their host halo.  Satellites are assigned an additional Gaussian dispersion, $\sigma(M)$, to mimic the virial velocities. 

Finally, the catalogues are subjected to the VIPERS survey selection described in Chapter \ref{chap:VIPERS}.  Apparent magnitude limits and the colour selection are imposed -- by subsampling galaxies to mimic the affect of the colour selection on $\bar n(z)$ in the latter case.  Spectroscopic slits are then assigned according to the VIPERS selection algorithm, in an identical manner to that applied to the parent photometry.  The photometric and spectroscopic angular masks (Samhain and Nagoya v7 respectively) are then applied.  The result of this process is realistic mocks that are similar as possible to the data.

\section{Measures of clustering}
\subsection{Two-point estimators}
If the primordial $\delta(\mathbf{x})$ is a Gaussian random field (as predicted by the simplest inflationary theories) the independent information is fully contained in the power spectrum, $P(\mathbf{k})$.  Defining the Fourier transform \citep{FourierT} as
\[
\tilde \delta(\mathbf{k}, a) = \frac{1}{V} \int d^3 r \ \delta(\mathbf{r}, a) \ \rm{e}^{-i \mathbf k \cdot \mathbf r},
\]
the power spectrum is defined by $\langle \tilde \delta(\mathbf{k}) \tilde \delta^{*}(\mathbf{k}') \rangle = (2 \pi)^3 \delta^3(\mathbf{k} - \mathbf{k}') P(\mathbf{k})$; here homogeneity has been assumed and $V$ is taken to be unity for the remainder of this work.  The power spectrum is a function solely of the wavenumber magnitude: $P(k)$, if the universe is also isotropic.  For the common case of discrete harmonic modes, e.g. when analysing a simulation with periodic boundaries, each mode is independent for a Gaussian field (when the Cosmological Principle holds) and has a random phase, $\phi \in [0, 2 \pi]$, together with a squared amplitude drawn from an exponential distribution.  This may be derived from 
\[
\text{Prob}(|\tilde {\delta}(\mathbf{k})|^2 \ > X) = \exp(\frac{-X}{P(k)}).
\]

Equivalently, in the continuum limit, this information is contained in the correlation function, 
\[
\xi(\mathbf{r}) = \langle \delta(\mathbf{x}) \delta(\mathbf{x} + \mathbf{r}) \rangle. 
\label{xi}
\]
The two constitute a Fourier transform pair:
\[
\xi(\mathbf{r}) = \int \frac{d^3 k}{(2 \pi)^3} \ |\tilde \delta(\mathbf{k})|^2 \ \rm{e}^{-i \mathbf{k} \cdot \mathbf{r}}.
\]
Although this equivalency is true of the expectation -- as obtained from an infinitely large volume, finite-volume estimators are not equivalent; both Fourier space and configuration space measurements are therefore of merit.  There are advantages in each case, but the most precise analyses to date \citep{Alam, Beutler_2016A} have shown that the power spectrum delivers the more stringent constraints; see the right panel of Fig. 8 of \cite{Alam} for example.  Moreover, the independent evolution of Fourier modes in linear theory results in a reasonably well-defined wavenumber beyond which the evolution is non-linear \citep{LittleWP}.  At a given separation, $\xi(r)$ is a weighted average of $P(k)$ over a range of wavenumbers and hence the transition is blurred in configuration space.  Modes do not evolve independently following the onset of non-linearity and phase correlations are introduced.  This results in a non-Gaussian field that is not fully defined by the power spectrum, but contains independent higher order correlations. 

Assuming both ergodicity and homogeneity, the ensemble average in eqn. (\ref{xi}) may be estimated with a volume average.  But the only direct observables of dark matter are in projection in practice, e.g cosmic shear or CMB lensing.  There is however evidence to suggest galaxies trace the matter linearly on large scales; in the Poisson sampling model \citep{KaiserElements}, the galaxy distribution is assumed to be a stochastic sampling of the dark matter density:  
\[
n(\mathbf{x}) = \bar{n}(\mathbf{x}) (1+\delta(\mathbf{x})),
\]
where $n(\mathbf{x})$ is the comoving number density; The mean, $\bar{n}$, of which may vary depending on evolution and survey selection.  Assuming the Cosmological principle, the ensemble average, $\langle \delta(\mathbf{x}) \delta(\mathbf{x} + \mathbf{r}) \rangle$, is a function of scalar separation only. The expected number of galaxy pairs between two small volumes is then given by 
\[
\langle N_1 N_2 \rangle = \bar n_1 \bar n_2 \left[ 1 + \xi(r) \right] dV_1 dV_2.   
\]
In practice, the Landy-Szalay estimator \citep{LandySzalay} has better statistical properties than computing $N_1 N_2$ directly.  With a random catalogue bounded by the survey geometry and incorporating the survey selection, i.e. $R_1 R_2 = \alpha^{-2} \bar n_1 \bar n_2 dV_1 dV_2$, this estimator is
\[
\hat \xi(r) = \frac{DD(r) - 2DR(r) + RR(r)}{RR(r)}.
\]
Here pair counts including randoms have been implicitly rescaled to account for the increased density; see page 522 of \cite{CP}.  
\begin{figure}
\centering
\includegraphics[trim={0 0 0 0.51cm}, clip, width=\textwidth]{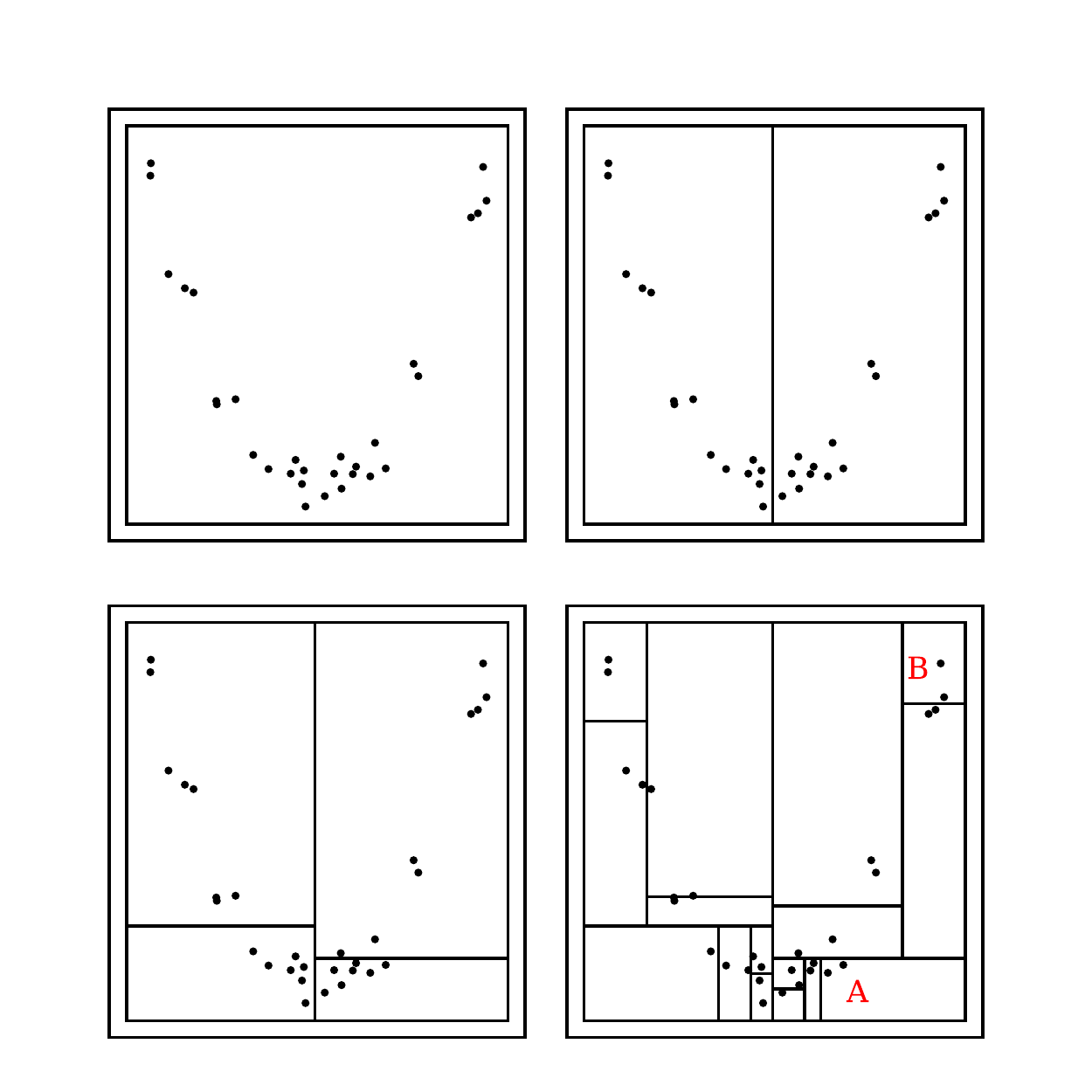}
\caption[A k-d tree graphic.]{A k-d tree groups a set of galaxies or randoms into subsets or `nodes' for efficient pair counting. The full catalogue or `root node' is divided into two `children' with each containing half the objects.  The process is then recursive: each child is further divided in two, this time along a different direction.  The direction of division cycles between $\{\mathbf{\hat{x}}, \mathbf{\hat{y}}, \mathbf{\hat{z}}\}$ and stops for a given child when further division results in $N<N_{\rm{min}}$ particles.  Following this division into nodes, large separation pairs, such as those between node A and node B, may be neglected on the basis of the minimum separation between the boundaries of A and B.  Adapted from \cite{astroML}.}
\label{fig:kdtree}
\end{figure}

A second useful tool for efficient pair counting is a k-d tree \citep{kdTree}, which groups the galaxies or randoms into subpopulations or `nodes'.  This allows for a number of large-separation pairs to be rapidly counted according to  
\[
DD(r \simeq |\mathbf{\bar x_1} - \mathbf{\bar x_2}|) \simeq N_{\rm{node}}(\mathbf{\bar x}_1) \times N_{\rm{node}}(\mathbf{ \bar x}_2),
\] 
where $\mathbf{\bar x}$ is the mean position of the particles in a node.  This is valid as errors on $\hat \xi(r)$ are sizeable on large scales and hence the binning in separation may be coarse.  This approach is also advantageous when only close pairs must be counted.  By estimating the minimum separation between nodes, pairs with a separation greater than a given separation may be ignored.  I make extensive use of this approach in Chapter \ref{chap:maskedRSD}, which is further described in the caption to Fig.~\ref{fig:kdtree}. 

\subsubsection{Tests on simple catalogues}
\label{haloCats}
To validate the pipeline for estimating both the power spectrum -- necessary for the RSD analysis in Chapter \ref{chap:VIPERS_RSD}, and the correlation function (applicable to Chapter \ref{chap:maskedRSD}) I created a number of mocks with known clustering and varying degrees of realism.  The first type assume the lognormal approximation; in this case, a Gaussian field is first generated and eqn. (\ref{delta_lnnormal}) is then applied.  By Poisson sampling the resulting density field a mock catalogue may be created.  The expected correlation function for these mocks is  
\begin{figure}
\centering
\includegraphics[width=\textwidth]{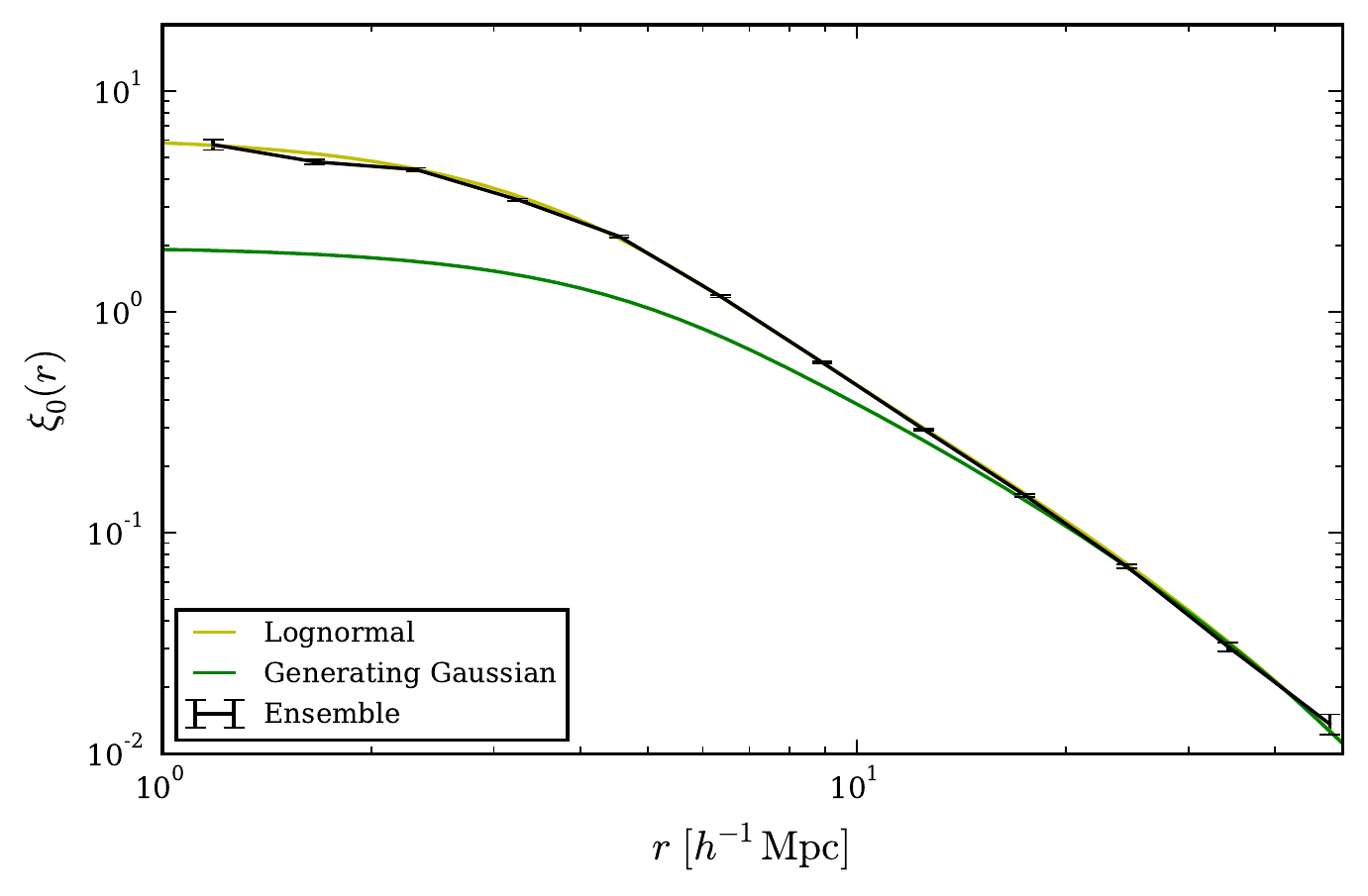}
\caption[A test of the $\xi_0(r)$ estimate with Poisson-sampled lognormal mocks.]{Mocks with a known isotropic correlation function were generated by Poisson sampling a lognormal density field.  This allows a test of the accuracy of $\hat \xi(r)$ to be performed.  Shown in the figure are the correlation function of the generating Gaussian (green), the expected autocorrelation of the lognormal mocks (yellow) and the measurement obtained by applying a simple $(DD/RR)$ estimator.  These results show the $\hat \xi(r)$ estimate to be accurate.}
\label{fig:lnnorm_real_xi}
\end{figure} 
\[
1 + \xi_{\rm{\ln}}(r) = \exp(\xi_g(r)).
\]
The results of pair counting this random catalogue and applying a simple $(DD/RR)$ estimator are shown in Fig. \ref{fig:lnnorm_real_xi}; this shows the kd-tree is being built correctly and that the an accurate $\hat \xi(r)$ estimate is made.  It was unclear how best to place these mocks into redshift space at the time (in order to repeat this test for an anisotropic $\xi(\mathbf s)$).  In hindsight this is clear -- the generating Gaussian corresponds to a rescaling of the velocity divergence field, eqn. (\ref{eqn:rho_ln}); one component of a curl-free velocity field with this divergence is required to add RSD (in the distant observer approximation).  This is simply a linear model and does not account for the fingers-of-God effect however.      

Instead, I created mocks with a known anisotropic clustering by assuming the halo model of large-scale structure (\citejap{SeljakHalomodel}, \citejap{PeacockDodds}).  Haloes of identical mass were first created and randomly distributed throughout the volume, which creates the shot noise spectrum described in \S \ref{sec:ShotNoise}.  Additional realism, such as incorporating a biasing model and a mass function are unnecessary for simply obtaining an anisotropic correlation function.  The correct two-halo power was added on large scales by displacing the haloes according to the Zel'dovich approximation, as discussed previously.  Satellites were then added to each host halo according to a NFW density profile, which incorporates the one-halo power on non-linear scales.   

The mocks were then placed into redshift space by assuming the distant observer approximation.  To do so, the $z$-component of the displacement must be simply multiplied by $(1+ f)$ prior to being added to the halo Lagrangian position \citep{White_Zeldovich}.  A velocity dispersion was included by adding a Gaussian random variable of width $2 \mpcoh$ to the $z$-component of each satellite position.  The clustering was then estimated in both real and Fourier space; the results of this are shown in Fig. \ref{fig:halomodel_mocks}.         
\begin{figure}
\subfloat{\includegraphics[width=\linewidth]{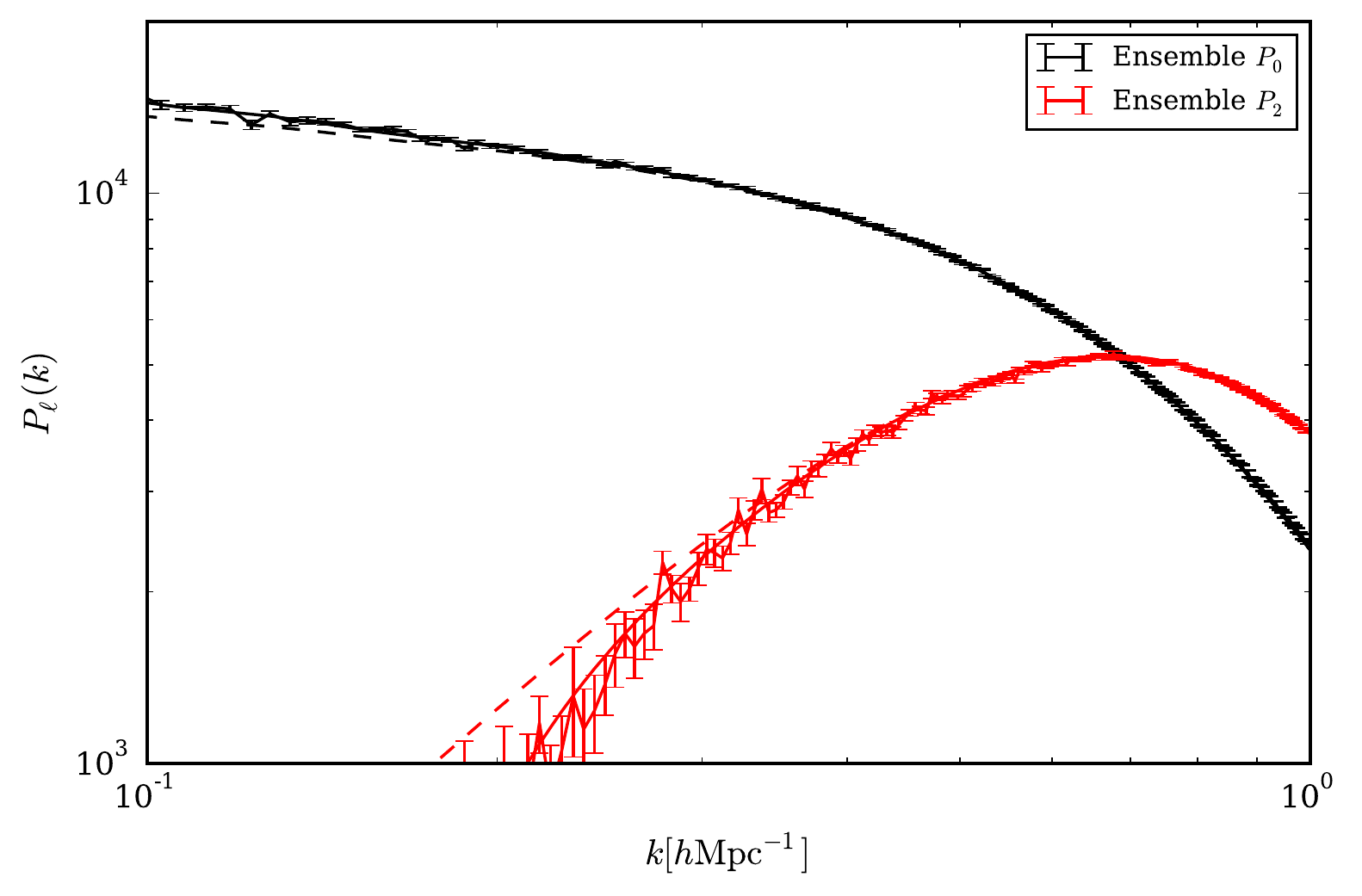}} \\ 
\subfloat{\includegraphics[width=\linewidth]{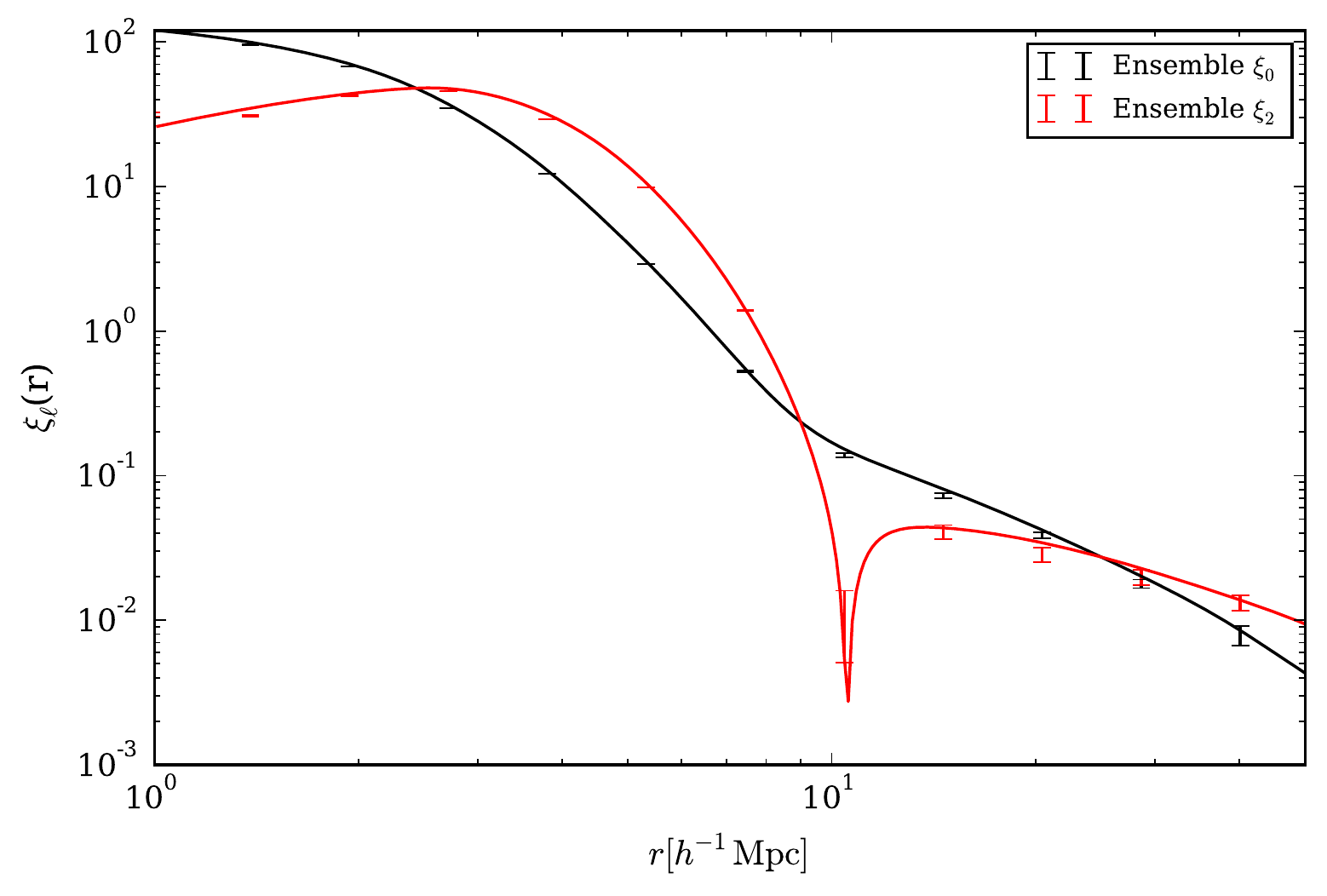}}     
\caption[Clustering tests with simplified halo model mocks.]{To ensure the accuracy of the power spectrum and correlation function estimates I created mocks with a known anisotropic clustering.  In this case, the halo model predictions for $(P_0, P_2)$ and $\xi(r)$ are shown (solid).  The difference between the dashed and solid lines show the affect of applying the Kaiser amplification to the two-halo term.  Error bars show the mean and standard error of the measurements obtained from a large number of mocks.  These results show the clustering estimates to be accurate.}
\label{fig:halomodel_mocks}
\end{figure}
\subsection{Practical power spectrum estimation}
\label{sec:FKP_pk}
In this section I outline a practical scheme for estimating the power spectrum of the VIPERS galaxy survey; this follows that presented in \cite{FKP}.  In particular, it includes how the finite volume and survey selection criteria should be accounted for.  

\subsubsection{The Feldman-Kaiser-Peacock (FKP) estimator}
I implement the Feldman-Kaiser-Peacock power spectrum estimator \citep[FKP,][]{FKP}, by using a large random catalogue bounded by the survey volume.  Following the discussion of a non-uniform, non-Poisson sampling estimator given in \S 2.3 of FKP, the best estimate of the selection-corrected number density is $n_g = \sum_g E(\mathbf{x}_g)^{-1}\delta^3(\mathbf{x} - \mathbf{x}_g)$.  The origin of the effective VIPERS sampling rate, $E(\boldsymbol{x}_g)$, is defined in Chapter \ref{chap:VIPERS}.  Correcting this effective sampling rate (ESR) by upweighting galaxies, rather than downweighting randoms, is necessary due to the (projected) density dependent VIPERS sampling.  The two-point covariance of this corrected field is
\[
\langle n_g n_g' \rangle = \bar n \bar n'(1 + \xi) + \bar n E^{-1} \delta^3(\mathbf r - \mathbf r'), 
\]
where $\bar n$ is the ESR corrected number density.  But assuming $\langle n_g n_g'\rangle \simeq \left (1 + \xi(|\mathbf r - \mathbf r'|) \right)$ is only valid for pairs with a transverse separation for which the sampling is constant.  This will be the case for only a limited number of VIPERS pairs; the effect of the sampling rate on the VIPERS clustering estimates is investigated in Chapter \ref{chap:VIPERS_RSD}.  

The Fourier transform of this field is
\[
\tilde F(\mathbf{k}) =  \frac{1}{\sqrt{N}} \sum_g \frac{w(\mathbf{x}_g) w_c(\mathbf{x}_g)}{E(\mathbf{x}_g)} e^{i \mathbf{k} \cdot \mathbf{x}_g} - \frac{\alpha}{\sqrt{N}} \sum_s w(\mathbf{x}_s) e^{i \mathbf{k \cdot \mathbf{x}_s}},
\label{eqn:FKP_estimator}
\]
where the clipping weights defined in Chapter \ref{chap:Clipping} have been included.  In this and later expressions, sums over $s$ represent a sum over a given randoms realisation.  The expected value of this estimator is given by 
\begin{align}
\langle |\tilde F(\mathbf k)|^2 \rangle = & \frac{1}{N} \int d^3r \int d^3r' w w' \bar n \bar n'
\xi(|\mathbf r - \mathbf r'|) e^{i \mathbf k \cdot (\mathbf r - \mathbf r')} \nonumber  \\ 
&+ \frac{1}{N} \int d^3 r \ \bar n w^2(\mathbf r) \left( E^{-1} + \alpha \right).
\label{eqn:expected_Fk}
\end{align}
Here, and in the following expressions, the effect of clipping is assumed to be incorporated in an effective model for $\xi(r)$ or $P(k)$ respectively.  

The optimal FKP weighting: 
\[
w(\mathbf{x}) \propto \frac{1}{1 + \bar{n}(\mathbf{x}) P_0},
\]
is applied in order to minimise the statistical error, which is due to both the finite-volume surveyed and Poisson sampling.  In the discrete case, the power spectrum amplitude of each mode is drawn from an exponential distribution on large scales and therefore the fractional error on a single mode estimate is unity.  The error on a binned measurement, obtained from the mean of $N$ available modes, is then 
\[
\left( \frac{\sigma_P}{P} \right ) = \frac{1}{\sqrt{N}}.
\]
This sampling (cosmic) variance is one source of error.  The second is due to the finite-number of galaxies surveyed, which are assumed to be a stochastic sampling of the density field.  Only the most massive galaxies are luminous enough to meet the apparent magnitude limit at large redshift and $\bar n(z)$ falls sharply in this regime: $\bar n P_0 \ll 1$.  Hence the Poisson error -- with a variance equal to the mean, is significant and the finite-volume variance is comparatively negligible.  Each galaxy should be weighted equally in this case.  In the opposite extreme, at the peak $\bar n$, the finite volume is the dominant source of error and hence each volume should be weighted equally.  These limits are clearly satisfied by this FKP weighting as the volume weighting is $\bar n w(\mathbf{x})$.  Note that the convolution introduced by the survey mask, see Chapter \ref{chap:maskedRSD}, has been neglected in deriving these weights. They should be normalised such that
\[
\int \bar n^2 w^2 d^3 x \mapsto \alpha \sum_s \bar{n} w^2 (\mathbf{r}_s) = 1, 
\]
which corrects for the scaling of the observed power with the (weighted) surveyed volume -- $V(w) \ P(k)$ is a constant.  I follow the convention of plotting the volume independent quantity, $PV$, rather than $P$ throughout and abbreviate the labelling to $P(k)$.  

In principle, the value of $P_0$ should be reassigned to the expected power for each mode prior to the estimate; in practice, I choose a fiducial value of $8000 (\mpcoh)^3$ as the results are commonly not sensitive to this choice; see \cite{BlakeGAMA} for example.  The conclusions given in later chapters would be unchanged given the likely gains from an optimal choice.  Moreover, the difference should be relatively small as, by definition, $\partial P / \partial w |_{w_{FKP}} =0$ and the modes most affected by weighting also reside near the peak of $P(k)$ -- therefore the amplitude will vary only a little for small changes in the wavelength.  The mean density of the random catalogue defines  $\alpha \equiv \bar n / \bar n_s = N_E/N_s$; $N_E= \sum_g E^{-1}$ and similarly $N_s$ are calculated for the low-$z$ and high-$z$ volumes independently.   

\subsubsection{Shot noise}
\label{sec:ShotNoise}
Assuming the galaxy population to be a Poisson sample of the continuous matter density, there is a contribution to the observed variance simply from the stochastic nature of galaxy formation.  The second term of eqn. (\ref{eqn:expected_Fk}) refers to this scale-independent shot noise: 
\begin{align}
P_{\rm{shot}} = \frac{1}{N} \int d^3 r \ \bar n w^2(\mathbf r) \left( E^{-1} + \alpha \right) &= \alpha \sum_s w^2 \left( E^{-1} + \alpha \right) \nonumber \\ 
&= \ \ \ \sum_{\rm{spec }} w^2 E^{-2} + \sum_s w^2 \alpha^2.
\end{align}
I calculate this shot noise correction from the spectroscopic galaxies, which have the most robust estimate of  $E(\mathbf{x})$.  This shot noise correction may be understood as follows: the Fourier coefficient of a weighted discrete set of galaxies is  
\[
\tilde \delta(\mathbf{k}) = \frac{\sum_i w_i \exp(i \mathbf{k} \cdot \mathbf{x}_i)}{\sum_j w_j }.
\]
In the absence of clustering: $\langle n_i n_j \rangle \propto (1 + \xi(|\mathbf{r}_i - \mathbf{r}_j|)) \simeq 1$ and the expected power spectrum may be evaluated by appealing to a set of infinitesimal microcells; these have an occupancy of either unity or zero. The expectation is then  
\[
\langle | \tilde \delta (\mathbf{k}) |^2\rangle = \frac{\sum_i w_i^2}{\left( \sum_j w_j \right )^2}. 
\]
This simply gives $PV = \bar n^{-1}$ for equal weights.  As the dark matter $\bar n$ must be much larger than the mean density of galaxies, this shot noise is normally subtracted in order to obtain an estimate in the $\bar n \mapsto \infty$ limit appropriate for dark matter.  There is also a contribution to the shot noise from the randoms as they are generated by a Poisson sampling of an unclustered density field; this must also be subtracted.
%When measuring the clipped power spectrum we do not include the clipping weights in the normalisation constraint or the shot noise correction.  This is justified for two reasons: both $N$ and $P_{\rm{shot}}$ correspond to volume averages and the volume affected by clipping is small ($\sim 5 \%$); galaxies are down weighted with clipping as we posit that they have resulted from a Poisson sampling of the linear density field, at the same sampling rate, 
%\[
%n = \bar{n}(1 + \delta) \mapsto \bar{n}(1 + \delta_0), 
%\]
%In fact, $P_{\rm{shot}}$ is expected to be reduced as the clipped density field is free of shot noise in the volume for which both the perturbation to the mass and to the Poisson sample satisfy $\delta \gg \delta_0$; in this case both fields map to $\delta_0$ irrespective of the sampling  \citep{Fergus}.  We neglect this correction due to the small volume to which this argument applies.

Once shot noise corrected, the expected value is given by the convolution of the galaxy power spectrum:
\[
P'(\mathbf{k}) \equiv \langle |\tilde F(\mathbf k)^2| \rangle = \int \frac{d^{3} q}{(2 \pi)^3}  P(\mathbf{q}) |\tilde W(\mathbf{k} - \mathbf{q})|^2,
\]
with the effective survey mask, $W(\mathbf{x}) = \bar{n} w$.  This is simply the volume weighting.  A new forward modelling approach for calculating this correction is presented in Chapter \ref{chap:maskedRSD}.
\label{sec:pk_estimate}

\subsubsection{Integral constraint correction}
It has been assumed in the previous derivation that the true density perturbations are simply multiplied by a weight, $\delta(\mathbf{x}) \times W(\mathbf{x})$. However this is not the case in practice: the mean density of selected galaxies must be estimated from the survey itself \citep{PeacockNicholson}.  As this estimate will be subject to fluctuations on wavelengths approaching the survey length (and larger), a further integral constraint correction must be included; this correction is derived in the following.  

An incorrect estimate of the mean introduces a DC shift: $\tilde \delta_{\cross}(\mathbf{k}) = \tilde \delta(\mathbf{k}) - A \delta^3(\mathbf{k} - \mathbf{k}')$ 
and therefore the estimated power is 
\[P_{\cross}(\mathbf{k}) = P(\mathbf{k}) - A^2 \delta^3(\mathbf{k}), 
\]
where the constant $A$ is to be determined.  This false mean field is subject to the convolution detailed above, which gives
\[
P'_{\cross}(\mathbf{k}) = P'(\mathbf{k}) - A^2 |W(\mathbf{k})|^2.
\]
If the survey volume is assumed to be a fair sample then the constraint $P'_{\cross}(\mathbf{0}) =0$ is enforced and, by solving for $A^2$, it follows that 
\[
P'_{\cross}(\mathbf{k}) = P'(\mathbf{k}) - \frac{|W(\mathbf{k})|^2}{|W(\mathbf{0})|^2} P'(\mathbf{0}).
\]
This expression predicts the net affect of the survey mask on the measured power; a new approach for calculating this correction is presented in Chapter \ref{chap:maskedRSD}.     

\subsubsection{Aliasing and Jenkins's folding}
\label{sec:Jenkins}
A practical estimate of the power spectrum requires the calculation of the discrete Fourier transform (DFT), e.g. with FFTW \citep{Johnson}.  This converts the \{$\sum_g$, $\sum_s$\} to a sum over the cells of a volume in which the survey is embedded.  This conversion requires a mass assignment scheme -- associating a given galaxy or random to either a single cell (Nearest Grid Point, NGP, \citejap{HockneyEastwood}) or a weighted contribution to a number of cells.  An assignment may be chosen that returns a smoothed estimate of the density field, which reduces aliasing of the DFT.  The grid-point estimates of the density field provide a sampled density field:
\[
\delta_s(\mathbf{x}) = \sum_{\mathbf{n}} \delta(\mathbf{x}) \ \delta^3(\mathbf{x} - \Delta \mathbf{n}),
\]
for $\mathbf{n}$ -- a three-vector of integers and a cell size of $\Delta$.  This results in a convolution in Fourier space and a copy of $\tilde \delta (\mathbf{k})$ is therefore replicated at each point of the reciprocal lattice: $\sum_{\mathbf{m}} \delta^3(\mathbf{k} - (2 \pi / \Delta) \mathbf{m})$.  If $\tilde \delta (\mathbf{k})$ is not bandwidth limited to $k < k_{\rm{N}}$, for a Nyquist frequency of $k_N = (\pi / \Delta)$, the measured power will be a superposition of the true power and leakage from copies evaluated at large $k$; see pg.~47 of \href{http://www.roe.ac.uk/japwww/teaching/fourier/fourier1415.pdf}{\tt http://www.roe.ac.uk/japwww/teaching/fourier/fourier1415.pdf} for more detail.  I implement the cloud-in-cell scheme \citep{HockneyEastwood}, which is equivalent to smoothing with a top-hat twice, and correct the measured $\tilde F(\mathbf{k})$ for the applied smoothing:
\[
\tilde F(\mathbf{k}) \mapsto \frac{\tilde F(\mathbf{k})}{\prod \limits_{i} \text{sinc}^2(\frac{\pi k_{i}}{2k_N})}. 
\]

Although aliasing has been reduced by the mass assignment, it may be removed entirely by Jenkins's folding of the embedding volume, $V_B = L^3$.  Consider the harmonic modes of $V_B$ and of an octant $V_T$; the harmonic modes of $V_T$ are the subset which are also periodic on $L/2$.  For this subset, the phase of the exponential in eqn. (\ref{eqn:FKP_estimator}) is invariant under 
`folding': 
\[
x_i \mapsto x_i \mod (L/2);
\]
this can be seen in Fig. \ref{fig:Jenkins_foldplot}.  Thus for any mode which is harmonic with respect to $V_T$, $\tilde F(\mathbf{k})$ may be determined after folding all galaxies and randoms into a single octant.  For a memory-limited FFT,  $\Delta$ may be halved and $k_N$ doubled with each fold and therefore aliasing may be removed from progressively smaller scales by repeated application.  The disadvantage of this approach is that it applies to only a limited subset of modes, which is made smaller by every fold, and this causes the statistical errors on large scales to increase.  To implement this method, the boundaries of $V_B$, $\mathbf{x}_g$ and $\mathbf{x}_s$ should be folded and otherwise the estimation of $\tilde F(\mathbf{k})$ proceeds as before (with respect to the new volume $V_T$).  Where aliasing is significant, above $k \simeq 0.6 \hompc$, I repeat the measurement after a single fold and use the corresponding values for $k>0.4 \hompc$.  Further discussion of this technique may be found in \cite{Jenkins} and \cite{Smith}.
\begin{figure}
\centering
\includegraphics[trim=5.5cm 3.5cm 5.cm 3.5cm, clip=true, width=0.6\textwidth]{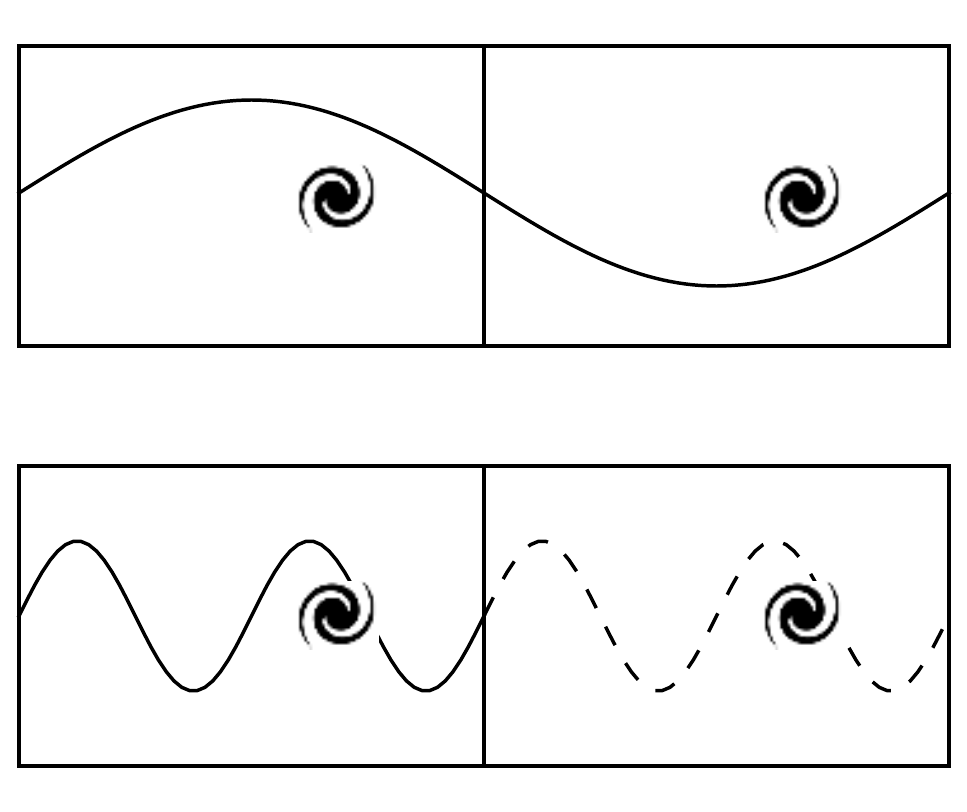}
\caption[An illustration of Jenkins's folding.]{An illustration of Jenkins's folding in one dimension.  Top: a harmonic mode of the embedding volume, $V_B$.  The mode phase differs when evaluated at the original and folded, $x \mapsto x \mod (L/2)$, positions.  Bottom: a harmonic mode of the folded volume, $V_T$.  For the subset of modes that are also harmonic with respect to the folded volume, $V_T$, the mode phase phase, $e^{i \mathbf{k} \cdot \mathbf{x}}$, is conserved.  In practice, I fold once along each of the three Cartesian dimensions.}
\label{fig:Jenkins_foldplot}
\end{figure}
\subsubsection{Multipole moments by regression}
Finally, as the DFT provides estimates of the power spectrum for a Cartesian lattice of modes non-linearly and often sparsely spaced in $\mu$, I estimate the multipole moments, $P_{\ell}(k)$, by linear regression.  This is as opposed to approximating the multipole decomposition as a Riemann sum, which one is prevented from doing by the non-linear spacing of modes in $\mu$.  Having binned the available modes in $k$ (with those in a given bin positioned at values $\mu_i$) I assume a truncated Legendre expansion: $P(k, \mu_i) = P_0 + P_2 L_2(\mu_i) + P_4 L_4(\mu_i)$.  This allows the coefficients $(P_0(k), P_2(k), P_4(k))$ to be determined by least-squares minimisation:
\begin{equation}
S = \sum_i \left[|\tilde F (k, \mu_i) |^2 - P_0 - P_2 L_2(\mu_i)  - P_4 L_4(\mu_i) \right]^2.
\end{equation}
At a minimum: $\partial S / \partial P_0 = \partial S /\partial P_2 = \partial S /\partial P_4 = 0$, which may be written as the matrix equation: 
\begin{equation}
\begin{pmatrix}
N & \sum Q_i & \sum H_i \\
\sum Q_i & \sum Q_i^2 & \sum Q_i H_i \\
\sum H_i & \sum Q_i H_i & \sum H_i^2 
\end{pmatrix}
\begin{pmatrix}
P_0 \\
P_2 \\
P_4
\end{pmatrix}
=
\begin{pmatrix}
\sum \ \ \ |\tilde F (k, \mu_i) |^2 \\
\sum Q_i |\tilde F (k, \mu_i) |^2  \\
\sum H_i |\tilde F (k, \mu_i) |^2 
\end{pmatrix}
,
\label{multipoles_byregression}
\end{equation}
and subsequently inverted to obtain $P_0(k)$, $P_2(k)$ and $P_4(k)$.  In this expression, $N$ is the number of modes contained in a given bin, $Q_i \equiv L_2(\mu_i)$, $H_i \equiv L_4(\mu_i)$ and the calculation is repeated for each bin independently.  I have confirmed that the monopole and quadrupole moments obtained when the expansion is truncated at the quadrupole term are indistinguishable from those obtained with a truncation at the hexadecapole.

\subsection{Maximum likelihood}
Having established the machinery required for making a robust estimate of the power spectrum it remains to determine how this measurement constrains the free parameters, $\boldsymbol \theta$, of viable theories, e.g. those with an evolving equation-of-state parameterised by $(w_0, w_a)$.  The best-fitting parameter set will minimise the chi-squared statistic:
\[
\chi^2 = (x_i - \langle x_i \rangle)\ C_{ij}^{-1} \ (x_j - \langle x_j \rangle), 
\]
given the available data and a model for predicting $\langle x (\boldsymbol \theta) \rangle$; here the Einstein summation convention is assumed.  From a Bayesian perspective, this corresponds to the `most probable' or maximum likelihood, $\mathcal{L}$ = $\Pi_i \ p(y_i) \propto \exp(- \chi^2/2)$, solution -- when the data are assumed to be drawn from a joint Gaussian distribution and flat priors are assumed.  Here $y_i$ is the linear combination of $x_i$ required to diagonalise the covariance; the covariance simply applies inverse variance weighting with respect to $y_i$.  See \cite{HeavensStats}, \cite{NorbergPeder} and \S 4 of \href{http://www.roe.ac.uk/japwww/teaching/astrostats/astrostats2012.pdf}{\tt http://www.roe.ac.uk/japwww/teaching/astrostats/astrostats2012.pdf}.

The inverse of the covariance (precision) matrix, $C_{ij}^{-1}$, must be estimated in order to calculate a $\chi^2$.  This may be done analytically on linear scales, providing that the dependence on the specific set of modes used in the regression is correctly accounted for.  I outline this linear theory calculation in the following section; this derivation neglects the Poisson sampling, which is negligible on large scales (i.e. $P(k) \gg \bar n^{-1}$ is assumed).  This calculation serves as a known case to which numerical estimates can be compared.  The covariance used in later chapters is obtained from the realistic mocks described previously as, realistically, the covariance is affected by the survey strategy, non-linearity and galaxy biasing. 

\subsubsection{Linear theory covariance and regression}
\begin{figure}
\subfloat{\centering \includegraphics[width=0.85\linewidth]{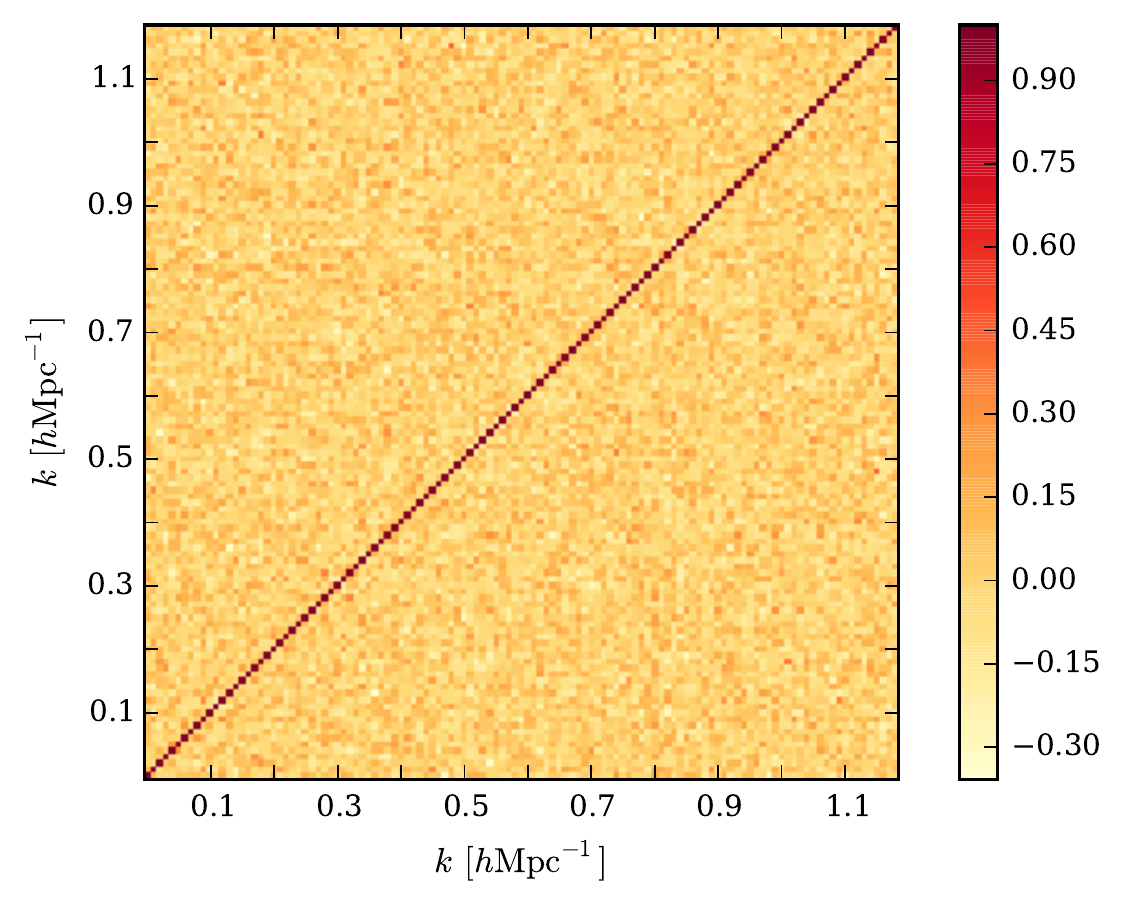}} \\ 
\subfloat{\centering \includegraphics[width=\linewidth]{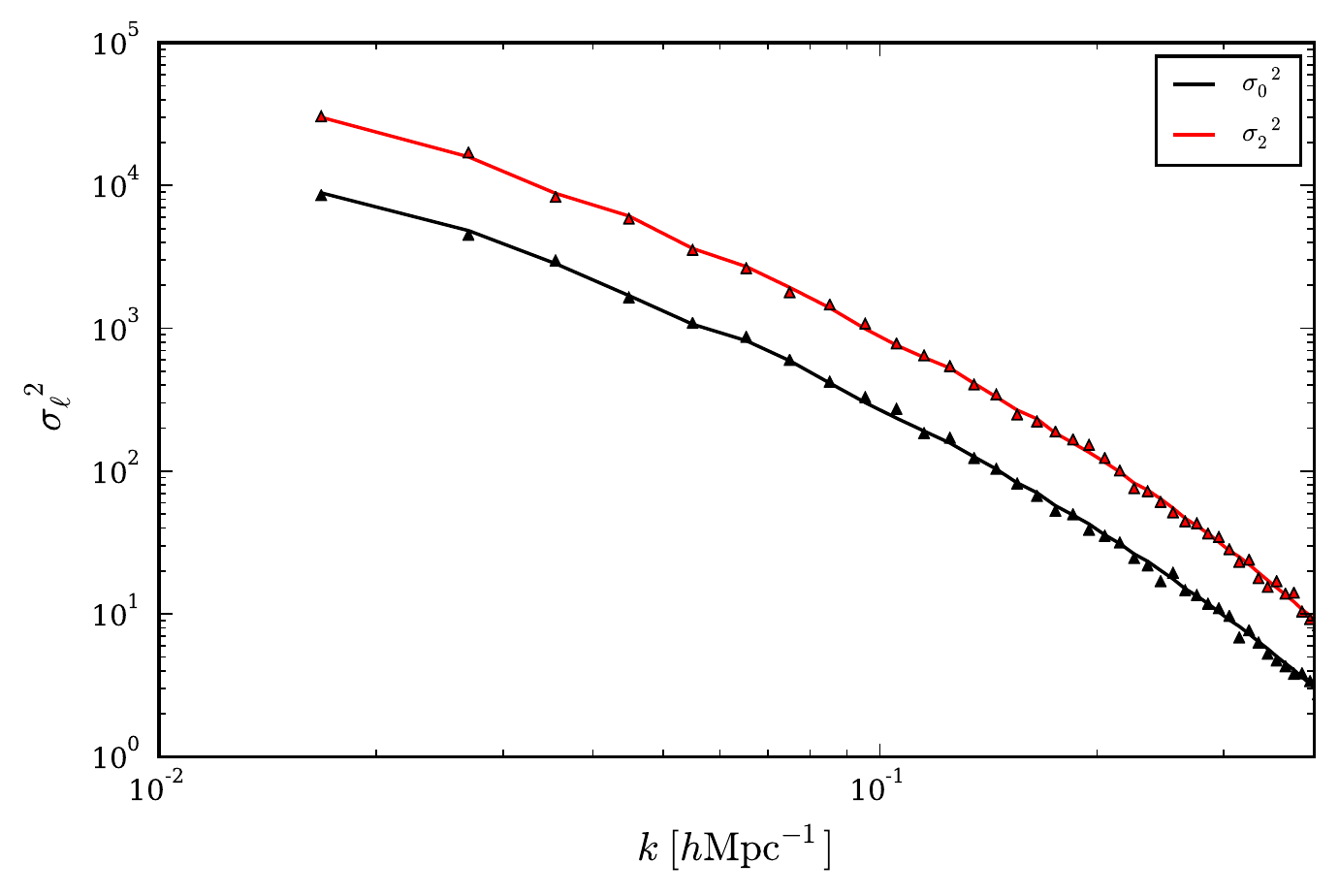}}     
\caption[A correlation matrix estimate for Gaussian random fields.]{Top: estimated correlation matrix of $P_0(k)$ for an ensemble of Gaussian random fields with a known power spectrum.  This correlation matrix should be diagonal, which is the case to within the statistical error.  Bottom: Predicted variances of $(P_0, P_2)$ -- black and red respectively, and their estimates (triangles).  The linear theory variance is dependent on the set of modes in the regression -- in particular, the distribution in $\mu$, as $P(\mathbf{k})$ is amplified  for those closer to $\boldsymbol{\hat \eta}$ and the variance is proportional to the mean.  These tests ensure the covariance estimate is performing well.}
\label{fig:GaussianCovariance}
\end{figure}
This section provides a linear theory calculation of the power spectrum errors, when the estimate is obtained by the regression of a predefined set of Fourier modes.  With this result, I ensured the machinery for estimating the covariance from the mocks was accurate; this same machinery could then be applied with confidence to more realistic simulations.

Neglecting a possible dependency between estimates of the regression parameters, $P_0(k)$ and $P_2(k)$, it follows that $P_0(\{ \tilde F_i \})$.  By further assuming that the realised power, $\tilde F_i(k)$, of each mode is independent -- valid for linear scales, the errors simply add in quadrature:  
\begin{equation}
\sigma_0^2 = \sum_i \left ( \frac{\partial P_0}{\partial |\tilde F_i|^2} \right )^2 \sigma^2_{\tilde F}(k, \mu_i).
\end{equation}
Here $\sigma_0^2$ is the variance of the $P_0$ estimate and $\sigma_{\tilde F}$ is the rms variation in $|\tilde F|^2$, which is given by $\sigma_{\tilde F} = \langle | \tilde F |^2 \rangle \equiv P(k, \mu_i)$ in linear theory.  With the inversion for $(P_0, P_2)^T$ given above, it follows that
\[
\frac{\partial P_0}{\partial |\tilde F_i|^2} = \sum Q_i^2 - Q_i \sum_j Q_j, \qquad 
\frac{\partial P_2}{\partial |\tilde F_i|^2} = - Q_i \sum_j Q_j + N Q_i.
\]
In which case, for $A = N \sum Q_i^2 - \left ( \sum Q_i \right )^2$,
\begin{eqnarray}
{\sigma_0}^2 =& \ \ \ A^{-1} \sum_i P^2(k, \mu_i) \left[ \sum Q_j^2 - Q_i \sum Q_j\right]^{2}, \nonumber \\ 
{\sigma_2}^2 =& A^{-1} \sum_i P^2(k, \mu_i) \left[ - \sum Q_j + N Q_i   \right]^{2}.
\end{eqnarray}
Note that only the $N$ independent modes should be included in this calculation, e.g. the hemisphere with $k_z>0$.  This result may be used to ensure the accuracy of the estimate of the covariance between $P_0$ and $P_2$ on different scales; this is shown in Fig. \ref{fig:GaussianCovariance}.

\subsection{Multipole covariance estimation}
\label{sec:Covariance}
Realistic mocks must be used to obtain a robust estimate of the power spectrum multipole moments as this ensures the effects of non-linearity, galaxy biasing and survey selection are properly accounted for.  Each simulation incorporates the expected statistical error -- resulting from sample (cosmic) variance and Poisson sampling, and therefore a large number are required for a converged estimate.  As the required number increases with the number of fitted data points \citep{TaylorCovariance, PercivalCovariance}, I reduce the latter -- computing the joint likelihood of $\{ P_0(k), P_2(k) \}$ to a given $k_{\rm{max}}$; this is as opposed to fitting the entire independent quadrant of $P(k, \mu)$, which would include the (poorly constrained) higher order multipoles.       

A maximum likelihood analysis of the observed multipoles requires both a model for the expected value, such as those described in Chapter \ref{chap:RSD}, and an estimate of the multipole moment covariance:
\[
C_{ij} = \Bigg \langle \Big ( P_{\ell}(k_i) - \big \langle P_{\ell}(k_i) \big \rangle \Big ) \Big ( P_{\ell}(k_j) - \big \langle P_{\ell}(k_j) \big \rangle \Big ) \Bigg \rangle,
\] 
where $(\ell, \ell') \in \left \{ 0, 2 \right \}$ and $k_i$ denotes the mean $k$ for modes in bin $i$.  This covariance matrix is non-diagonal for three reasons: the survey mask, non-linear structure formation and the multipole moment decomposition.

The survey mask introduces covariances on all scales; in Fourier space, the masked density field is a convolution: $\tilde \delta \mapsto \tilde \delta * \tilde W$, which ensures that the $|\tilde F(\mathbf{k})|^2$ measured for neighbouring modes are given by weighted sums over the same matter fluctuations, $|\tilde \delta(\mathbf{k})|^2$.  Inevitably then, $|\tilde F(\mathbf{k})|^2$ is correlated amongst those modes that are separated in scale by less than the extent of the window, $|\tilde{W}(\mathbf{k})|^2$.  

On large scales, these matter fluctuations evolve independently but this is no longer the case on small scales due to the development of non-linearity for $\delta \simeq 1$. This non-linear mode-coupling leads to a significant covariance between the measured multipole moments on small scales (\citejap{Meiksin}, \citejap{RomanCov}).  Lastly, a covariance between the estimates, $\hat P_0(k_i)$ and $\hat P_2(k_i)$, is introduced by the multipole decomposition, eqn. (\ref{multipoles_byregression}).  As the monopole and quadrupole are separated by weighted sums over $|\tilde F(\mathbf{k})|^2$ -- the observed values of which correspond to a given statistical realisation, there will inevitably be cross-talk between the derived values for each.  

Although analytic approximations exist for the covariance, or trispectrum, introduced by the onset of non-linearity (see \cite{HaloModelReview} and references therein) it is difficult to also include the effects of the selection and survey mask analytically.  As a result, it is necessary to simulate the observed volume for a number of realisations, which possess the expected sampling variance and shot noise, and subject these to the survey strategy. Following this, the covariance may be obtained by analysing each mock individually and calculating the correlation between the estimates. 

I measure $P_{\ell}(k)$ for each of the 306 W1 \& W4 VIPERS v7 mocks; the correlation matrix of these measurements is shown in Fig. \ref{fig:MultipoleCovariance}.  An increased covariance on small scales, attributable to non-linearity, is clearly apparent together with that on large scales -- due to the survey mask.  The multipole moment estimates can also be seen to be correlated on the same scale.  The statistical error of this estimated covariance is likely to be small with this large number of mocks and hence I do not consider a shrinkage estimator \citep{PopeSzapudi} or a Hartlap correction \citep{Hartlap}.  However, I ensure that the conclusions of Chapter \ref{chap:VIPERS_RSD} are robust to the covariance matrix estimate.
\begin{figure}
\centering
\includegraphics[width=0.85\linewidth]{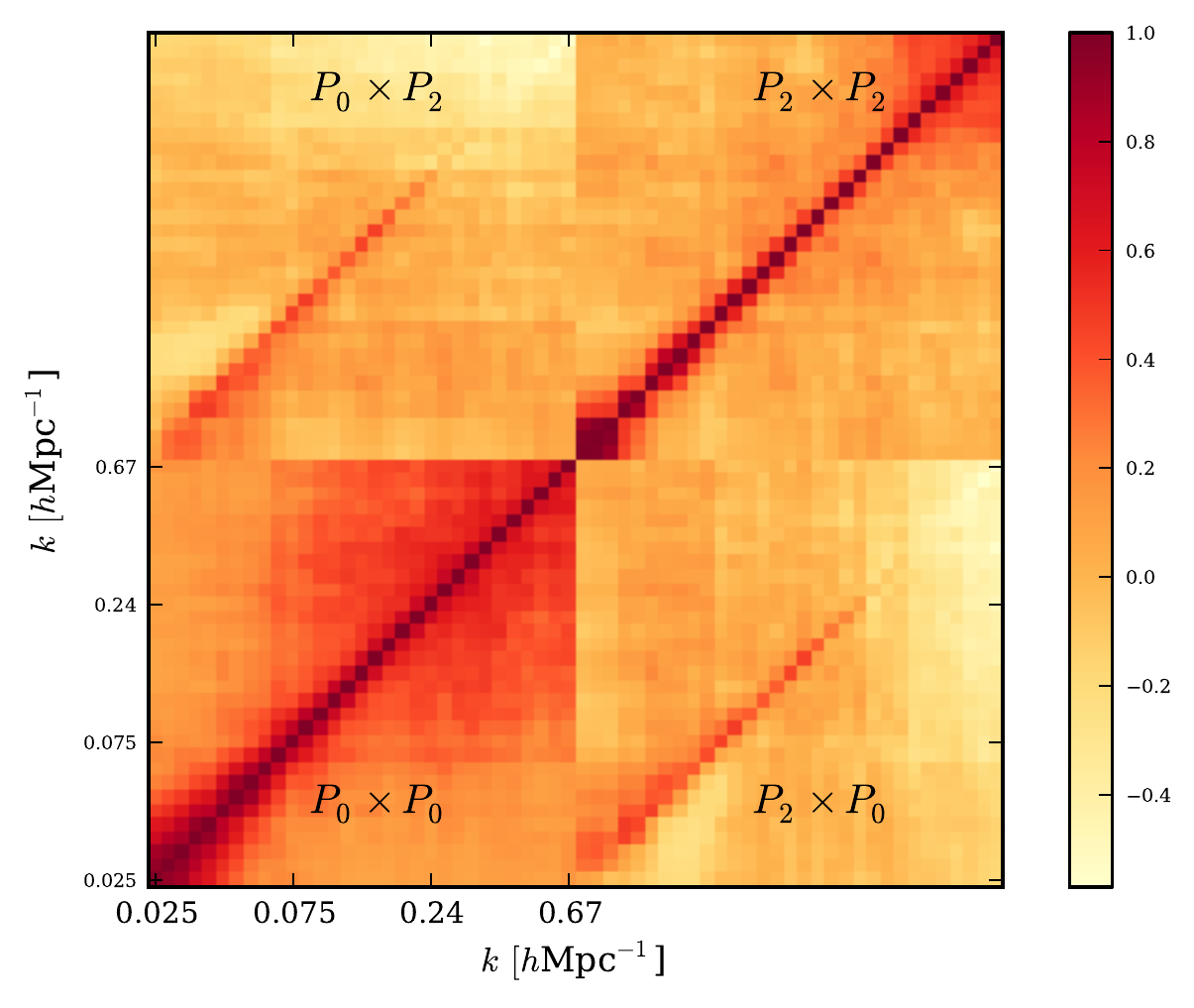}
\caption[A correlation matrix estimate for realistic VIPERS mocks.]{Estimated correlation matrix for the $P_0(k)$ and $P_2(k)$ moments.  This is obtained from the $306$ VIPERS v7 mocks previously described in the text.  An increased covariance on small scales, attributable to non-linearity, is clearly apparent, together with that on large scales -- due to the survey mask convolution (see Chapter \ref{chap:maskedRSD}).  The multipole moment estimates on the same scale can also be seen to be correlated.  If this covariance is not been  accounted for the maximum likelihood parameters and errors would be biased; note that this analysis follows the common practice of neglecting the model dependence of the covariance.  See \cite{Covariance_cosmologyDependence} for greater detail.}
\label{fig:MultipoleCovariance}
\end{figure}
\end{chapter}
\begin{chapter}{The VIPERS galaxy redshift survey}
\label{chap:VIPERS}
This chapter provides a detailed account of the VIMOS Public Extragalactic Redshift Survey (VIPERS), which provides the $z \simeq 0.8$ galaxy sample used to test gravity in Chapter \ref{chap:VIPERS_RSD}.  This discussion focuses on the characteristics of the survey relevant to the accuracy of clustering estimates, e.g. the criteria for spectroscopic selection and the angular mask, and details the methods used to minimise any resulting biases.  Further discussion of VIPERS may be found in \cite{Vipers} and \cite{sylvainClustering}.
\begin{figure}
\centering
\subfloat{\includegraphics[height=1.5\linewidth]{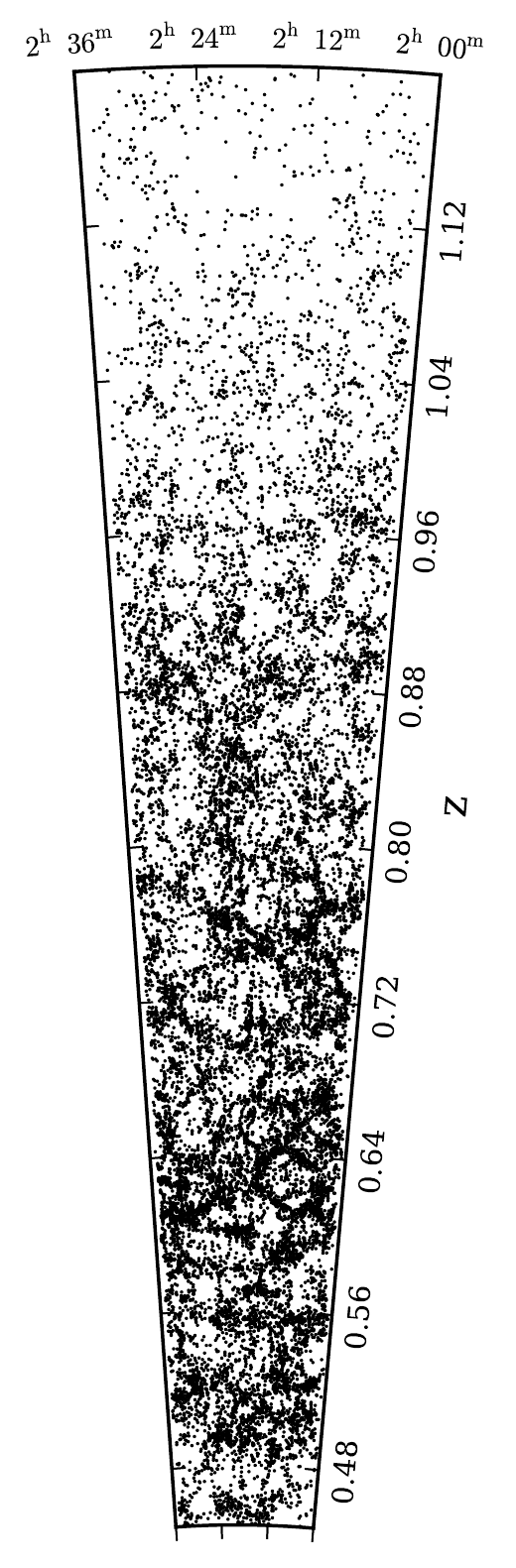}} 
\subfloat{\includegraphics[height=1.5\linewidth]{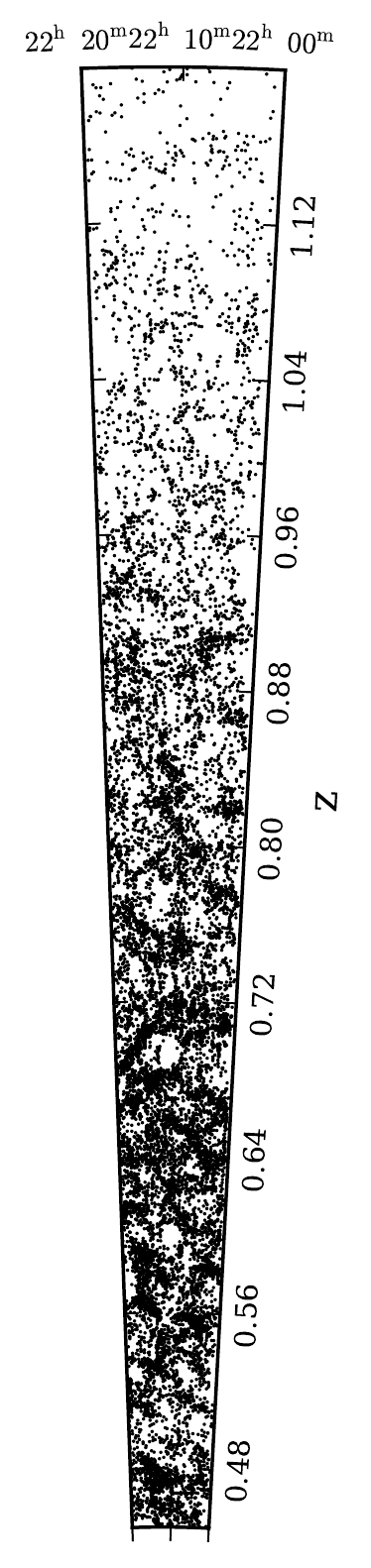}}
\caption[Light-cones for the VIPERS W1 and W4 fields.]{Light-cone plots for the VIPERS W1 and W4 fields.  These comprise a total of 88,901 galaxies with a median redshift:$\bar z = 0.8$.  Clearly apparent is the significantly reduced number density beyond $z \simeq 0.96$; only the most luminous sources meet the apparent magnitude limit at this distance.  This sample of large-scale structure is used to test gravity on cosmological scales with RSD in Chapter \ref{chap:VIPERS_RSD}.  To do so requires the affect of the survey selection on the power spectrum to be very well understood.} 
\label{fig:lightcone}
\end{figure}

\section{Survey details}
VIPERS is a recently completed spectroscopic survey comprising 88,901 galaxies selected from the optical photometry of the \href{http://terapix.iap.fr/cplt/T0006-doc.pdf}{Canada-France-Hawaii Telescope Legacy Survey-Wide} (CFHTLS-Wide).  The excellent seeing quality of this dataset and large number of photometric bands allows for efficient star removal, such that the stellar contamination is only $\simeq 3 \%$.  With an apparent magnitude limit of $i'_{AB}< 22.5$, VIPERS spans a total of 24 deg$^2$ divided between the W1 \& W4 CFHTLS fields.  Those galaxies selected for spectroscopy have satisfied a simple, yet robust, $u g r i$ colour-colour selection that removes $z<0.5$ interlopers while remaining $98 \%$ complete for $z>0.6$; see Fig. 5 of \cite{Vipers} for the completeness-redshift relation.  This selection criterion has been calibrated using pre-existing spectroscopic redshifts from the partially overlapping VVDS-Wide survey (Fig. 3 of \citejap{Vipers}).  Due to this informed colour selection, VIPERS galaxies largely occupy the redshift range: $0.6 < z < 1.2$; under half of a purely magnitude-limited sample would lie in this range.  

The survey volume, $5 \times 10^7 (\mpcoh)^3$, ensures that VIPERS replicates the statistical significance attained by the Two-degree Field Galaxy Redshift Survey (\citejap{Colless1}, \citejap{Colless2}) but at a significantly higher median redshift, $ \langle z \rangle =0.8$.
\begin{figure}
\centering
\includegraphics[width=0.75\textwidth]{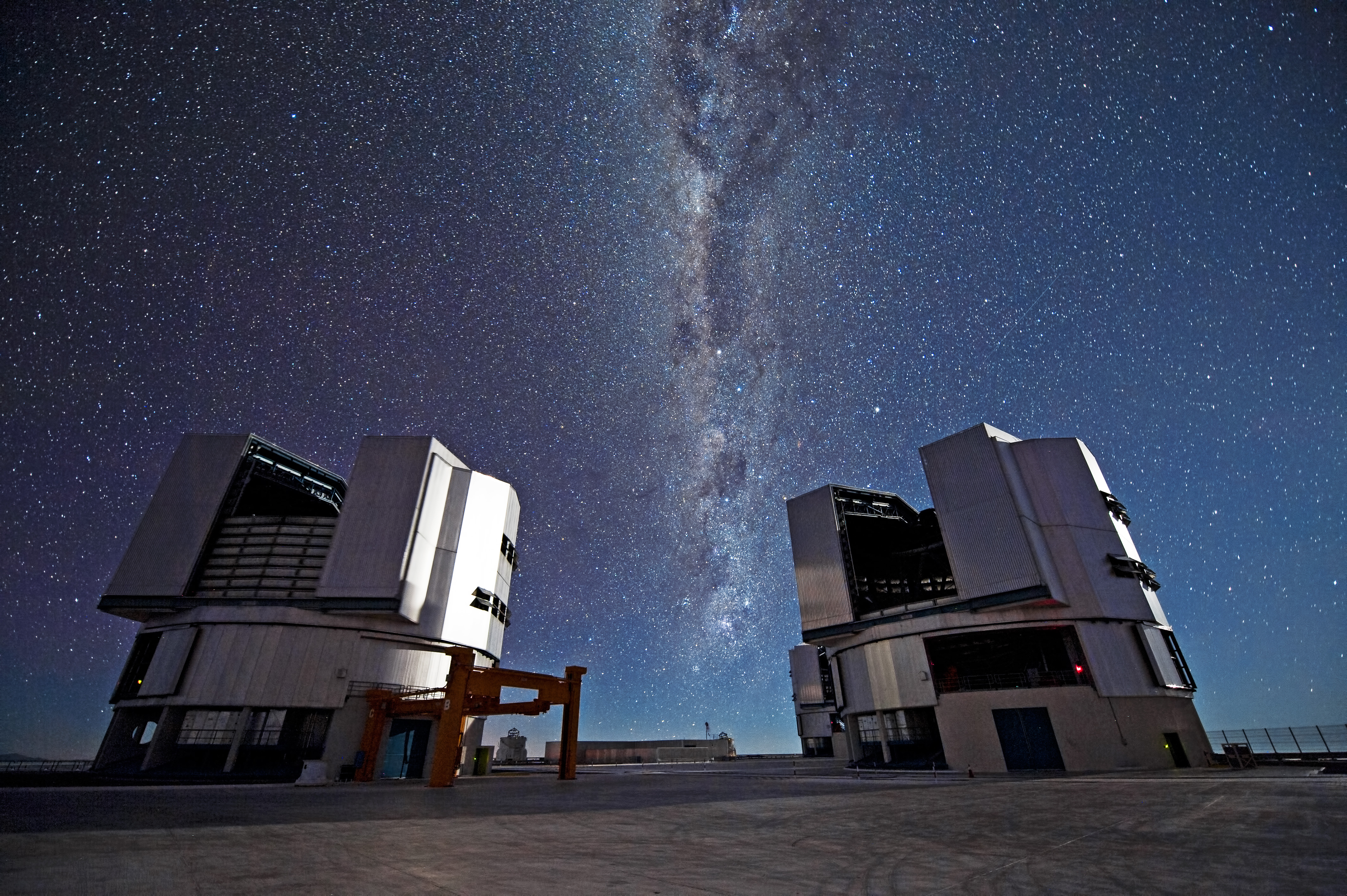}
\caption[The ESO Very Large Telescope.]{The ESO Very Large Telescope (VLT): the VIMOS spectrograph at the VLT provides the spectra necessary for determining the radial position (in redshift space) of a sample of $\simeq 90, 000$ galaxies.} 
\label{fig:VLT}
\end{figure}
The VIPERS spectra were collected by the VIMOS multi-object spectrograph at the ESO Very Large Telescope, which is shown in Fig. \ref{fig:VLT}, and have a wavelength coverage of $5500-9500$\r{A} at moderate resolution, $R=210$.  Secure redshifts obtained from these spectra are accompanied by a confidence flag.  This analysis includes only those satisfy $(2 \leq f_z < 10) \ | \ (12 \leq f_z < 20)$; these are secure at the $98 \%$ confidence level and comprise 64\% of the total redshifts.  The typical redshift error of this sample is $\sigma_z = 4.7(1+z) \times 10^{-4}$.  This estimate is obtained from a limited number of galaxies that were reobserved due to overlapping pointings.  In this work, I analyse the Nagoya v7 dataset, which will subsequently become available at \href{http://vipers.inaf.it}{\tt{http://vipers.inaf.it}}. 
\begin{figure}
  \begin{adjustbox}{addcode={\begin{minipage}{\width}}{
      \caption[The VIPERS v7 angular footprint.]{The angular footprint of the Nagoya v7 \& Samhain mask relevant to this work.  The VIMOS spectrograph is comprised of four quadrants separated by a central cross in which spectra cannot be obtained.  The surveyed area is overlaid by gaps in the coverage as a result.  There are further gaps due to failed pointings and imperfections in the parent photometry -- including stellar masks and corrupted or missing bands.     
      }
      \label{fig:VIPERS_footprint}
      \end{minipage}}, rotate=90, center}
      \includegraphics[scale=1.3]{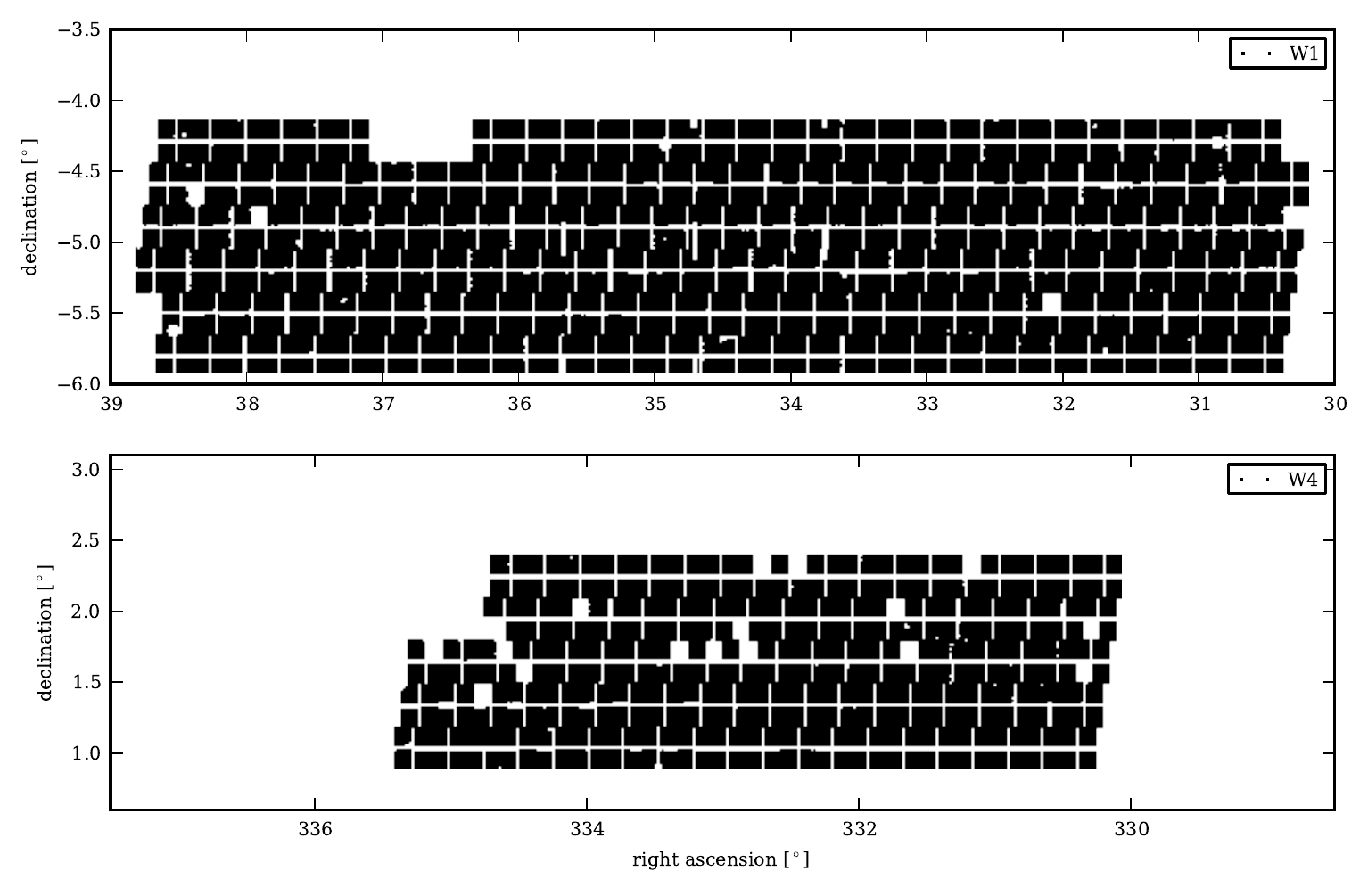}
  \end{adjustbox}
\end{figure}
\label{sec:footprint}
\subsection{Angular selection function}
The design of the VIMOS spectrograph gives VIPERS a distinct angular footprint across the sky; this is shown in Fig. \ref{fig:VIPERS_footprint}.  The surveyed area is comprised of multiple rows of pointings of a single VLT unit, each of which surveys a $\simeq 218$ arcmin$^2$ area divided between four quadrants which are separated by a central cross in which spectra cannot be obtained.  
\begin{figure}
\centering
\includegraphics[width=0.75\textwidth]{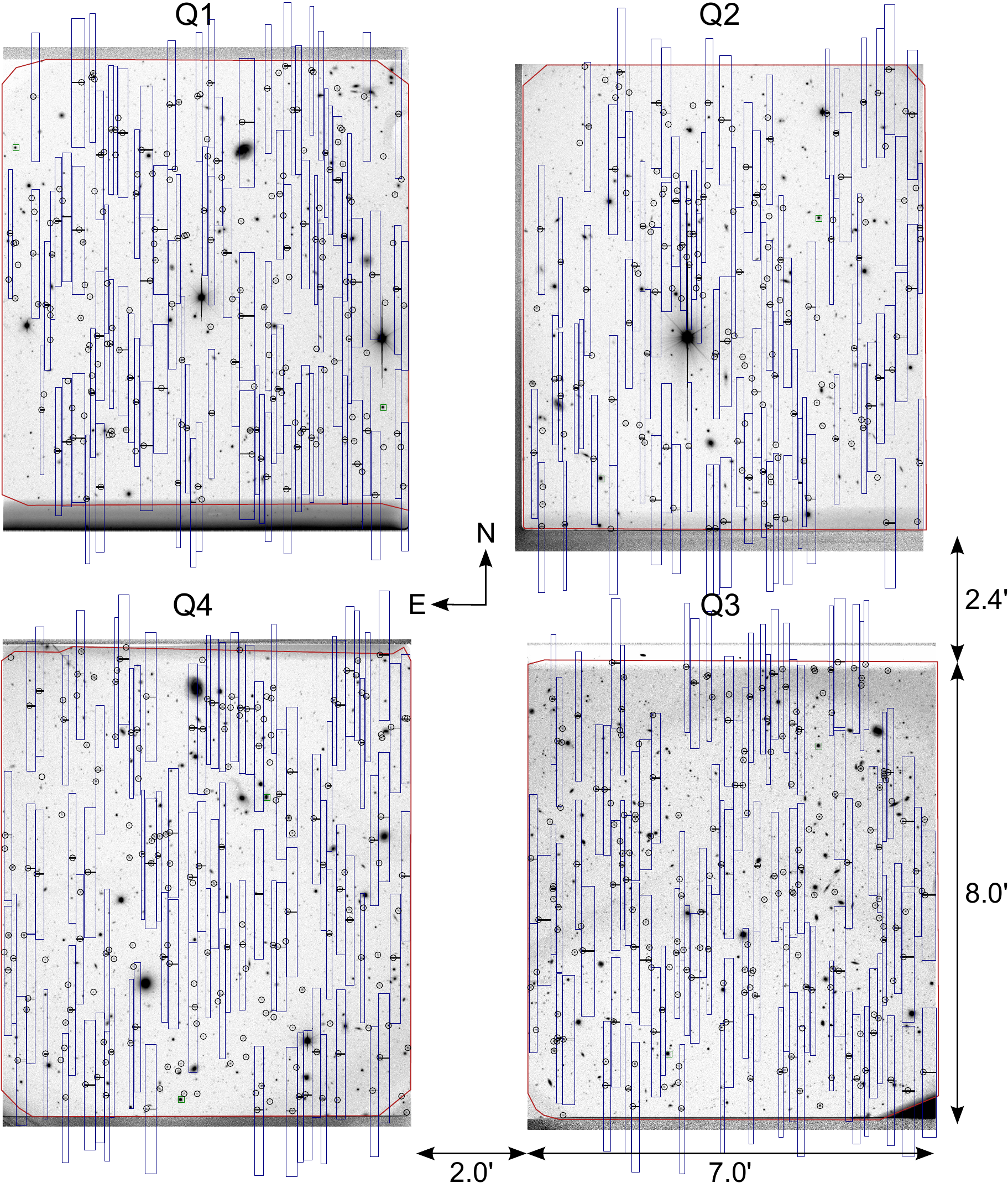}
\caption[A graphic illustrating the VIPERS slit assignment.]{A single pointing of the VLT is comprised of four quadrants, which are separated by a central cross in which spectra cannot be obtained.  An example area of the parent photometry (CFHTLS-WIDE) is overlaid across each.  The spectroscopic slits assigned by the SPOC algorithm \citep{Bottini} can be seen in each case.  The spectra are prevented from overlapping to avoid confusion in the redshift estimation.  As spectra are dispersed in declination, this results in the prominent column alignments seen in the $\simeq 400$ assigned slits.  In particular, the effective sampling rate will be low where the projected density is high. The projected density of the spectroscopic galaxies is effectively homogeneous as a result.  This density dependent selection may bias clustering estimates if not properly corrected for.} 
\label{fig:slits}
\end{figure}
As the survey is completed in a single pass -- to maximise the volume surveyed, these gaps overlay the entirety of the sky coverage.  The resulting footprint is given by the  Nagoya v7 angular mask \citep{Vipers}.  In addition to this is a second mask, `Samhain', which accounts for imperfections in the parent photometry stemming from the removal of the areas dominated by diffraction patterns around bright stars and extended extragalactic sources in particular.  The final angular mask for this analysis corresponds to the union of the Nagoya v7 and Samhain masks. 

The restricted number of slits per VIMOS quadrant requires the selection of a subsample of galaxies in the parent photometry.  The algorithm used for this selection \citep[SPOC, ][]{Bottini} maximises the number of slits that may be simultaneously placed on targets.  The rate of spectroscopic selection (SSR) is $\simeq 40 \%$ on average, but is highly non-random.  The VIPERS spectra are dispersed along the declination direction and consequently slits are more widely separated in declination (to prevent overlapping).  Therefore the selection rate drops significantly in  areas of large (projected density).  This angular anisotropy and density dependence has the potential to significantly bias the power spectrum measurement and I ensure this can be robustly corrected for in Chapter \ref{chap:VIPERS_RSD}.

\begin{figure}
\centering
\includegraphics[width=0.75\textwidth]{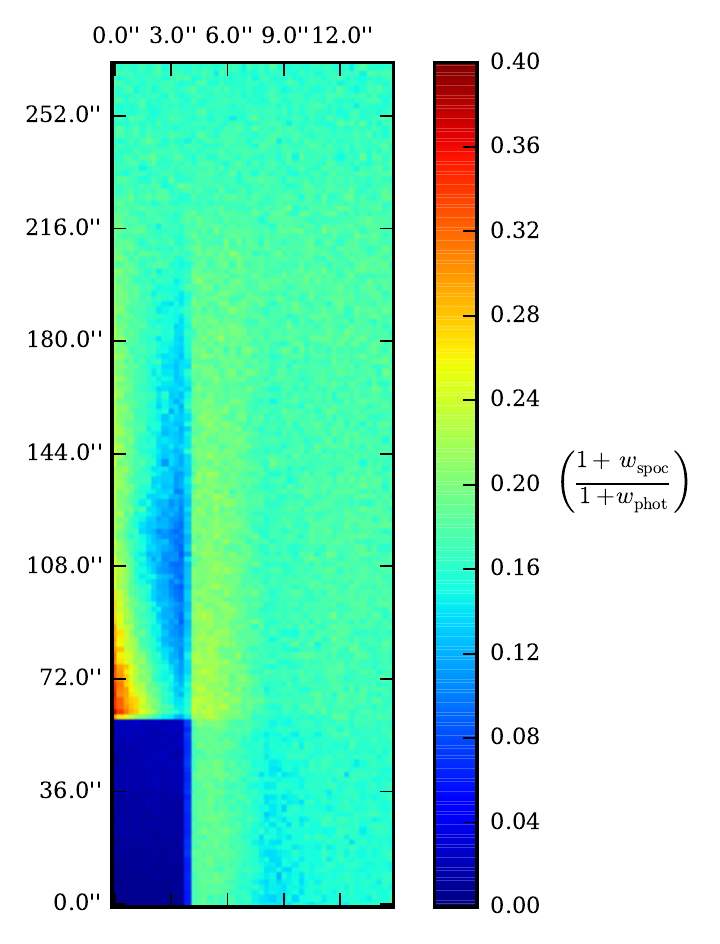}
\caption[Dependence of the angular clustering on the VIPERS selection.]{This figure shows the ratio of the pair counts: $1+w(\theta)$, where $w(\theta)$ is the angular correlation function, measured in the parent and spectroscopic mock catalogues.  The two differ as the latter is selected from the photometry by a slit assignment algorithm (SPOC) described in the text.  There is a hard exclusion zone as SPOC prevents the overlap of spectra; this is clearly apparent on the bottom left.  This region roughly corresponds to the maximum length achievable with a pair of just-touching slits in any direction.  This effect is analagous to the fibre collisions in the BOSS survey; see \cite{FiberCollision} and \cite{Beutler}.  There is an enhanced probability just above this zone to compensate for this, which results in the prominent column arrangement highlighted earlier.  Clearly there is significant anisotropy and an annuli averaged weight, of the type advocated in \cite{sylvainClustering} -- e.g. Fig. 3, is far from ideal.  Moreover, this shows the selection to vary on a scale much smaller than a VIMOS quadrant.  I therefore apply a local weighting scheme based on a slit-like ($180'' \times 3''$) shape.} 
\label{fig:spoc_wtheta_ratio}
\end{figure}
I correct for the SPOC selection by upweighting the spectroscopic galaxies in the density field estimation; the alternative of downweighting the expected number is excluded by the density dependence of the selection.  I therefore define a local target sampling rate (TSR) for each spectroscopic galaxy of 
\[
\label{eqn:TSR}
T(\boldsymbol{\theta}_{\rm{spec}}) = \left \langle \frac{\delta_{\text{spec}}}{\delta_{\text{phot}}} \right \rangle.
\] 
Here $\delta_{\text{phot}}(\boldsymbol \theta)$ is an estimate of the projected overdensity, which is obtained by a Delaunay Tessellation (\citejap{DelaunayTessellation}, \citejap{DTFE}) of the target population (in the parent photometry); similarly, $\delta_{\text{spec}}(\boldsymbol \theta)$ is obtained by using the spectroscopic sample as the point set.  Delaunay Tessellation is employed to further reduce the statistical noise in low density regions.  An average of this ratio over the effective `shadow' of the spectra -- approximately a $180'' \times 3''$ rectangle (as shown in Fig. \ref{fig:spoc_wtheta_ratio}), is then computed for each spectroscopic galaxy.  These weights more accurately account for the local density dependence of the sampling than the quadrant based approach advocated in \cite{sylvainClustering}.  Note that no knowledge of the photometric redshift is used when calculating the weighting; this choice is proven to be sufficient in Chapter~\ref{chap:VIPERS_RSD}.  

\label{sec:ESR}
Further incompleteness results from an inability to assign a secure redshift to a given spectra -- due to poor observing conditions or the lack of strong emission lines (in particular OII), and is especially important for the faint galaxies that just meet the apparent magnitude limit.  The spectroscopic success rate (SSR) is empirically defined as the ratio of the number of redshifts assigned to number of spectra observed (on a per quadrant basis).  The product of the target sampling rate (TSR) and the spectroscopic success rate (SSR) defines an effective sampling rate (ESR), which is denoted by $E(\boldsymbol \theta)$.  

\subsection{Radial selection function} 
\label{sec:radialSelection}
Both the imposed colour selection (CSR) and apparent magnitude limit have significant influence in determining the completeness-corrected $\overline{n}(z)$; this is shown in Fig. \ref{fig:nbar}.  There is a strong suppression of the number density for $z \leq 0.5$ due to the colour selection (as intended) while at higher redshift only the most luminous sources can be observed.
\begin{figure}
\centering
\includegraphics[width=\textwidth]{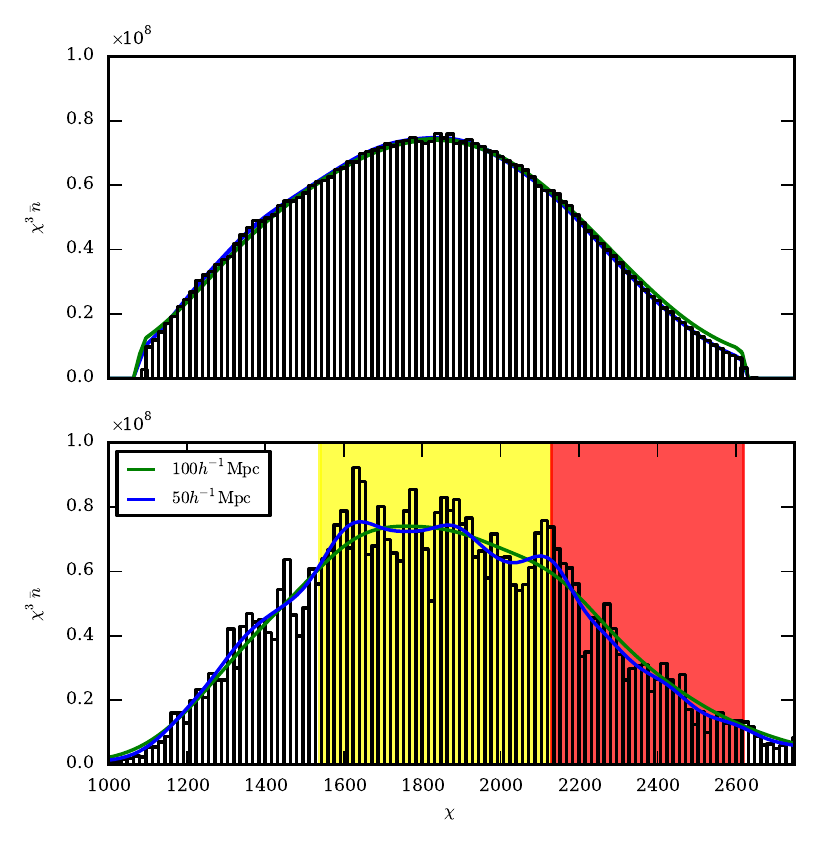}
\caption[The VIPERS v7 radial selection function.]{The VIPERS v7 radial selection function.  Top: mock average of the ESR-corrected $\bar n(z)$, prior to and following smoothing with kernels of $100$ and $50 \mpcoh$ (green and blue respectively).  The low-$z$ slice (yellow) has a greater sample variance and requires a larger $\simeq 100 \mpcoh$ kernel.  I find a kernel of this size biases the estimate at large redshift.  There is greater volume at high redshift however and a finer $50 \mpcoh$ suffices. Statistical errors in the $\bar n(z)$ estimate are reflected in the multipole moment covariance and hence propagated to the derived parameter constraints.  Bottom: similar format but for the VIPERS v7 data.  Subsequent chapters will show results for a partition of the surveyed volume into four subvolumes.  These are defined by the permutations of the low and high $z$ slices, $0.6<z<0.9$ and $0.9<z<1.2$ respectively, and the two fields, W1 \& W4.} 
\label{fig:nbar}
\end{figure}
As the surveyed area is relatively modest, it is crucial that the radial selection be separated from the intrinsic radial clustering.  Failing to do so will bias the power spectrum estimate on large scales.  

I separate the radial selection from the intrinsic clustering with a two-fold approach: firstly, the contribution of intrinsic clustering is reduced by estimating a joint-field average of $\overline{n}(z)$.  I subsequently smooth this estimate with a Gaussian, assuming the radial selection to be constant on scales smaller than the kernel.  This kernel is reflected at the observed redshift boundaries when smoothing, to ensure the tails are not biased by the initial zero padding.  The low-$z$ slice (yellow) has a smaller volume and is therefore more sensitive to sample variance.  A relatively large kernel of $100 \mpcoh$ is appropriate in this case.  I find that a kernel of this size introduces a bias in the $\bar n(z)$ estimate at high$z$ in the mocks; this is shown by the top panel.  But the sample variance is reduced for the larger volume  high$-z$ slice (red) and I find $50 \mpcoh$ to be sufficient.  Statistical errors in the $\bar n$ estimate will be reflected in the power spectrum covariance and therefore propagated to the parameter constraints.  

I assume a given subvolume to be a fair sample and normalise the smoothed $\bar n(z)$ such that $\int \bar n \ d^3x = N_E$, where $N_E$ is the ESR-corrected counts in a given subvolume -- specified by a field and redshift slice.  This imposes an integral constraint correction that is corrected for in the modelling.  This correction is detailed in Chapter \ref{chap:maskedRSD}; I also outline how the radial smoothing and joint-field estimate should be included in this correction.  The bottom panel shows the ESR-corrected $\bar n(z)$ for the data prior to and following smoothing.  A consequence of the joint-field $\bar n(z)$ estimate is that a covariance between the power spectrum measured in each field is introduced.  I detail an investigation of the sensitivity of the results to the $\bar n$ estimate in Chapter~\ref{chap:VIPERS_RSD}.

\subsection{Complementary VIPERS analyses}
\label{sec:CompVIPS}
Together with investigating galaxy formation, evolution and biasing, constraining the linear growth rate with redshift-space distortions is a central goal of the VIPERS survey.  As a result there are, and have been, numerous VIPERS RSD analyses that possess both important similarities and differences.  In particular, the complimentary analysis of \cite{sylvainClustering} finds
\[
f \sigma_8(0.8) = 0.47 \pm 0.08,
\]
based on the correlation function of the VIPERS PDR-1 dataset \citep{Vipers}.  A goal of my own analysis is to ensure that a Fourier-based analysis of the VIPERS v7 dataset yields consistent results.  
This consistency provides a reassurance that systematic effects are well controlled in both analyses, as the $\bar n(z)$ estimate, survey mask and non-linearity affect the power spectrum and correlation function in distinct ways.  The conclusions of my own analysis are compared to those of \cite{sylvainClustering} in Chapter \ref{chap:VIPERS_RSD}. 

In addition to \cite{sylvainClustering}, Pezzotta et al. (in prep.) and Rota et al. (in prep.) perform clustering analyses of the v7 data used in this work, in configuration and Fourier space respectively.  The former has done extensive work in confirming the accuracy of the ESR weights used in this analysis (in addition to that detailed in Chapter \ref{chap:VIPERS_RSD}).  As an independent power spectrum analysis, which constrains the $\Lambda$CDM parameters held fixed in this analysis, comparison with Rota et al. has allowed numerous tests to be carried out that ensure the robustness of both analyses.
\end{chapter}
\begin{chapter}{Masked redshift-space distortions}
\label{chap:maskedRSD}
In this work I reformulate the forward modelling of the redshift-space power spectrum multipole moments for a masked density field, as encountered in galaxy redshift surveys.  Exploiting the symmetries of the redshift-space correlation function, I provide a masked-field generalisation of the Hankel transform relation between the multipole moments in real and Fourier space.  Using this result, I detail how a likelihood analysis requiring computation for a broad range of desired $P(k)$ models may be executed $10^3-10^4$ times faster than with other common approaches, together with significant gains in spectral resolution.  I present a concrete application to the complex angular geometry of the VIPERS PDR-1 release and discuss the validity of this technique for finite-angle surveys.  This work is reproduced from \cite{maskedRSD}, which is the more polished and complete work.
Sections \S \ref{sec:smoothed_icc} and \S \ref{sec:ap_masking} are original and will comprise a section of Wilson et al. (in prep.)

\section{Introduction}
\label{sec:intro}
For a Fourier-based analysis of a galaxy redshift survey, the imprint of the survey geometry is commonly the largest systematic difference between the observed large-scale power spectrum and that predicted by fundamental physics.  This difference is of increased importance in a redshift-space distortions (RSD) analysis as the principal observable is the power spectrum anisotropy, which is equally sensitive to the density field and survey mask.  The density field appears anisotropic when inferred from observed redshifts as the radial component of the peculiar velocity field introduces an additional Doppler shift.  These peculiar velocities are a consequence of the formation of large-scale structure via gravitational collapse and the magnitude of this effect is therefore dependent on the effective strength of gravity on cosmological scales \citep{Guzzo}.  A measurement of the observed anisotropy by future surveys such as Euclid \citep{Euclid} will therefore provide a stringent test of modified gravity theories.

The goal of such surveys is to achieve a statistical error of $\simeq 1 \%$ on the logarithmic growth rate of density fluctuations, which requires systematic uncertainties to be very well understood.  This work proposes a new approach for the forward modelling of the systematic change in the galaxy power spectrum due to the survey mask.  Because statistical noise is amplified by deconvolution, a forward modelling of this effect as part of a likelihood analysis is the logical approach; but with a broad range of theoretical models to consider it is important to do so efficiently.  I show that the symmetries of the redshift-space correlation function make it possible to do so in a manner that offers a greater physical insight, together with significant gains in speed and resolution when compared to other common methods.
 
In the usual approach, the density field as observed by a redshift survey is `the infinite sea of density fluctuations' multiplied by a mask $W(\mathbf{x})$ that accounts both for the survey geometry and a local weighting, which may include incompleteness corrections or FKP weights \citep{FKP}:
\[
\delta(\mathbf{x}) \mapsto \delta(\mathbf{x}) \ W(\mathbf{x}).
\]
This multiplication in configuration space results in a convolution in Fourier Space: $\tilde \delta' (\mathbf{k}) = \tilde \delta (\mathbf{k}) * \ \tilde W(\mathbf{k})$ which, in the absence of phase correlations between the density field and the mask, is also true of the observed power \citep{PeacockNicholson}:
\[
\label{P_cnvld}
P'(\mathbf{k}) = \int  \frac{d^3 q}{(2 \pi)^3}  \ P(\mathbf{k} - \mathbf{q}) \ | \tilde W(\mathbf{q})|^2.
% \langle |\tilde \delta ' (\mathbf{k})|^2 \rangle
\]
Different Fourier modes are uncorrelated due to statistical homogeneity: $\langle \tilde \delta(\mathbf{k}) \tilde \delta^*(\mathbf{k'}) \rangle = (2 \pi)^3 \delta^3(\mathbf{k} - \mathbf{k}') \, P(\mathbf{k})$ is diagonal but this is no longer true of $\langle \tilde \delta'(\mathbf{k}) \tilde \delta' {}^*(\mathbf{k'}) \rangle $ \citep{Hamilton06}.  However, the diagonal term is the quantity that contains the principal cosmological information and it is the systematic change in shape of this function that I seek to calculate.  I adopt a convention in which $P(k)$ has units of volume and exploit the independence of $P'(\mathbf{k})$ on the phases of the density field by presenting various tests based on Gaussian random fields.  

This convolution alters both the amplitude and shape of the observed power spectrum with respect to that of the true field.  I assume the amplitude of the observed power has been suitably corrected, 
\[
P^{\rm{obs}}(\mathbf{k}) \mapsto  \left(  \int d^3 x  \ W^2(\mathbf{x}) \right)^{-1} P^{\rm{obs}}(\mathbf{k}),
\label{ampcorr}
\]
and address solely the change of shape in the forward modelling.  Here $P^{\rm{obs}}(k)$ differs from $P'(k)$ due to the integral constraint correction discussed in \S\ref{sec:intcor}.

As first described by \cite{Kaiser}, $P(\mathbf{k})$ is anisotropic about the line-of-sight when the radial comoving position of a galaxy is inferred from a measured redshift.  On large scales, this anisotropy is dependent on the infall (outflow) rate of galaxies into (out of) over (under) densities and hence is sensitive to the strength and therefore theory of gravity on cosmological scales \citep{Guzzo}.  With an additional large-$k$ suppression due to the virialised motions of galaxies in groups and clusters, a commonly assumed model for the power spectrum is the dispersion model, which combines the Kaiser anisotropy factor with a `fingers-of-God' damping:
\[
P(\mathbf{k}) = \frac{(1 + \beta \mu^2)^2}{1 + \frac{1}{2} k^2 \sigma_p^2 \mu^2} P_{\text{g}}(k).
\]
Here $P_g(k)$ is the real-space galaxy spectrum, $\mu = \mathbf{\hat{k}} \cdot \hat{\boldsymbol \eta}$ for a unit vector $\hat{\boldsymbol \eta}$ lying along the line-of-sight, $\sigma_p$ is an empirical pairwise dispersion (in this case for a Lorentzian damping model) and $\beta$ is the ratio of the logarithmic growth rate of density fluctuations to the linear galaxy bias.  For an outline of the approximations underlying this model see \cite{Cole94} and \cite{Cole}.  Taking $\hat{\boldsymbol \eta}$ as the polar axis, the azimuthal symmetry and $\mu^2$ dependence of RSD allows $P(\mathbf{k})$ to be distilled into a series of multipoles of even order in $\ell$:
\[
\label{legenSeries}
P(\mathbf{k}) = \sum_{\ell =0}^{\infty} P_{\ell}(k) L_{\ell}(\mu).
\]
Here $L_{\ell}$ is a Legendre polynomial of order $\ell$; these form a complete basis for $-1 \leq \mu \leq 1$.  The monopole and quadrupole modes are given by $L_0 = 1$ and $L_2 = \frac{1}{2} (3\mu^2 -1)$ respectively.  

Although I have presented the symmetries I exploit in the context of the dispersion model they are inherent to all models that assume the validity of the `distant observer' approximation -- when the variation of line-of-sight across the survey is neglected and subsequently both $\hat{\boldsymbol \eta}$ and $\mu$ are well defined.  However, this is only valid for surveys of relatively small solid angle or for RSD analyses that are restricted to pairs of small angular separation; I comment on the validity of this approach for finite-angle surveys in \S\ref{wide-angle}.

Despite being a physically well-motivated approximation, this dispersion model fails to incorporate the more subtle aspects of RSD.  More developed models that apply appropriate corrections include those by \cite{Scoccimarro} and \cite{Taruya} -- see \citejap{sylvainModels} for further details.  The former relaxes the assumption of the linear theory relations between the overdensity and velocity divergence fields on the largest scales surveyed, which yields an effective Kaiser factor that is typically calibrated with numerical simulations \citep{Jennings}.  However, the ansatz proposed by Scoccimarro continues to neglect the physical origin of the fingers-of-God damping -- the virialised motions of galaxies are sourced by the same velocity field responsible for linear RSD.  \cite{Taruya} apply further corrections that more accurately account for the correlation of this non-linear suppression with the velocity divergence field.  Despite these shortcomings, the dispersion model is sufficient for illustration as these more developed models continue to satisfy the symmetries I exploit.

It is necessary to be able to compute the multipole moments for the assumed RSD model in order to implement this approach.  These are quoted for the Kaiser-Lorentzian model in Chapter \ref{chap:RSD}.  In the following section I present the main result of this chapter: a masked-field generalisation of the known Hankel transform relation between the multipole moments in real and Fourier space, which allows for the rapid prediction of $P'_{\ell}(k)$ by 1D FFT.  This may be used to correct the one-loop power spectrum for example, which has recently been shown to be rapidly computable using similar means \citep{Schmittfull}.

\section{Power spectrum multipoles for a masked density field}
\subsubsection{Outline of the method}
\label{sec:multipoles}
I start with the required convolution:
\[
\label{convo}
P' (\mathbf{k}) =   \ \int  \frac{ d^3 q}{(2 \pi)^3} \ P(\mathbf{q}) \ |\tilde W(\mathbf{k} - \mathbf{q})|^2 . 
\]
The simplest methods to evaluate this integral are by approximating it as a Riemann sum or by the application of the convolution theorem and 3D FFTs.  These approaches share a number of disadvantages for inclusion in a likelihood analysis (especially in the common case of a pencil-beam geometry): 
\begin{enumerate}
\item The broad extent of the $|\tilde W(\mathbf{k})|^{2}$ kernel when the mask is narrow along one or more dimensions in real space, such as for the pencil beam geometry common to $z \simeq 1$ surveys.  In this case a Riemann sum is prohibitively slow as the addition of a large number of non-negligible terms is required.  This calculation must then be repeated for each mode for which $P'(\mathbf{k})$ is desired.
\item Systematic errors introduced by the use of a FFT for the estimation of $|\tilde W(\mathbf{k})|^{2}$ -- not least due to memory limited resolution and the subsequent aliasing effect; see \S\ref{sec:W2delta} for further discussion on this. 
\item The estimation of the multipole moments of $P'(\mathbf{k})$:  the modes available from a FFT lie on a Cartesian lattice and as such are irregularly spaced in $\mu$.  It is therefore invalid to calculate the multipole moments by approximating the multipole decomposition, eqn. (\ref{multipoleDecomp}), as a Riemann sum and instead linear regression must be used.  I find that the time required for this decomposition is at least comparable to that spent on a 3D FFT estimate of $P'(\mathbf{k})$.  As it is necessary to perform this calculation for a range spanning many decades in wavenumber, a large number of modes and hence a computationally expensive 3D FFT is required.
\item The necessary calculation of $P'_{\ell}(k)$ for each of a broad range of models in a likelihood analysis, which has led to recent approaches utilising pre-computed lookup tables to optimise the calculation, e.g. \cite{Blake}.  While this is an equally rapid approach, this `mixing matrix' technique has limited portability -- for a given matrix, the model power spectra must be provided in predefined wavenumber bins and there is a hard ceiling to the highest order $P_{\ell}'(k)$ that may be calculated before a new matrix is required.  In contrast to this, I show that the multipole moments of the mask autocorrelation function allow for the calculation of $P_{\ell}'(k)$ to any order with limited restrictions on the wavenumber range and resolution. 
\end{enumerate}

In this work I present a reformulation that predicts $P'_{\ell}(k)$ directly.  This allows for a rapid implementation that requires only a small number of 1D FFTs per model and hence achieves a significantly greater spectral resolution, which minimises the number of FFT based artefacts. To begin with, I first generalise the known Hankel transform relation between $P_{\ell}(k)$ and $\xi_{\ell}(\Delta)$ to a masked field by making use of the symmetries of the redshift-space correlation function. 

From the convolution theorem, the autocorrelation functions of both density field, $\xi(\boldsymbol \Delta)$, and mask, $Q(\boldsymbol \Delta)$, multiply to give the masked autocorrelation:
\[
\label{overautocorr}
\xi'(\mathbf{\Delta}) = \xi (\mathbf{\Delta}) \ Q(\mathbf{\Delta}).
\]
Note that this masked autocorrelation is equally sensitive to the anisotropy of the redshift-space density field and the survey mask.  Here I have introduced both $\xi(\mathbf{\Delta})$ and $Q(\mathbf{\Delta})$ as the inverse Fourier transforms of $P(\mathbf{k}$) and $|\tilde W(\mathbf{k})|^{2}$ respectively:
\[
Q(\boldsymbol \Delta)  = \int d^3 x \ W(\mathbf{x}) \ W(\mathbf{x} + \boldsymbol \Delta) = \int \frac{d^3 k}{(2\pi)^3} \ |\tilde W(\mathbf{k})|^{2} \ e^{i \mathbf{k} \cdot \boldsymbol \Delta}. 
\label{eqn:Qdef}
\]
Spherical coordinates present a natural coordinate system for RSD in which the physical symmetries may be best exploited. A coordinate transformation may be achieved by expanding the plane wave in spherical waves using the Rayleigh plane wave expansion:
\[
\label{Rayleigh}
e^{-i  \mathbf{\Delta} \cdot \mathbf{k}} = \sum_{p=0}^{\infty} (-i)^{p} (2p + 1) \ j_p(k \, \Delta) L_p(\hat{\mathbf{\Delta}} \cdot \mathbf{\hat{k}}),
\]
see equation (B3) of \cite{Cole94}.  Here $j_{p}(k \, \Delta)$ represents a spherical Bessel function of order $p$.  Conventionally, the chosen observables are the multipole moments of $P(\mathbf{k})$ despite the convolution:
\[
\label{multipoleDecomp}
P'_{\ell}(k) = \frac{(2\ell + 1)}{2} \int d(\hat{\mathbf{k}} \cdot \hat{\boldsymbol \eta}) \int \frac{d \phi_k}{(2 \pi)} P'(\mathbf{k}) \ L_{\ell} (\mathbf{\hat{k}} \cdot \hat{\boldsymbol \eta}); 
\]
here $\phi_k$ denotes the azimuthal coordinate of $\mathbf{k}$. With equations~(\ref{overautocorr})~--~(\ref{multipoleDecomp}), I find
\[
P'_{\ell}(k) = (-i)^{\ell} (2 \ell + 1) \int d^{3} \Delta  \ j_{\ell}(k \Delta) \ \xi' (\mathbf{\Delta})  L_{\ell} (\mathbf{\hat{\Delta}} \cdot \hat{\boldsymbol \eta}).
\label{noangleavg}
\]
Here I have used the identity presented by equation (A11) of \cite{Cole94}:
\[
\frac{(2\ell + 1) }{2}\int d(\mathbf{\hat{k}} \cdot \hat{\boldsymbol \eta} ) \int \frac{d \phi_k}{(2 \pi)} L_{\ell}(\mathbf{\hat{k}} \cdot \hat{ \boldsymbol \eta}) \ L_{\ell'} (\mathbf{\hat{k}} \cdot \mathbf{\hat{\Delta}} ) = \delta^{K}_{\ell \ell'} \ L_{\ell} (\mathbf{\hat{\Delta}} \cdot \hat{\boldsymbol \eta}).
\label{eqn:LegendreOrthogonality} 
\]
This ensures only the $p=\ell$ term survives from the Rayleigh plane wave expansion.  

In the distant observer approximation, when an expansion of $P(\mathbf{k})$ such as eqn. (\ref{legenSeries}) is valid, $\xi(\boldsymbol \Delta)$ may be similarly decomposed and the expression may be further simplified:
\begin{align}
P'_{\ell}(k) = 4 \pi &(-i)^{\ell} \left ( \frac{2 \ell + 1}{2q + 1} \right ) \nonumber \\ 
		& \times A_{\ell, \ell'}^{q} \int \Delta^2 d\Delta \ \xi_{\ell'} (\Delta) \ Q_q (\Delta) \ j_{\ell}(k \Delta).
\label{cnvldpk}
\end{align}
I assume the Einstein summation convention over the repeated \textit{dummy} indices in this equation.  Here the multipole moments of the mask autocorrelation function have been defined as 
\[
Q_{q}(\Delta) = \left ( \frac{2q + 1}{2} \right ) \int d(\mathbf{\hat{\Delta}} \cdot \hat{\boldsymbol \eta}) \ \int \frac{d\phi_{\Delta}}{(2 \pi)} \ Q(\mathbf{\Delta}) \ L_{q}(\mathbf{\hat{\Delta}} \cdot \hat{\boldsymbol \eta}). 
\label{eqn:Qelldef}
\]
Finally, Legendre polynomials have been used as a basis in $\mathbf{\hat{\Delta}} \cdot \hat{\boldsymbol \eta}$: 
\[
L_{\ell}(\mathbf{\hat{\Delta}} \cdot \hat{\boldsymbol \eta}) L_{\ell'}(\mathbf{\hat{\Delta}} \cdot \hat{\boldsymbol \eta}) = \sum^{\min(\ell, \ell')}_{q=0} A_{\ell, \ell'}^{q} \ L_{q}(\mathbf{\hat{\Delta}} \cdot \hat{\boldsymbol \eta}).  
\]
The product of two Legendre polynomials has been derived in \cite{PSP:1736752}; this result:
\[
L_{\ell} L_{\ell'} = \sum_{p=0}^{\min(\ell, \ell')} \frac{G_{\ell-p} G_p  G_{\ell'-p}}{G_{\ell+\ell'-p}} \left ( \frac{2\ell + 2\ell' -4p +1}{2\ell + 2\ell' -2p +1} \right) L_{\ell+\ell' -2p},
\label{Bailey}
\]
where 
\[
G_p = \frac{1 . 3 . 5 \dots (2p-1)}{p!} \equiv \frac{2^p (\frac{1}{2})_p}{p!}, \text{ and } \ell \geq \ell',
\]
may be used to obtain the $A_{\ell, \ell'}^q$ coefficients.

The structure of this equation is clear as the multipole moments in configuration and Fourier space are known to form a Hankel transform pair:
\begin{eqnarray}
\label{Hankelpair}
P_{\ell}(k) &=& 4 \pi (-i)^\ell \int \Delta^2 d\Delta \ \xi_{\ell}(\Delta) \, j_{\ell}(k\Delta).
\end{eqnarray}
Equation (\ref{cnvldpk}) generalises eqn. (\ref{Hankelpair}) to the case of a masked density field and is a primary result of this chapter.  In the masked case, the Hankel transform relation is preserved as $\xi_\ell(\Delta)$ is replaced by an effective $\xi'_{\ell}(\Delta)$ defined below.

This is a relation of significant practical importance as a Hankel transform may be evaluated in a single 1D FFT \citep[FFTlog,][]{Hamilton}.  It is then possible to quickly transform between P$_{\ell}(k)$ and $\xi_{\ell}(\Delta)$ -- a fact I utilise to achieve both a $10^3-10^4 \times$ speedup and increased spectral resolution relative to a 3D FFT approach.  As a 3D FFT is $\simeq 3 N^2 \times$ slower than a 1D FFT of mesh size $N$, (for given time) there is the potential for an improvement of $\simeq 10^6$ in speed, a similar gain in resolution, or a balance may be struck between the two.  As an additional benefit, when the problem is approached in real space, there is no need to embed the survey in an effective volume and subsequently there is no resolution limit imposed by a fundamental mode.  A direct comparison of the resolution achieved is impractical as the optimised method I implement computes $P'_\ell(k)$ for logarithmically spaced intervals in $k$, as opposed to the linear spacing of the 3D FFT.  This is in itself a benefit due to the many decades of wavenumber over which $P'_{\ell}(k)$ is desired.  Consequently, the minimal amount of required memory allows aliasing associated with the 1D FFT to be confined to wavenumbers of no practical interest.  

\subsection{Practical details}
\label{practical}
Given that $\xi_{\ell}(\Delta)$ may be rapidly computed for an assumed RSD model, how is $P'_{\ell}(k)$ to be calculated?  In this approach, first the $Q_{q}(\Delta)$ are precomputed from a random catalogue bounded by the survey geometry; this is described further in \S\ref{sec:W2delta}.  At each point in parameter space $\xi_{\ell}(\Delta)$ can be found by Hankel transformation (1D FFT) of $P_{\ell}(k)$ for $\ell=0,2,4, \cdots $.  For the monopole, the necessary linear combination is formed:
\[
\xi_0'(\Delta) = \xi_0 Q_0 + \frac{1}{5} \xi_2 Q_2 + \frac{1}{9} \xi_4 Q_4 + \frac{1}{13} \xi_6 Q_6 + \cdots,  
\label{monoxi}
\]
and an inverse Hankel transform (1D FFT) computes $P'_0$.

This relation is obtained by calculating the $A_{\ell, \ell'}^{q}$ coefficients, which account for the weighted volume average given by eqn. (\ref{noangleavg}).  To illustrate this point, consider a simple case in which both density field and mask are composed of solely quadrupole terms.  Following eqns. (\ref{overautocorr}) \& (\ref{Bailey}), the $\mu$ dependence of the product is a linear combination of monopole, quadrupole and hexadecapole terms, $(1/5) L_0 + (2/7) L_2 + (18/35) L_4$, such that $\xi'_{2} = (2/7) \ \xi_2 Q_{2}$.  In this case the masked monopole and quadrupole can be seen by inspection.  More generally, a contribution to $P'_\ell(k)$ is generated by all $(q, \ell')$ terms of $Q_q$ and $\xi_{\ell'}$ due to the angular dependence of $L_{\ell} L_{\ell'}$ -- not least by leakage of the monopole to the quadrupole via $Q_{2}$.  By calculating the $A_{2, \ell'}^{q}$ coefficients an explicit expression for the lowest order terms in $\xi_2'(\Delta)$ may be given:
\begin{eqnarray}
\xi_2'(\Delta) = \xi_0 Q_2 &+ \ \xi_2&\left( Q_0 + \frac{2}{7} Q_2 + \frac{2}{7} Q_{4} \right)  \nonumber \\
 		  &+ \ \xi_4&\left( \frac{2}{7} Q_2 + \frac{100}{693} Q_4 + \frac{25}{143} Q_{6} \right)   \nonumber \\ 
		  &+ \ \xi_6&\left( \frac{25}{143} Q_4 +  \frac{14}{143} Q_6 + \frac{28}{221} Q_8 \right) \nonumber \\
		  &+ \cdots &.
\label{quadxi}
\end{eqnarray}
Equations (\ref{monoxi}) \& (\ref{quadxi}), together with the FFTlog implementation of eqn. (\ref{Hankelpair}) and its inverse, suffice to calculate $P'_0(k)$ and $P'_2(k)$.  As $\xi_{\ell}(\Delta)$ is non-zero for even $\ell$ only there is no dependence on $Q_q(\Delta)$ for odd numbered q.  One might expect this expression to include $Q_q$ terms of arbitrarily high order; this is not the case as $L_{2} L_{\ell'}$ has only three non-zero Legendre coefficients for given $\ell'$.  I return to a test of the convergence rate of these expressions with respect to $\xi_{\ell}$ in \S \ref{ConvergenceTest}. 

The required coefficients may be calculated to increasingly higher order to obtain a given precision, however the measurement of $Q_{\ell}$ becomes progressively noisier for $\ell \gg 1$; I comment on an approach to this problem in \S\ref{sec:W2delta}.  On large scales, $\Delta \gg 1$, the series will be truncated by the Kaiser limit: $\xi_{\ell}(\Delta) = 0$ for $\ell > 4$ \citep{Hamilton92}.  On small scales the series will commonly be restricted by $Q_q(\Delta)$, which are negligible compared to $Q_0$ for $\Delta < 10$; see  Fig. \ref{WindowMultipoles}. It is clear that of order 10 one-dimensional FFTs are required for the prediction of $P_0'(k)$ \& $P_2'(k)$ if both $\xi$ and $Q$ are well approximated by terms for which $\ell$ and $q \leq 6$.

Although the logical approach is to measure the multipole moments of the mask using a large random catalogue in the usual manner (see \S \ref{sec:W2delta}) they may instead be obtained by FFT as in other common approaches.  In this case the real space moments $Q_{q}(\Delta)$ may be estimated by
\begin{eqnarray}
Q_{q}(\Delta)&=&i^{q} (2q +1) \int \frac{d^{3}k}{(2 \pi)^3} |\tilde{W}(\mathbf{k})|^{2} \ j_{q}(k \, \Delta) \ L_{q}(\mathbf{\hat{k}} \cdot \hat{\boldsymbol \eta}) \nonumber, \\
&\approx& i^{q} (2q +1) \sum |\tilde{W}(\mathbf{k}) |^2  \ j_{q}(k \, \Delta) \ L_{q}(\mathbf{\hat{k}} \cdot \hat{\boldsymbol \eta}).
\end{eqnarray}
Here the sum is restricted to the modes available with a 3D FFT.  To obtain this result I have made use of eqn. (\ref{eqn:Qdef}), replaced plane with spherical waves and used eqn. (\ref{eqn:LegendreOrthogonality}).  I tested this last approximation and confirmed the validity of these expressions with the following illustrative example.   

\subsection{A simple validity test}
\label{validity}
To test the validity of this approach I generated a set of 5000 realisations of anisotropic Gaussian random fields with a $(1 + \mu^2/2)P_g(k)$ power spectrum, to which I applied a common 3D `mask' and measured the resultant $P'_{\ell}(k)$.  Sn independent Gaussian field was generated for the mask; the power spectrum was chosen for convenience of illustration to have the same $P(k)$ as the density field.  This allows for the calculation of $Q_{\ell}(\Delta)$ by Hankel transformation of the known $P_{\ell}(k)$.  A comparison between that `observed' and the prediction of this method is shown in Fig. \ref{ConvldMultipoles}.  This simple test shows the validity of this approach, with excellent agreement found between the predictions and the realisations. 
\begin{figure}
\centering
\includegraphics[width=\textwidth]{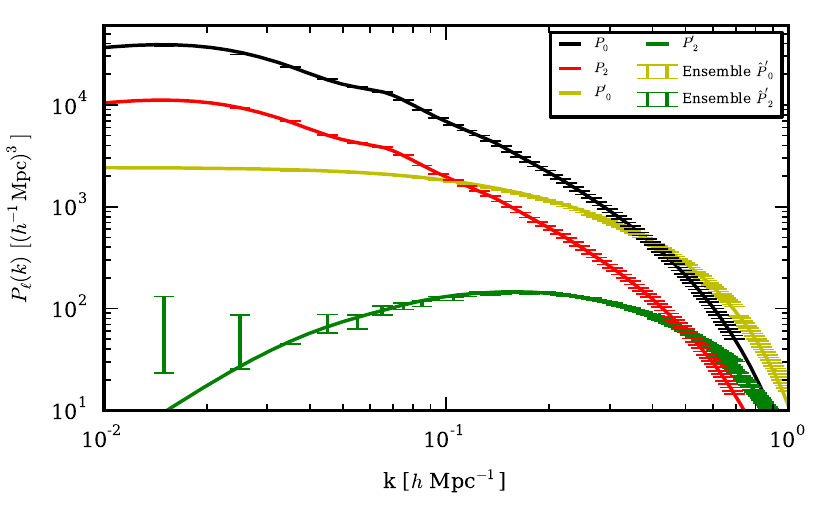}
\caption[A test of the accuracy of the predictions for a Gaussian density field.]{A comparison of the power spectrum multipole moments for a masked density field predicted by this approach and that of 5000 realisations for an illustrative test case comprised of a mask and density field given by 3D Gaussian random fields.  A $(1 + \mu^2/2)P_g(k)$ model was assumed for the mask and realisations, with $P_{\text{g}}(k)$ taken to be a non-linear $\Lambda$CDM power spectrum for biased tracers smoothed with a sphere $3 h^{-1} \emph{Mpc}$ in radius.  The step in the standard error of the mean at $k=0.7 h \emph{Mpc}^{-1}$ corresponds to replacing the results with those obtained from a box of half the size.  This allows for a higher mesh resolution to be obtained and thus avoids the spurious increase of power due to aliasing from the finite FFT grid.}
\label{ConvldMultipoles}
\end{figure}

\section{Obtaining $Q_{q}(\Delta)$ with a pair counting approach}
\label{sec:W2delta}
In this section I outline a method for obtaining $Q_{q}(\Delta)$ by pair counting a random catalogue of constant number density that is bounded by the survey geometry.  I also comment on the artefacts introduced by accounting for the mask with FFT based approaches. 

Consider the available distinct pairs of a random catalogue of constant number density, $\overline n_s$, that populates two volumes, $dV_1$ and $dV_2$.  Following the application of weights, $W(\mathbf{x})$, the weighted pair count is given by 
\[
RR(\mathbf{x}_1, \mathbf{x}_2) = \frac{1}{2} \, \overline{n}_s^2 dV_1 dV_2 W(\mathbf{x_1}) W(\mathbf{x_2}). 
\]
If $dV_2$ is taken to be centred on $\boldsymbol x_2 = \boldsymbol x_1 + \boldsymbol \Delta$ and the total such pairs over the surveyed volume, $\int dV_1$, are counted then  
\[
RR^{\mathrm{tot}}(\boldsymbol \Delta) = \frac{1}{2} \overline n_s^2 Q(\boldsymbol \Delta) dV_2.
\]
By first applying a $(2q+1) L_q(\boldsymbol{\hat \Delta} \cdot \boldsymbol {\hat \eta})/2$ weighting to each pair and taking $dV_2$ as a narrow shell of width d$(\ln \Delta)$ centred on $\boldsymbol x_1$, the total number of weighted pairs summed over the surveyed volume is
\[
\overline{RR}^{\mathrm{tot}}_{q}(\Delta) = \frac{1}{2} \overline{n}_s^2 . 2 \pi \Delta^3 d(\ln \Delta) Q_q(\Delta).
\]
Here $\overline{RR}^{\mathrm{tot}}_{q}$ corresponds to distinct pairs binned by separation in logarithmic intervals and each pair has been weighted by
\[
(2q +1) L_{q}(\boldsymbol{\hat \Delta} \cdot \boldsymbol{\hat \eta}) W(\mathbf{x}_1) W(\mathbf{x}_2)/2.  
\]
Significantly,  $\overline{RR}^{\mathrm{tot}}_{q}$ differs only in amplitude from $Q_q(\Delta)$.  For each $q$, the counts should be similarly rescaled such that $\overline{RR}^{\mathrm{tot}}_{0}/\Delta^3 \mapsto 1$ for $\Delta \ll 1$.  This is because the renormalisation in eqn. (\ref{ampcorr}) is equivalent to enforcing $Q_0(\boldsymbol 0) = 1$.

A real-space pair counting approach has a number of advantages compared to an estimate of $|\tilde{W}(\mathbf{k})|^2$ by FFT:
\begin{enumerate}
\item  Large volume suveys are challenging with a memory limited FFT due to the large volume required to embed the survey, which enforces a small fundamental mode and hence a small Nyquist frequency.  If the survey has small scale angular features then $|\tilde{W}(\mathbf{k})|^2$ at large $k$ may be large and the aliasing introduced by coarse binning may be significant.  
\item The integral constraint correction presented in \S \ref{sec:intcor} requires a robust estimation of $|\tilde{W}_{\ell}(k)|^2$ for $k \ll 1 \hompc$. 
A FFT is imprecise for this estimate due to the limited number of modes available in this regime.  In contrast, a Hankel transform of the pair counts yields a much higher resolution estimate and hence a more robust correction may be made; this is shown quite clearly in Fig. \ref{windowkmultipoles}.  This is discussion further in \S \ref{sec:intcor}.
\item The required pair counting is performed prior to the likelihood analysis and is easily optimised with the use of a k-d tree or a similar technique.  For $\Delta < 10 \mpcoh$, the relatively small number of pairs gives a noisy estimate but this regime can be rapidly remeasured with a higher density by decreasing the maximum separation of nodes that are to be included.
\end{enumerate}

It is often the case that an additional weighting is applied to the surveyed volume (rather than simply a binary geometric factor).  For the commonly used FKP estimator \citep{FKP} a further $\bar n(\mathbf{x})/(1 + \bar n(\mathbf{x}) P_0)$ weighting is required, which may be seen by contrasting eqn. (2.1.6) of \cite{FKP} with eqn. (\ref{convo}).  The best approach in this case is to generate a random catalogue with the same radial distribution function as the survey; each pair should then be weighted by
\[
(2q + 1) L_{q}(\boldsymbol{\hat \Delta} \cdot \boldsymbol{\hat \eta})/ \left [ 2 (1+ \bar n_1 P)(1 + \bar n_2 P) \right ].
\]

\section{VIPERS: an application to a realistic survey geometry}
\begin{figure}
\centering
\includegraphics[width=\textwidth]{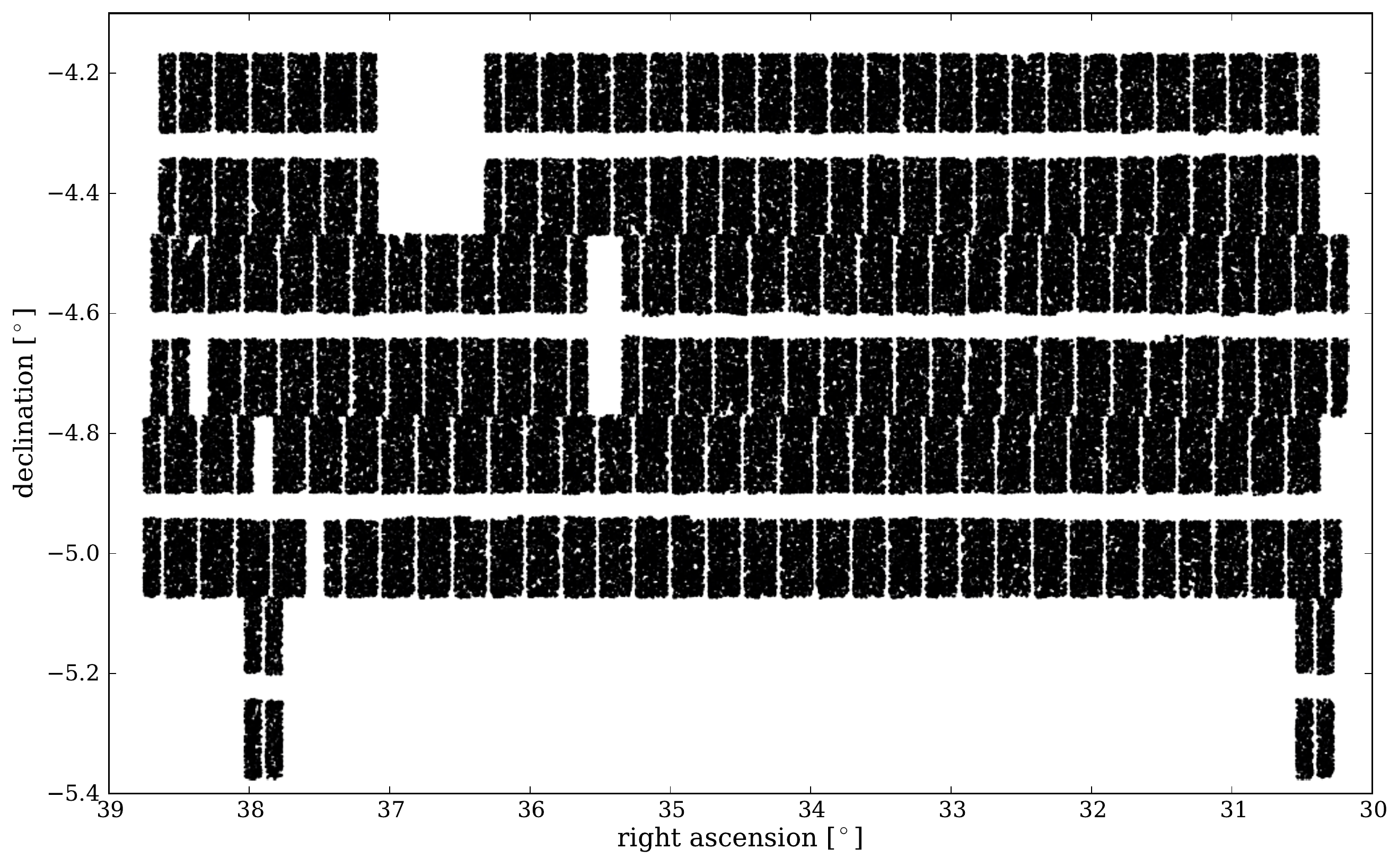}
\caption[Angular footprint of the VIPERS PDR-1 release.]{The angular footprint of the VIMOS spectrograph across the W1 field of the VIPERS PDR-1 release \citep{Vipers}.  This angular selection represents an idealised VIPERS mask in which the spectroscopic success rate is unity and the target success rate is binary across the sky; see \protect \cite{sylxi} for further discussion. The final VIPERS v7 footprint is shown in \S \ref{sec:footprint}.}
\label{angularfootprint}
\end{figure}

This section presents a concrete application to a realistic test case: the W1 field of the VIPERS PDR-1 release \citep{Vipers}.  VIPERS is a large spectroscopic survey that has measured approximately 100,000 galaxies in the redshift range $0.5 < z < 1.2$.  Despite a significant volume,~$\simeq 5 \times 10^7 (\mpcoh)^{3}$, and a high sampling rate,~$\simeq 40 \%$, VIPERS is afflicted by a complicated angular selection as shown in Fig.~\ref{angularfootprint}.  The survey footprint is made up of rows of pointings, each of which is comprised of four quadrants separated by a central cross in which spectra cannot be obtained.  This is in order to obtain the maximum possible volume and hence only a single pass is performed.  
\begin{figure}
\centering
\includegraphics[width=0.9\textwidth]{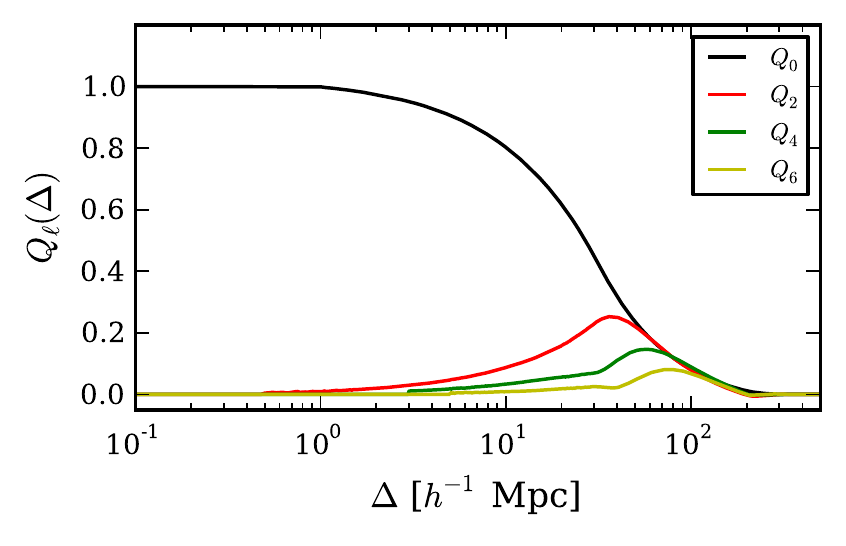}
\caption[Multipole moments of the VIPERS PDR-1 mask autocorrelation]{Multipole moments of the VIPERS PDR-1 mask autocorrelation function for the idealised angular selection shown in Fig. \ref{angularfootprint} and $0.7<z<0.8$.  It is clear from this figure that the anisotropy of the VIPERS mask is relatively insignificant below $\simeq 10 h^{-1} \emph{Mpc}$.}
\label{WindowMultipoles}
\end{figure}

This disjointed network presents a challenging test case for modelling the effect of the survey mask.  However, it should be noted that the convolution is dependent only on the mask autocorrelation function -- effectively the mask is self-smoothed and hence the masked power spectrum is relatively insensitive to sharp angular features.  Fig. \ref{WindowMultipoles} shows the lowest order moments, $Q_q(\Delta)$, of the mask autocorrelation for $0.7<z<0.8$.  It is clear from this figure that the anisotropy of the VIPERS mask is relatively insignificant below $\simeq 10 \mpcoh$.
\begin{figure}
\centering
\includegraphics[width=0.9\textwidth]{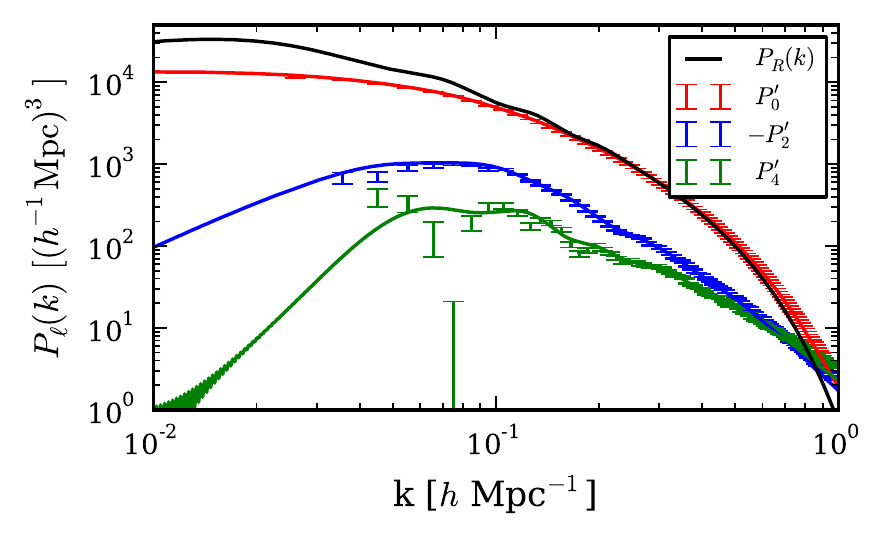}
\caption[$P_{\ell}(k)$ for isotropic Gaussian fields subject to the PDR-1 mask]{Observed multipole moments for a set of Gaussian realisations with an \textit{isotropic} $P_g(k)$ to which a common mask has been applied.  The mask is equivalent to that of the VIPERS W1 field for $0.7<z<0.8$. The observed quadrupole and hexadecapole moments result solely from the anisotropy of the survey mask, which is given explicitly for $P_2(k)$ in eqn. (\ref{quadxi}).  Note that the predictions (solid) are in excellent agreement with that observed.}
\label{VIPERS_cnvld_hexpk}
\end{figure}

Fig. \ref{VIPERS_cnvld_hexpk} shows the accurate prediction of the observed $P'_{0}$(k) and $P'_{2}$(k) for a more realistic test case comprised of the application of the VIPERS PDR-1 mask to an isotropic Gaussian field. This illustrates the generation of higher order $P'_\ell(k)$ simply due to the anisotropy of the survey mask. 

\subsection{A convergence test for VIPERS PDR-1}
\label{ConvergenceTest}
Fig. \ref{fig_convergence} shows a test of the convergence rate of the expressions for $\xi_{\ell}'(\Delta)$ given by equations (\ref{monoxi}) and (\ref{quadxi}).  In this figure, $M'_{p}$ denotes the predicted monopole power spectrum for the masked density field when the expansion is truncated at $\xi_{p}(\Delta)$ (inclusive) and similarly for the quadrupole, $Q'_{p}$.  I analyse the VIPERS PDR-1 mask and assume $(\beta, \sigma_p) = (0.5, 5.0 \mpcoh)$, which corresponds to a conservative choice of $\sigma_p$ and therefore the series should converge relatively slowly.  The Kaiser model is recovered for small $\sigma_p$ and therefore $\xi_{\ell'}(\Delta) = 0$ for $\ell'>4$.  It is clear from this figure that the inclusion of the hexadecapole term is sufficient for obtaining subpercent precision on the masked multipoles.  The fractional error of the quadrupole diverges at $k=0.8 \hompc$ as $Q_6'$ passes through zero at this point.  To calculate the higher order multipole moments I make use of an implementation of \cite{DeMicheli}, which enables the first $N$ multipole moments to be obtained with a 1D FFT of size $N$ for each wavenumber, $k$.
\begin{figure}
\centering
\includegraphics[width=0.9\textwidth]{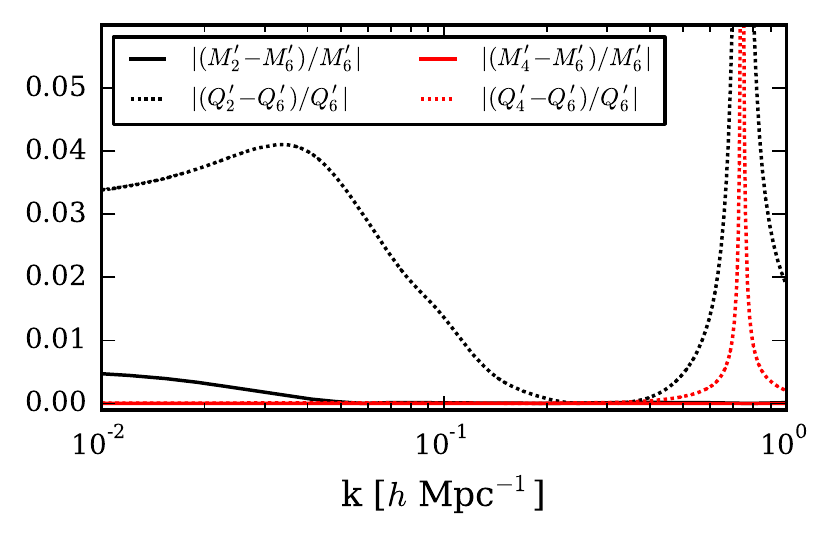}
\caption[A convergence test of the expansion with respect to $\xi_{\ell}(r)$.]{This figure illustrates the convergence rate of $P'_{\ell}(k)$ with respect to the expansion in $\xi_{\ell}'(\Delta)$ given by eqns. (\ref{monoxi}) \& (\ref{quadxi}).  Here $M'_{p}(k)$ denotes the predicted monopole power spectrum for the masked density field when the expansion is truncated at $\xi_{p}$ (inclusive).  Similarly, $Q'_{p}$ represents the quadrupole.  I analyse the VIPERS PDR-1 geometry and assume $(\beta, \sigma_p) = (0.5, 5.0 h^{-1} \emph{Mpc})$, which corresponds to a conservative choice of $\sigma_p$.  For $\sigma_p \ll 1 h^{-1} \emph{Mpc}$ the Kaiser model is recovered and the series formally converges with $\xi_4$.  It is clear from this figure that the inclusion of the hexadecapole terms is sufficient for obtaining subpercent precision on the masked multipoles.  There is a divergence in the fractional error of the quadrupole at $k = 0.75 h \emph{Mpc}^{-1}$ as $Q_6'(k)$ passes through zero at this point.}
\label{fig_convergence}
\end{figure}

\section{Integral constraint correction}
\label{sec:IntConCorr}
\begin{figure}
\centering
\includegraphics[width=0.9\textwidth]{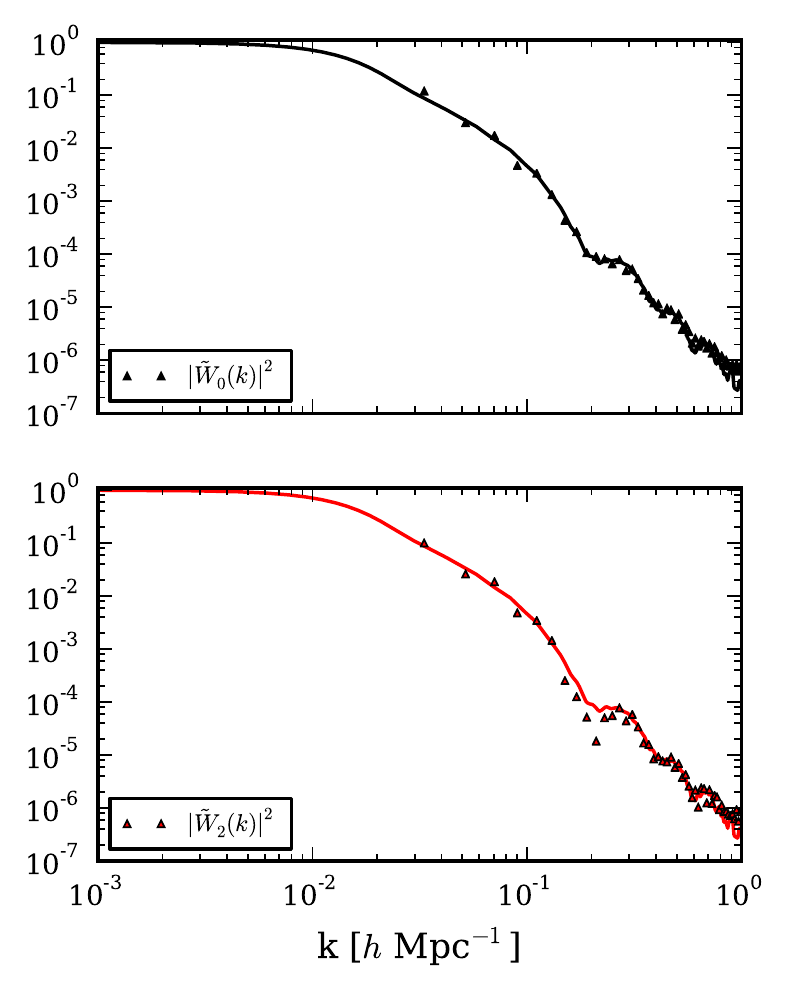}
\caption[A $|\tilde W_{\ell}(k) |^2$ estimate by Hankel transform of the pair counts.]{A comparison of $|\tilde W_{0}(k) |^2$ and $|\tilde W_{2}(k) |^2$ for the VIPERS W1 mask.  This is obtained by a Hankel transform of the pair counts (solid) and by 3D FFT (triangles).  The FFT measurement is coarsely binned to suppress the statistical noise.  In contrast, the Hankel Transform is independent of the fundamental period of any embedding volume.  Note that where the integral constraint correction is largest, $k \ll 1 h \emph{Mpc}^{-1}$, there are very few FFT modes to make a robust estimate of $|\tilde W_{\ell}(k) |^2$.}
\label{windowkmultipoles}
\end{figure}
\begin{figure}
\centering
\includegraphics[width=0.9\textwidth]{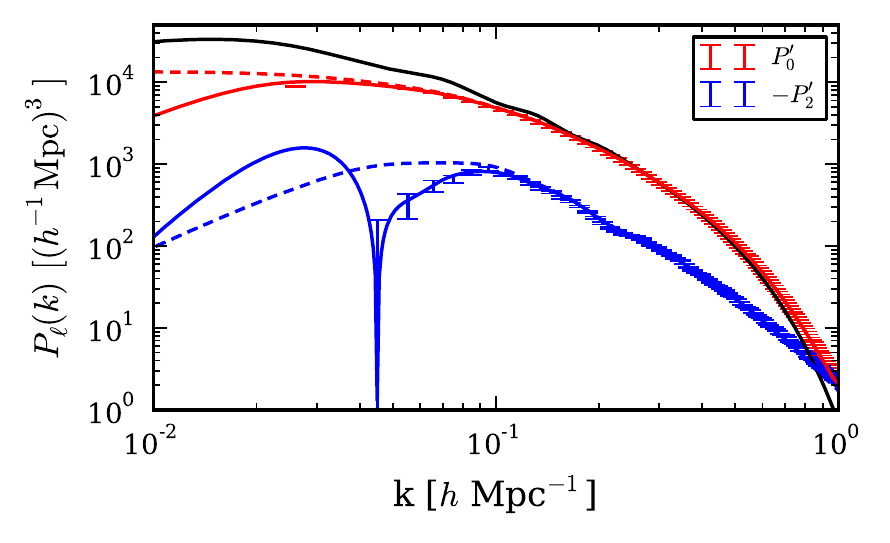}
\caption[The integral constraint correction for the PDR-1 $P_{\ell}(k)$.]{This figure repeats Fig. \ref{VIPERS_cnvld_hexpk} but further assume that the surveyed volume is a fair sample of the density field. This introduces the integral constraint correction discussed in \ref{sec:intcor}.  The monopole and quadrupole predictions are shown both prior to (dashed) and following (solid) correction.  Clearly there is a significant impact on the observed quadrupole, but this could be reduced in a more realistic VIPERS analysis by making a joint-field estimate of $\bar n$.}
\label{VIPERShex_intcor}
\end{figure}
\label{sec:intcor}
An additional integral constraint correction is required for a masked density field: the background number density $\bar n (z)$ is estimated from the finite survey volume and may differ from the true value due to clustering on wavelengths approaching the survey size.  The assumption that the survey volume is a fair sample enforces the constraint: $\tilde \delta(\mathbf{0}) = 0$.  As this false mean $\delta$ is subject to the linear convolution detailed above, a convolved spike centred on $\mathbf{k} = \mathbf{0}$ contributes to the observed power. To ensure $P^{\text{obs}}(\mathbf{0})=0$, the observed power must be given by \citep{PeacockNicholson}
\[
P^{\text{obs}}(\mathbf{k}) = P'(\mathbf{k}) -  |\tilde{W} (\mathbf{k}) |^2 \ P'(\mathbf{0}). 
\]
Here $|\tilde{W}(\mathbf{k}) |^2$ is rescaled such that $|\tilde{W}(\mathbf{k}) |^2 \mapsto 1$ as $k \mapsto 0 \hompc$.  This clearly equates to
\[
P^{\text{obs}}_{\ell}(k) = P'_{\ell}(k) -  P'_{0}(0)  \ |\tilde{W} _{\ell}(k) |^2 
\]
in the distant observer approximation.  In this case, the rescaling is such that $|\tilde{W}_{0}(0) |^2 = 1$.  Given a measurement of  $Q_{\ell}(\Delta)$, $|\tilde{W}_{\ell}(k) |^2$ may be obtained by Hankel transformation.  This allows for a much higher resolution measurement than with a 3D FFT and is free of the fundamental mode of an embedding volume.  Fig. \ref{windowkmultipoles} shows a comparison of $|\tilde{W}_{\ell}(k) |^2$ when obtained via these contrasting approaches.  Note that the FFT estimate is coarsely binned in order to suppress the statistical noise present.

Fig. \ref{VIPERShex_intcor} repeats Fig. \ref{VIPERS_cnvld_hexpk} but assumes the mean density over the survey volume is a fair sample.  This introduces the integral constraint correction discussed above.  The impact on the quadrupole is significant, but in a more realistic analysis this could be reduced by estimating $\bar n(z)$ with a joint-field estimate and over a larger redshift range. 

\subsection{Including a smoothed joint-field $\bar n(z)$ estimate}
\label{sec:smoothed_icc}
It is often the case in redshift surveys that discontiguous areas of the sky are surveyed.  Sample variance in the estimated radial selection may be minimised in this case by making a joint-field estimate.  This is not reflected in the previous derivation as the $\langle \delta \rangle = 0$ constraint must be enforced over the joint volume, rather than each field individually.  The required correction to the power is
\[
P^{\rm{obs}}_{\ell}(k) = P'_\ell(k) - \left( \frac{\Omega_i}{\Omega_J}\right) P'_J(0) \ |\tilde W_\ell(k)|^2,  
\]
in this case.  This is an original result and is obtained by generalising \cite{PeacockNicholson}.  This expression correctly accounts for the joint-field estimate of $\bar n(z)$.  A given field is denoted by $i$ and has a solid angle of $\Omega_i$.  Joint field quantitites are denoted by $J$ and $P'(k)$ denotes the amplitude corrected power spectrum, as before.  I adopt this correction for the RSD analysis detailed in Chapter \ref{chap:VIPERS_RSD}.  Further details may be found in Wilson~et~al.~(2016,~in prep.). 

The intrinsic radial clustering may be suppressed in the $\bar n(z)$ estimate with a radial Gaussian smoothing, which assumes the survey selection is smooth below the kernel scale.  This approach may be accounted for in the integral constraint correction as the affect of a radial smoothing on the power spectrum is simply the damping term in the dispersion model.  When determining $P'_J(0)$ in this case, the appropriate model for $P(k)$ is
\[
\left( \frac{P}{P_g} \right ) = (1 + \beta \mu^2)^2 \exp (- \sigma_{n_z}^2 k^2 \mu^2).
\]
Here $\sigma_{n_z}$ is the scale of the smoothing kernel, which is either $100 \mpcoh$ or $50 \mpcoh$ in the analysis detailed in later chapters.  The additional damping from the non-linear RSD will be negligible on these scales.  I neglect this small correction in later chapters due to the sizeable VIPERS errors on large scales.  

\section{Masking and the Alcock-Paczy\'nski effect}
\label{sec:ap_masking}
The forward modelling required for the masked power spectrum is  
\[
P'_*(\mathbf{k'}) = \int \frac{d^{3} q'}{(2 \pi)^3}  P'(\mathbf{k'} - \mathbf{q'}) |\tilde{W'}(\mathbf{q'})|^2.
\]
when the AP effect is considered; i.e all quantities are in the fiducial cosmology for a mask defined in redshift and angle. The AP mode remapping discussed in Chapter \ref{chap:RSD} can then be applied to $P'(\mathbf q') \mapsto P(\mathbf q)$, as there exists a physical model for the AP distortion-free power spectrum, e.g. the Kaiser-Lorentzian model.  However, the undistorted mask is not known apriori -- the choice of a fiducial cosmology may have increased (or decreased) the anisotropy.  However, this integral only requires the mask in the fiducial cosmology and this may be measured.  Forward modelling then proceeds as in the AP distortion-free case -- with $P'(\mathbf q')$ obtained with a mode remapping of $P(\mathbf q)$.  Incidentally, a useful approach may be to simplify the mask modelling by assuming a fiducial cosmology in which the mask is less anisotropic.

\section{Complete impact of the VIPERS mask}
All necessary mask corrections for the VIPERS PDR-1 W1 release are summarised in Fig.~\ref{intcor}.  This figure assumes an intrinsically anisotropic density field with a $(1 + \mu^2/2)P_g(k)$ power spectrum, such that $P_0(k) = (7/6) P_g$ (black line) and $P_2(k) = (1/3) P_g$ (blue long dashed line) in the absence of the survey mask.  This corresponds to a quadrupole-to-monopole ratio with an effective value of $\beta \simeq (1/4)$ in the Kaiser model.  A breakdown of the individual components is also plotted; firstly, the isotropic component of the density field yields a distorted monopole, ${P'}_0^m$, due to the mask and creates a significant quadrupole, ${P'}_2^m$.  This leakage may mitigate the intrinsic quadrupole depending on the relative anisotropy of the density field and mask.  The integral constraint correction is also applied to the unmasked power for illustration; the result of this is denoted by $P_{0}^{\rm{IC}}$ and $P_{2}^{\rm{IC}}$.  The combined correction predicted the impact of the PDR-1 mask on the observed multipoles.  This is shown by the solid lines and is in excellent agreement with that measured. 

\section{Validity for finite-angle surveys}
\label{wide-angle}
This approach assumes the validity of the distant observer approximation -- when  the variation of the line-of-sight across the survey is assumed to be negligible.  In this case, the redshift-space $P(\mathbf{k})$ possesses the symmetries outlined in the introduction; if this assumption is relaxed and widely separated pairs are included in the analysis then the redshift-space $\xi(\boldsymbol \Delta)$ is dependent on the triangular configuration formed by a given pair and the observer \citep{SphericalRSD}.  As there remains a statistical isotropy about the observer, any such configuration may be rotated into a common plane \citep{Szalay97} following which the remaining degrees of freedom are vested solely in the triangular shape.  Alternative parametrisations of this configuration are possible and therefore there is an ambiguity in the definition of the `line-of-sight'.  One possibility is to define $\boldsymbol \eta$ as that bisecting the opening angle; the triangle is then fully defined by $\mu$, $\Delta$ and the opening angle, $\theta$; see Fig. 1 of \citejap{Yoo}.  

It is clear that the finite-angle redshift-space $\xi(\boldsymbol s)$ does not possess the symmetries that have been exploited in the distant observer limit, $\theta \mapsto 0 \deg$.  This method must therefore be applied with some care to surveys with a large sky coverage.  While this is currently a limitation, the median redshift of future surveys will be considerably larger and the modal opening angle of pairs separated by the BAO scale will be $\simeq 4-6 \deg$, as compared to the $\simeq 20 \deg$ of current surveys (Y15).  In fact, Fig. 7 of Y15 shows that the systematic error introduced by assuming the distant observer approximation in the modelling is negligible for both Euclid and DESI -- provided the redshift evolution of the density field and bias is correctly accounted for.  In any case, a practical perspective is to accept that any bias introduced by using an approximate model, such as the dispersion model, may be calibrated with numerical simulations and a correction applied to the final data analysis.  Finite-angle effects are merely one instance where this approach may be taken.   

\section{Conclusions}
The effect of the survey mask represents the largest systematic difference between the observed large-scale power spectrum and that predicted by fundamental physics.  This work presents a new forward modelling approach for predicting the redshift-space power spectrum multipole moments in light of this effect.  By exploiting the symmetries of the redshift-space correlation function in the distant observer approximation, I derive a masked-field generalisation of the known Hankel transform relation between the multipole moments in real and Fourier space.  As a Hankel Transform may be computed with a 1D FFT, this implementation is $10^3 - 10^4 \times$ faster than other common approaches and achieves a higher spectral resolution that minimised FFT-based artefacts.  These advantages are especially relevant for large volume surveys with sharp angular features. 

I describe and validate an approach for obtaining the required multipole moments of the mask autocorrelation function, $Q_q(\Delta)$, by pair counting a random catalogue bounded by the survey geometry.  By accounting for the mask in real space, rather than Fourier space, a more robust integral constraint correction may be made.  This approach allows for greater physical insight into the impact of the mask anisotropy and, with a kd-tree or similar, the calculation may be conveniently optimised.   Although other approaches have achieved similar speeds, e.g. the `mixing matrix' approach of \cite{Blake}, this formulation suggests a physically motivated compression of the mask into the multipole moments of the autocorrelation function, $Q_q(\Delta)$.  With these in hand, $P_{\ell}'(k)$ may be calculated to arbitrary order with limited restrictions on the wavenumber range and resolution.  This is in contrast to the mixing matrix, which achieves a compression by restricting the allowed modelling to a limited number of $P_{\ell}'(k)$ for a predefined set of wavenumber bins. 

A concrete application to the VIPERS PDR-1 W1 survey geometry is presented; I show that the power spectrum multipole moments can be accurately predicted in this case, which serves as a proof of principle for this approach.  Amongst other corrections, a significant quadrupole is generated by the quadrupole component of the survey mask, which has been noted by other surveys \citep{Beutler}.  Although this method is limited by assuming the distant observer approximation, and must be applied with care to large solid angle surveys, this issue will be mitigated by the larger median redshift of future surveys.  In any case, any small error introduced by finite-angle effects may be calibrated with numerical simulations and a subsequent correction applied to the final data analysis. 

The machinery I have constructed should prove valuable in the RSD analyses of future galaxy redshift surveys such as VIPERS, eBOSS, DESI and Euclid, and allow for the systematic error due to the mask to be rapidly corrected.  This serves as an important step towards providing robust constraints on modified gravity theories based on the linear growth rate of density fluctuations. 
\begin{figure}
\centering
\includegraphics[width=\textwidth]{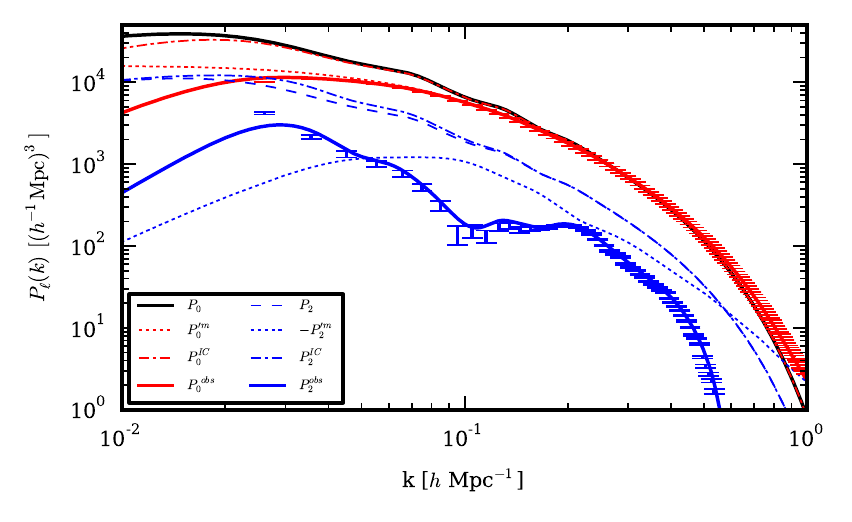}
\caption[A breakdown of all $P_{\ell}(k)$ corrections due to the PDR-1 mask.]{This figure provides a breakdown of all necessary mask corrections for the VIPERS PDR-1 W1 mask.  An anisotropic density field with a $(1 + \mu^2/2)P_g(k)$ power spectrum is assumed, such that $P_0(k) = (7/6) P_g$ and $P_2(k) = (1/3) P_g$ in the absence of the mask.  Firstly, the isotropic component of the density field yields a distorted monopole, ${P'}_0^m$, due to the mask and contributes significantly to the quadrupole, ${P'}_2^m$.  This leaked component may mitigate the intrinsic quadrupole depending on the relative anisotropy of the density field and mask.  The integral constraint correction is applied to the unmasked power for illustration; the result of this is denoted by $P_{0}^{\rm{IC}}$ \& $P_{2}^{\rm{IC}}$.  The combined correction predicts the complete impact of the VIPERS mask on the monopole and quadrupole.  This is shown by the solid lines and is in excellent agreement with the measurements.}
\label{intcor}
\end{figure}

\section{Derivation of eqn. (\ref{eqn:LegendreOrthogonality})}
This section provides the identities required for the derivation of the orthogonality relation:
\[
\frac{(2\ell + 1)}{2} \int d(\hat{\mathbf{k}} \cdot \hat{\boldsymbol \eta}) \int \frac{d \phi_k}{(2 \pi)} L_{\ell}(\hat{\mathbf{k}} \cdot \hat{\boldsymbol \eta}) L_{\ell'} (\hat{\mathbf{k}} \cdot \hat{\boldsymbol \Delta})  = \delta^{K}_{\ell \ell'} L_{\ell}( \hat{\boldsymbol \Delta} \cdot \hat{\boldsymbol \eta}). \label{B1}
\]
 These identities are reproduced from Appendix (B3) of \cite{Beutler}.  Starting with the definition of the Legendre polynomials in terms of spherical harmonics:
\[
L_{\ell}(\hat{\mathbf{k}} \cdot \hat{\boldsymbol \eta}) = \frac{4 \pi}{(2 \ell +1)} \sum_{m = -\ell}^{\ell} Y_{\ell m}(\hat{\mathbf{k}}) Y^{*}_{\ell m}(\hat{\boldsymbol \eta}),
\]
and using the orthogonality relation:
\[
\int d(\hat{\mathbf{k}} \cdot \hat{\boldsymbol \eta}) \int d \phi_k \, Y_{\ell m}(\hat{\mathbf{k}}) Y^{*}_{\ell' m'} (\hat{\mathbf{k}}) = \delta^{K}_{\ell \ell'} \delta^{K}_{m m'},
\]
eqn. (\ref{B1}) follows after a few simple steps.

\end{chapter}
\begin{chapter}{VIPERS: geometric and growth rate tests of gravity}
\label{chap:VIPERS_RSD}
\section{Synopsis}
Motivated by difficulties in reconciling the current cosmic acceleration with the vacuum energy density predicted by Quantum Field Theory, I investigate the consistency of the expansion history and linear growth rate with that predicted by General Relativity, when a \cite{Planck} fiducial cosmology is assumed.  Specifically, I constrain the anisotropy of the redshift-space power spectrum with the completed VIPERS v7 census of large-scale structure at $z \simeq 0.8$, which is sensitive to both $D_AH(z)$ and $f \sigma_8(z)$. This Fourier approach complements the VIPERS PDR-1 correlation function analysis by \cite{sylvainClustering}.  The observed consistency of these two approaches provides a reassurance that systematic biases, including those due to non-linear evolution and galaxy bias, are smaller than the statistical error.  

To do so, I measure the optimally weighted \citep{FKP} monopole and quadrupole power spectrum having corrected for the (projected) density dependent VIPERS sampling.  A maximum likelihood calculation is performed by forward modelling the observed signal -- including the effect of the survey mask, non-linear RSD and the Alcock-Paczy\'nski effect \citep{AP}.  A robust estimate of the power spectrum covariance is made using realistic mock catalogues to ensure an accurate likelihood calculation.  Stringent tests of possible systematic biases are carried out, again with the aid of realistic mocks.

I find that the predictions of General Relativity \cite{EinsteinGR}, for a fiducial cosmology compatible with \cite{Planck}, remain consistent with both the inferred expansion history, as quantified by
\[
F_{\rm{AP}} \equiv \frac{(1+z)}{c} D_A H, 
\]
and the linear growth rate.  I place constraints of:
\begin{align}
&f \sigma_8(0.76) = 0.44 \pm 0.04, \nonumber \\
&f \sigma_8(1.05) = 0.28 \pm 0.08, 
\end{align}
on the latter at 68\% confidence; the errors are obtained from the mock-to-mock scatter.  The dependence of these $f \sigma_8$ estimates on the assumed priors will be investigated in future work.  This corroborates the conclusions reached by \cite{sylvainClustering} -- see \S~\ref{sec:CompVIPS}.  These results are a consistency test of GR, defined to be the field equation given by eqn. (\ref{eqn:EinsteinField}), as the linear growth rate is assumed to be scale independent, which limits the applicable actions, and error estimates are obtained assuming this law.

The low-$z$ joint-field constraint is obtained by fitting to a maximum wavenumber: $k_{\rm{max}} = 0.8 \hompc$ and represents a $1.1 \sigma$ deviation from GR; the high-$z$, a $2.4 \sigma$ deviation.  I find GR to be successfully recovered from the mocks with 68\% confidence in both cases; surprisingly, as this is deep into the non-linear regime.  This is consistent with \cite{sylvainClustering}, who found a similar model to be unbiased to $s_{\rm{min}} = 6 \mpcoh$ at 17\% precision; this corresponds to $k_{\rm{max}} = 1.05 \hompc$.  This is also likely the case for this work as, even when including these very small scales, the fractional error remains greater than $9 \%$.  The improvement of this VIPERS v7 constraint is principally driven by the increased volume -- \cite{sylvainClustering} analysed 68\% of the area used in this work.

The quoted redshifts are the mean of the weighted radial selection (including both FKP and ESR weights) for the two radial slices: $0.6<z<0.9$ and $0.9<z<1.2$.  I do not quote the pair-weighted redshift as the difference betwen the two should be much smaller than the above errors; e.g there is only a 7\% difference in $f \sigma_8(z)$ between the quoted redshift for the low-$z$ and high-$z$ slices. 

After marginalising over the anisotropic AP distortion, with the scale dilation factor set to unity \citep{PadmanabhanWhite}, the preferred values are
\begin{align}
&f \sigma_8(0.76) = 0.31 \pm 0.10, \nonumber \\
&f \sigma_8(1.05) = -0.04 \pm 0.26.
\end{align}
This shows no compelling evidence that a $\Lambda$CDM expansion history in a \cite{Planck} cosmology is disfavoured.  In this case, the errors are derived from the width of the data posterior.  The implications of the degeneracy between these two sources of anisotropy and their consistency with General Relativity are explored in \S \ref{sec:AP_degen}. 

\section{Modelling overview}
\label{sec:RSD_modellingOverview}
This section provides a concise summary of the forward modelling of the observed $P_{\ell}(k)$ and states the sections in which a more detailed account may be found.  In particular, this includes the model assumed for both linear and non-linear redshift-space distortions and the necessary corrections for the survey mask and the Alcock-Paczy\'nski (AP) effect.

This AP distortion is introduced by the necessary assumption of a fiducial cosmology when measuring a two-point clustering statistic as a function of comoving separation.  Differences between the assumed cosmology and the truth can cause both a radial scale dilation and introduce additional anisotropy.  When fitting models with a varying expansion history, the power spectrum should be remeasured for each point in parameter space in principle -- after galaxy positions have been recomputed.  As this would be unfeasibly slow, the power spectrum is measured once in practice.  The AP distortion introduced by this fiducial choice is then forward modelled for each point in the parameter space.  Ensuring this additional AP anisotropy is small provides evidence that the fiducial expansion history -- specifically $D_A H(z)$, is close to the truth.  This is a valuable consistency check for galaxy clustering analyses and provides late-time constraints on the expansion history.  Moreover, exhibiting the inherent degeneracy between the AP anisotropy and RSD should be a given for any transparent analysis.  To build a practical algorithm that delivers informative constraints, I adopt an approximate approach: fitting for the AP distortion parameters with $P_g$ fixed to that in the fiducial cosmology.  See Chapter \ref{chap:RSD} for further discussion of the validity and limitations of this approach.     

The foundation of the forward modelling is the multipole moments in the Kaiser-Lorentzian model, which are given in \S \ref{eqn:KL_multipoles}.  This relatively simple model assumes the linear symmetry: $P_{\delta \delta} = P_{\delta \theta} = P_{\theta \theta}$, which is broken to some extent on even the largest scales surveyed, but I prove that this and any additional assumptions are sufficiently valid by ensuring $f \sigma_8$ can be correctly recovered from the mocks.  In particular, the distant observer approximation is assumed throughout as finite-angle effects are negligible for the relatively modest surveyed area, $24 \rm{\ deg}^2$, of VIPERS.  See \S \ref{sec:finiteAngle} and \cite{Yoo} for further detail.

Given a position in the 4D parameter space: $(f \sigma_8, b \sigma_8, \sigma_p, \epsilon)$, the monopole, quadrupole and hexadecapole power spectrum multipole moments ($P_0, P_2$ and $P_4$ respectively) are calculated assuming either a linear theory or non-linear \citep[Halofit-2, ][]{halofit2} prescription for $P_g(k)$;  the latter accounts for the enhanced growth of the power spectrum for $k > 0.1 \hompc$.  These multipole moments are converted to configuration space by Hankel transformation (FFTlog, \citejap{Hamilton}) to rapidly correct for the AP distortion and the survey mask.  The AP effect is corrected for according to eqn. (\ref{eqn:AP_multipoles}) and the survey mask is corrected for according to \cite{maskedRSD}, which is reproduced in Chapter \ref{chap:maskedRSD}.  To do so requires the multipole moments of the mask autocorrelation function, which are shown in Fig. \ref{fig:Qmultipoles}; intermediate corrections are calculated to hexadecapole.  The corrected correlation function multipole moments are then transformed to Fourier space, again by FFTlog.  With these in hand, the likelihood is computed with the estimated covariance, which is further described in \S \ref{sec:Covariance}.  The fiducial cosmology for this analysis is given by eqn. (\ref{eqn:fiducial}) 
\begin{figure}
\centering
\subfloat{\includegraphics[width=\linewidth]{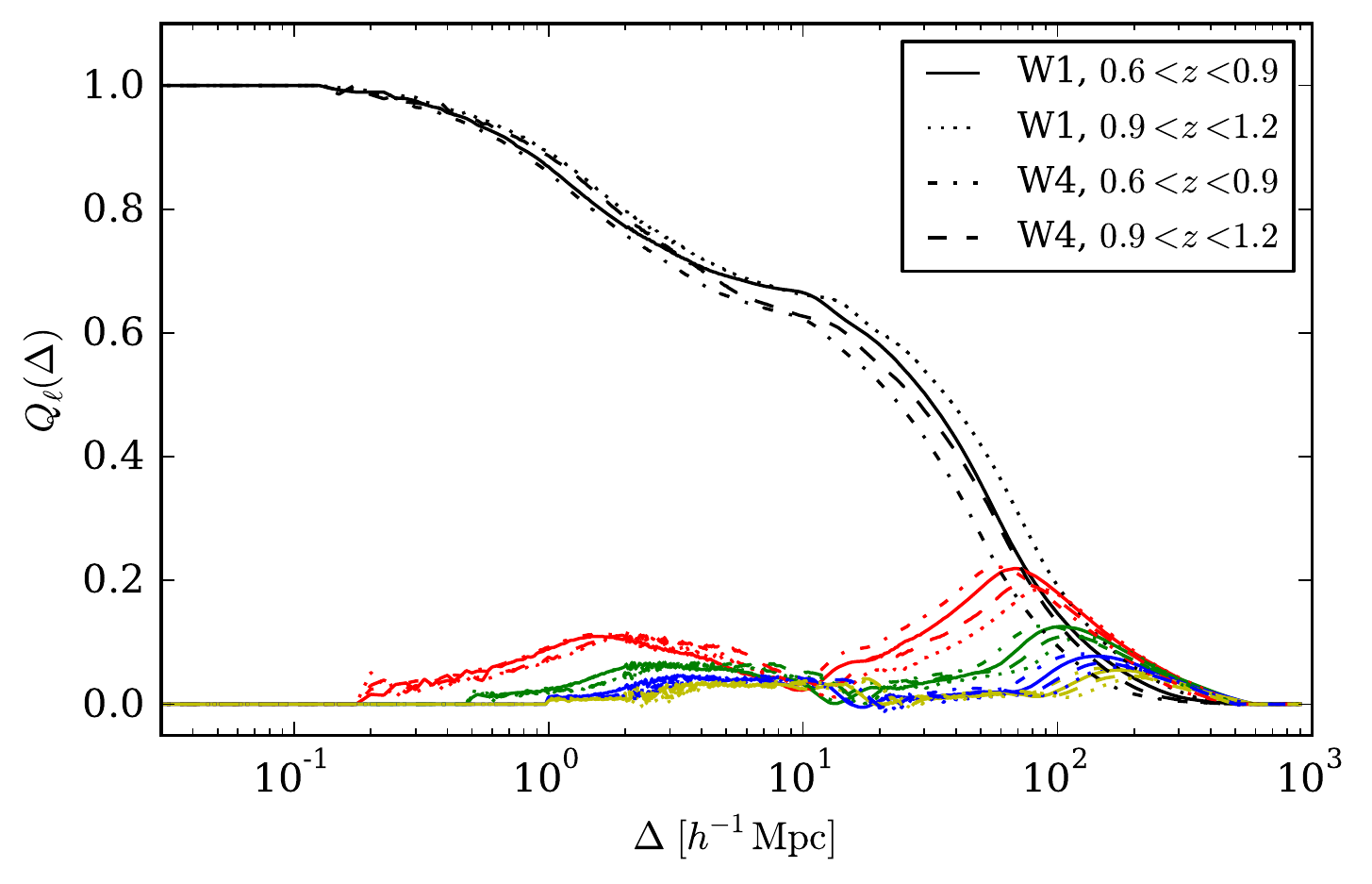}} \\
\subfloat{\includegraphics[width=\linewidth]{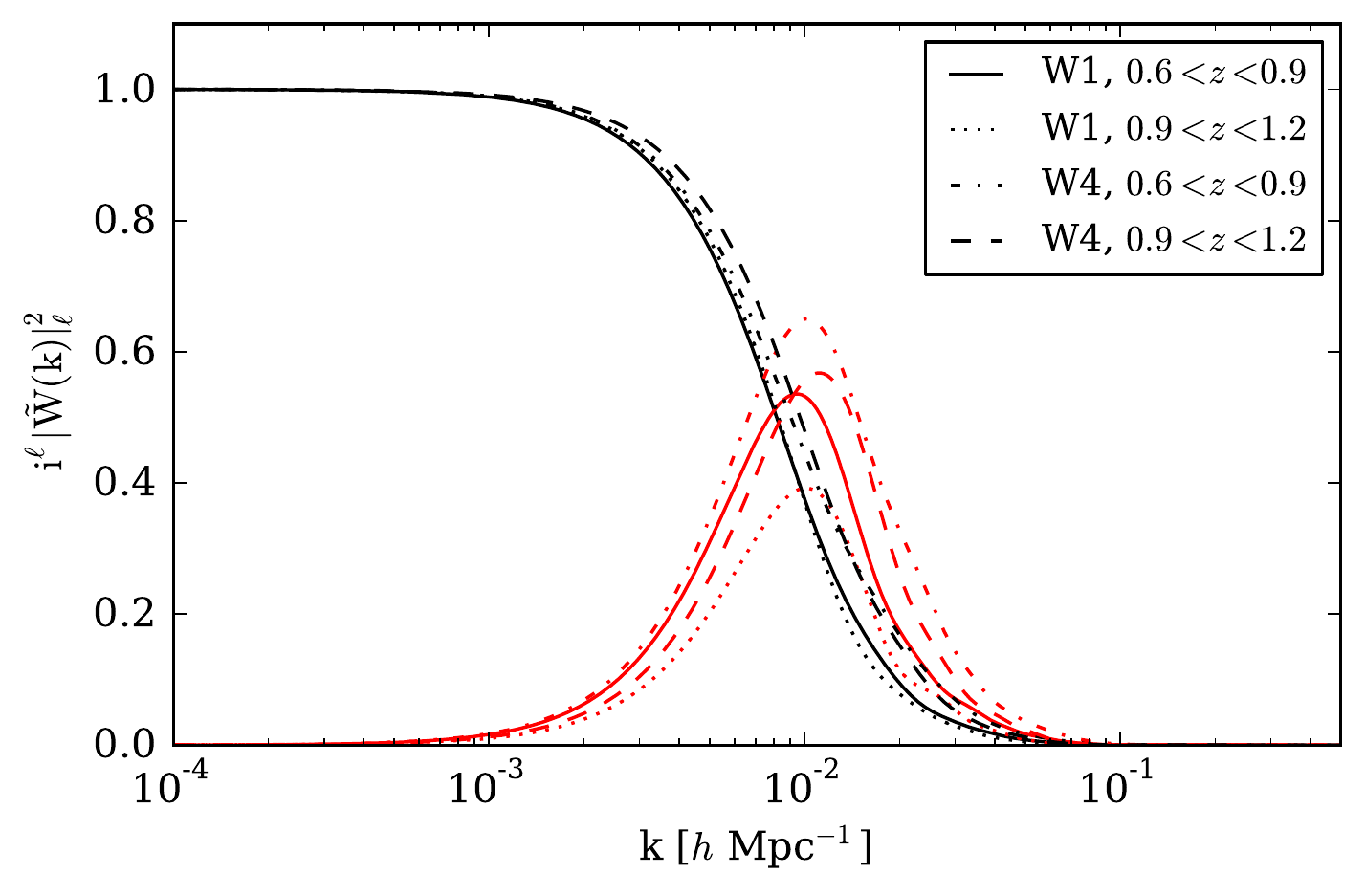}}
\caption[Multipole moments of the VIPERS v7 mask autocorrelation function.]{Top: multipole moments of the mask autocorrelation function, $Q_{\ell}(\Delta)$, for the four subvolumes of the VIPERS v7 survey.  These correspond to the possible permutations of the W1 and W4 fields ($10.692$ \& $5.155$ \emph{deg}$^2$ respectively) and the low-$z$ and high-$z$ slices.  The largest subvolume contains the pairs with the largest separation as expected and has the smallest quadrupole (red) and higher order moments (green, blue and yellow respectively).  This scaling suggests the method of \cite{maskedRSD} should be readily applicable to still larger surveys.  Bottom: the Fourier space multipole moments, which are required for the integral constraint correction discussed in \S \ref{sec:IntConCorr}.}
\label{fig:Qmultipoles}
\end{figure}
\begin{table}
\centering
  \begin{tabular}{lllllllll}
  \hline
    \toprule 
      $z$ interval 
      & \multicolumn{4}{c}{W1} \\
            & $N_{\rm{gal}}/10^4$ 
            & $N_{\rm{ESR}}/10^4$
            & $V/(h^{-1}\rm{Gpc})^3$ 
            & $V_{\rm{eff}}/(h^{-1} \rm{Gpc})^3$  \\
            \midrule
                $0.6 - 0.9$ & 2.82 & 7.37 & $6.5 \times 10^{-3}$ & $6.4 \times 10^{-3}$ \\
                $0.9 - 1.2$ & 0.74 & 2.21 & $9.1 \times 10^{-3}$ & $7.6 \times 10^{-3}$ \\
            %\bottomrule
\end{tabular}
    \begin{tabular}{lllllllll}
    \toprule 
      $z$ interval 
      & \multicolumn{4}{c}{W4} \\
            & $N_{\rm{gal}}/10^4$ 
            & $N_{\rm{ESR}}/10^4$
            & $V/(h^{-1}\rm{Gpc})^3$ 
            & $V_{\rm{eff}}/(h^{-1} \rm{Gpc})^3$  \\
            \midrule
                $0.6 - 0.9$ & 1.41 & 3.76 & $3.1 \times 10^{-3}$ & $3.1 \times 10^{-3}$ \\
                $0.9 - 1.2$ & 0.33 & 1.00 & $4.4 \times 10^{-3}$ & $3.6 \times 10^{-3}$ \\
            \bottomrule
\end{tabular}
\caption[Subvolume properties of the VIPERS v7 data release.]{Subvolume properties: shown are the number of galaxies, $N_{\rm{gal}}$, that meet the criteria on redshift selection and security, $(2 \leq f_z \leq 10) \ | \ (12 \leq f_z \leq 20)$ -- where $f_z$ denotes the redshift flag; these redshifts are secure with 98\% confidence.  $N_{\rm{ESR}}$ is the effective number following the ESR correction and the difference reflects the $\simeq 30 \%$ average VIPERS sampling.  The volume of each partition (for the Nagoya v7 \& Samhain angular mask and in the fiducial cosmology) are also provided. The FKP weighted effective volume is given by the volume integral of the two-point volume weighting, $[\bar n P_0 / (1 + \bar n P_0 )]^2$ (\citejap{FKP}, \citejap{Tegmark06}) and quantifies the expected precision.  This is quoted for the ESR corrected $\bar n(z)$ and assuming a fiducial value of $P_0 = 8000 (h^{-1} \rm{Mpc})^3$.}
\label{table:vol_stats}
\end{table}

\section{A VIPERS v7 measurement of the growth rate}
Chapter \ref{chap:RSD} described the anisotropy of the redshift-space power spectrum and how this may be used to constrain the growth rate of large-scale structure.  This section presents a new power spectrum measurement of the completed VIPERS v7 data together with numerous statistics that quantify its anisotropy.  The derived constraints on $f \sigma_8$ and $F_{AP}$ are presented, for both the data and mock catalogues; the latter prove the method to be robust and provide the expected statistical errors.

\subsection{Measures of the power spectrum anisotropy}
\begin{figure}
\centering
\includegraphics[width=\textwidth]{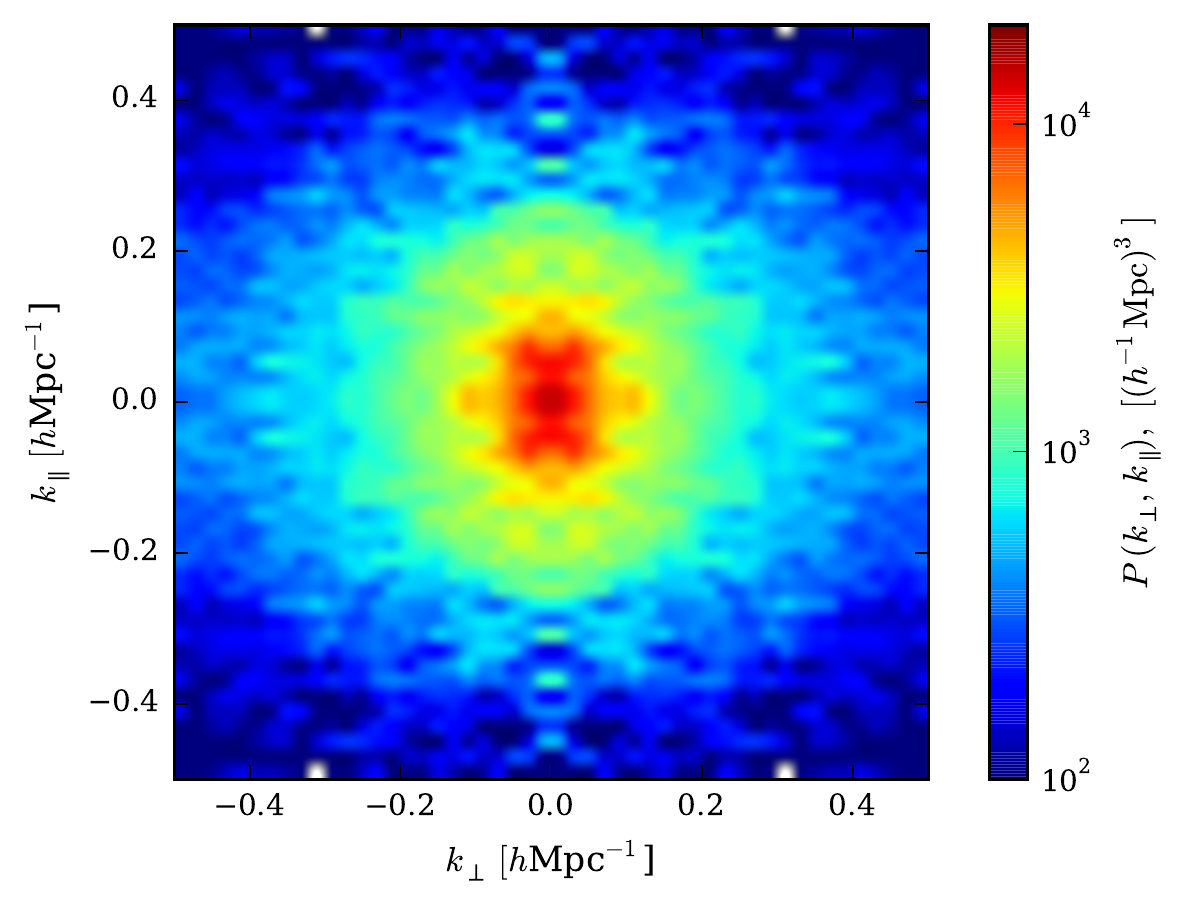}
\caption[A VIPERS v7: W1 measurement of $P(k_{\perp}, k_{\parallel})$ for $0.6<z<0.9$.]{The measured 2D power spectrum, $P(k_{\perp}, k_{\parallel})$, for the VIPERS v7: W1, $0.6<z<0.9$ dataset.  The independent quadrant contains $25 \times 25$ bins, which have been smoothed with a centripetal Catmull-Rom spline \citep{Matplotlib}.  The power spectrum anisotropy is clearly evident on large scales, $k<0.1 \hompc$, due to both linear redshift-space distortions and the survey mask anisotropy \citep{maskedRSD}.  On smaller scales, the non-linear velocity dispersion of galaxies in groups and clusters suppresses the Kaiser amplification;  See Fig. 1 of \cite{ColeFourierOmega} and discussion therein.  The remainder of this analysis focuses on extracting the most statistically significant measures of the scale-dependent anisotropy in order to constrain the growth rate of density fluctuations.}
\label{fig:vipers_data_2dpk}
\end{figure}
Fig. \ref{fig:vipers_data_2dpk} shows the 2D redshift-space power spectrum of the VIPERS v7: W1, $0.6<z<0.9$ dataset.  This is measured according to \S \ref{sec:FKP_pk}, following the correction of the (projected) density dependent ESR.  With the survey embedded in a $(800 \mpcoh)^3$ cubic volume, which has fundamental mode of $(2 \pi/800 \mpcoh) = 7.85 \times 10^{-3} \hompc$, a $256^3$ FFT was computed.  This results in a  cell size of $3.125 \mpcoh$ and a Nyquist frequency of $1.01 \hompc$.  The power measurements shown for $k > 0.4 \hompc$ are obtained by 
halving the cell size with a single Jenkins's fold; this doubles the Nyquist frequency.  Further detail of Jenkins's folding may be found in \S \ref{sec:Jenkins}.  I have ensured the results are robust to these choices.

To combine Jenkins's folding and the FKP estimator for a non-trivial survey mask the measured power spectrum must be differenced with that from a large random catalogue.  When estimating the power spectrum for a given mock I reassign the radial position of each random to one consistent with that particular joint-field $\overline n(z)$ estimate, as is the case for the data.  This ensures that the mocks analysis is as similar as possible to that for the data.  The covariance matrix estimate then includes the variance resulting from the necessity of estimating the radial selection from the survey itself.  As a Gaussian smoothed, ESR corrected, radial selection is estimated, the radial positions of randoms are drawn from the probability density: $P(r) = r^2 \ \overline n(r) dr$ based on the cumulative distribution -- see pg. 288 of \cite{TheBible}.  To do so requires inverting the cumulative distribution, which I achieve by reversing the arguments of a cubic-spline interpolation.  The alternative of drawing redshifts from those available in each mock or dataset (with the addition of Gaussian errors) is excluded by the non-discrete ESR weights.                        

An accurate likelihood calculation requires a sufficiently converged covariance estimate \citep{TaylorCovariance, PercivalCovariance}.  This is more easily obtained by reducing the number of data points; therefore I retain only the most statistically significant estimates of the scale-dependent anisotropy: the monopole and quadrupole moments up to a given $k_{\rm{max}}$.  This is discussed further in \S \ref{sec:Covariance}.  The mean power spectrum multipoles for an ensemble of mocks are shown in Fig. \ref{fig:mocks_multipoles}; these mocks have been described in detail in \S \ref{sec:mocks}.  The reasonable agreement between the model and mean measurement inspires confidence in the analysis method -- in particular, in the mask correction and assumed RSD model.  This statement is supported by a quantitative analysis that follows, in which the inferred posteriors on $f \sigma_8$ for the mocks are compared to the known GR expectation.  

The VIPERS v7 data multipole moment measurements are shown in Fig.~\ref{fig:data_multipoles}.  Significant deviations from the model are to be expected in this case, due to the large errors expected for $k < 0.1 \hompc$.  The application of a simple dispersion model may questioned s the smaller scales are determined by both real-space and redshift-space non-linearity.  The validity of this model is proven by ensuring the known $f \sigma_8$ can be correctly recovered from the mocks.  This is shown to be the case in the following section.          
\begin{figure}
\centering
\includegraphics[width=\textwidth]{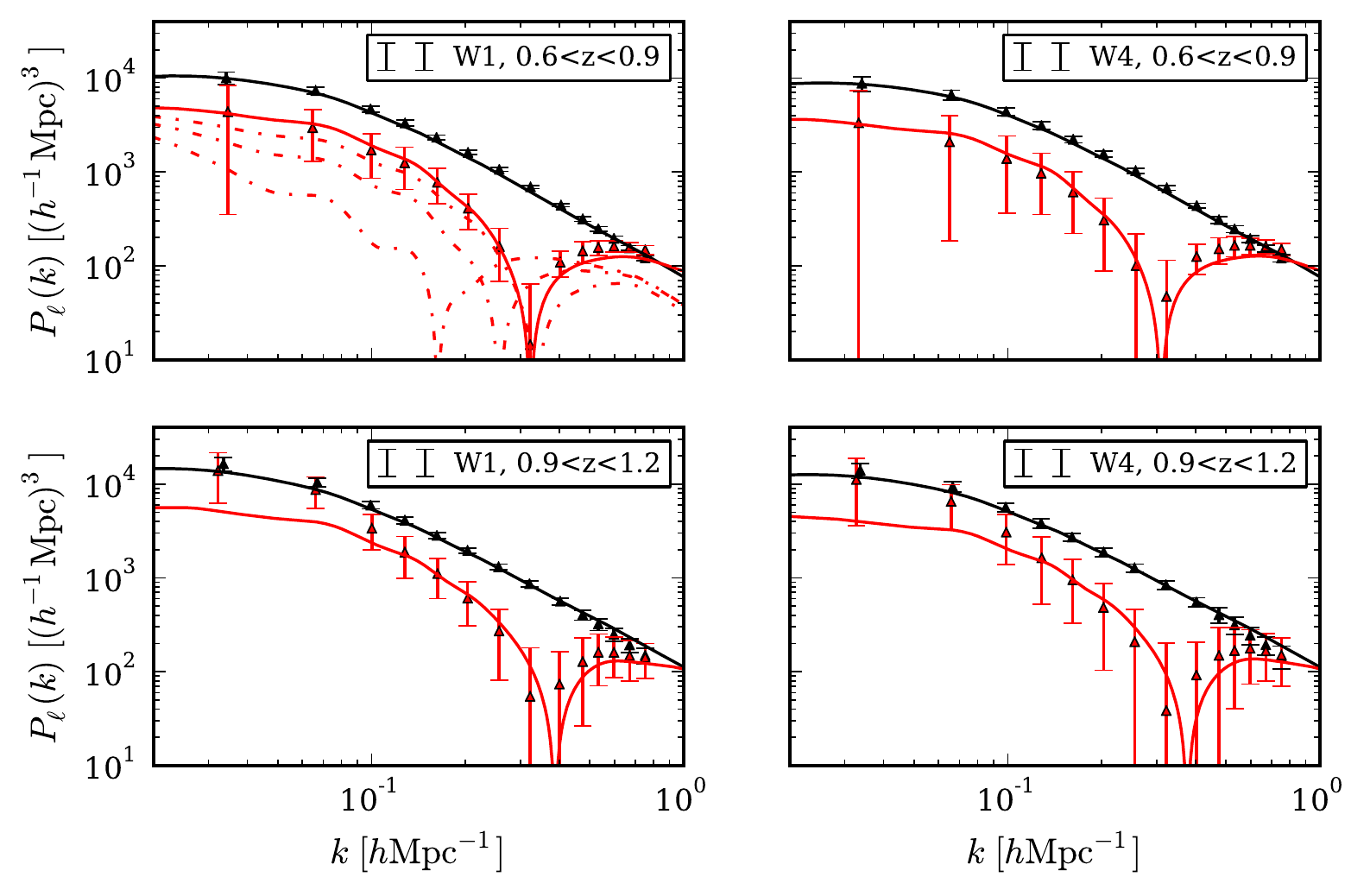}
\caption[Mean $P_0(k)$ \& $P_2(k)$ mock estimates and their expectation.]{Shown are the mean $P_0(k)$ and $P_2(k)$ moments (black and red respectively) of the 306 VIPERS mocks when split by subvolume; errors show the mock-to-mock scatter.  The mask-corrected dispersion model is overplotted for a non-linear $P_g(k)$ given by Halofit-2 \citep{halofit2}.  The cosmology assumed in the mock construction defines $f \sigma_8$ for the mean redshift and $(b \sigma_8, \sigma_p)$ have been roughly estimated by inspection.  The top left panel shows additional models for $f \sigma_8 = \{ 0.15, 0.25, 0.35\}$ and the $\Lambda$CDM expectation of $f \sigma_8 (0.75) = 0.49$.  There is little evidence for a statistically significant bias and the model provides an acceptable fit to even non-linear scales.}
\label{fig:mocks_multipoles}
\end{figure}

\begin{figure}
\centering
\includegraphics[width=\textwidth]{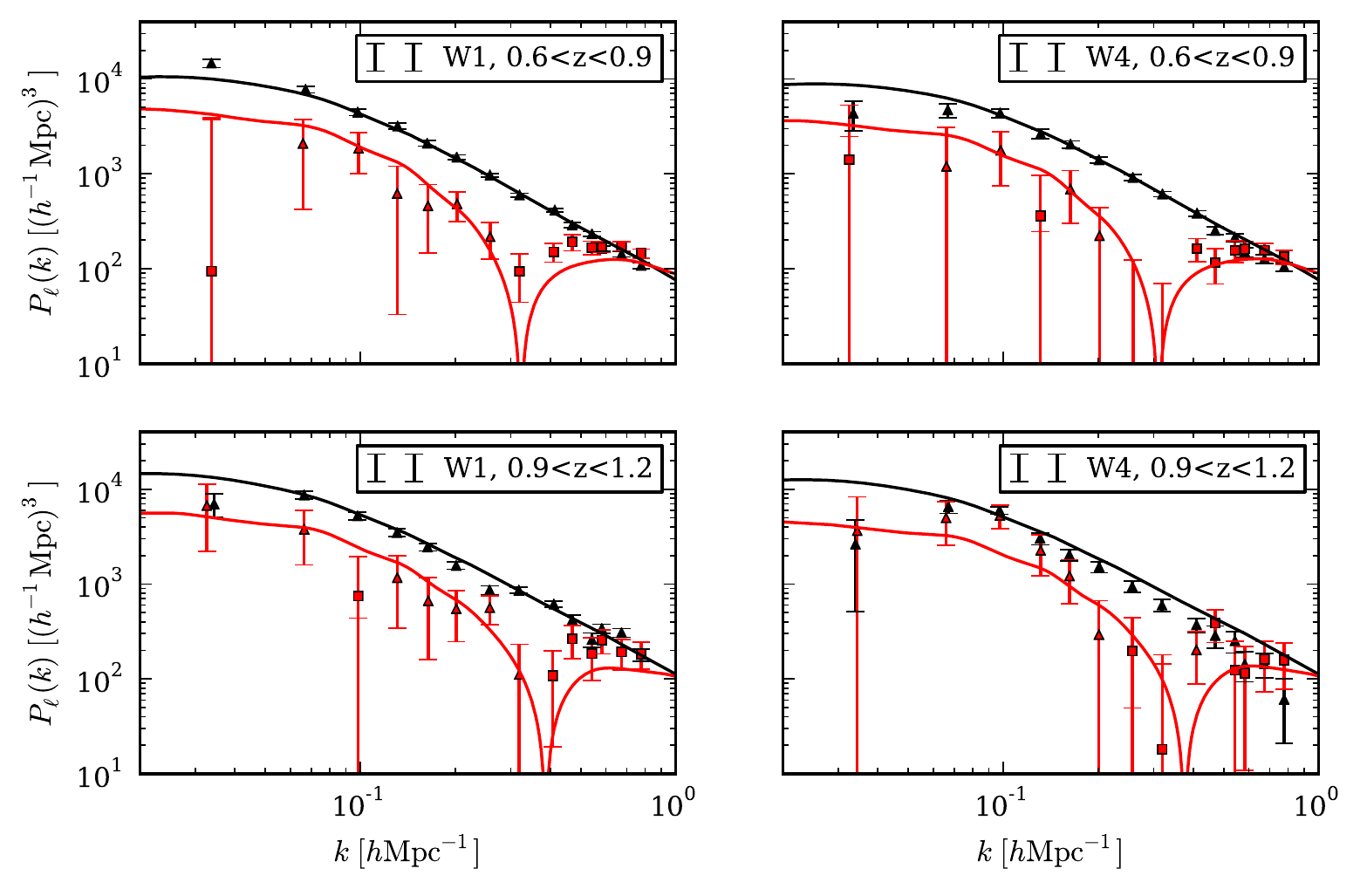}
\caption[Estimates of $P_0(k)$ \& $P_2(k)$ from the VIPERS v7 data.]{Same format as Fig \ref{fig:mocks_multipoles} but the VIPERS v7 data measurements are shown in this case; squares denote negative amplitudes and error bars replicate the mock-to-mock scatter.  The models shown are identical to those previously.  Given the sizeable errors and the large covariance on small scales, there is reasonable agreement shown for each subvolume -- certainly up to mildly non-linear scales.  The W1 high-$z$ slice is seemingly discrepant in isolation and I calculate the significance of the combined measurements in \S \ref{sec:StatSig}.}
\label{fig:data_multipoles}
\end{figure}

\subsection{Derived constraints on $f \sigma_8(z)$}
\begin{figure}
%\pagenumbering{gobble}
\centering
\subfloat{\includegraphics[width=\linewidth]{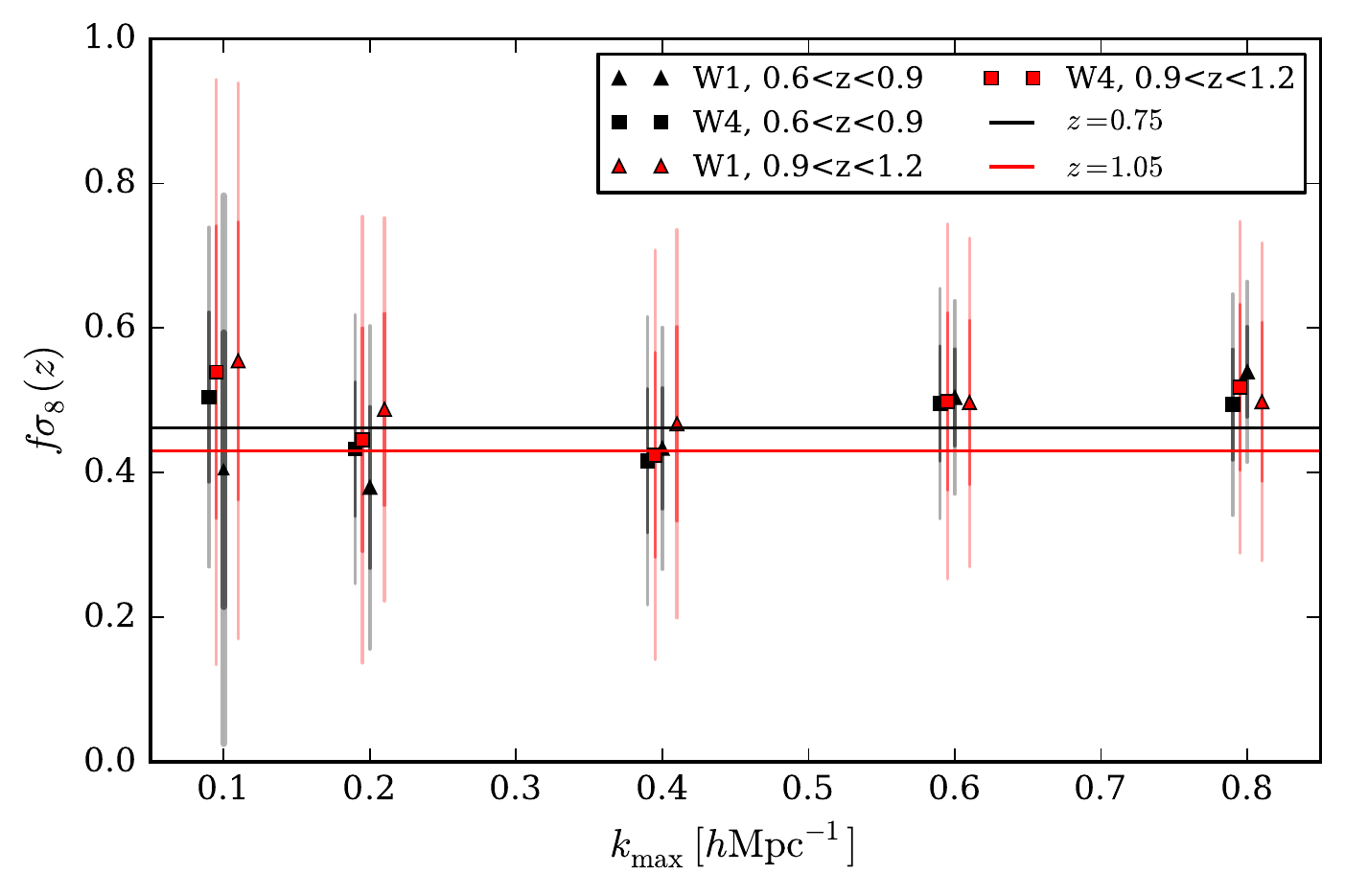}} \\
\subfloat{\includegraphics[width=\linewidth]{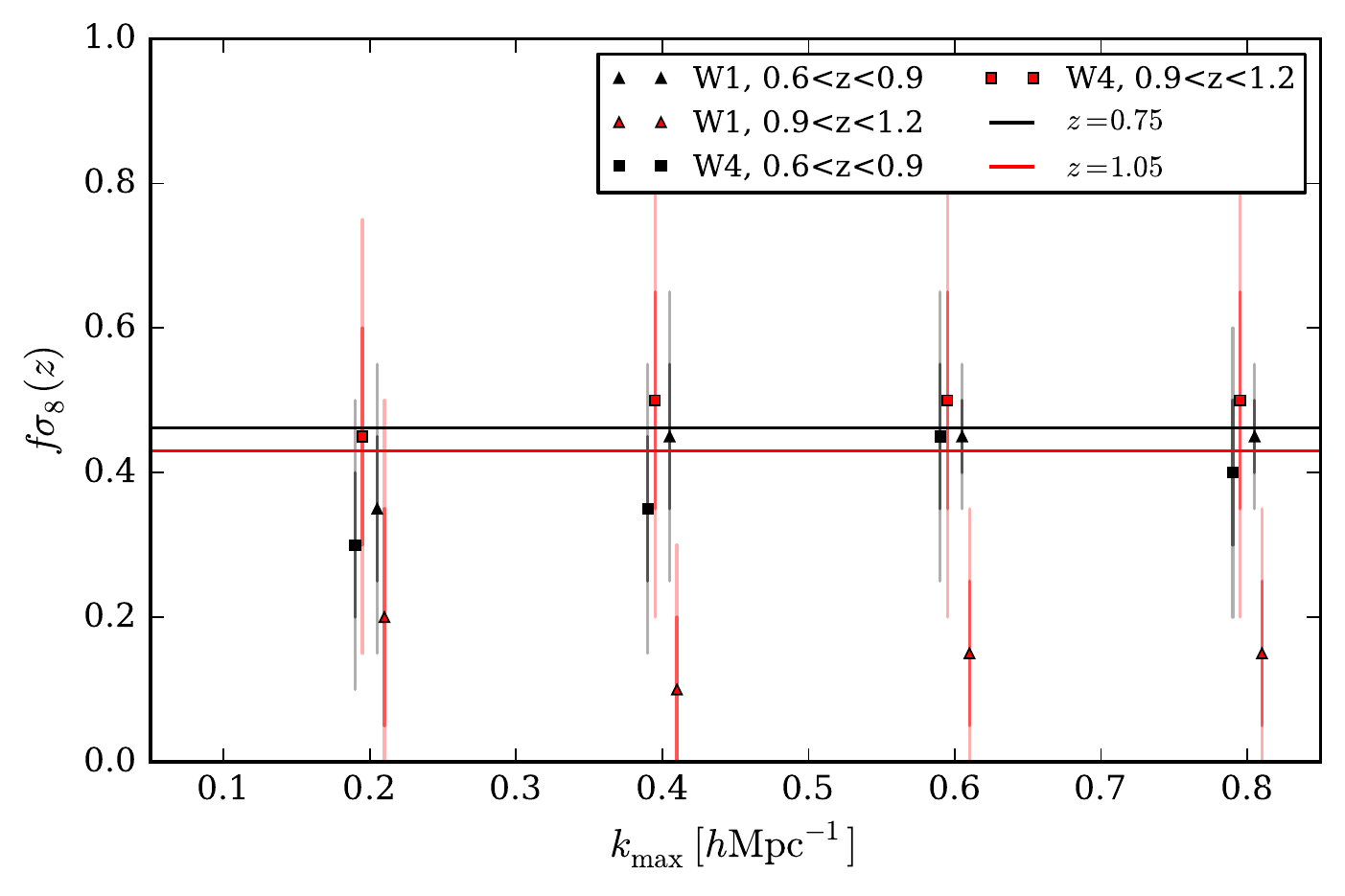}}
\caption[Confidence intervals on $f \sigma_8$ for the VIPERS mocks]{Top: 68 and 95\% confidence limits on $f \sigma_8$ for the mocks.  The dispersion model is seemingly unbiased up to $0.8 h \emph{Mpc}^{-1}$ when marginalised over $(b \sigma_8, \sigma_p)$ -- surprisingly, as this is deep into the non-linear regime.  This is partly due to the $\simeq 20 \%$ errors, which only moderately change beyond $0.2 h \emph{Mpc}^{-1}$.  The multipole covariance and $(f \sigma_8, \sigma_p)$ degeneracy must therefore reduce the small-scale constraining power.  A larger $f \sigma_8$ is favoured for larger $k_{\rm{max}}$ although the shift is small.  Bottom: Same format for the VIPERS v7 data.  Good agreement with GR is shown for every subvolume except the W1 high-$z$ slice.  This is seemingly discrepant in isolation.}
\label{fig:fsig8_kmax}
%\pagenumbering{arabic}
\end{figure}
\begin{figure}
\centering
\subfloat{\includegraphics[width=\linewidth]{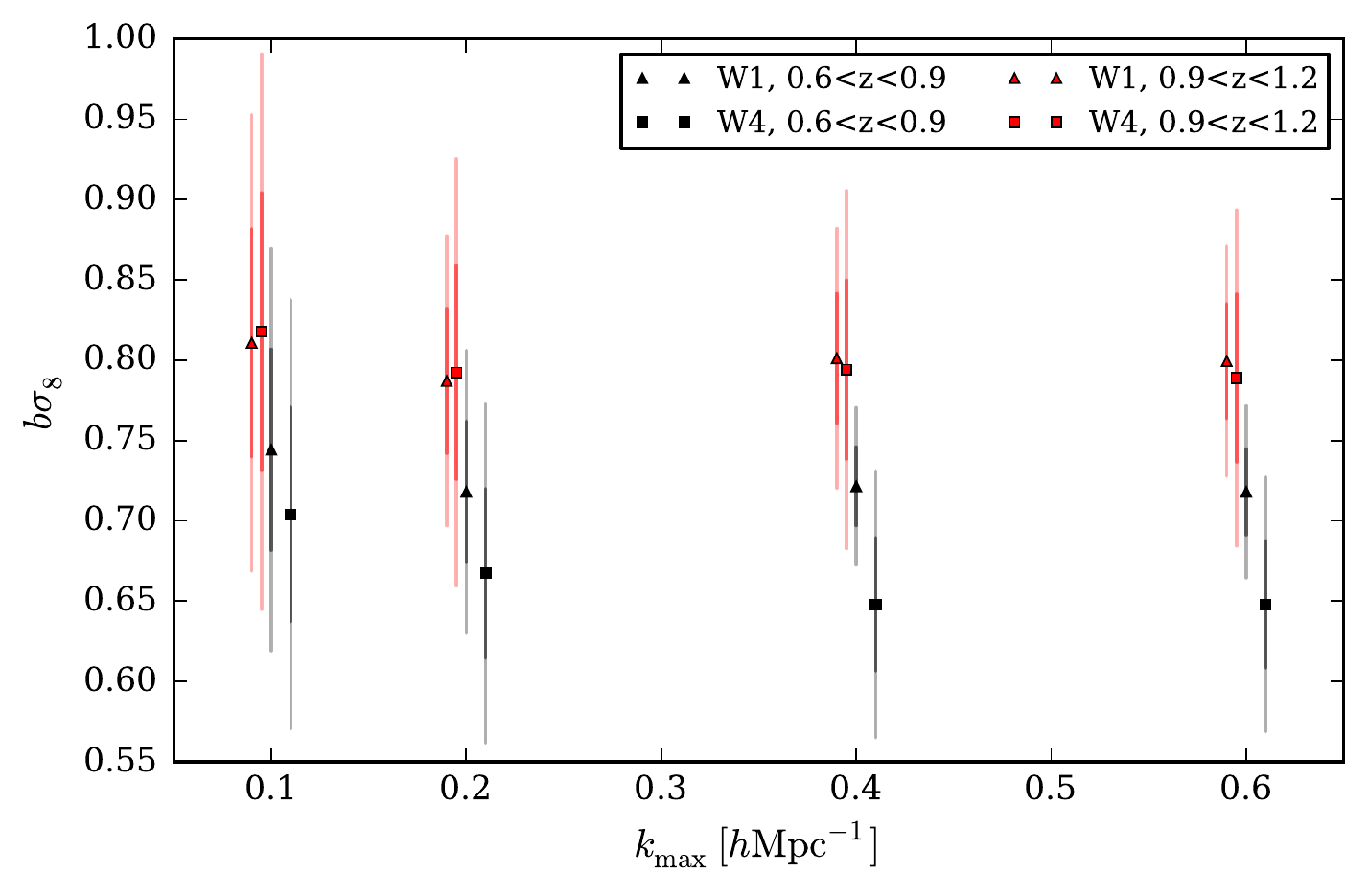}} \\
\subfloat{\includegraphics[width=\linewidth]{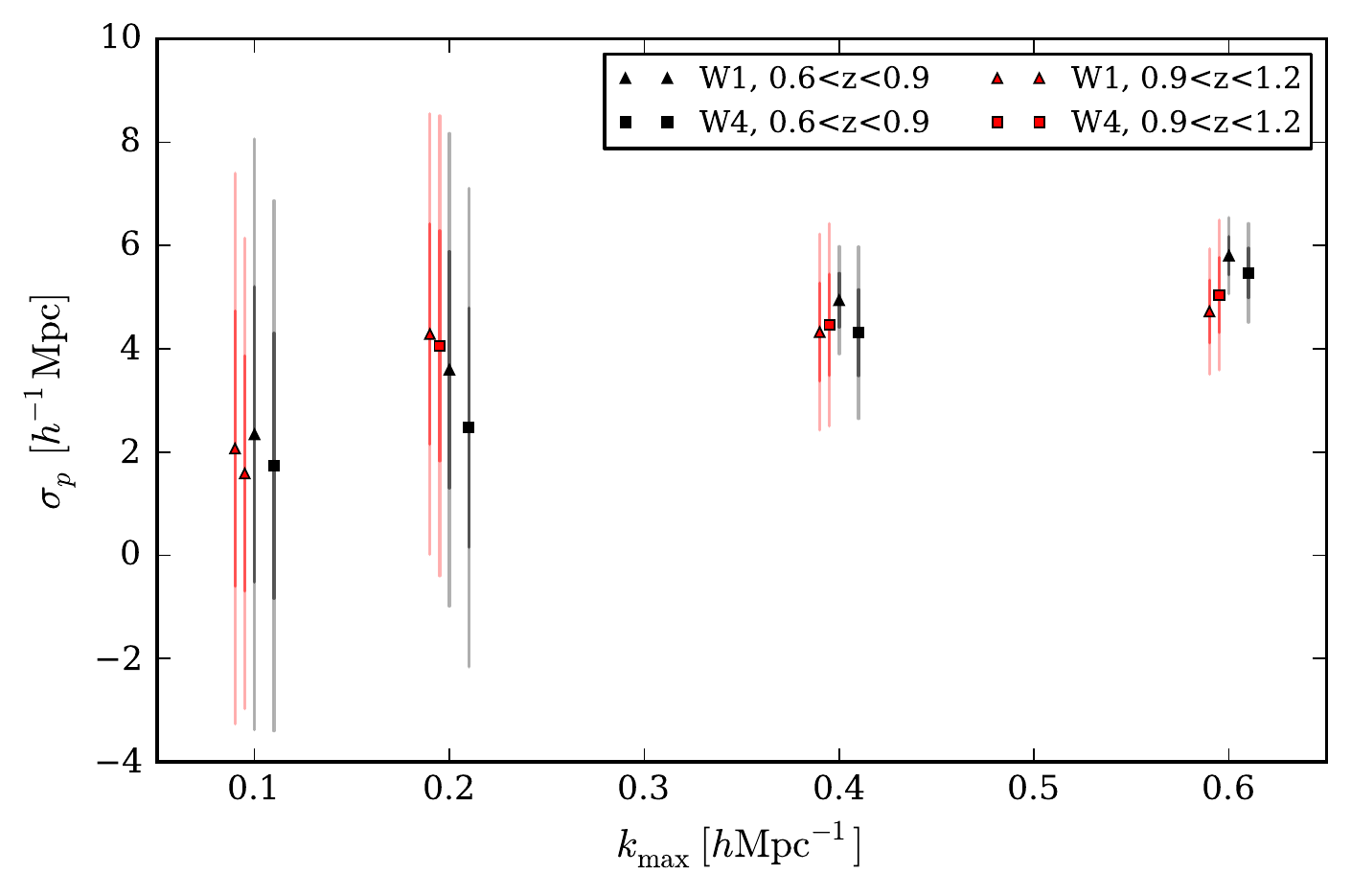}}
\caption[Confidence intervals on $b \sigma_8$ and $\sigma_p$ for the VIPERS mocks.]{Top: $b \sigma_8(z)$ posteriors for the VIPERS mocks as a function of the minimum fitting scale, $k_{\rm{max}}$.  Bottom: derived posteriors on $\sigma_p$ when marginalised over $f \sigma_8$ and $b \sigma_8$.  $\sigma_p$ is assumed to be positive definite within the prior range and the 68\% interval is to be interpreted as showing the standard deviation only (as opposed to intervals of cumulative probability).}
\end{figure}
Assuming a known expansion history and therefore a negligible AP distortion, I derive posteriors on $f \sigma_8(z)$ having assumed $(\alpha, \epsilon) = (1.0, 0.0)$ and conservative flat priors of  
\begin{alignat}{3}
0.05 &\leq f \sigma_8 \leq 0.80,  \notag \\
0.05 &\leq b \sigma_8 \leq 1.05,  \notag \\
0.00 &\leq \sigma_p/(h^{-1} \rm{Mpc}) \leq 6.0, 
\label{eqn:priors}
\end{alignat}

The likelihood is computed for a $16^3$ grid spanning a 3D parameter space defined by these boundaries.  To so so, I first `prewhiten' the covariance \citep{NorbergStats} -- rescale the covariance to the correlation matrix.  Using the pre-whitened matrix eases the danger of roundoff error in the necessary eigenvalue, $\Lambda_i$, and eigenvector estimation,  which is a problem because the covariance matrix elements span many decades in wavenumber. This is achieved by $\mathbf x_i \mapsto \mathbf{x}_i/C_{ii}$; no sum on $i$ is implied and $\mathbf x$ is the data vector to be fitted -- the binned $P_0$ and $P_2$ estimates at a number of wavenumbers. 

Following prewhitening, if the eigenvectors (principal components) of the correlation matrix are $\mathbf e_i$ then new variables with a diagonal correlation may be defined: $\mathbf y_j = \mathbf e_j \cdot \mathbf x$; the expectation of each is trivially: $\langle \mathbf y_j \rangle = \mathbf e_j \cdot \langle \mathbf x \rangle$.  The $\chi^2$ for a given realisation of $\mathbf y$ is then easily calculated with the appropriate covariance, $\diag(\Lambda_1, \Lambda_2, \cdots, \Lambda_N)$.  As the results shown here satisfy a convergence test of the estimated covariance, see \S \ref{sec:Covariance}, I do not consider shrinkage estimators \citep{PopeSzapudi} or singular value decomposition \citep{TheBible}.  The latter would proceed by removing the least informative $\mathbf y _j$ -- that with the smallest variance (eigenvalue), from the $\chi^2$ calculation.  See \S 3.3 of \cite{NorbergPeder}, \S 2.6 of \cite{TheBible} or \cite{MultivariateAnalysis} for further detail.
\begin{figure}
\centering
\includegraphics[width=\textwidth]{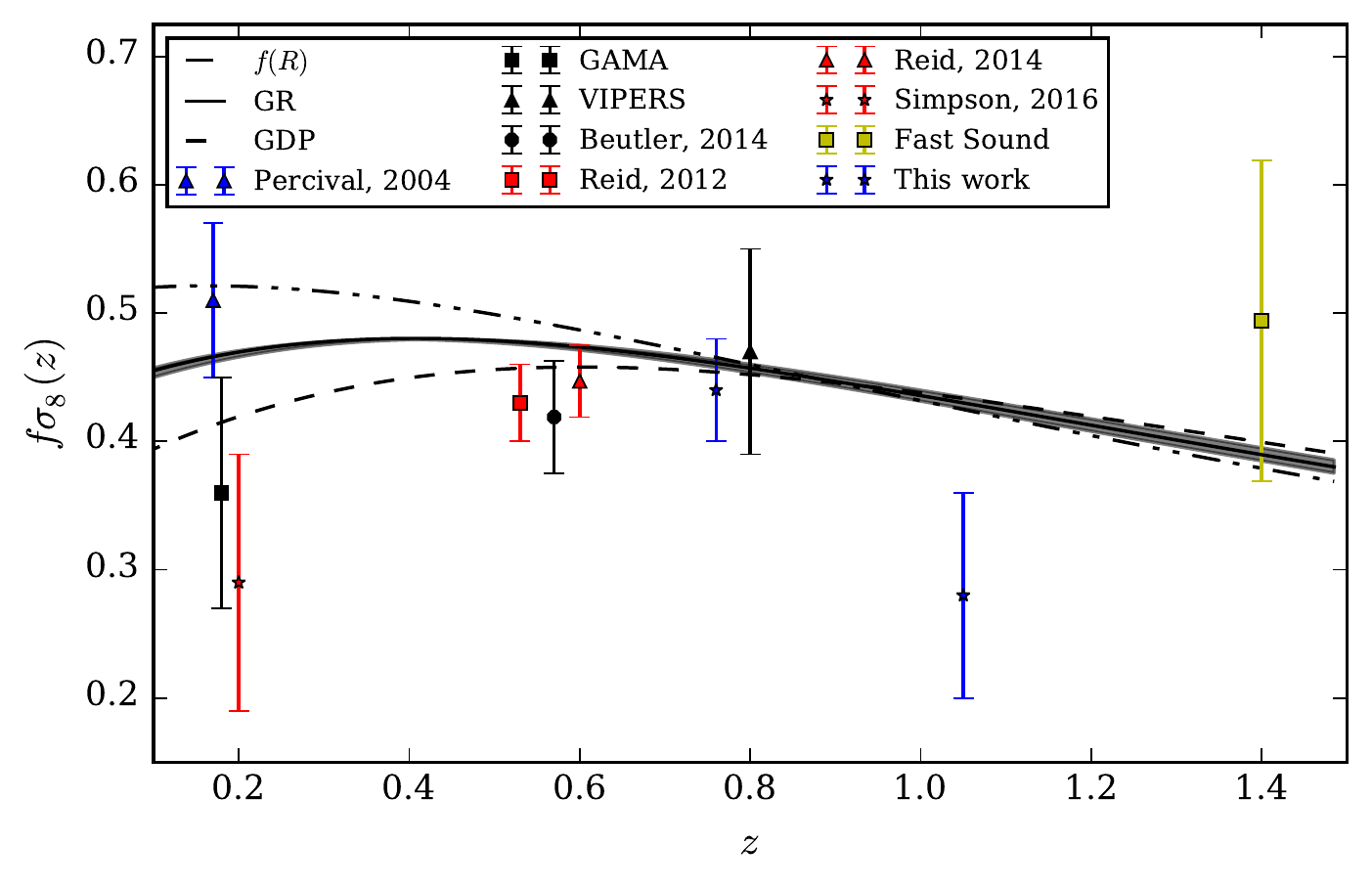}
\caption[Constraints on $f \sigma_8(z)$ from VIPERS v7 and recent surveys.]{The VIPERS v7 constraint on $f \sigma_8(z)$ placed by this work (blue star) and by other recent surveys.  The GR prediction for a \cite{Planck} fiducial cosmology is shown (black) with an associated errorband due to the finite precision of the $\Lambda$CDM parameter estimates -- see the text for further discussion.  The predictions of the \cite{DGP} model (dot-dashed) and the \cite{HuSawicki} $f(R)$ model are also shown; these span the range $0.4 < \gamma < 0.7$.  These measurements represent a slightly unlikely realisation, but the measurements of future surveys (eBOSS, DESI, Euclid) will be much more conclusive.  An introduction to RSD and constraints on the linear growth rate is given in Chapter \ref{chap:RSD}.}
\label{fig:fsig8_z}
\end{figure}

The resulting posteriors on $f \sigma_8(z)$ from measurements of both the mocks and VIPERS v7 data are shown in Fig. \ref{fig:fsig8_kmax}.  The mocks analysis shows no statistically significant deviations from General Relativity and recovers the input value with $68-95 \%$ confidence.  Three subvolumes of the data are also in very good agreement with the GR expectation, but the W1 high-$z$ slice is clearly discrepant in isolation.  I analyse the significance of the combined measurements in \S \ref{sec:StatSig}.

\subsection{Goodness-of-fit}
From a Bayesian perspective, the VIPERS v7 dataset will always update the degree-of-belief or credence of a model or parameter of interest -- predominantly $f \sigma_8(z)$ in RSD analyses.  But it is good practice to ensure that the maximum-likelihood model really does provide a satisfactory fit to the data.  The most common statistic for doing so is the \smash{reduced-$\chi^2$} \citep{TheBible}.

In this case the measurements are assumed to be normally distributed about the expectation, which is a good approximation for bins containing a large number of modes given the central limit theorem (but note that large-$k$ modes are not independent).  The distribution of the reduced-$\chi^2$ statistic, $(\chi^2/ \nu)$, is given by eqn. (176) of
\url{http://www.roe.ac.uk/~jap/teaching/astrostats.html}; see \cite{JasperWall} for an indepth discussion.  Here $\nu$ is the number of degrees of freedom -- equal to the number of data points minus the number of fitted parameters.  Fig. \ref{fig:absChi2} shows a comparison of the reduced-$\chi^2$ distribution of the $f \sigma_8$ estimes for the VIPERS mocks with this expectation, for two values of $k_{\rm{max}}$.           
\begin{figure}
\centering
\includegraphics[width=\textwidth]{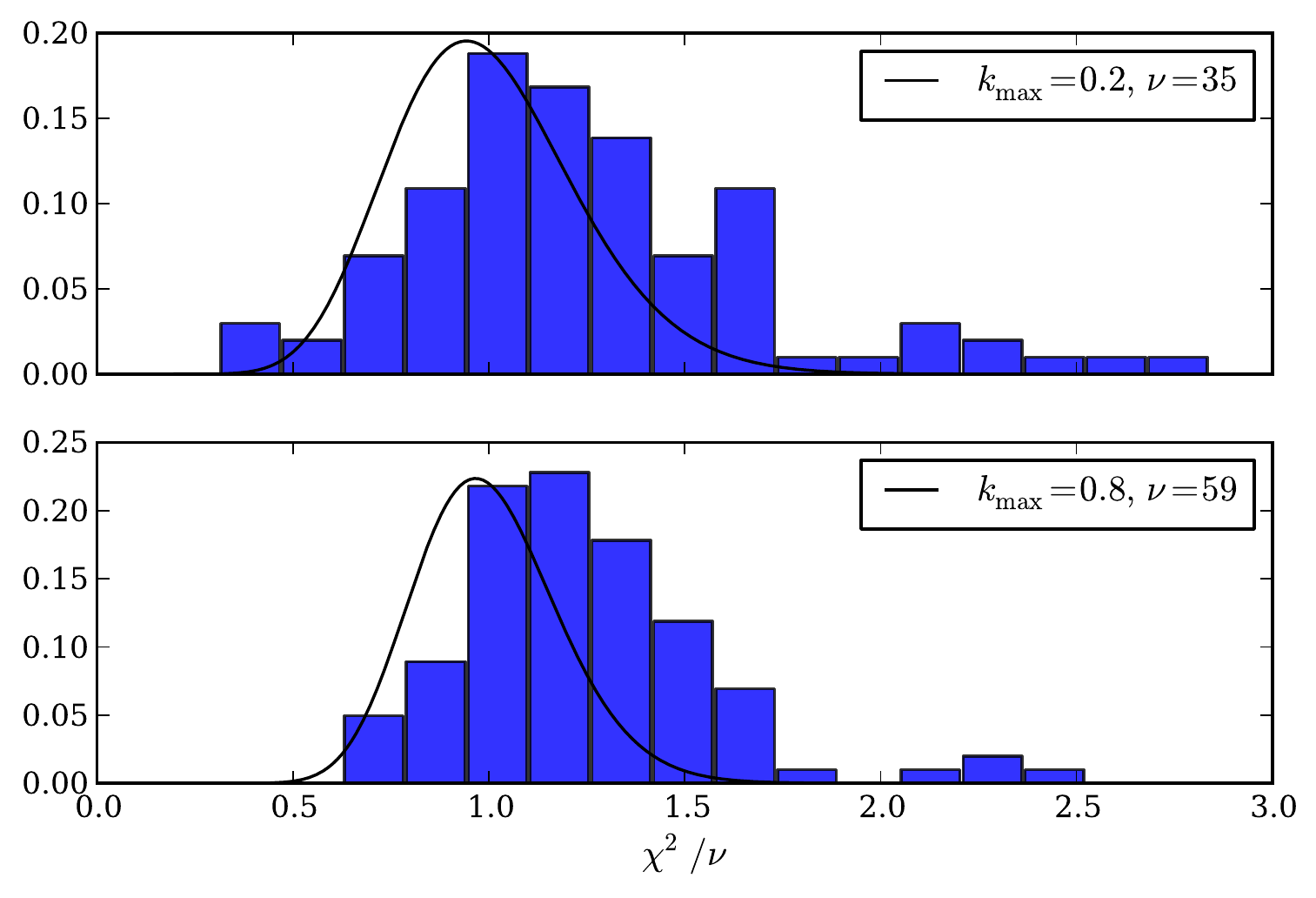}
\caption[A goodness-of-fit test of the mocks measurements.]{It is clear from the plots above that the assumed model provides an acceptable fit and recovers an unbiased $f \sigma_8(z)$ estimate.  This is sufficient from a Bayesian perspective, the VIPERS v7 data will always update our credence of $f \sigma_8$.  But it is good practice to see how well the maximum likelihood model actually fits the data.  This can be quantified by the reduced-$\chi^2$, which both tests whether the errors are normally distributed and accurately estimated.  The reduced-$\chi^2$ distribution of the $f \sigma_8$ estimates from the mocks is shown, for two $k_{\rm{max}}$ values; the expectation for each $\nu$ \citep{JasperWall} is overlaid.  Clearly there is some discrepancy, as a larger reduced-$\chi^2$ is favoured and there is a long tail for $k_{\rm{max}} = 0.2 h \emph{Mpc}^{-1}$; this warrants further investigation, especially for surveys that achieve a greater precision.  If non-Gaussian errors are present, robust parameter errors may still be derived from the measured scatter between the mocks.}
\label{fig:absChi2}
\end{figure}

\section{Tests of systematics}
This section records the results of various systematic tests of the method, which were performed with the realistic VIPERS mocks.  Arguments for the validity of additional modelling assumptions are also given; these are based on indirect evidence if not tested directly. 

\subsection{Potential biases due to the $\bar n(z)$ estimate.}
The VIPERS v7 radial selection function of the mocks and data can be seen in Fig. \ref{fig:nbar}; the estimation method is described in the accompanying text.  It is clear from this figure that a smoothed $\bar n(z)$ estimate, which reduces the intrinsic radial clustering, can recover an unbiased estimate.  However, this joint-field $\bar n (z)$ results in power spectrum measurements for the W1 and W4 fields that are dependent -- assuming the intrinsic covariance (given by $\xi (r)$ for scales close to the homogeneous limit, $r \gg 1 \mpcoh$) is negligible.  This effect is difficult to quantify as the two fields of each VIPERS mock may not be extracted from a unique volume, i.e. the same dark matter field may be used for more than one field or mock.  Therefore it is left to future analyses to ensure that this is a negligible systematic.

Fig. \ref{fig:pk_ell_nbar} shows the result of the necessity of a radial selection function estimate on the power spectrum multipole moments.  Further discussion is to be found in the accompanying caption, but the main conclusions are: the difference is within the expected sample variance; this difference is somewhat accounted for by the integral constraint correction and the effect is likely to be greater for the high-$z$ slice.  Ultimately, the method is validated as $f \sigma_8$ can be successfully recovered from the mocks.  Further evidence for this may be found in Rota et al. (in prep.).  

An investigation of any resulting bias in $f \sigma_8(z)$ is shown in Fig. \ref{fig:fsig8_truenbar}. It may be concluded from this figure that any systematic shift is within the 68\% error and that GR can still be successfully recovered; this is despite of the necessity of estimating $\bar n (z)$ from the survey itself.
%\afterpage{%
%\thispagestyle{empty}
\begin{figure}
\centering
\subfloat{\includegraphics[width=\linewidth]{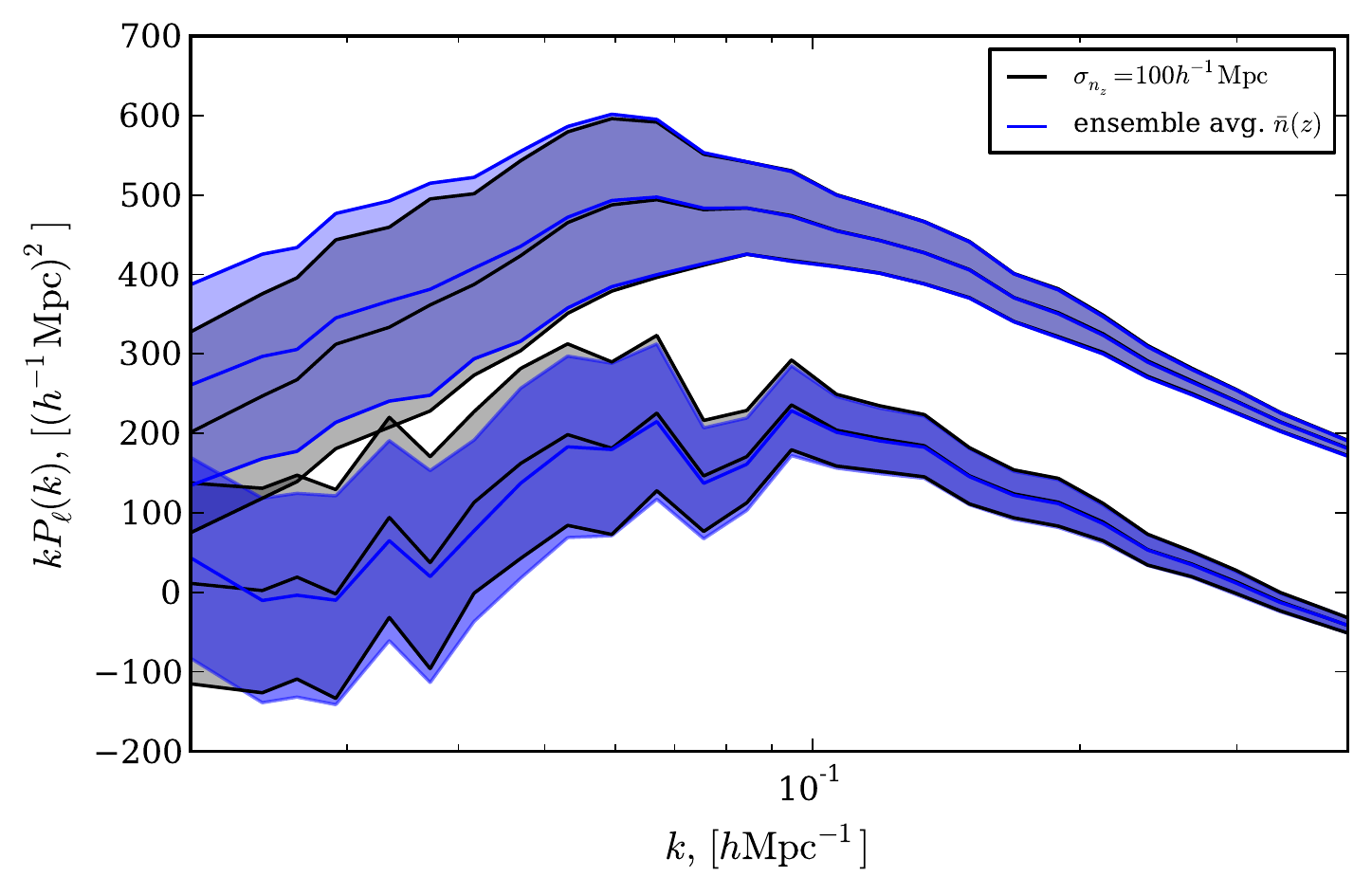}} \\
\subfloat{\includegraphics[width=\linewidth]{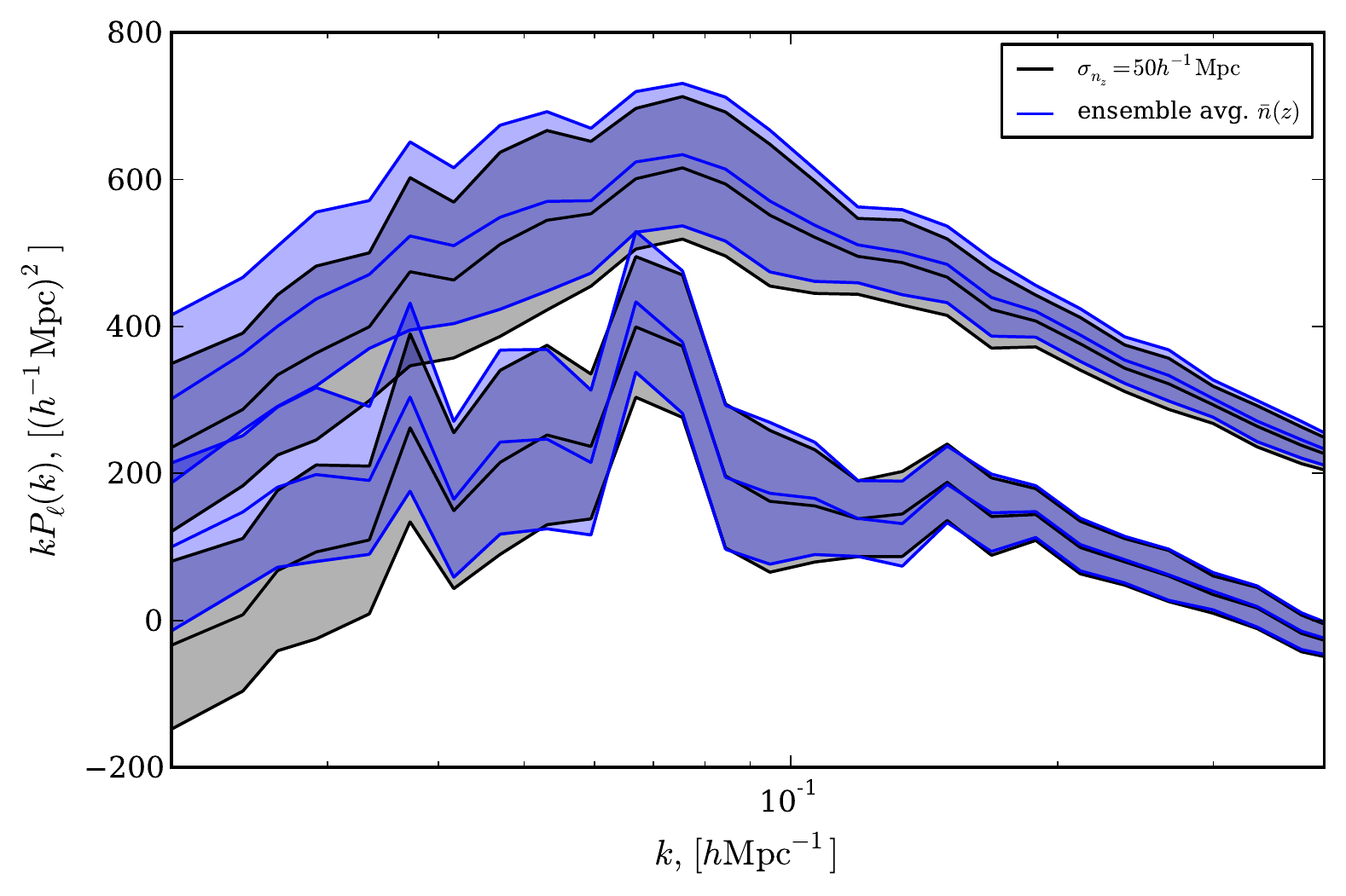}}
\caption[Mean $P_{\ell}(k)$ of the mocks when $\bar n(z)$ is known.]{Top: Mean $P_{\ell}(k)$ measurements of 26 W1 low-$z$ mocks when the ESR corrected $\bar n(z)$ is taken to be the mock-average (blue) and a per-mock estimate (black).  Estimating $\bar n(z)$ from the survey itself suppresses large-scale clustering for $k<0.06 h$\emph{Mpc}$^{-1}$.  This suppression is somewhat accounted for by the integral constraint correction (see \S \ref{sec:intcor}) and well within the statistical error.  This suggests this method is adequate.  There is seemingly only a small change in the expected variance and, as the quadrupole is slightly increased, a larger $\beta$ will be inferred on average.  Bottom: same format, but for the high-$z$ slice.  There is a greater effect in this case -- $P_0$ is underestimated up to $k=0.4 h$\emph{Mpc}$^{-1}$.  Surprisingly, the quadrupole recovers the true value on larger scales despite this being a radial (anisotropic) systematic.  Ultimately, the bias remains within the statistical error and $f \sigma_8$ may still be successfully recovered.}
\label{fig:pk_ell_nbar}
\end{figure}
%\clearpage
%}
\begin{figure}
\centering
\includegraphics[width=\textwidth]{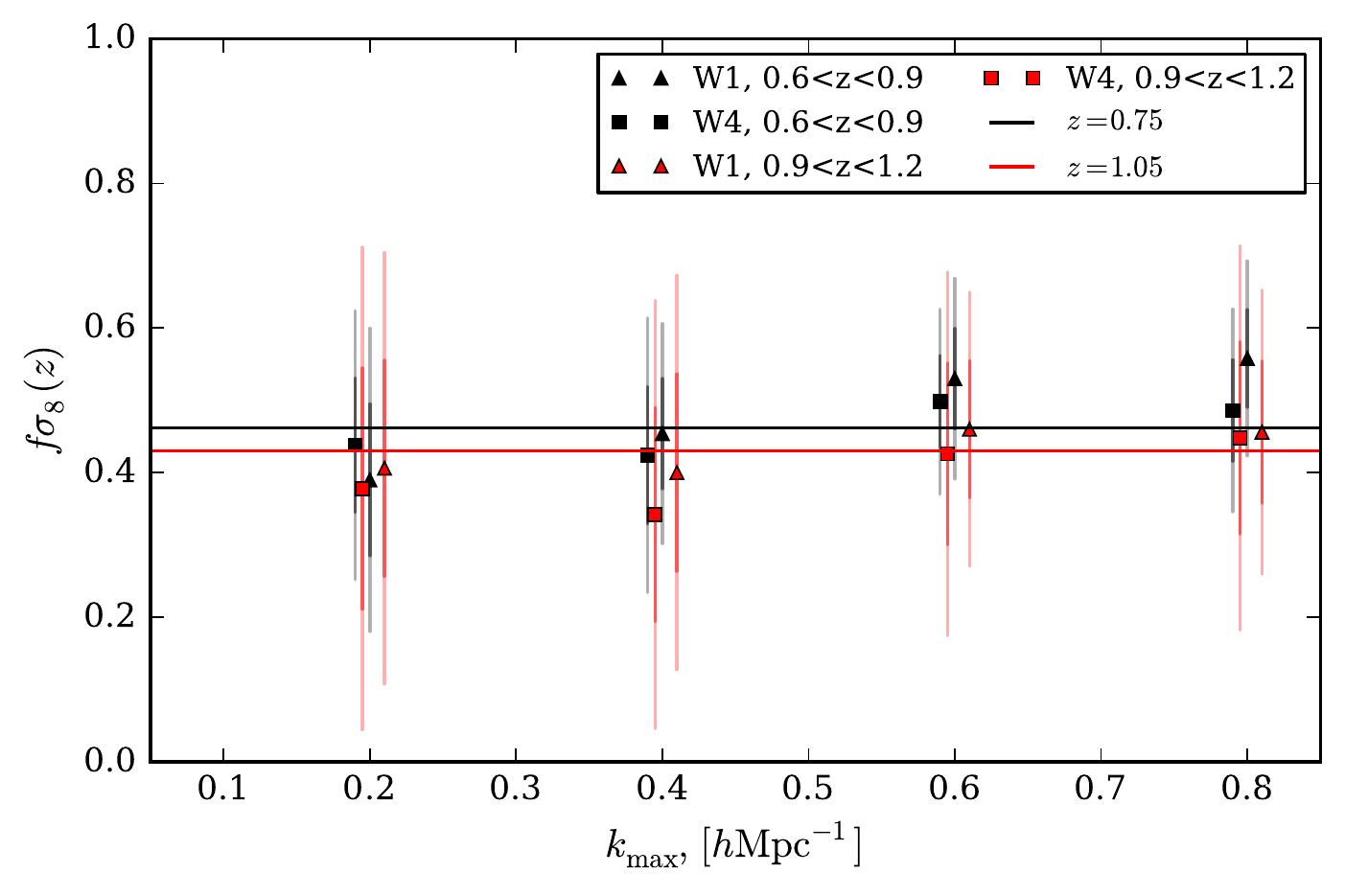}
\caption[Confidence intervals on $f \sigma_8$ for the mocks when $n(z)$ is known.]{Confidence intervals on $f \sigma_8$ for the mocks when the multipole moments are given by the mock-average $\bar n(z)$; the estimated covariance for a per-mock estimate of $n(z)$ was retained in this estimate.  When compared to the top panel of Fig. \ref{fig:fsig8_kmax}, this figure shows that a larger $f \sigma_8$ value is inferred when a per-mock $\bar n(z)$ estimate is made and there is a larger effect for the low-$z$ slice.  This systematic bias remains within the statistical error.  Ultimately, $f \sigma_8$ is proven to be successfully recovered despite the necessity of estimating $\bar n(z)$ from the survey itself.}
\label{fig:fsig8_truenbar}
\end{figure}

\subsection{Potential biases due to the incompleteness correction}
A significant completeness correction has been applied to the spectroscopic sample to obtain the $f \sigma_8$ estimate -- an upweighting of a factor-of-three is required to recover the target sample $\bar n$ and correct for the scale-dependent bias.  The applied scheme should be highly accurate given the magnitude of this correction.  This must be the case as $f \sigma_8$ has been shown to be successfully recovered from the mocks.  Further evidence for this is provided by Pezzotta et al. (in prep.), who analyse the sensitivity of the correlation function to the ESR and its correction.  I do not consider further tests of the ESR correction on this basis.  But one avenue for improvement is the inclusion of a photometric redshift posterior in the weighting scheme; this redshift information has been neglected in the weights we apply to date. 
%
%$\implies$ Fig - Multipole moments of parent, spoc-mocks and with ESR.
%
\subsection{Potential biases due to the assumed model}
Three specific assumptions have been made for the redshift-space power spectrum model: the real-space spectrum, the linear RSD factor and the non-linear damping term; the validity and alternatives to these choices are discussed below.  The successful recovery of $f \sigma_8$ from the mocks provides evidence that each assumption is valid for the precision of the VIPERS v7 dataset, but I have not tested their validity directly.  It is unlikely that alternatives could provide better constraints as the $f \sigma_8$ constraints shown above are seemingly robust to $k = 0.8 \hompc$.  This is further evidenced by a relatively small change in the obtained $f \sigma_8$ error when modes greater than $0.2 \hompc$ are included. 

\textit{Real-space model}:  The real-space power spectrum is modelled according to a non-linear Halofit-2 \citep{halofit2} prescription.  Given the good agreement in shape between the redshift-space monopole of this model and the mocks, the scale-dependence of galaxy bias -- from the preferential occupancy of galaxies in the most massive haloes, must be small on the relevant scales (or at least well modelled by a combination of misestimating $P_g(k)$ and a RSD damping term).  In either case the $f \sigma_8$ estimate obtained is unbiased.  Thus scale-dependent bias is less of a concern, even though this may have been thought to be a significant effect in the HOD mocks a priori.

\textit{Linear theory RSD}: the VIPERS monopole errors are approximately a factor-of-two at $k=0.05 \hompc$ (see Fig. \ref{fig:data_multipoles}).  The expected deviations from the linear theory symmetry: $P_{\delta \delta} = P_{\delta \theta} = P_{\theta \theta}$ (see Fig. \ref{fig:sdlt_Pdd_Pdt_Ptt}) are negligible in this case.  Therefore assuming the simple Kaiser model should be a sufficiently valid approximation.  

\textit{Non-linear theory RSD}:  Perhaps the greatest uncertainty in the assumed model is in the choice of the non-linear damping -- the dispersion model is itself an approximation.  But there must be limited sensitivity to this choice as the expected $f \sigma_8$ error changes only slightly when modes larger than $k=0.2 \hompc$ are included.  Further support for the applied damping is given by Fig. \ref{fig:mocks_multipoles}, which shows that the model provides an adequate fit to the zero-crossing point and large-$k$ shape of $P_2(k)$.  To some extent this is by construction; in creating the mocks a mass-dependent Gaussian variable was added to the velocity to mimic virialised motions, which results in a Lorentzian form for a mass averaged observable \citep{Sheth_exp}.  However, spectroscopic errors have also been added, which have comparable magnitude at these redshifts ($\sigma_{z} = 141(1+z) \rm{km \ s}^{-1}$ as opposed to $\simeq 300 \rm{km \ s}^{-1}$ for the intrinsic dispersion), hence a Lorentzian form is far from inevitable.  This illustrates a potential danger of over-fitting the mocks, which will be a greater concern for future surveys.    

\subsection{Potential biases due to the covariance estimation}
The convergence of the estimated covariance matrix is a common worry \citep{Hartlap, TaylorCovariance, PercivalCovariance} and many analyses attempt to increase the rate of convergence with a `shrinkage' estimator \citep{PopeSzapudi} or a similar technique, e.g. \cite{sylvainClustering}.  This VIPERS PDR-1 analysis was able to make use of only 28 mocks however, as opposed to the 306 available for this work.  The top panel of Fig. \ref{fig:CovarianceConvergence} shows the covariance estimate to be sufficiently converged with respect to the number of mocks included; consistent results are obtained when the covariance is estimated using only half of the available mocks.  Again, the GR expectation is successfully recovered for a range of $k_{\rm{max}}$.  

The bottom panel of the same figure shows the results of assuming a diagonal covariance (when using all of the available mocks).  The inferred $f \sigma_8(z)$ values are overestimated compared to the non-diagonal case, but, perhaps surprisingly, the estimates again recover the expected value with 68\% confidence.  This is a reassurance that the impact of the covariance, or lack thereof, is correctly accounted for.  This perhaps reflects the relative independence of the quadrupole estimates; effectively only neighbouring $P_2$ estimates are correlated and there is only a weak correlation between $P_0$ and $P_2$ estimates; this was shown previously in Fig. \ref{fig:MultipoleCovariance}.   
\begin{figure}
\centering
\subfloat{\includegraphics[width=\linewidth]{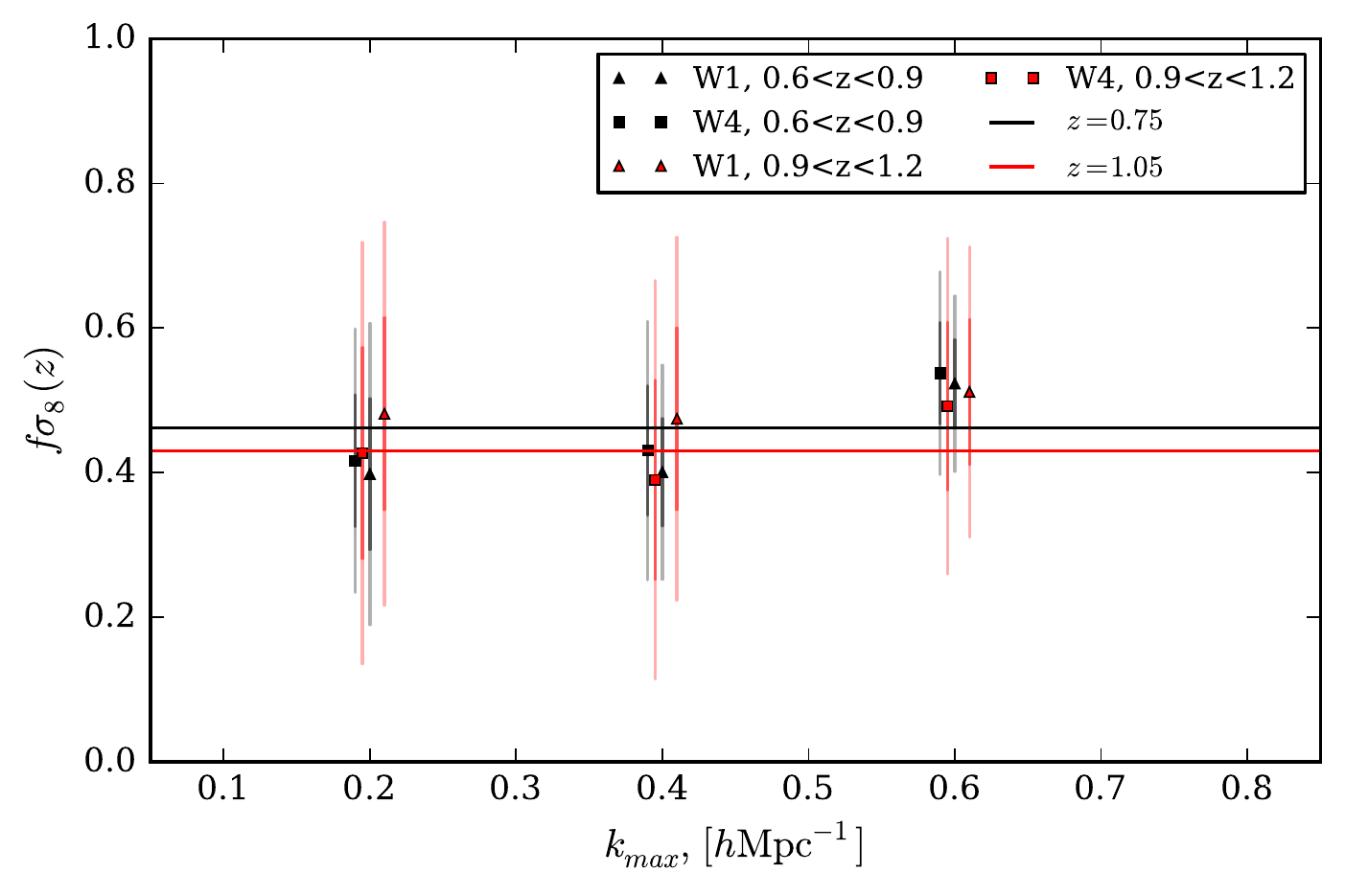}} \\
\subfloat{\includegraphics[width=\linewidth]{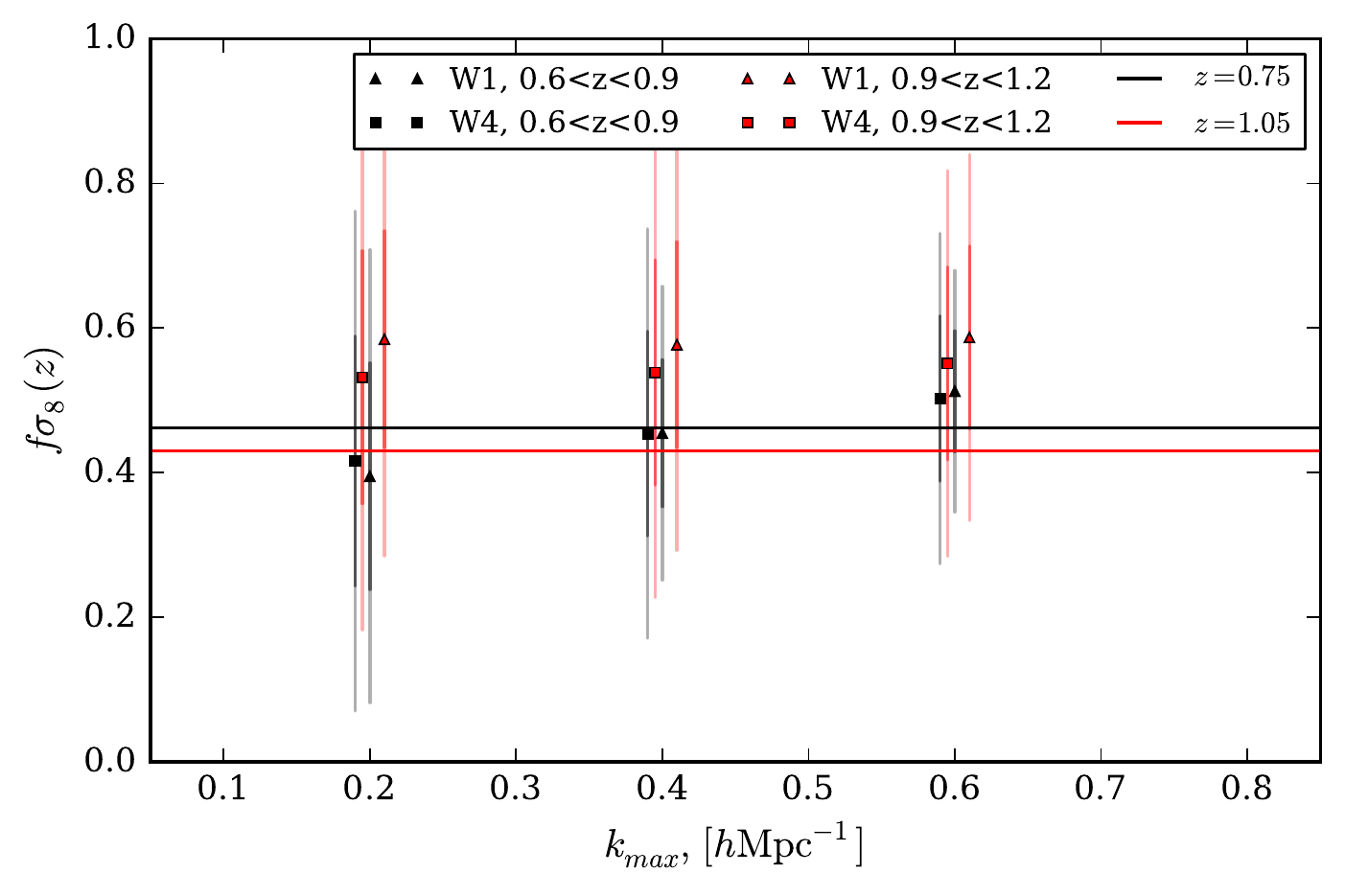}}
\caption[Impact of the covariance matrix convergence on $f \sigma_8$.]{This figure shows the robustness of the $f \sigma_8$ estimates to the convergence of the covariance matrix.  The top panel shows those obtained from using only half the mocks in the covariance estimate.  These remain consistent with both the GR expectation and the estimates from the covariance of 306 mocks.  Therefore the covariance seems to be sufficiently converged.  The bottom panel shows the result of assuming a diagonal covariance in the $f \sigma_8$ estimate and using all of the mocks.  The expected GR value is recovered even in this case.  This perhaps reflects the weak covariance of $P_2$ and $P_0$ on the same scale.  See Fig. \ref{fig:MultipoleCovariance}}
\label{fig:CovarianceConvergence}
\end{figure}

\section{Statistical significance of the VIPERS v7 measurements}
\label{sec:StatSig}
In order to calculate the combined statistical significance of the VIPERS v7 subvolume constraints the expected evolution of $f \sigma_8(z)$ must first be accounted for by defining:
\[
y = \frac{f \sigma_8(z)}{\langle f \sigma_8(z) \rangle}.
\]
For General Relativity ($\gamma=0.545$) and a \cite{Planck} fiducial cosmology -- $\sigma_8 = 0.82$ in particular, the expected values are 
\[
\langle f \sigma_8(z) \rangle = 
   \begin{cases}
   0.462 \pm 0.006 \qquad \text{for } z=0.75, \\
   0.430 \pm 0.005 \qquad \text{for } z=1.05,
   \end{cases}
\]
where $z$ is the mean weighted redshift for each slice.  These errors are derived from \cite{Planck} (TT + lowP + lensing constraints), which state
\begin{align}
\sigma_8 &= 0.8149 \pm 0.0093, \nonumber \\
\Omega_m &= 0.308 \pm 0.012.
\end{align}
%\begin{figure}
%\centering
%\includegraphics[width=0.75\textwidth]{Om_sig8_fid.pdf}
%\caption{Constraints on the $(\Omega_m, \sigma_8)$ plane from recent cosmic shear analyses and Planck.  This figure illustrates the lack of degeneracy between the $\Omega_m$ and $\sigma_8$ Planck estimates.  Accordingly, I calculate the error in the expected $f \sigma_8$ -- from uncertainties in the baseline cosmology, by adding in quadrature the fractional error in $f$, when varying $\Omega_m$ within the 68 \% Planck confiddence limits, to the similar fractional error in $\sigma_8$.  Reproduced from Fig. 2 of \cite{DES_shearCosmology}.}
%\label{fig:sig8_Om.pdf}
%\end{figure}
%\noindent
Assuming these to be independent, the errors are propagated to errors on $f \sigma_8$ via the \cite{LinderCahn} approximation.  As there is no significant degeneracy in these Planck estimates (see Fig. 2 of \citejap{DES_shearCosmology} for example) I add in quadrature the fractional error in $f D_+(z)$ as $\Omega_m$ is varied within the 68\% confidence limits to the fractional error in $\sigma_8$.  This modelling error is shown in Fig. \ref{fig:fsig8_z}, which  illustrates that uncertainties in the $\Lambda$CDM parameters are comparable to $0.4 < \gamma < 0.7$ at $z \simeq 0.8$.  Therefore more precise constraints on the $\Lambda$CDM parameters are required to constrain gravity at $z \simeq 0.8$, where a greater and more linear volume is available.  
As this modelling error is much smaller than the VIPERS v7 statistical error I do not include it when calculating the significance, eqn. (\ref{eqn:chi2_GRsignificance}).

By similarly rescaling the errors, $\sigma \mapsto \sigma / \langle f \sigma_8 \rangle$, the best-fitting $y$ for the four measurements satisfies
\[
y_{\rm{best}} = \frac{\sum_{\rm{subvols}} w_i y_i}{\sum_i w_i}.
\]
I apply inverse variance weighting, $w_i = (1/ \sigma_i ^2)$, with respect to the renormalised errors.  I calculate a $\chi^2$ statistic:
\[
\chi^2 = \sum_{\rm{subvols}} \frac{(y - y_{\rm{best}})^2}{\sigma^2},
\label{eqn:chi2_GRsignificance}
\]
that assumes the four subvolumes to be independent.  The significance may then be judged on the basis of the $\chi^2$ probability.  This is an approximation; there will be a small covariance introduced by the joint-field estimate of $\bar n(z)$ in addition to the negligible intrinsic correlation given by $\xi(r)$ for $r \gg 1 \mpcoh$.  The magnitude of this field-to-field covariance is difficult to quantify as the two fields of a given VIPERS mock may not have been drawn from an independent dark matter distribution.  The $\chi^2$ values as a function of $k_{\rm{max}}$ are given in Table \ref{table:fsig8_sig} for $\nu = 3$ degrees-of-freedom -- four subvolumes and one fitted parameter, $y_{\rm{best}}$.

From these results it is clear that the combined likelihood is of marginal significance in requiring a deviation from GR.  Within a frequentist interpretation, the results are expected at least once-in-twenty realisations, which is far short of the `five sigma metric' often employed for claiming a detection. 
\begin{table}
\centering
\begin{tabular}{|p{0.2\textwidth} | p{0.2\textwidth} | p{0.2\textwidth} | p{0.2\textwidth}|}
\hline
\hline
$k_{\rm{max}} [ \hompc ]$ & $y_{\rm{best}}$ & $\chi^2$ & $\text{min. } P(\nu=3)$ \\
\hline
0.2 & 0.718 & 1.55 & $0.50$ \\
\hline
0.4 & 0.732 & 7.40 & 0.05 \\
\hline
0.6 & 0.899 & 6.77 & 0.05 \\
\hline
0.8 & 0.881 & 6.63 & 0.05 \\
\hline
\end{tabular}
\caption[VIPERS v7 data: statistical significance of deviations from GR.]{The combined statistical significance of the$y = f \sigma_8(z) / \langle f \sigma_8(z) \rangle$ values for the four VIPERS subvolumes.  The minimum probability that $\chi^2$ exceeds the observed value is denoted by $\emph{min.} P(\nu=3)$, which is obtained from Table A2.6 of \cite{JasperWall}.  The observed $\chi^2$ for $k_{\rm{max}} = \{ 0.4, 0.6, 0.8\}$ are significant at the $5-10 \%$ level on this basis; hence the W1 high-$z$ volume is marginally discrepant from GR when considered as the set.  As there is also a small `look-elsewhere' effect -- the $\chi^2$ is recalculated for a number of $k_{\rm{max}}$ values, this significance is further reduced.}
\label{table:fsig8_sig}
\end{table}

\section{The geometric and growth rate degeneracy}
\label{sec:AP_degen}
The percent level \cite{Planck} constraints on the $\Lambda$CDM parameters are much more precise than most other cosmological surveys.  But relying on a single observable leads to the possibility of unaccounted for systematics; there is some evidence that Planck is in tension with both local measures of $H_0$ \citep{Riess16} and weak gravitational lensing \citep{Heymans}.  Although this could also be due to extensions to the vanilla $\Lambda$CDM model.  Galaxy redshift surveys can constrain the expansion history with the Alcock-Paczy\'nski (AP) effect \citep{AP}. This is most robustly performed by using baryon acoustic oscillations (BAO) as a standard ruler.  Assuming a fiducial cosmology differing from the true one introduces further anisotropy, dependent on $D_AH(z)$, and a isotropic dilation of scale,  determined by $D_A^2H(z)$.  In the following section I explore the extent to which the VIPERS v7 anisotropy constrains $D_AH(z)$; a more detailed introduction to the AP effect is given in Chapter \ref{chap:RSD}.  

As a 4D parameter space is required when marginalising over the anisotropic AP distortion -- with the scale dilation factor set to unity \citep{PadmanabhanWhite}, I estimate the posterior by Markov Chain Monte Carlo (MCMC) with my own implementation of the Metropolis-Hastings algorithm -- based on the description in \cite{HeavensStats}.  To ensure the first steps in the chain are unbiased by the starting position, which may reside in a region of low likelihood -- subsequent jumps are then strongly directed towards the (hyper)volume of peak likelihood and hence the MCMC is not `stationary', I follow standard practice and consider an initial chain length as unrepresentative.  I remove the first 5000 elements of the chain due to this `burn-in' phase.  The (symmetric) proposal function is chosen to be a multi-variate Gaussian with a diagonal covariance of $\diag(0.01, 0.05, 0.05, 0.50)$.  This corresponds to approximately 1\% error in $\epsilon$, where $k_{\perp} = \alpha^{-1} (1 + \epsilon) k'_{\perp}$ and $k_{\parallel} = \alpha^{-1} (1 + \epsilon)^{-2} k'_{\perp}$ for a true and fiducial cosmology denoted by unprimed and primed respectively, and 10\% errors in $(f \sigma_8, b \sigma_8, \sigma_p)$.  Typical chain lengths are $10^6$ elements long and require around three hours to complete.

\subsection{Parameter degeneracies}
This section investigates the degeneracies between the free parameters of the AP-distorted Kaiser-Lorentzian model.  The degeneracy between the AP and RSD sources of anisotropy is shown in Fig. \ref{fig:VIPERS_ruiz_data}.
\begin{figure}
\centering
\subfloat{\includegraphics[width=.65\linewidth]{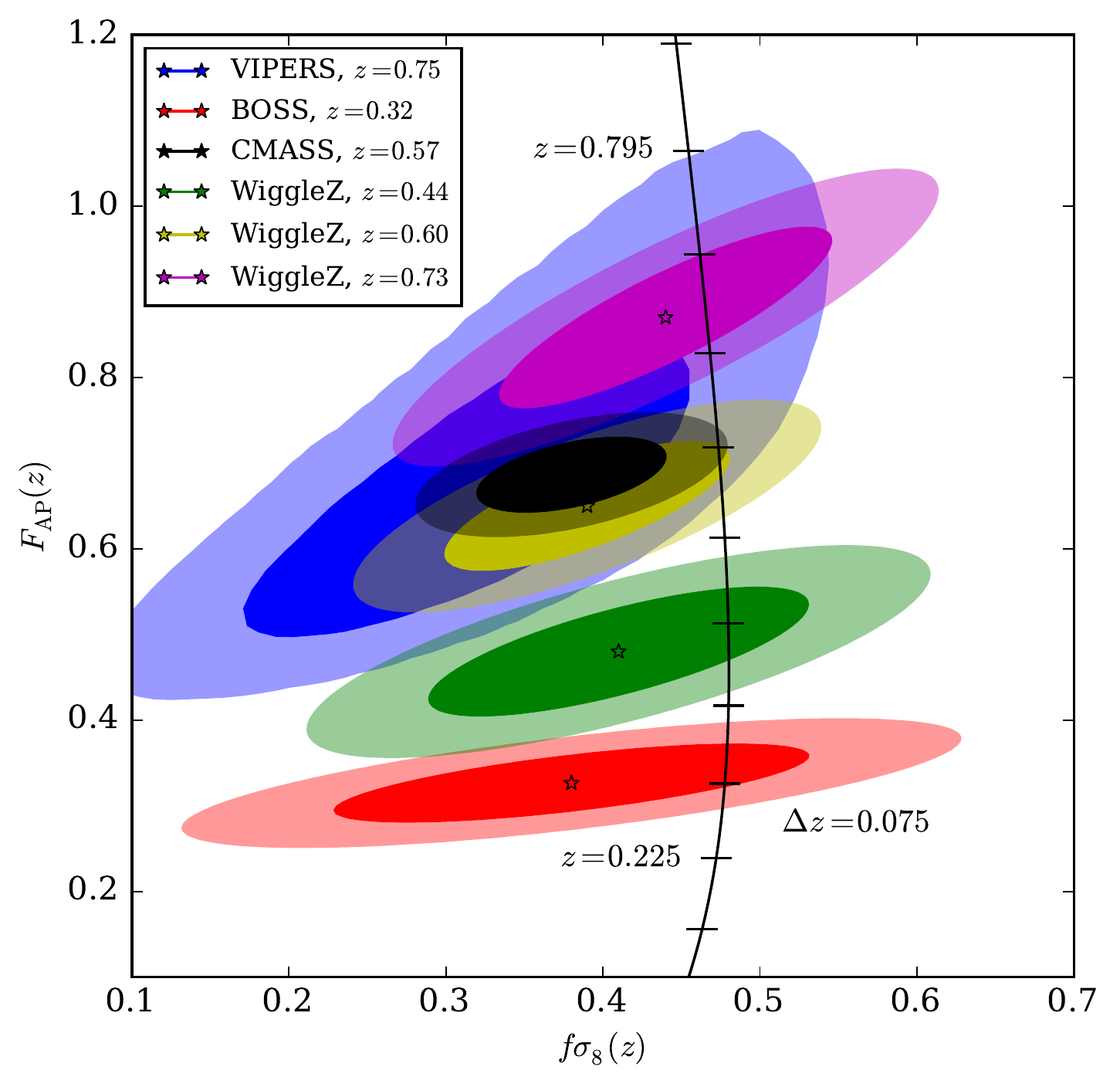}} \\
\subfloat{\includegraphics[width=.65\linewidth]{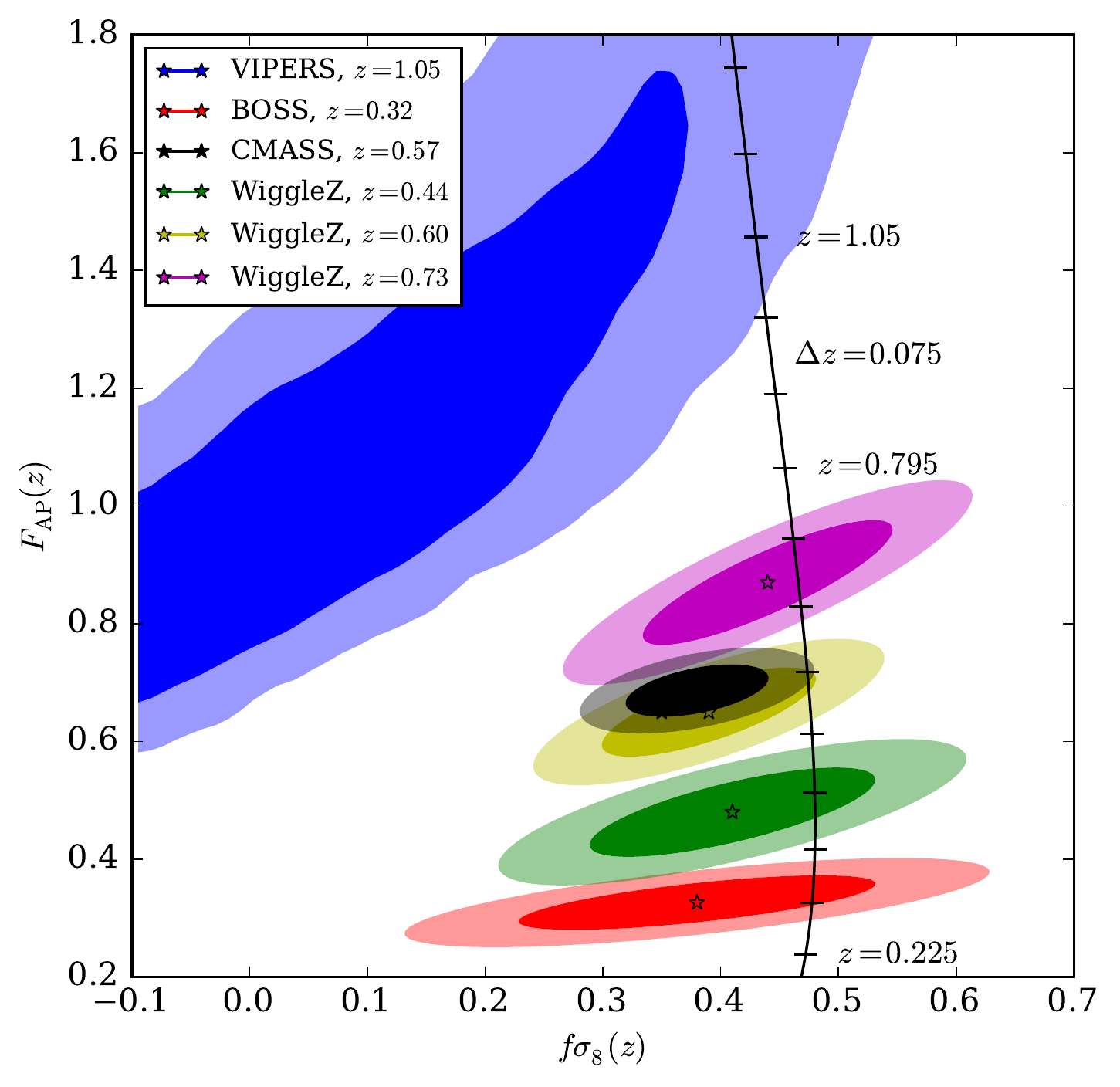}}  
\caption[VIPERS v7 data posteriors on $(f \sigma_8, F_{AP})$.]{The VIPERS v7 joint-field posteriors on $(f \sigma_8, F_{AP})$; The low-$z$ and high-$z$ slices are shown by the top and bottom panel respectively.  These constraints are obtained using the combined likelihood of $P_0(k)$ and $P_2(k)$ for $k_{\rm{max}}=0.8 h \emph{Mpc}^{-1}$.  The parameters $(b \sigma_8, \sigma_p, \epsilon)$ have been marginalised over and $\alpha$ is set to unity.  This approach is similar to that of WiggleZ \citep{Blake_ExpansionGrowth}, which differs as $\sigma_p$ is unmarginalised and the 2D $P(k)$ is used to compute $\chi^2$ in that case.  The range of $F_{\rm{AP}}$ spans $\epsilon \in [-0.15, 0.5]$ and therefore the \cite{PadmanabhanWhite} approximation should be valid.  The expected values are $\langle F_{\rm{AP}}(z) \rangle = (0.95, 1.46)$ in the fiducial cosmology, for $z=(0.75, 1.05)$ respectively. No significant deviation from the GR expectation is apparent, but the errors are sizeable.}
\label{fig:VIPERS_ruiz_data}
\end{figure}
In this case, rather than assuming the fiducial expansion history is the truth, the VIPERS v7 data is used to separate the two effects based on their distinct scale and angular dependence.  Clearly there is some ambiguity in doing so given the significant measurement errors but this serves to illustrate the inherent degeneracy and likely increase in the $f \sigma_8(z)$ error. 

The approach I take is an approximation for two reasons; firstly, I marginalise over only the anisotropic AP distortion and fix the scale dilation to unity. This is similar to that of WiggleZ \citep{Blake_ExpansionGrowth}, which differs as there is no marginalisation over a damping term and the 2D $P(\mathbf{k})$ is fitted in that case.  Secondly, the real-space power spectrum, $P_g(k)$, is fixed to that in the fiducial cosmology -- see \S \ref{sec:AP}.  Setting $\alpha$ to unity is logical in this case as this will have a greater dependence on the shape of the real-space power spectrum.  Conclusions may be drawn only on the consistency with GR and the fiducial expansion history under these assumptions.  If there is an inconsistency, this does not necessarily mean that GR is refuted.  Fig. \ref{fig:VIPERS_ruiz_data} shows that the results are consistent, but the sizeable errors limit the strength of this conclusion.        

When the anisotropic AP distortion is separated from the RSD anisotropy on the basis of the distinct scale and angular dependence, the VIPERS v7 data place constraints of
\begin{align}
f \sigma_8(0.76) &= 0.31 \pm 0.1, \\
f \sigma_8(1.05) &= -0.04 \pm 0.26,
\end{align}
for $k_{\rm{max}} =0.8 \hompc.$  This is roughly a factor-of-three increase in the error and remains consistent with the constraints obtained when the expansion history is assumed.  Here the quoted errors are obtained from the width of the data posterior; this is shown in Fig. \ref{fig:VIPERS_eps_marginalised_fsig8}. Fig. \ref{fig:fsig8_ap} shows no systematic bias  when this method is applied to the mocks.
\begin{figure}
\centering
\includegraphics[width=\textwidth]{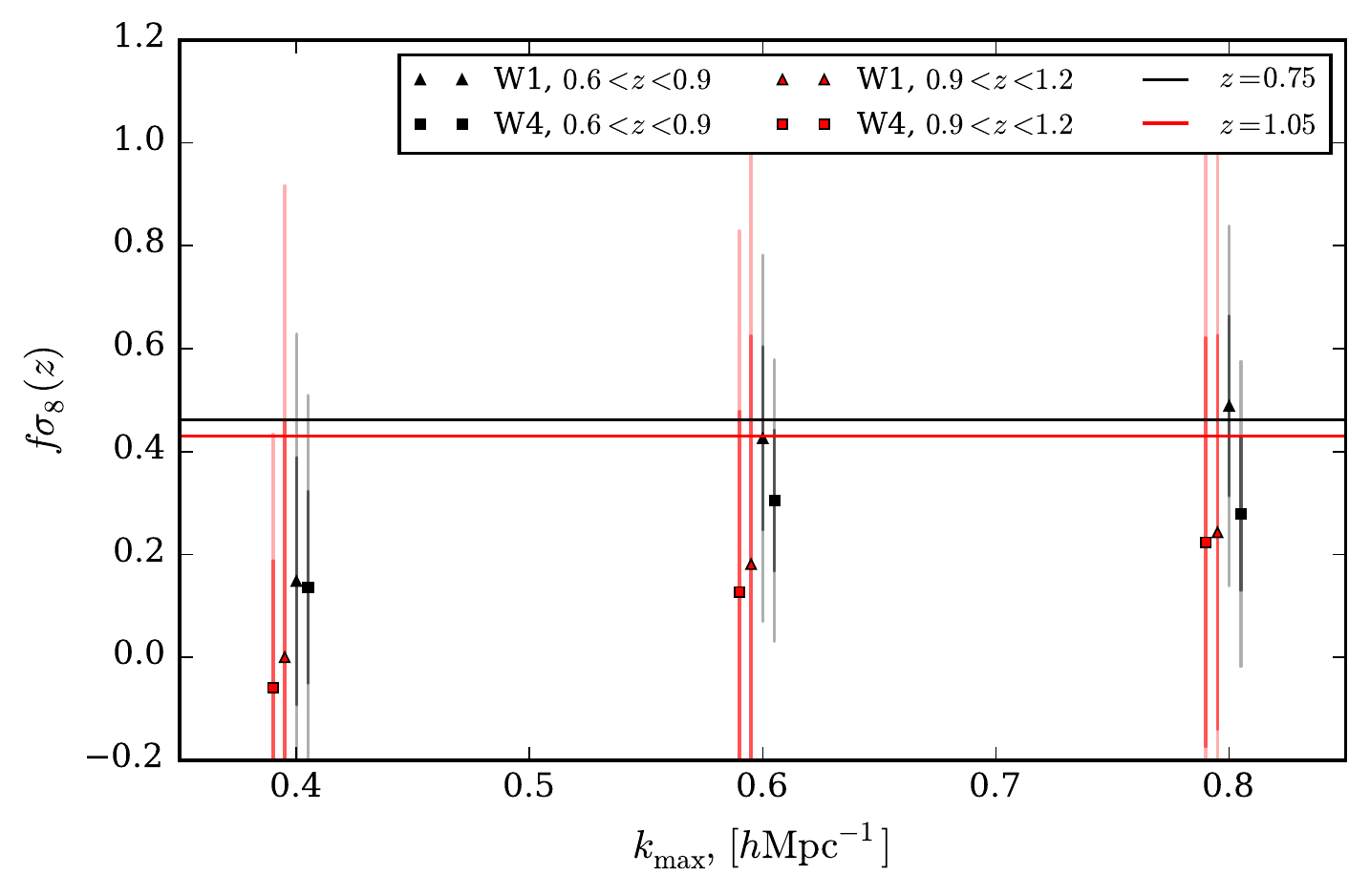}
\caption[Mocks: Confidence levels on $f \sigma_8$ after marginalising the anisotropic AP effect.]{Constraints on $f \sigma_8(z)$ after marginalising $(b \sigma_8, \sigma_p, \epsilon)$ for 26 mocks.  Shown are the symmetric 68\% and 95\% confidence limits, which are roughly a factor-of-three greater than if the expansion history is assumed apriori.  The trend of increasing $f \sigma_8$ with increasing $k_{\rm{max}}$ is again apparent, but there is no significant deviation from GR up to $k_{\rm{max}}=0.8 h \emph{Mpc}^{-1}$.}
\label{fig:fsig8_ap}
\end{figure}
\begin{figure}
\centering
\subfloat{\includegraphics[width=.7\linewidth]{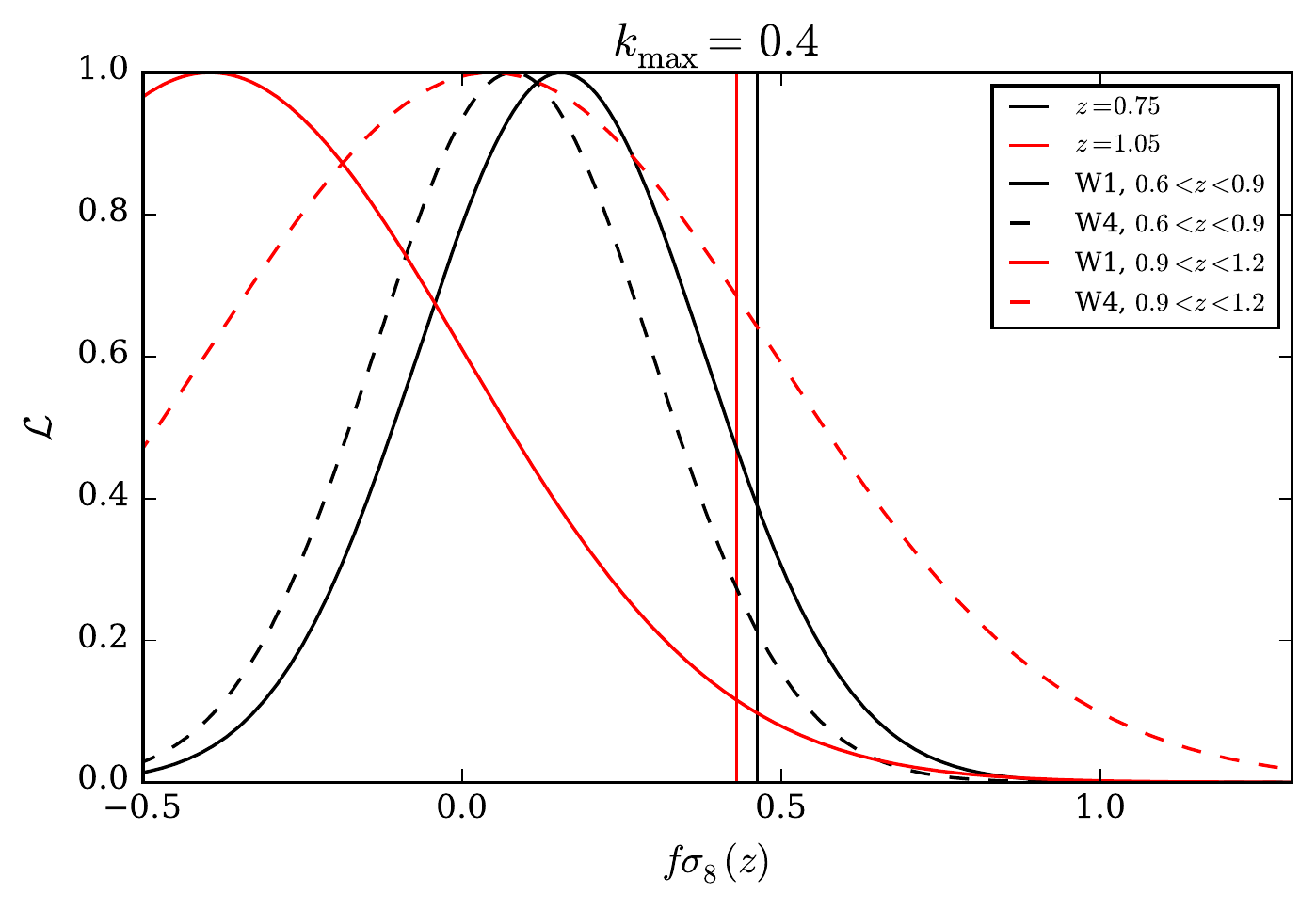}} \\
\subfloat{\includegraphics[width=.7\linewidth]{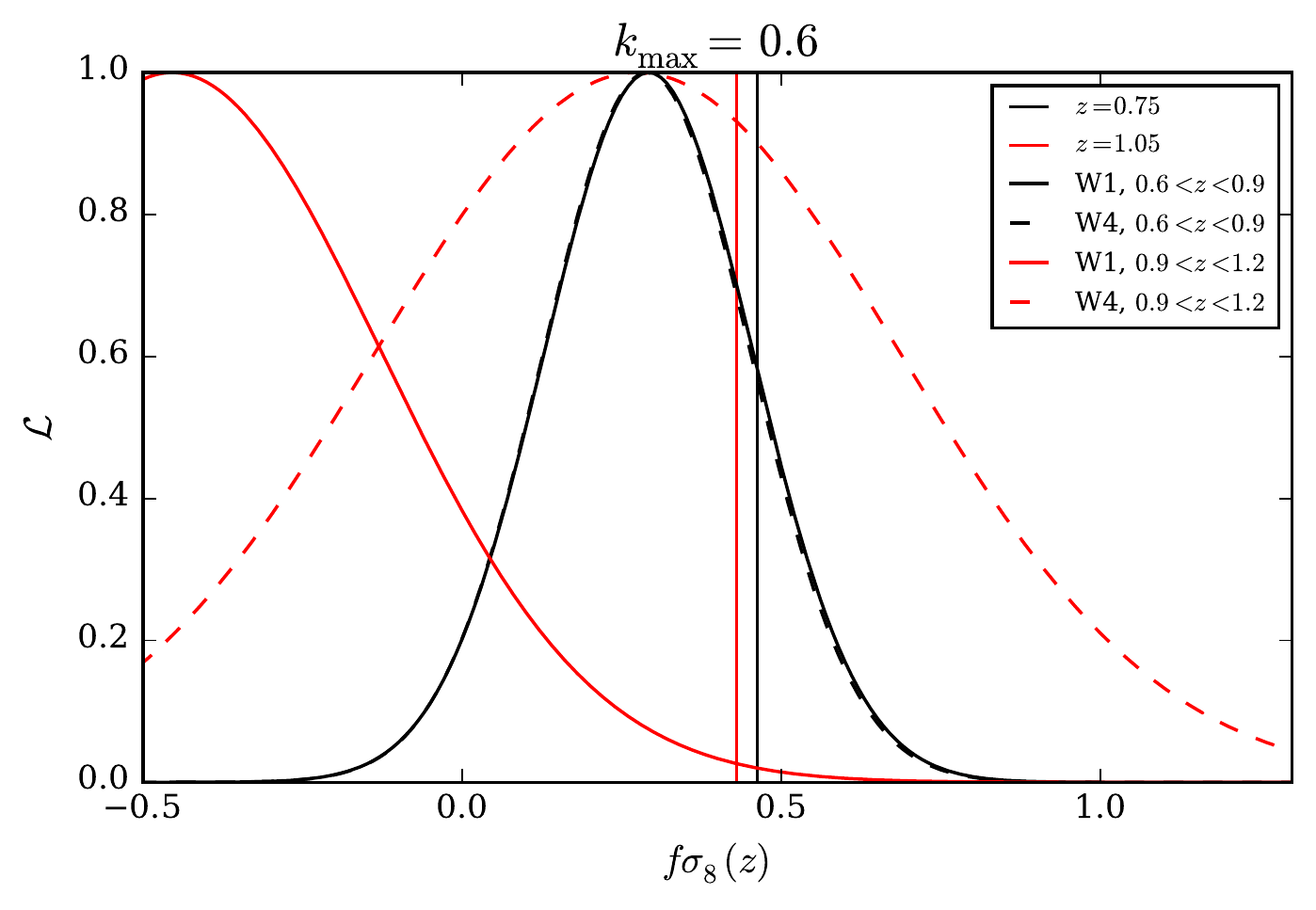}} \\    
\subfloat{\includegraphics[width=.7\linewidth]{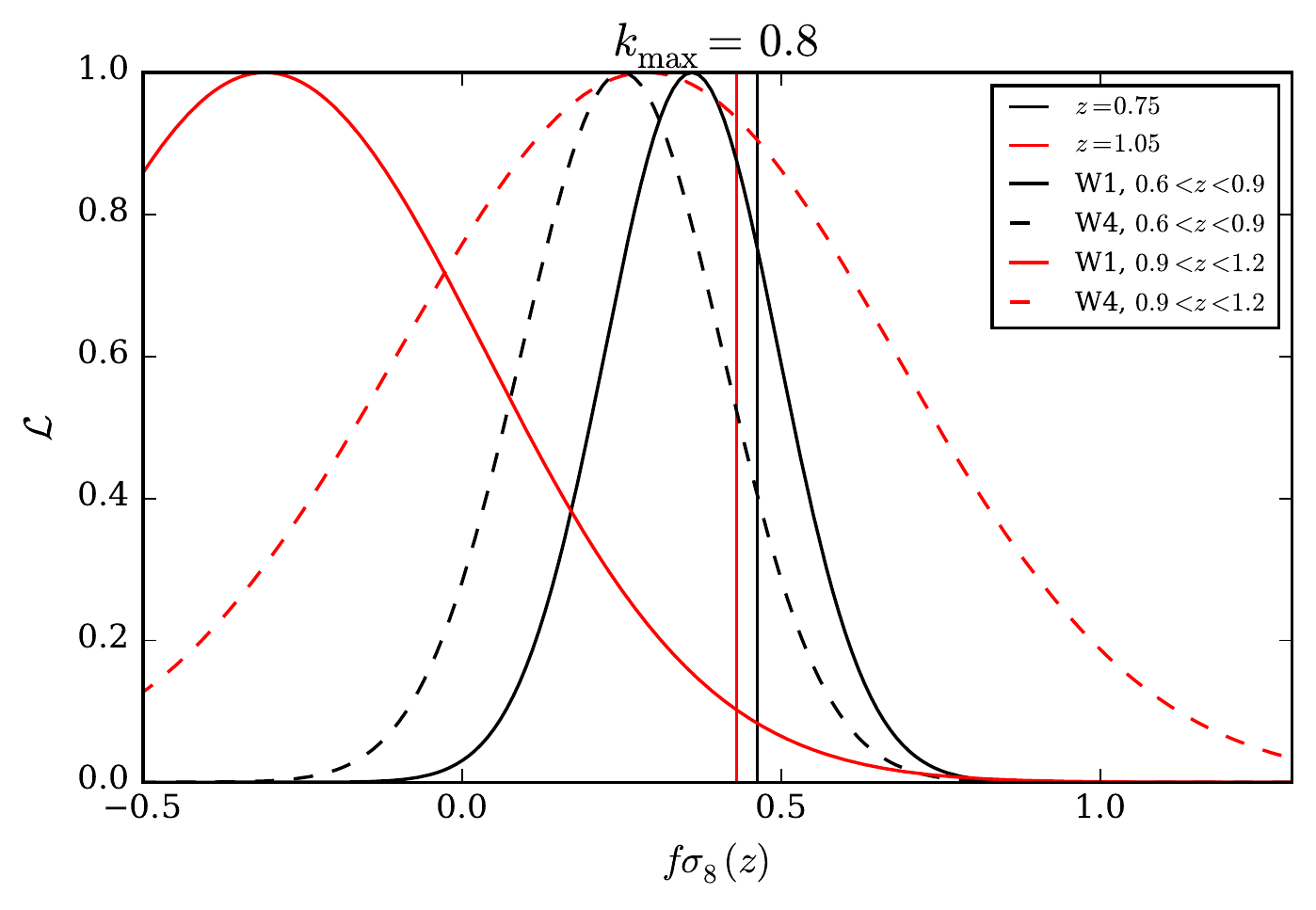}}
\caption[Data: Confidence levels on $f \sigma_8$ after marginalising the anisotropic AP effect.]{The VIPERS v7 dataset constraints on $f \sigma_8(z)$ when marginalised over the anisotropic AP distortion.  The $f \sigma_8$ estimates are robust to the choice of $k_{\rm{max}}$ -- peak shifts are less than the posterior width, as was the case when the expansion history was assumed.  The precision is degraded by roughly a factor-of-three compared to the known expansion history case.}
\label{fig:VIPERS_eps_marginalised_fsig8}
\end{figure}

A peculiar aspect of the results shown so far is that small-scale modes do not further constrain the anisotropy; the error achieved is effectively fixed from $k=0.2 \hompc$. The strong covariance on small scales goes some way to explaining this, but the correlation of neighbouring $P_2(k)$ measurements and between $P_0(k)$ and $P_2(k)$ is small (this is in itself a surprise due the leakage of power caused by the survey mask) although this may simply reflect the larger $P_2(k)$ errors.  Another possible explanation may be a possible degeneracy between $f \sigma_8$ and the other nuisance parameters.  I investigate whether this is the case in Figures \ref{fig:degeneracies.pdf} and \ref{fig:degeneracies2.pdf}.   
\begin{figure}
\centering
\subfloat{\includegraphics[width=0.65\linewidth]{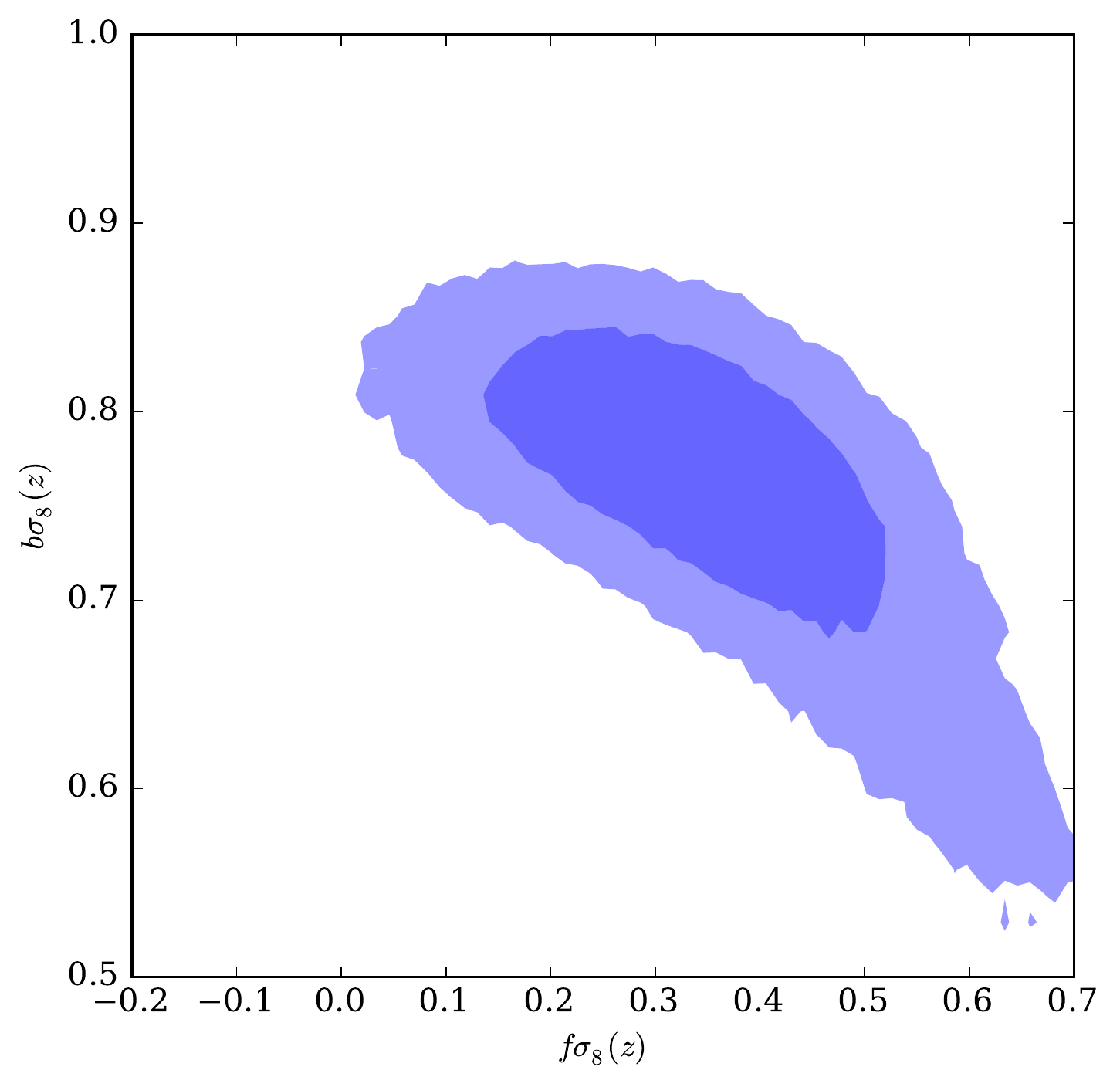}} \\
\subfloat{\includegraphics[width=0.65\linewidth]{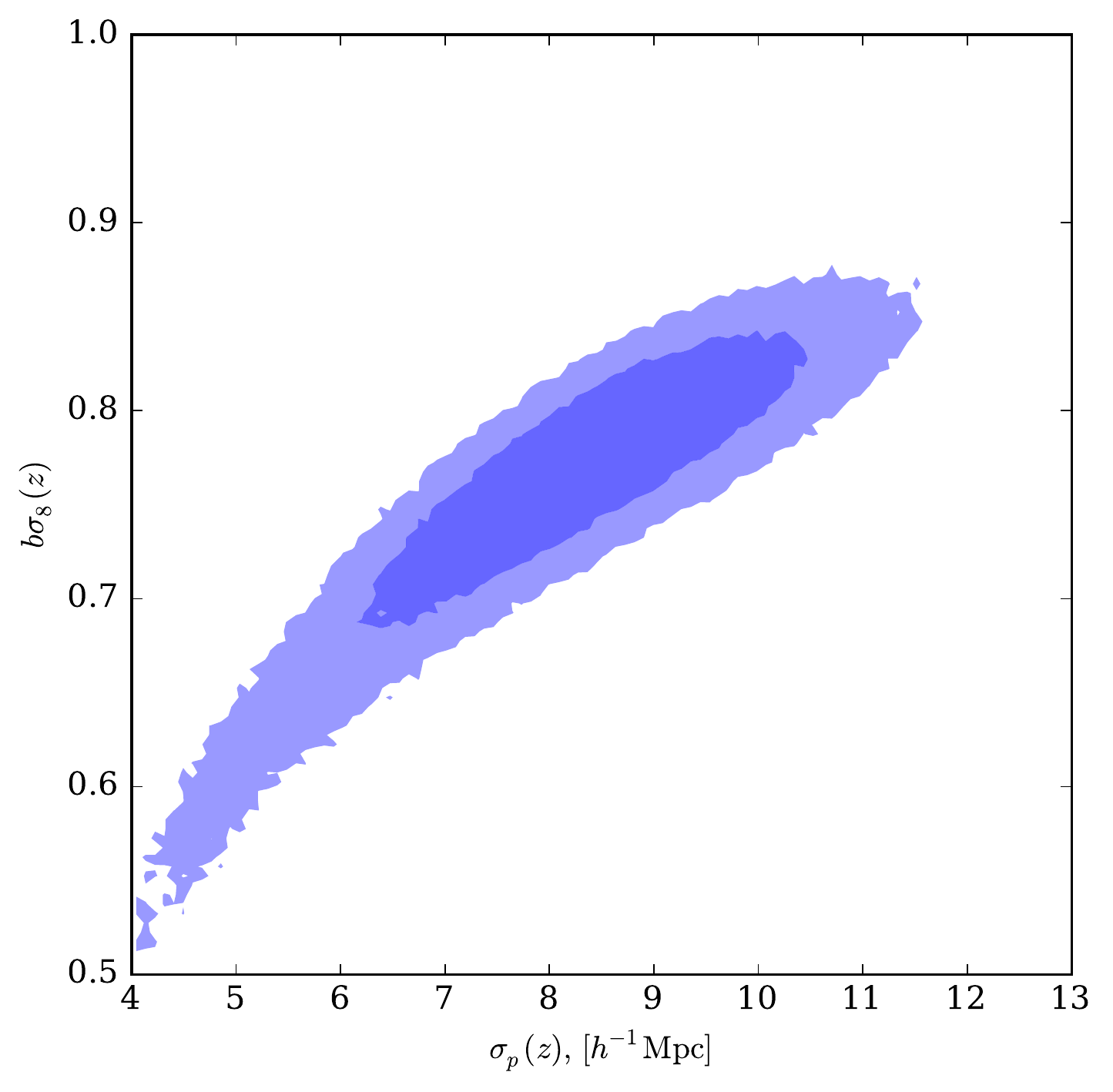}}
\caption[Parameter degeneracies of the Kaiser-Lorentzian model.]{Parameter degeneracies of the Kaiser-Lorentzian model after marginalising $\epsilon$.  This figure shows the 68 and 95\% confidence levels placed by the VIPERS v7: W1, $0.6 \leq z \leq 0.9$ subvolume for $k_{\rm{max}} =0.8 h \emph{Mpc}^{-1}$.  There is a significant degeneracy in both cases; one could solve this problem by determining $b \sigma_8(z)$ from the angular clustering of the parent photometry, which is independent of RSD.  Alternatively, $\sigma_p$ may be fixed to that in the mocks, $\simeq 6 \mpcoh$. The conditional error on $b \sigma_8$ is then $0.025$, as opposed to $0.067$ when $\sigma_p$ is marginalised -- a 60\% decrease.  But, while not inconsistent, this $\sigma_p$ value is one standard deviation from the peak likelihood.}
\label{fig:degeneracies.pdf}
\end{figure}
\begin{figure}
\centering
\subfloat{\includegraphics[width=0.65\linewidth]{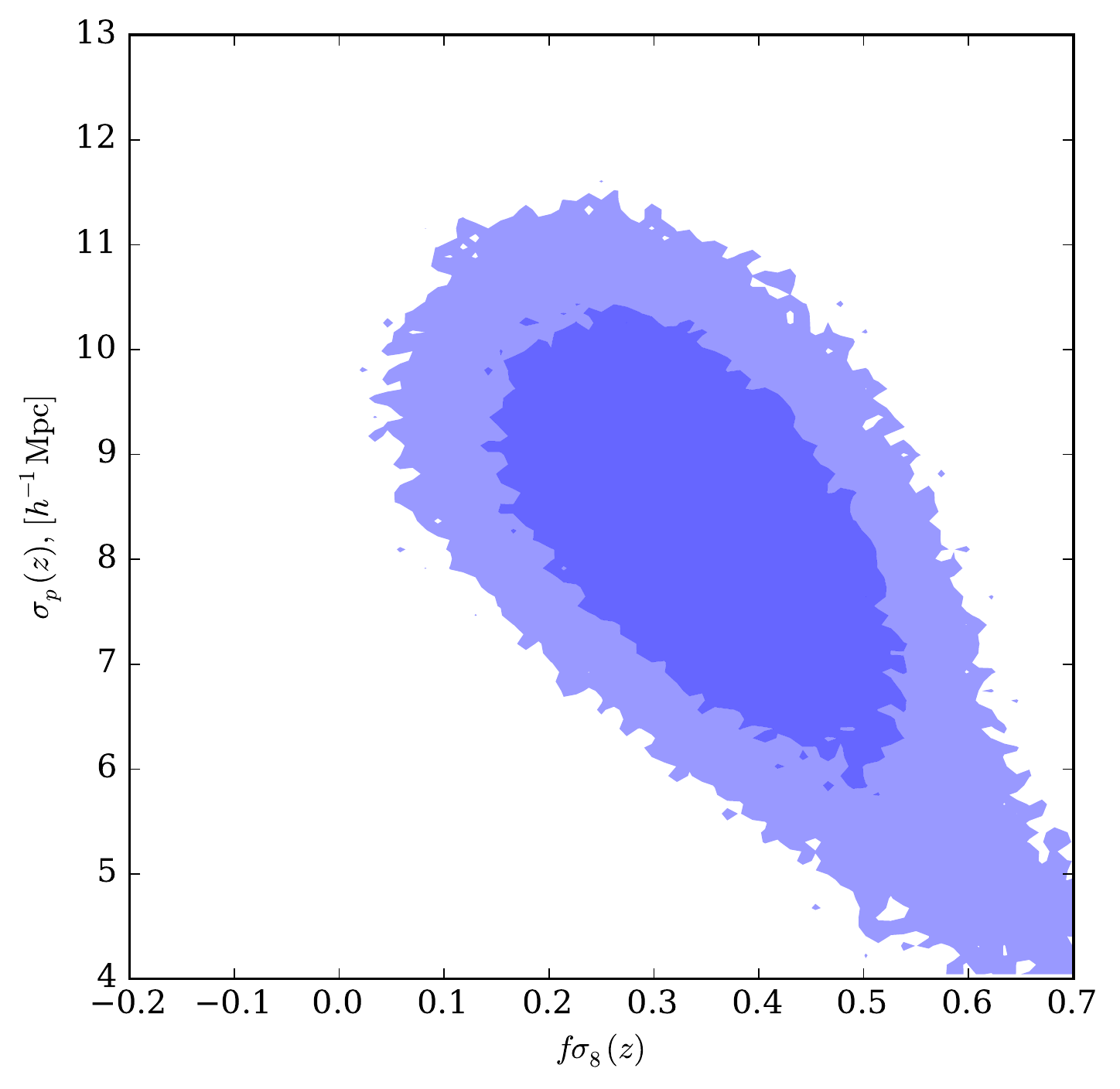}}
\caption[$(f \sigma_8, \sigma_p)$ degeneracy after marginalising the anisotropic AP effect.]{Same format as the previous figure but for the $(f \sigma_8, \sigma_p$ degeneracy in this case.  The degeneracy is modest, but stronger for large $f \sigma_8$.  This accounts somewhat for the surprisingly small constraining power of the small-scale measurements.  Another explanation is the strong covariance between $P_0$ estimates on neighbouring scales and modest covariance between $P_0$ \& $P_2$ and neighbouring $P_2$ estimates.}
\label{fig:degeneracies2.pdf}
\end{figure}

\section{Future work}
\begin{figure}
\includegraphics[width=0.9\linewidth]{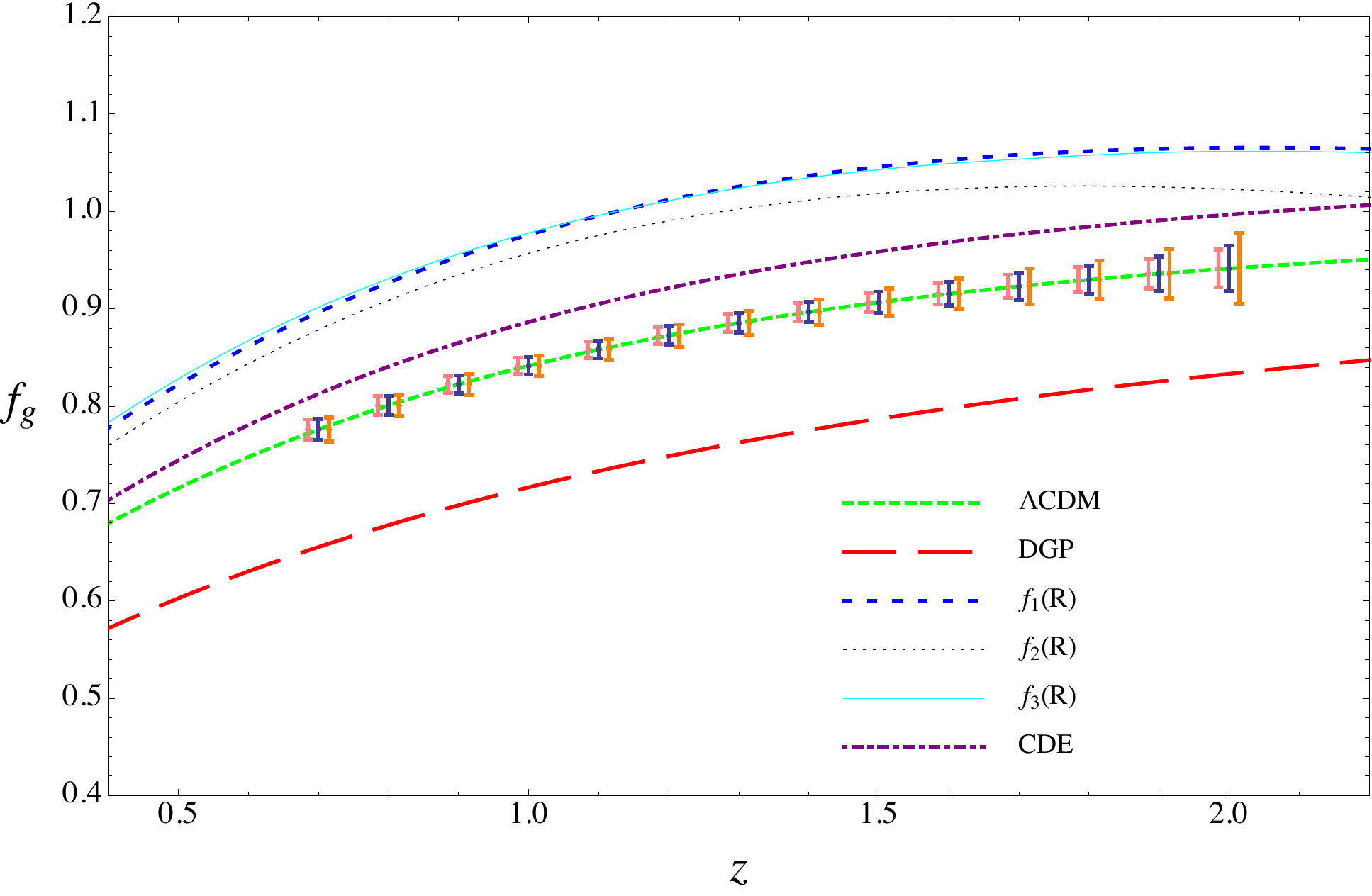} 
\caption[Expected Euclid constraints on $f \sigma_8(z)$.]{Expected Euclid constraints on $f(z)$;  the models shown include a coupled Dark Energy-Dark matter (CDE) model, the \cite{DGP} model and the \cite{HuSawicki} $f(R)$ model; these have been discussed previously.  Note that $f$ is plotted rather the independent observable for RSD analyses, $f \sigma_8(z)$.  Presumably this assumes that $\sigma_8$ may be determined by independent surveys.  Reproduced from Fig. 1.16 of \cite{EuclidForecast}}
\label{fig:Euclid}
\end{figure}
The clear next step is to apply this method to a greater volume by exploiting forthcoming surveys such as eBOSS, DESI, Euclid and WFIRST.  Smaller errors will certainly go some distance in excluding a range of posited modified gravity models, but degeneracies will remain.  With this greater precision comes a increased number of sources of systematic bias and each of these will have to be investigated and minimised.  This is not a distinct point; even though more precise measurements may be obtained without a greater volume, these larger surveys allow for more precise systematic tests -- e.g. by subdividing the sample according to absolute magnitude or colour, and thereby ensure the results are robust to galaxy bias and environmental dependence, for example.  The limited statistical significance of subpopulations in current surveys reduces the effectiveness of such tests.        

Further consistency tests of the method must be developed given the likelihood of future analyses being systematics dominated.  For instance, as is commonly the case, I neglect any checks of consistency of the data with the HOD framework -- beyond recovery of the power spectrum and, by construction, the angular clustering used to constrain the HOD parameters \citep{sylvainModels}.  Moreover, the HOD parameter errors have not been propagated to the mocks nor the $f \sigma_8(z)$ constraints.  Given the dependency of further systematic checks (e.g. for survey selection biases) on the mocks it is clear that ensuring these are sufficiently realistic must be the foundation of any robust analysis.  This should include investigating whether alternatives to the HOD formalism, e.g. sub-halo abundance matching, yields consistent results.  Current uncertainty in our understanding of galaxy formation is reflected by a range of available models.  It is clear that more stringent tests of gravity with large-scale structure will require increasingly accurate models of galaxy formation.    

In addition to uncertainties in the formation of galaxies in dark matter haloes and subhaloes, there remain subtleties in the physics of galaxy velocities; this issue was discussed in Chapter \ref{chap:RSD} -- see Fig. \ref{fig:ReidBias} in particular.  When RSD analyses with percent-level precision are available this will likely require extensive investigation \citep{Reid2.5}.  The central galaxy is presumed to reside in the densest part of the halo in \cite{Reid2.5} and therefore the strongly non-linear physics of baryonic feedback will be crucial in determining the magnitude of this effect.  

A further neglect in current RSD analyses is the model dependence of the error estimation, which is quantified by the covariance matrix.  In particular, the robustness of Fig. \ref{fig:fsig8_z} and Fig. \ref{fig:Euclid} to the assumed gravity model; these figures are derived from mocks that assume General Relativity.  Intuitively, models with a larger $G_{\rm{eff}}$ should be more non-linear and hence have a greater covariance.  For a given volume then, the $f \sigma_8$ error should increase.  This is not a simple problem to address as a gravitational theory must be assumed to constrain the HOD parameters, e.g. with the observed angular correlation function.  There is therefore likely to be a degeneracy between the assumed gravity model and the HOD occupancy.  This degeneracy will have an impact on both the expected clustering and expected covariance.  In the worst case scenario, the entire RSD analysis would have to be carried out on a per-model basis in order to fully exclude models similar to GR.  

Assuming the error estimates are accurate, any method that may lead to smaller errors without requiring a larger volume are an obvious avenue for future research.  These include the application of more involved analytic models, e.g. \cite{Okumura} and references therein, or extending the validity of linear models by downweighting non-linear structures prior to the RSD analysis.  I explore a \cite{Fergus} approach for tackling the root-cause of non-linearity directly in the next chapter.  Similar motivation has also kindled the recent interest in the analysis of voids.  Finally, the possibility of a cosmic variance-free linear growth rate measurement with multiple tracers is an enticing prospect \citep{McDonaldSeljak, BlakeGAMA, Abramo}.  Whether this is a viable approach when placed under further scrutiny is uncertain, but further investigation is warranted nonetheless. 

Given that there are models that make degenerate predictions for both the expansion and growth rate history, it will be important to utilise all available observables.  This is eloquently summarised by:~\begin{quote}
``It can scarcely be denied that the supreme goal of all theory is to make the irreducible basic elements as simple and as few as possible without having to surrender the adequate representation of a single datum of experience'',  A. Einstein, 1933.
\end{quote}
\noindent
Given the great successes of constraints on the expansion history with Type-Ia supernovae and BAO, a near-term focus of cosmology should be on constraining the linear growth history as precisely; redshift-space distortions and tomographic weak gravitational lensing are the two approaches with the greatest potential in this respect. However, large-scale structure constraints on the expansion history could be further improved if the AP distortion of pairs with a finite angular separation and large comoving separation could also be included; this is a potentially fruitful avenue for further research.  In addition, as the linear RSD of voids has been shown to not include a hexadecapole term \citep{CaiTaylorPeacock}, the AP and RSD degeneracy may be broken by the addition of a void-galaxy clustering analysis to the conventional analyses.  In the (potentially much) longer term the propagation of gravitational waves may also stringently constrain modified gravity theories \citep{LombriserTaylor}.
\end{chapter}
\begin{chapter}{Clipping: a local transform of the overdensity}
\label{chap:Clipping}
The development of non-linear dynamics greatly complicates the interpretation of the redshift-space galaxy distribution.  Linear theory is sufficient at early times or on large scales and allows for analytical predictions to be made; see Chapter \ref{chap:Basics}, \cite{PeeblesBook} or \cite{CP}.  But once the rms $\delta(\mathbf{x})$ is greater than unity it is necessary to resort to approximate models, which includes perturbation theory (\citejap{Taruya},  \citejap{Okumura}), the halo model (\citejap{SeljakHalomodel}, \citejap{PeacockSmith}) and empirical comparison to N-body simulations (\citejap{Smith}, \citejap{TinkerRSD_simulations}).  As the covariance matrix must also be known precisely, an analysis that includes non-linear scales is significantly more involved.  A common approach is therefore to limit the analysis to linear scales, $r \gsim 20 \mpcoh$, which, while robust, discards much of the available information.

In contrast, this chapter explores the \cite{Fergus} approach: tackling the root cause of non-linearity by downweighting the most overdense volumes prior to the RSD analysis.  This is simple in real space, as the non-linear tail of the overdensity may be removed with a remapping of \citep{GAMA_Clipping}:
\begin{eqnarray}
\label{eqn:clipTransform}
\text{g:}
            \begin{cases}
            \ \delta \mapsto \delta \quad \text{ if  } \delta \leq \delta_0, \\
            \ \delta \mapsto \delta_0 \quad \text{otherwise};
            \end{cases}
            \label{eqn:clipping_transform}
\end{eqnarray}
\begin{figure}
\centering
\subfloat{\includegraphics{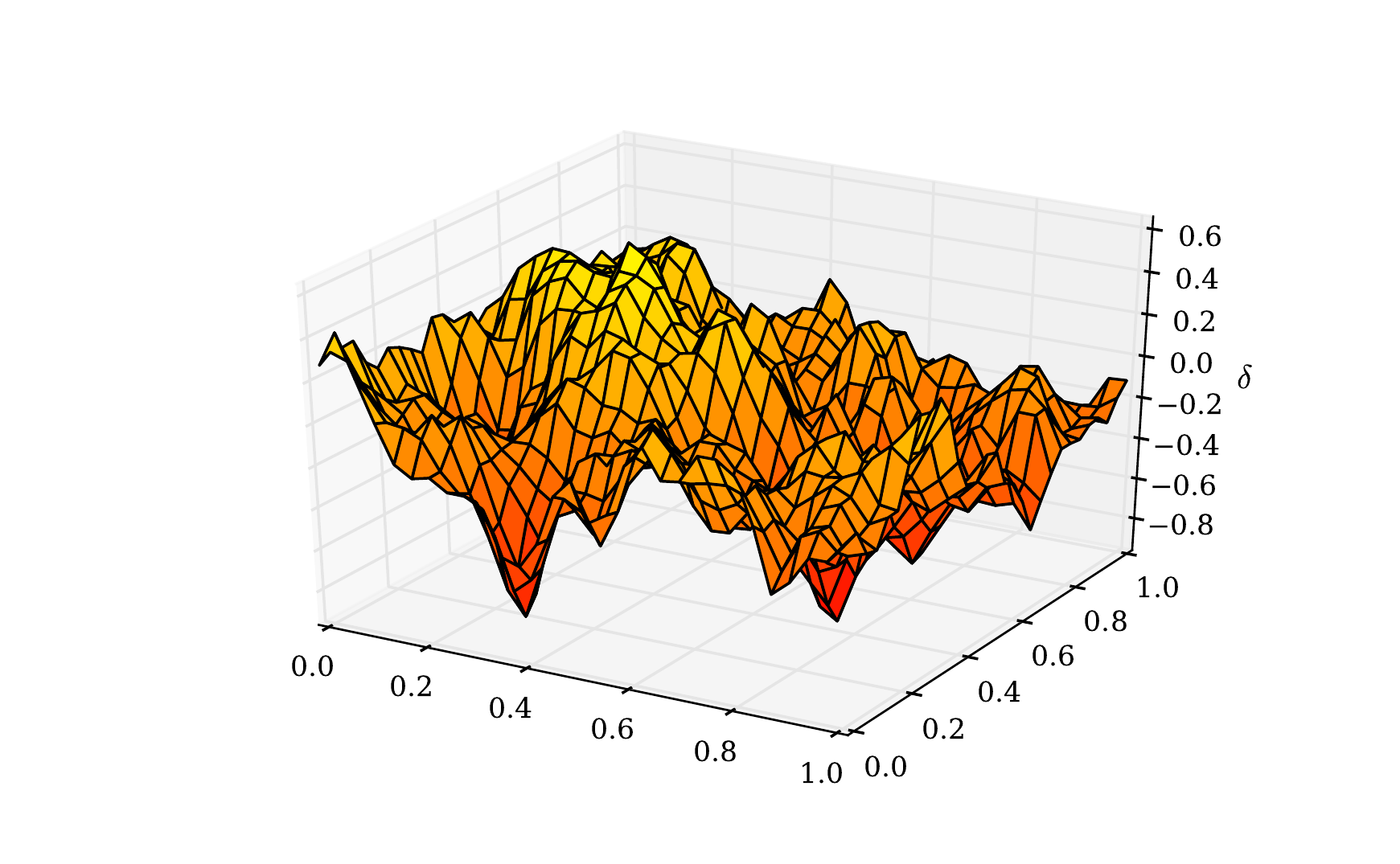}} \\
\subfloat{\includegraphics{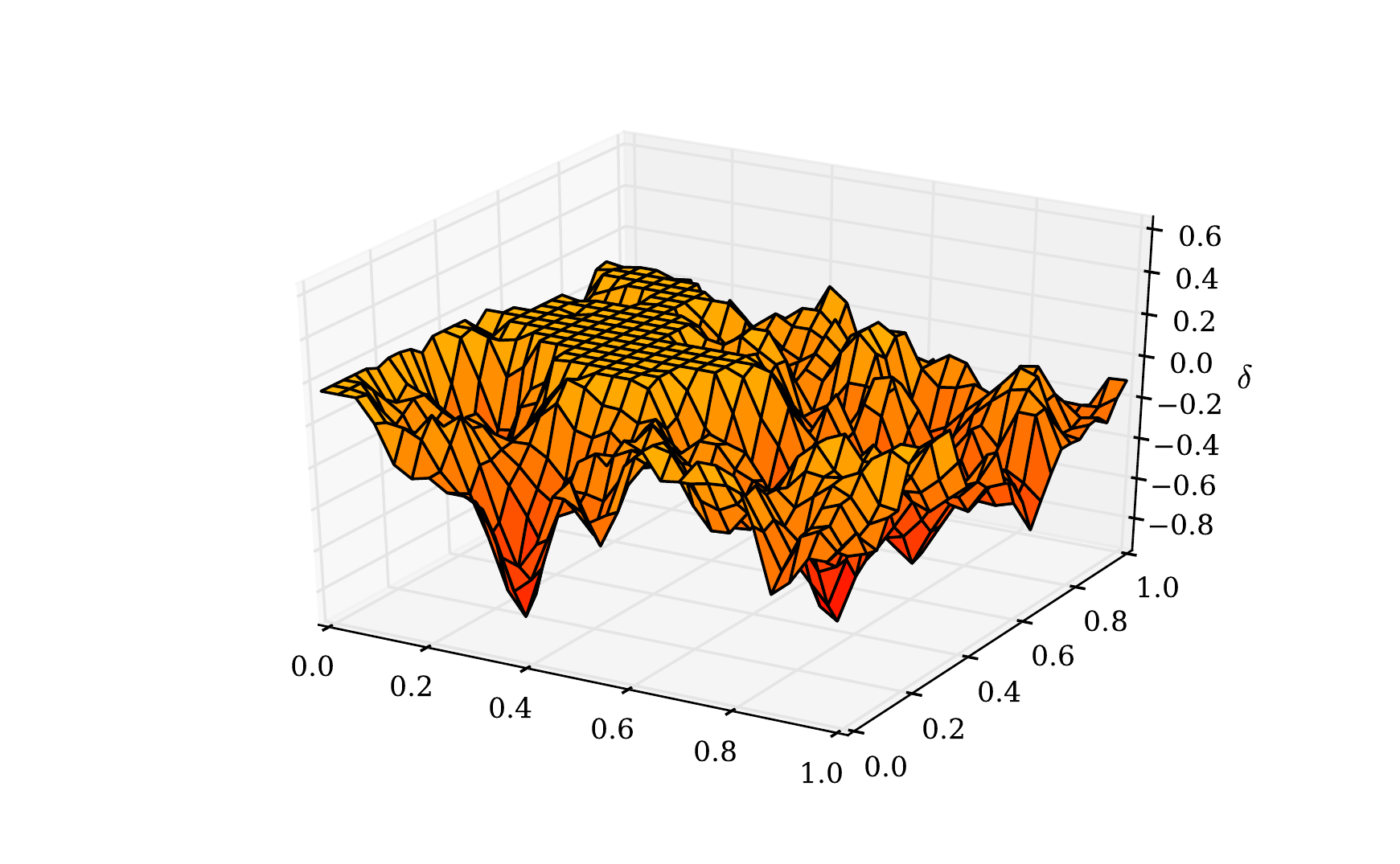}}  
\caption[An analogy to clipping]{A analogy of clipping: if the overdensity field is represented by distance a horizontal plane at $\delta=0$, clipping corresponds to a saturation at a pre-defined threshold, $\delta_0$.  This results in the Table mountain-like ranges shown.}
\label{fig:clipped}
\end{figure}
\noindent
this is equivalent to a saturated estimate of the overdensity field.  The positive tail, $\delta > 0$, is specifically targeted as gravitational instability drives $p(\delta)$ to be highly skewed with a long tail to large densities.  

Clipping has been shown to recover the linear prediction for the real-space bispectrum \citep{clippedBispectrum} and proven to be effective in a GAMA RSD analysis \citep{Fergus}.  The optimal threshold was shown to be that which halves the power spectrum amplitude in this work.  One might naively presume that a still smaller threshold would yield a yet more linear volume, but there is a fundamental floor: the regime $\delta \simeq -1$ is highly non-linear as linear growth is forbidden by the positive-definite condition on the density.  This property will become significant in shaping $P(k)$ for a very small $\delta_0$.  In addition, the reduced effective number density -- due to upweighting a small subsample of galaxies, leads to strictly clipped fields becoming progressively more shot noise dominated.  

In addition to non-linear dynamics, we observe galaxies as opposed to the mass directly; this hinders the potential of RSD measurements.  Although a complete understanding of galaxy formation is a fundamental goal of cosmology, this has proven to be a complicated field to address.  For many years it has been known that galaxies linearly trace the mass only on the largest scales.  With galaxies being intrinsically more clustered than the total mass as galaxy formation is biased to overdense environments, i.e. to the most massive haloes; see Fig. \ref{fig:tinker_biasedclustering} for instance.  The physics of galaxy formation includes a wide range of processes spanning many magnitudes in scale, e.g. active galactic nuclei feedback and supernovae -- see \cite{CP} and a complete understanding is likely to be take many decades.  Non-linear galaxy bias is therefore likely to remain a significant uncertainty, which leads to the restriction of RSD to the only largest scales, $r \gsim 20 \mpcoh$.  Further motivation for clipping is then provided by galaxy bias models in which the galaxy overdensity is assumed to be a local and deterministic function of that of the mass:
\[
\delta_g(\mathbf{x}) = \sum_{n=0}^{\infty} \frac{b_n}{n!} \ \delta^n.
\]
In such models, linear bias will clearly be recovered in the volume where $|\delta| \lsim 1$.  However, these are unrealistic approximations, not least because galaxy formation is stochastic in nature.  But certainly the simplification of galaxy formation in underdense environments -- where ram-pressure stripping is absent for instance, is a plausible proposition.  

Further motivation for clipping is provided by the plethora of `shielded' modified gravity models.  These revert to GR in the large (or rapidly changing) curvature regime in order to circumvent the stringent constraints placed by solar system tests; see \cite{ModGravReview} for a review.  The signatures of modified gravity theories can therefore be amplified by focusing on the underdense volumes isolated by clipping \citep{LombriserClipping}.  This has led \cite{White2016} to advocate the addition of density-weighted clustering statistics such as clipping to the conventional set of clustering analyses.

Clipped power spectra may be obtained for a range of thresholds, $\delta_0$.  As these are obtained from a largely overlapping volume there is a strong covariance between them, but the combination represents a higher-order statistic than the power spectrum.  The small-scale density field is highly non-Gaussian and it is far from obvious than the information content of the bispectrum is second to only the power spectrum; especially as realistic mocks would be required to quantify this property for small scales.  Therefore the constraints of a clipped analysis are potentially competitive with that of the bispectra, e.g. \cite{GilMarin}.  This provides further motivation for the work described in this chapter.     

Although potentially advantageous, clipping is strongly correlated with the density field by construction and modelling the clipped power spectrum is not simply equivalent to an effective survey mask.  But only a small percentage of the volume, $\simeq 5 \%$, is typically affected and therefore the net modelling may be simplified (as the non-linearity of densities and velocities and galaxy biasing has been suppressed).  There is a significant reduction of small-scale power following clipping despite this restriction to a small volume, which reflects the disproportionate power from large haloes.  This has motivated similar local transforms, e.g. a lognormal approach \citep{Neyrinck}.  Care must be taken to ensure that the clipped fields do sufficiently recover the linear theory prediction and ultimately these methods may only be proven to be robust and effective by analysing realistic mock galaxy catalogues.

The following describes an estimator of a clipped density field and provides approximate models for the expected clipped power spectrum.  The effectiveness of this method for constraining $f \sigma_8$ is investigated using mock catalogues in \S \ref{sec:clipping_mocks}.

\section{An effective weight definition}
\label{sec:clip_weights}
The clipping weights require an estimate of the overdensity, $\delta$;  I calculate this for each cell of a lattice spanning the surveyed volume according to 
\[
\bar{n} (1 + \delta_i) \ dV = \sum_{g} \frac{1}{E},
\]
where $\bar{n}(\mathbf{x}_i)$ is the ESR corrected number density (see \S \ref{sec:ESR} for the definition of ESR), $dV$ is the cell volume and the sum is over the galaxies assigned to cell $i$ by nearest grid point.  This estimate will be shot noise dominated for small cells; to reduce this noise I smooth the field with a Gaussian and take $2 \mpcoh$ as the fiducial smoothing scale.  This follows the approach taken by \cite{GAMA_Clipping}.  The homogeneous constraint, $\left \langle 1+\delta \right \rangle = 1$, is then restored for the smoothed field by rescaling $(1 + \delta)$ in amplitude, such that volumes with $\delta=-1$ are unaltered.  If this field exceeds a saturation value, $\delta_i > \delta_0$, each galaxy in cell $i$ is assigned a weight:
\[
w_c(\mathbf{x}_i) = \frac{(1 + \delta_0)}{(1 + \delta_i)},
\label{eqn:clip_weight}
\]
and $w_c = 1$ otherwise.  I investigate the effectiveness of a range of $\delta_0$ in \S \ref{sec:v7_application}, but leave the dependence on smoothing scale to future work. 

Having defined effective weights that achieve a clipped estimate of the overdensity field, I outline plausible models for the expected two-point statistics of such fields in the following section.  An implicit assumption of this work is that clipping a Poisson sampled, biased and non-linear density field resembles clipping a linear density field as the linear volume is amplified by the weighting.  

\section{Modelling the clipped power spectrum}
The models for the clipped power spectrum derived in the following predict two generic, intuitive, effects: the reduced variance of the clipped field results in a suppression of the power spectrum amplitude and the sharp transition at of the density at $\delta_0$ increases the relative small-scale power.  A precise calculation of this spectral distortion would require detailed assumptions for the non-linear growth and bias that I hope to circumvent with clipping; but as only an approximately linear and Gaussian volume is retained, it may be sufficient to assume a clipped Gaussian field; the clipped power spectrum for a Gaussian field was rederived in \cite{GAMA_Clipping}.  This derivation is reproduced in the following section in order to define a notation and provide a basis for an original derivation of the `clipped lognormal' model in \S \ref{sec:lognormal}.

\subsection{Spectral distortion of a clipped Gaussian field}
Using Price's relation \citep{Price} for the autocorrelation function of a local transformation of a Gaussian field, $g(\delta_G)$, the clipped correlation function is given by  
\begin{equation}
\left \langle \frac{\partial \xi_c}{\partial \xi} \right \rangle = \int d1 \int d2 \ p(1, 2, r) \ \partial_1 g \ \partial_2 g.
\end{equation}
Here $1$ denotes $\delta(\mathbf{x_1})$, i.e. $\int d1$ is a succinct notation for $\int d\delta_1$ and similarly $\partial_1 g \equiv \partial g /\partial \delta_1$.  For the clipping transform defined in eqn. (\ref{eqn:clipping_transform}), $\partial g = 0$ for $\delta \geq \delta_0$, which limits the range of integration, and unity otherwise; there is a discontinuity at the threshold.  In this expression, $p(1, 2, r)$ is the scale-dependent two-point probability of a Gaussian random field: 
\begin{equation}
p(1, 2, r) = \frac{1}{2 \pi \sqrt{1 - \rho^2}} \ \exp \left( \frac{2 \rho \delta_1 \delta_2 - \delta_1^2 - \delta_2^2}{2 \sigma^2(1 - \rho^2)} \right ),
\end{equation}
for the rescaled correlation function, $\rho(r) = \xi(r) / \xi(0)$.  Applying \href{http://mathworld.wolfram.com/MehlersHermitePolynomialFormula.html}{Mehler's polynomial formula} for the normalised variables $x = \delta_1 / \sqrt{2} \sigma$ and $y= \delta_2 / \sqrt{2} \sigma$ results in  
\begin{equation}
p(x, y, r) = \frac{e^{-(x^2 + y^2)}}{2 \pi \sigma^2} \ \sum_{n=0}^\infty \frac{H_{n}(x) H_n(y)}{n !} \left( \frac{\rho}{2} \right)^n.
\end{equation}
Where the Hermite polynomials, $H_n(x)$, are defined by 
\begin{equation}
H_n(x) = (-1)^n e^{x^2} \frac{d^n}{dx^n}(e^{-x^2}).
\end{equation}
In this form, $p(x, y, r)$ is explicitly separable and Price's relation reduces to the square of a one-dimensional integral:
\begin{equation}
\left \langle \frac{\partial \xi_c}{\partial \xi} \right \rangle = \frac{1}{2 \pi \sigma^2} \sum_{n=0}^{\infty} \frac{\rho^n}{2^n n!} \ \left[ \int_{-\infty}^{\delta_0} d1 \ e^{-\delta_1^2/(2 \sigma^2)} \ H_n \left (\frac{\delta_1}{\sqrt{2} \sigma} \right )\right ]^2. 
\end{equation}
The integrand is odd for symmetric limits and odd $n$ and hence only even $n$ terms are non-zero.  As the clipping transform I have defined is asymmetric, all $n$ must be retained.  Using the definition of the Hermite polynomials given above, the integrand is a total derivative and the integral has a solution of \citep{GAMA_Clipping}:
\[
\xi_c(\mathbf{r}) = \mathcal{A}^2 \xi(\mathbf{r}) + \sigma^2 \sum_{n=1}^{\infty} \left [ \frac{\xi (\mathbf{r})}{\sigma^2}\right ]^{n+1} C_n(u_0).
\label{eqn:clipped_xi}
\]
The integration constant is zero due to the large-scale homogeneous limit, $\xi(r) \mapsto 0$ for $r\gg1$.
The spectral distortion is linked the the amplitude suppression, $\mathcal{A}$, in this Gaussian model.  Both are conveniently parameterised by the normalised threshold, $u_0 = \delta_0/(\sqrt{2} \sigma)$.  For an amplitude suppression of 
\begin{equation}
\label{eqn:clip_amp}
\mathcal{A} = \frac{1}{2} \left [ 1 + \erf(u_0) \right ],
\end{equation}
the spectral distortion coefficients are 
\[
C_{n}(u_0) = \frac{H^2_{n-1}(u_0)}{2^n \pi (n+1)!} e^{-2u_0^2}.
\]
This derivation closely resembles that for the correlation function of biased galaxy clusters \citep{KaiserClusters}. 

In the distant-observer approximation, $\xi_s(\mathbf{s})$ may be expanded in a Legendre series; the largest non-zero term is $\ell=4$ in the Kaiser limit.  Following \cite{maskedRSD}, the lowest order terms of the spectral distortion corrections are 
\begin{align}
\label{eqn:clip_xi}
\xi^{c}_{0} =&   \mathcal{A}^2 \xi_0 + \frac{C_1}{\sigma^2} \left( \xi_0^2 + \frac{\xi_2^2}{5} + \frac{\xi_4^2}{9}  + \frac{\xi_6^2}{13} + \mathcal{O}(\xi_8^2) \right) + \mathcal{O}(C_2), \nonumber \\
\xi_{2}^c =& \mathcal{A}^2 \xi_2 + \frac{C_1}{\sigma^2} \left(  2 \xi_0\xi_2 + \frac{2}{7} \xi_2^2 +\frac{4}{7} \xi_2 \xi_4  + \frac{100}{693} \xi_4^2 + \frac{50}{143} \xi_4 \xi_6 + \frac{14}{143} \xi_6^2 + \mathcal{O}(\xi_8) \right) \nonumber \\ &\qquad + \mathcal{O}(C_2), \nonumber \\
\xi^{c}_4 =& \mathcal{A}^2 \xi_4 + \frac{C_1}{\sigma^2} \bigg( \frac{18}{35} \xi_2^2 + 2 \xi_0 \xi_4 + \frac{40}{77} \xi_2 \xi_4 \nonumber + \frac{162}{1001} \xi_4^2 + \frac{90}{143} \xi_2 \xi_6 \nonumber \\ & \qquad \qquad \qquad \ \quad + \frac{40}{143} \xi_4 \xi_6 + \frac{252}{2431} \xi_6^2 + \mathcal{O}(\xi_8) \bigg) + \mathcal{O}(C_2).
\end{align}
Higher order $C_n$ terms were found to only be required for sub-percent precision in \cite{Fergus}.  A comparison of this approximation with the complete correction is shown in Fig. \ref{fig:rapid_clippedmultipoles}.  Computing the complete correction by 3D FFT is prohibitively slow as the spectral distortion must be computed for a range of models as part of the likelihood analysis.  Accordingly, I approximate the correction with this series expansion, which is computed using FFTlog.  Further arguments for a similar approximation are provided in Chapter \ref{chap:maskedRSD}.

The normalised threshold, $u_0$, must be determined to calculate the spectral distortion.  In the Gaussian model, an estimate may be made from the amplitude suppression between the clipped and unclipped $P_0(k)$ on large scales -- where the spectral distortion is negligible.  I assume the variance, $\sigma^2$, is given by $\xi(0)$ for the model of interest.  This spectral distortion correction then fits into the wider context of the forward modelling, which is described in \S \ref{sec:RSD_modellingOverview}.
\begin{figure}
\subfloat{\includegraphics[width=\linewidth]{pk_powerlaw_trunc2.pdf}} \\
\subfloat{\includegraphics[width=\linewidth]{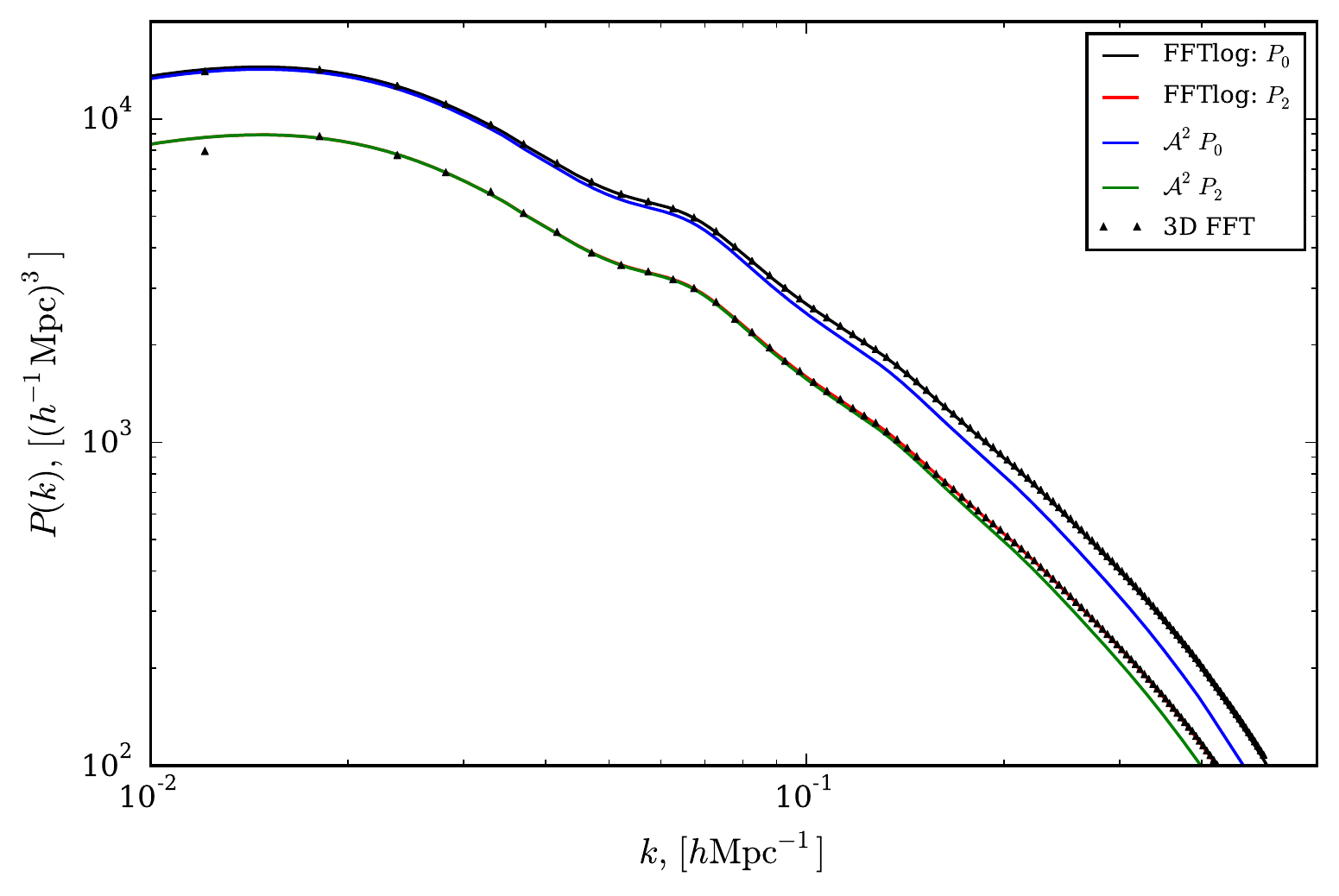}}
\caption[Equivalency of the 3D FFT and FFTlog clipped $P_{\ell}(k)$.]{Top: a comparison of the clipped $\xi_{\ell}$ obtained by 3D FFT and an approximate series expansion, which is computed by a Hankel transform (1D FFT).  The latter approach is rapid enough that an emulator technique, as utilised by \cite{Fergus}, is not required.  This approach also mitigates problems of resolution and aliasing associated to the memory limited 3D FFT; see \cite{maskedRSD} for further discussion.  Bottom: once $\xi_{\ell}^c$ has been computed it may be transformed to Fourier space by FFTlog without loss of accuracy; this figure proves this to be the case. }
\label{fig:rapid_clippedmultipoles}
\end{figure}

\subsection{Spectral distortion of a clipped lognormal field}
\label{sec:lognormal}
\begin{figure}
\centering
\includegraphics[scale=0.75]{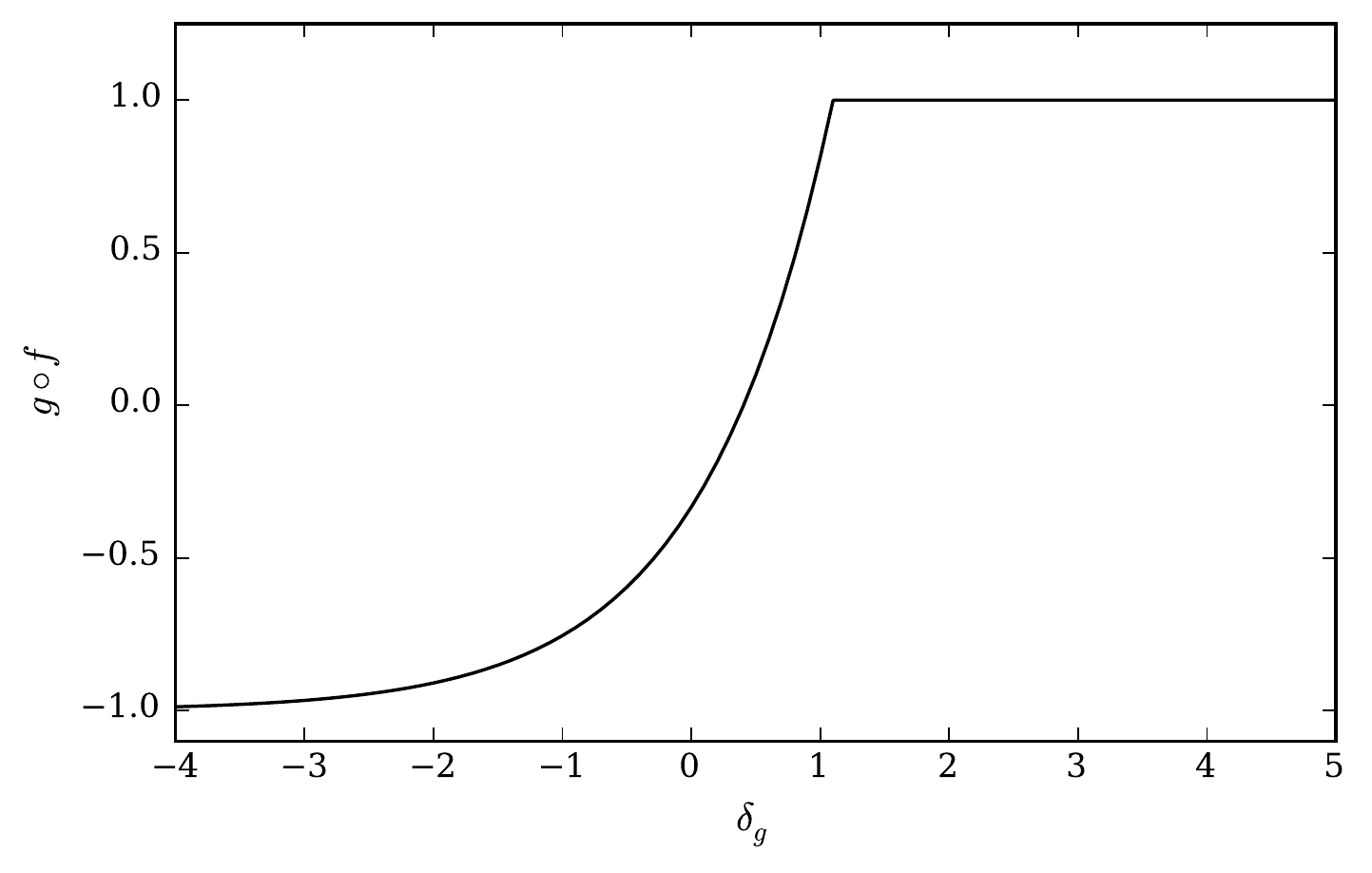}
\caption[The effective mapping of the clipped lognormal model]{The clipped lognormal model is equivalent to applying a composite map, $g \circ f$, to a Gaussian field.  Here $g$ is defined by eqn. \ref{eqn:clipping_transform} and $f$ is given by \ref{eqn:lognormal_transform}.  This figure illustrates this composite map.  As for the clipped Gaussian case, the flat gradient places an upper limit of $\delta_G=1$ on the integration range.}
\label{fig:clippedlognormal_compositemap}
\end{figure}
The power spectrum of a clipped lognormal field (see \S \ref{sec:lognormal} for an introduction to the lognormal model) may be derived from the correlation function obtained when a composite map, $g \circ f$ -- i.e first $f$ and then $g$, is applied to a Gaussian field.  Where $f$ is given by  
\begin{equation}
f: \exp(\delta_G - \frac{\sigma_G^2}{2}) - 1;
\label{eqn:lognormal_transform}
\end{equation}
the $\exp(-\sigma_G/2)$ factor arises from imposing the homogeneous constraint, $\langle 1 + \delta \rangle = 1$.  In this clipped lognormal case, Price's relation yields  
\begin{equation}
\left \langle \frac{\partial \xi_{c}}{\partial \xi} \right \rangle = \int_{-\infty}^{\delta_G} d1 \int_{-\infty}^{\delta_G} d2 \ p(1, 2, r) \ \partial_1 (g \circ f) \ \partial_2 (g \circ f).
\end{equation}
The range of integration is restricted by the effective threshold, $\delta_G$, applied to the generating Gaussian.  This satisfies
\begin{equation}
\delta_G = \frac{\sigma_G^2}{2} + \text{ln}(1 + \delta_0).
\end{equation} 
Within the integration limits: $g \circ f \mapsto f$; see Fig. \ref{fig:clippedlognormal_compositemap}.  Consequently,
\begin{equation}
\left \langle \frac{\partial \xi_c}{\partial \xi} \right \rangle = 2 \sigma_G^2  e^{-\sigma_G^2} \int dx \int dy \ p(x, y, r) \ e^{\sqrt{2} \sigma_G x} \ e^{\sqrt{2} \sigma_G y}.
\end{equation}
This relation is again separable if Mehler's formula is used. As a result: 
\begin{equation}
\left \langle \frac{\partial \xi_c}{\partial \xi} \right \rangle = e^{-\sigma_G^2} \sum_{n=0}^{\infty} \frac{\rho^n}{\pi 2^n n!} \ I_n^2,
\end{equation}
where $I_n$ is defined by 
\begin{equation}
I_n \equiv (-1)^n \int^{u_0}_{-\infty} e^{\sqrt{2} \sigma_G x} \left( \frac{d^n}{dx^n} e^{-x^2} \right) dx,
\end{equation}
for $u_0 = \delta_G / \sqrt{2} \sigma_G$.  The $e^{\sqrt{2} \sigma_G x}$ term is absent in the Gaussian case and the integrand is a total derivative.  Whereas, for the clipped lognormal model, an integration by parts leads to a recurrence relation: 
\begin{equation}
I_n = \sqrt{2} \sigma_G I_{n-1} - e^{\sqrt{2} \sigma_G u_0} H_{n-1}(u_0) \ e^{-u_0^2},
\end{equation}
which has a series solution of
\begin{equation}
\xi_c = \sum_{n=0}^{\infty} \frac{\xi^{n+1}}{(n+1)!} \ J_n^2 + \text{cnst}.
\end{equation}
The integration constant is again zero due to the large-scale homogeneity constraint, $\xi(r) \mapsto 0$ for $r \gg 1 \mpcoh$.  The $J_n$ coefficients are a renormalisation of $I_n$ and defined by 
\begin{equation}
J_n = \frac{e^{-\sigma_G^2/2}}{2^{n/2} \sqrt{\pi} \sigma_G^n} I_n. 
\end{equation}
These satisfy the recurrence relation:
\begin{equation}
\label{eqn:J_recurrence}
J_n = J_{n-1} - \frac{e^{-(u_0 - \sigma_G/ \sqrt{2})^2}}{2^{n/2} \sqrt{\pi} \sigma_G^n} H_{n-1} (u_0),
\end{equation}
with a first term of 
\begin{equation}
\label{eqn:J_firstterm}
J_0 = \frac{1}{2} \left (1 + \erf \left (u_0 - \frac{\sigma_G}{\sqrt{2}} \right) \right ).
\end{equation}
The lognormal limit should be recovered when $u_0 \mapsto \infty$; in this case:
$\erf(u_0 - \sigma_G/\sqrt{2}) \mapsto 1$, $J_0 = 1$ and $J_n = J_{n-1}$.  Consequently,
\begin{equation}
\xi_{\text{c}} \mapsto \sum_{n=0}^{\infty} \frac{\xi^{n+1}}{(n+1)!} = \text{exp}(\xi) - 1 \equiv \xi_{\text{ln}},
\end{equation}
as should be the case.  

Using the first term and recurrence relation for the clipped lognormal correlation function, equations (\ref{eqn:J_firstterm}) and (\ref{eqn:J_recurrence}) respectively, the Fourier equivalent may be calculated by Hankel transforming $\xi_0(r)$.  Fig. \ref{fig:clippedlognormal_powerspectrum} shows this prediction for a range of $\delta_0$, when parameterised by the amplitude suppression.  To validate this prediction I created and clipped a number of lognormal mocks; the results of this are shown as triangles in the figure.  A test of the accuracy of this model for predicting the affect of clipping on a real-space HOD cube is shown in Fig. \ref{fig:clippedlognormal_amplitudeHOD}.
\begin{figure}
\centering
\includegraphics[width=\textwidth]{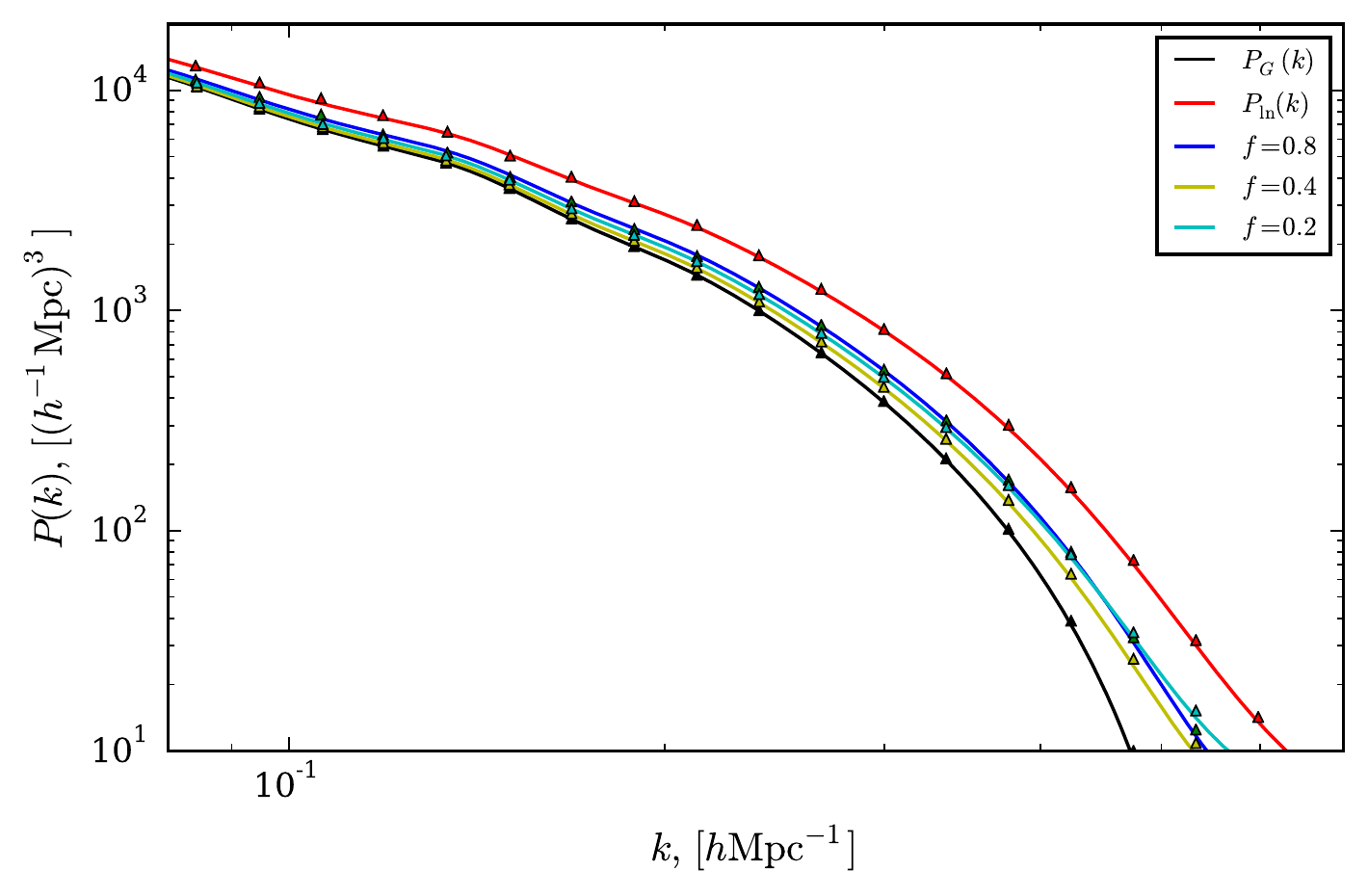}
\caption[Validation of the clipped lognormal model.]{Shown are the clipped lognormal prediction for $P_0(k)$ (solid) and the estimate from clipping a number of lognormal mocks (triangles) for various $\delta_0$; these are parameterised by the equivalent amplitude suppression, $f$.  The power spectrum of the generating Gaussian field (black) is assumed to be isotropic; the resultant lognormal power is shown in red.  The first order correction is simply an amplitude suppression, $\xi^c = f^2 \xi$, as in the Gaussian case.  The amplitude has been renormalised by $f^2$ in each case; The values of $f=\{0.8, 0.4, 0.2\}$ are equivalent to $\delta_0=\{2.864, 0.285, -0.289\}$ respectively.  Note that the relative small-scale power is increased by further clipping, as in the Gaussian case.}
\label{fig:clippedlognormal_powerspectrum}
\end{figure}
\begin{figure}
\centering
\includegraphics[width=\textwidth]{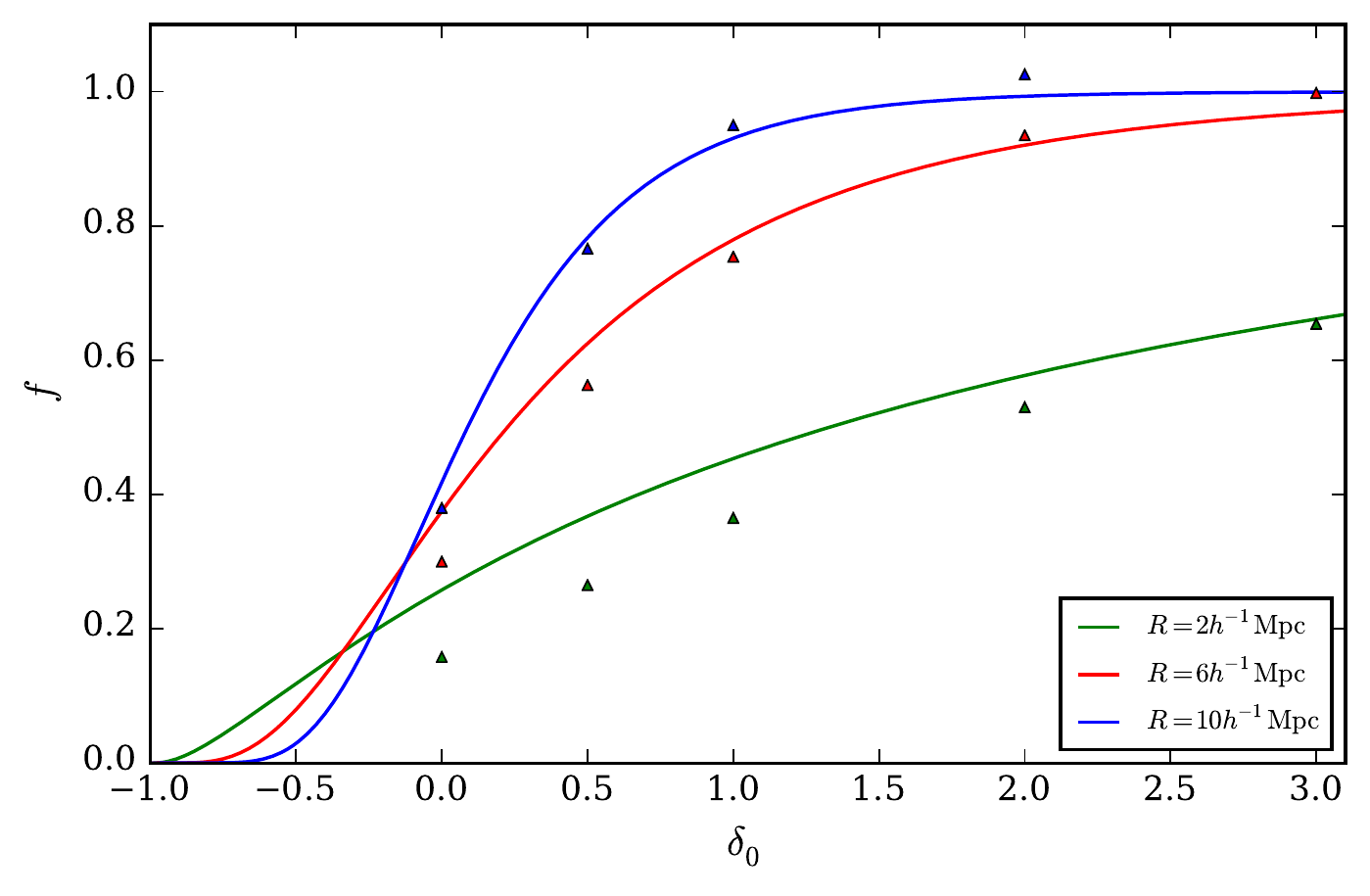}
\caption[Comparison of the predicted amplitude suppression for a HOD cube.]{A comparison of the power spectrum amplitude suppression for a real-space HOD cube (triangles) and the clipped lognormal model (solid).  This suppression is found by regression for $k \leq 0.06 h \emph{Mpc}^{-1}$ and is therefore is not affected by the spectral distortion.  The clipping weights are defined for a smoothed field with coherence length $R$.  For large  smoothing, $R=10 h^{-1} \emph{Mpc}$, the variance of the generating Gaussian can be estimated from the HOD cube; this is $\sigma_G = 0.41$.  For reduced smoothing, $R=2h^{-1} Mpc$, empty cells with $\delta=-1$ invalidate this estimate.  By simply fitting the model (green) it is clear that the clipped lognormal is not a good model for the suppression.  This is to be expected as on small-scales the velocity field is no longer linear; hence the lognormal model should be a poor approximation.}
\label{fig:clippedlognormal_amplitudeHOD}
\end{figure}

The redshift-space correlation function is anisotropic (see Chapter \ref{chap:RSD} and references therein).  \cite{GAMA_Clipping} advocate the prediction of the clipped redshift-space correlation function by separately applying Price's theorem to each $\mu$ slice of $\xi_s(\mathbf{s}, \mu)$. This results in eqn. (\ref{eqn:clipped_xi}).  The motivation for exponentiating the Lagrangian Gaussian field, eqn. (\ref{eqn:lognormal_transform}), is a real-space property that is not valid for the redshift-space density field.  It is therefore unclear that similarly exponentiating an anisotropic Gaussian field will yield a redshift-space model of any physical merit.  Accordingly, the remainder of this chapter focuses on applications of the clipped Gaussian model to the VIPERS v7 dataset.  But this model may be applicable when clipping other fields,  e.g. the weak lensing convergence field, which has been shown to be approximately lognormal \citep{lognormal_convergence}.   

\section{Clipping the VIPERS v7 data}
\label{sec:v7_application}
This section details the results of clipping the VIPERS v7: $0.6<z<0.9$ W1 and W4 mocks and data.  I investigate if a linear theory redshift-space power spectrum is sufficient for obtaining an unbiased estimate of $f \sigma_8$ with clipping; the expansion history is assumed to be known a priori.  Given the involved modelling of RSD -- predictions of both the effect of non-linearity (on densities and velocities) and galaxy bias are required, internal consistency checks such as ensuring the more-linear volume targeted by clipping delivers consistent results are of merit.  The validity of simple linear models is also potentially extended by applying a local density transform.  Moreover, it has been shown by \cite{LombriserClipping} that clipping can strongly amplify deviations of the matter power spectrum in shielded modified gravity models \citep{ModGravReview} from that in $\Lambda$CDM.  This motivates the addition of density-marked clustering statistics to the set of conventional analyses \citep{White2016}.  As the VIPERS v7 low-$z$ volume was shown to deliver the most stringent constraints in Chapter \ref{chap:VIPERS_RSD} I focus on this volume to begin with.  This low-$z$ volume contains approximately 80\% of the galaxies in the v7 sample and therefore most of the statistical power.  Moreover, the density field is better sampled in this volume and will not be prone to the shot noise artefacts likely to be present at high redshift.  An application to the full dataset is left to future work.

\subsection{Modelling summary}
The forward modelling is the same as that described in Chapter \ref{chap:VIPERS_RSD}, but $P_g(k)$ is taken to be the linear theory prediction and a Gaussian model for the spectral distortion, eqn. (\ref{eqn:clip_xi}), is assumed.  The magnitude of this distortion is determined by $u_0$, which is obtained from the amplitude suppression, $\mathcal{A}$, of the large-scale monopole prior to and following clipping.  This suppression is corrected for prior to the likelihood calculation and  therefore the $b \sigma_8$ parameter is physically meaningful.  I assume a  three dimensional parameter space, $(f \sigma_8, b \sigma_8, \sigma_p)$, with flat conservative priors of 
\begin{alignat}{3}
0.05 &\leq f \sigma_8 \leq 0.80, \notag \\
0.05 &\leq b \sigma_8 \leq 1.00, \notag \\
0.00 &\leq \sigma_p/(h^{-1} \rm{Mpc}) \leq 6.0.
\end{alignat}
These are similar to the unclipped case, which is given by eqn. (\ref{eqn:priors}).  The pairwise dispersion, $\sigma_p$, is expected to be reduced when clipping and therefore the maximum value is restricted to provide a better sampling of the space with the $16^3$ likelihood grid.  The allowed range of $b \sigma_8$ has been increased given the variation due to the amplitude suppression; the resolution, d(b$\sigma_8$) = 0.06, should remain sufficient.  The dependence of the posteriors on the assumed priors will be the subject of future work.  The power spectrum estimation is as detailed in \S \ref{sec:pk_estimate} for the clipping weights given by eqn. (\ref{eqn:clip_weight}).
\subsection{Results from the mocks}
\label{sec:clipping_mocks}
\begin{figure}
\centering
\includegraphics[width=\linewidth]{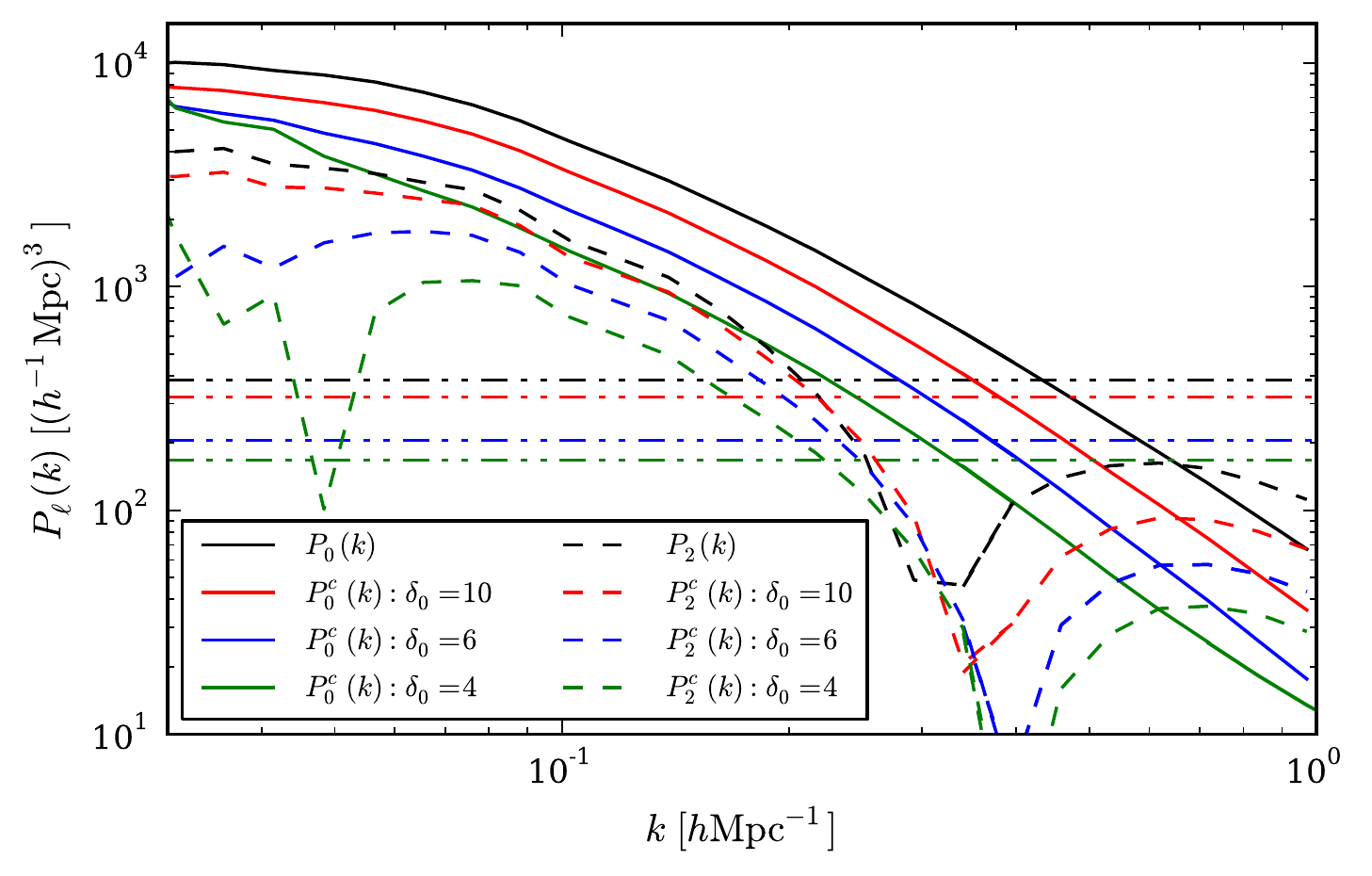}
\caption[$P_{\ell}(k)$ for clipped VIPERS v7: W1, $0.6<z<0.9$ mocks]{$P_{\ell}(k)$ for clipped VIPERS v7 W1: $0.6<z<0.9$ mocks as a function of $\delta_0$.  It is evident that the leading-order ``correction'' predicted by the Gaussian and lognormal models is valid as there is a significant reduction in amplitude -- $P(k)$ is halved for $\delta_0=6$. There is also a relative decrease in the small-scale amplitude despite the expected spectral distortion.  This is consistent with a more linear density field, which is corroborated by the shape of the quadrupole.  The quadrupole zero-crossing point can be seen to migrate to larger $k$, which is equivalent to a decrease in $\sigma_p$.  This is to be expected given the reduced virialised motions in the more linear volume targeted.  Despite these advantages, it is clear that clipping a survey with a realistic selection introduces large-scale features that are not predicted by simple models; there is an upturn in $P_0$ and a corresponding downturn in $P_2$ for strict clipping and $k < 0.05 h \rm{Mpc}^{-1}$.}
\label{fig:Clipped_pk}
\end{figure}
Fig. \ref{fig:Clipped_pk} shows the results of clipping the VIPERS v7 mock catalogues.  It is evident that the leading-order `correction' predicted by the Gaussian and lognormal models is valid as there is a significant reduction in amplitude -- $P(k)$ is halved for $\delta_0=6$. There is also a relative decrease in the small-scale amplitude despite the expected spectral distortion, which is consistent with a more linear density field.  This is corroborated by the shape of the quadrupole; a migration of the quadrupole zero-crossing to larger $k$ is apparent.  This is equivalent to a decrease in $\sigma_p$, which is to be expected given the reduced virial velocities in the more linear volume targeted.  Despite these advantages, it is clear that clipping a survey with a realistic selection introduces large-scale features that are not predicted by simple models -- there is an upturn in $P_0$ and a corresponding downturn in $P_2$ for strict clipping and $k < 0.05 h \rm{Mpc}^{-1}$.  To obtain a shot noise estimate for each $\delta_0$, Jenkins's thrice folded results were used to measure the amplitude of the (evidently flat) power spectrum in the range $1.5<k<2.5 h \rm{Mpc}^{-1}$.  These shot noise estimates are shown in the figure by the dot-dash lines and have been used to correct the monopole estimates (solid).

\begin{figure}
\centering
\includegraphics[width=0.85\linewidth]{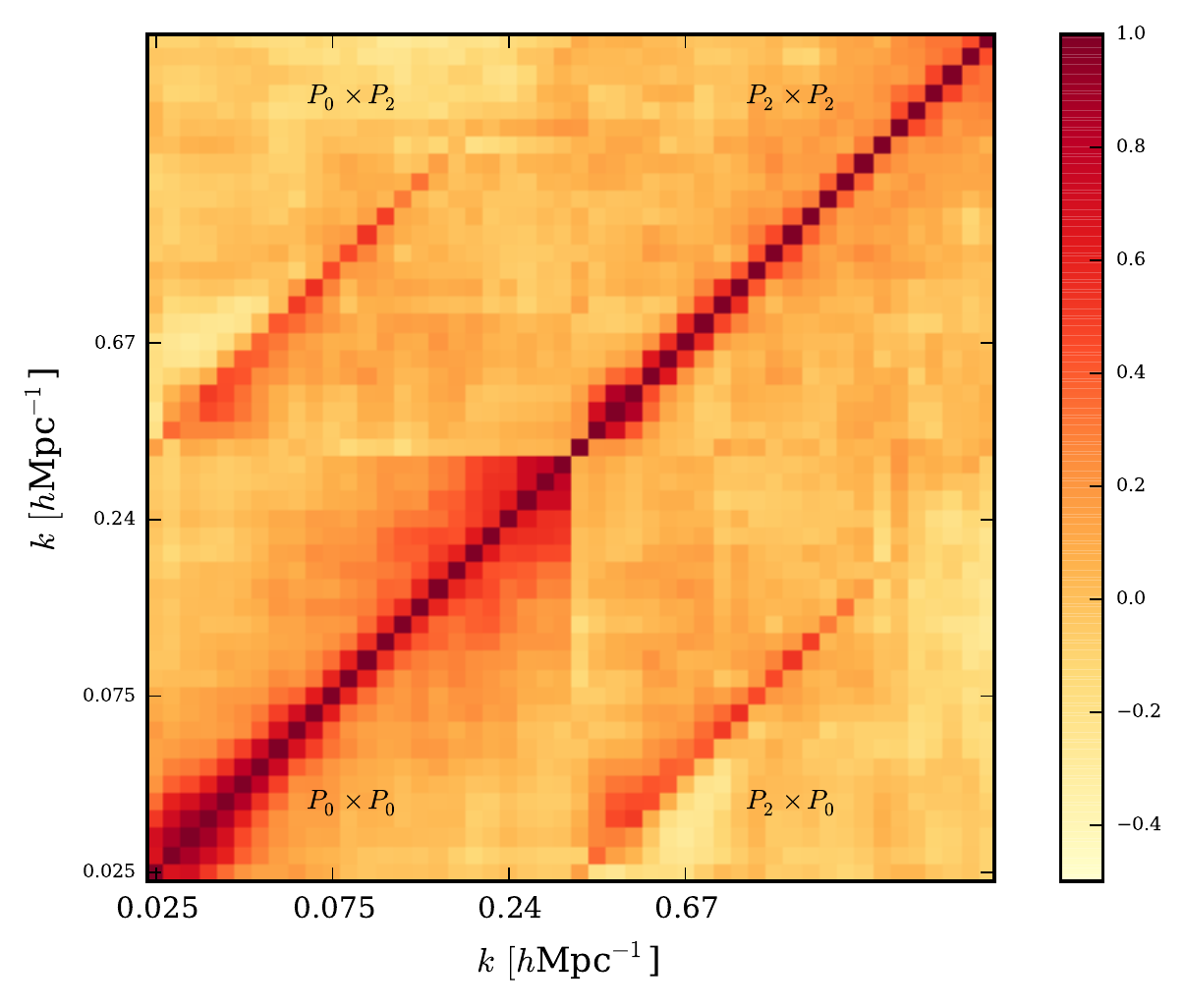}
\caption[Correlation matrix for a density field clipped at $\delta_0 = 10$.]{Correlation matrix of the power spectrum multipole moments for VIPERS v7 mocks clipped at $\delta_0 =10$.  When contrasted with the unclipped case, Fig. \ref{fig:MultipoleCovariance}, it is evident that clipping has significantly reduced the small-scale correlation.  This suggests a more linear density field and that tighter constraints on $f \sigma_8$ may be obtained for the same $k_{\rm{max}}$.  These constraints may be further improved if the linearised density field can also be modelled to a larger $k_{\rm{max}}$.  This is shown to be the case in Fig. \ref{fig:Clipped_fsig8}.  See Fig. 11 of \cite{ClippingCovariance} for a similar comparison.}
\label{fig:Clipped_MultipoleCovariance}
\end{figure}
An effective linearisation of the clipped density field is also reflected by the correlation matrix; this is shown in Fig. \ref{fig:Clipped_MultipoleCovariance}.  When compared to the unclipped case, Fig. \ref{fig:MultipoleCovariance}, there is clearly a significant decrease in the small-scale correlation.  As off-diagonal elements are also reduced, this is tending towards the diagonal case predicted by linear theory (in conjunction with the simplest inflationary models).  As alluded to previously, further decreasing $\delta_0$ beyond $\delta_0=4$ is likely to raise additional problems rather than linearise the density field further.     

The estimates of $f \sigma_8$ obtained from the clipped VIPERS v7 mocks are shown in Fig.~\ref{fig:Clipped_fsig8}.  These are modelled according to a dispersion model with $P_g(k)$ taken to be the linear theory prediction.  Simultaneously fitting the non-linear monopole and quadrupole shape with linear theory yields a biased estimate of $f \sigma_8 \simeq 0.2$ (irrespective of $k_{\rm{max}}$) without clipping, $\delta_0=1000$.  But with progressively stricter clipping the bias is reduced and ultimately removed by a threshold of $\delta_0=4$.  This provides a final calibration of the method as there is no other means by which an appropriate value of $\delta_0$ can be decided apriori.   

Alternatively, rather than choosing a single unbiased threshold, more information may be obtained from a combined constraint based on a number of thresholds.  This approach is equivalent to a higher-order density-dependent clustering statistic.  As large $\delta_0$ estimates are significantly biased it is necessary to calculate and apply calibration factors -- the ratio between the mock average estimate and the expectation for a given $\delta_0$ and $k_{\rm{max}}$.  The mocks also allow for the covariance between the calibrated estimates (due to the largely overlapping volume) to be obtained.
As an illustrative case, the correlation matrix of the calibrarted $f \sigma_8$ estimates for the W1 field and $k_{\rm{max}} = 0.4 \hompc$ is 
\[
\begin{pmatrix}
\label{eqn:fiducial}
&1.00 \
&0.89 \ 
&0.83 \ 
&0.63 \\
&0.89 \
&1.00 \
&0.91 \
&0.67 \\
&0.83 \ 
&0.91 \ 
&1.00 \
&0.69 \\
&0.63 \
&0.67 \
&0.69 \
&1.00
\end{pmatrix}.
\]
Here rows and columns are ordered according to $\delta_0=\{ 4, 6, 10, 1000\}$.  The combined constraint is largely determined by that with the least bias, $\delta_0=4$, as the errors scale with the calibration factor.  This is reassuring because calibrating a significantly biased estimate is unwise -- not least because it is unknown whether it is an additive or multiplicative bias a priori.  However, these biases would be reduced with a more realistic model and the combined constraint would be more informative as a result;  this is left to future work.  The increased correlation of the W4 estimates yields a close to singular matrix and therefore there is no more information to be gained from the combination.  I apply this method to the data in the following section.     
\begin{figure}
\centering
\includegraphics[width=\linewidth]{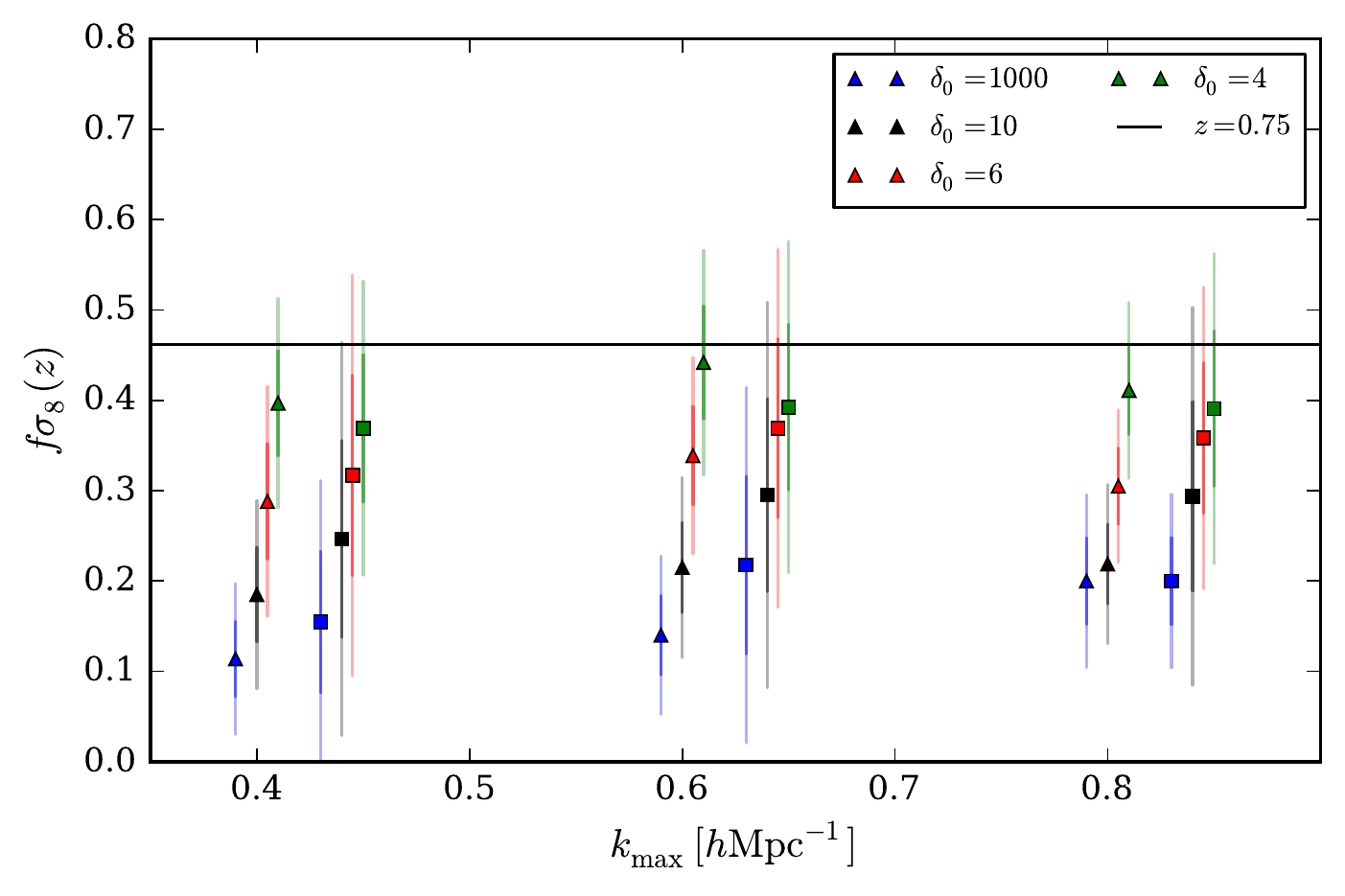}
\caption[Constraints on $f \sigma_8$ for the clipped VIPERS v7 mocks.]{Constraints on $f \sigma_8$ from the clipped VIPERS v7 W1 (triangles) and W4 (squares) mocks for $0.6<z<0.9$; constraints are shown for a range of thresholds: $\delta_0 \in \{ 1000, 10, 6, 4\}$.  A dispersion model is assumed with $P_g(k)$ taken to be the linear theory prediction.  The reduced small-scale amplitude of the linear $P_g(k)$ leads to a significant bias without clipping ($\delta_0 =1000$) as the zero-point of $P_2(k)$ cannot be fit simultaneously; but progressively stricter clipping leads to an unbiased estimate.  This is consistent with the results of  \cite{GAMA_Clipping} -- see their Fig. 4.  The combined constraint from a number of thresholds represents a higher-order density-dependent clustering statistic if the covariance is correctly accounted for.}  
\label{fig:Clipped_fsig8}
\end{figure}

\subsection{Results from the data}
\label{sec:clipping_data}
Fig. \ref{fig:data_clipped_pk} shows the shot noise corrected spectra of the VIPERS v7: $0.6<z<0.9$ W1 and W4 subvolumes when clipped at $\delta_0 = \{1000, 10, 6, 4 \}$ (black, red, blue and green respectively).  The mean measurement of the mocks (solid) is overplotted in each case and the shot noise estimates are given by the dot-dash lines. Reasonable agreement between the data and the mocks can be seen; there is some evidence for a reduced dispersion in the data given the zero-crossing of the quadrupole and a reduction in small-scale power.  Significantly, the dependence of the amplitude on $\delta_0$ seems to be well matched by the mocks.  This inspires confidence in the method given the necessary calibration on the mocks.
\begin{figure}
\centering
\includegraphics[width=\linewidth]{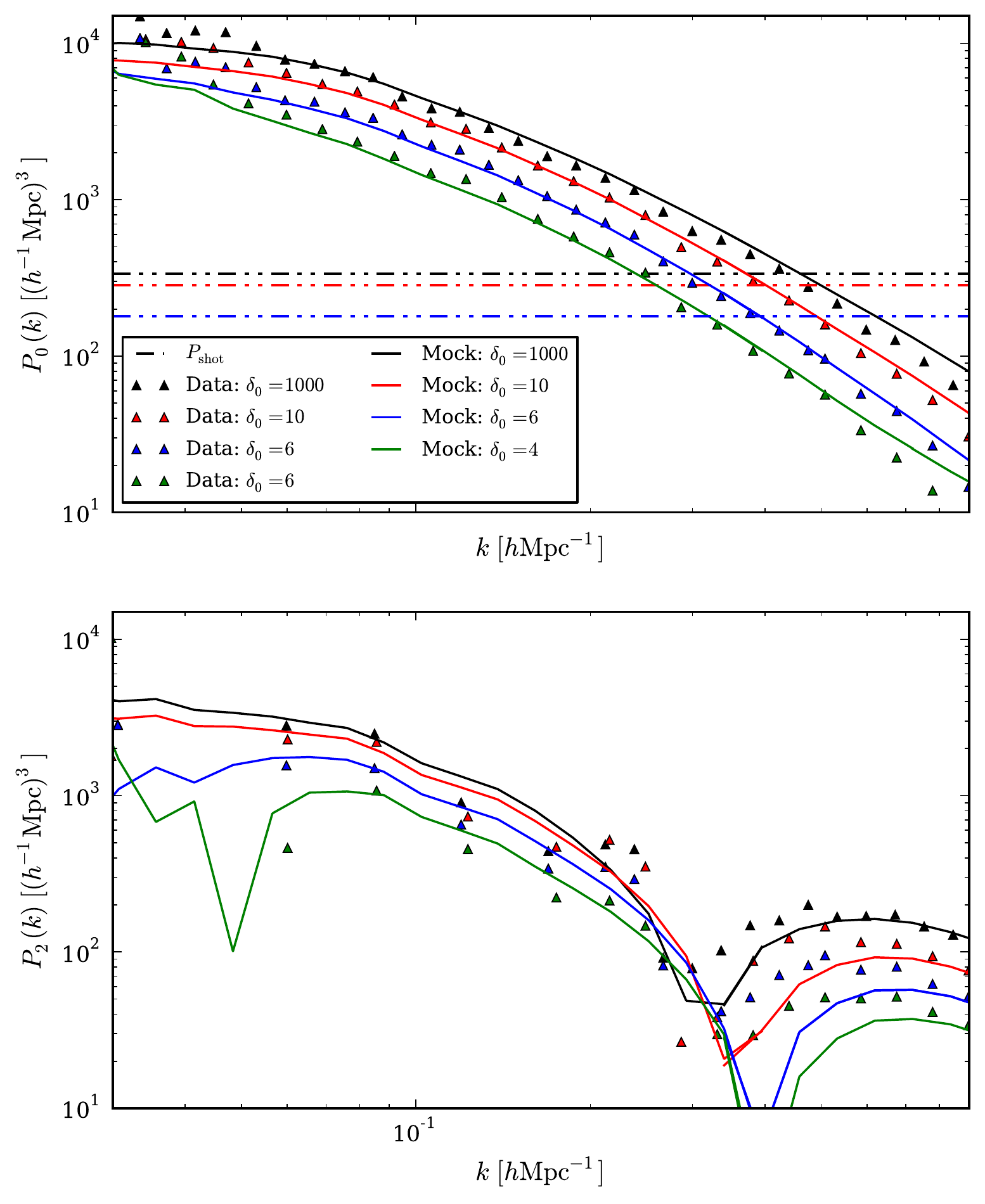}
\caption[$P_{\ell}(k)$ for the clipped VIPERS v7: W1, $0.6<z<0.9$ data]{Shown in this figure are the VIPERS v7: W1, $0.6<z<0.9$ power spectra when clipped at $\delta_0 = \{ 1000, 10, 6, 4\}$ (black, red, blue and green respectively).  The mean measurement of the mocks is overplotted (solid) for each case and the shot noise estimates are given by the dot-dash lines.  Reasonable agreement between the data and mocks can be seen; there is some evidence for a reduced dispersion in the data given the zero-crossing of the quadrupole and a reduction in the small-scale power.  Significantly, the dependence of the amplitude on $\delta_0$ seems to be well matched by the mocks.  This inspires confidence in the method given the necessary calibration on the mocks.}
\label{fig:data_clipped_pk}
\end{figure}

An $f \sigma_8$ estimate can be obtained by clipping at a range of thresholds, but these are highly correlated due to the overlapping volume.  Moreover, Fig. \ref{fig:Clipped_fsig8} has shown the estimates to be significantly biased for large $\delta_0$.  This bias can be corrected by defining calibration factors using the mocks, which also allow for the covariance to be estimated.  The maximum likelihood values for $f \sigma_8$ from the W1 low-$z$ slice are shown in Table \ref{table:fig8_combclip} for the combined constraint provided by all of $\delta_0= \{ 1000, 10, 6, 4\}$.  The $k_{\rm{max}} = 0.4 \hompc$ error represents a $\simeq 16\%$ decrease on the error of the most precise single-threshold ($\delta_0=4$) estimate in this case.  This is robust to the most biased $\delta_0$ estimate as the errors for each threshold scale with the calibration factor.
\begin{figure}
\centering
\includegraphics[width=\linewidth]{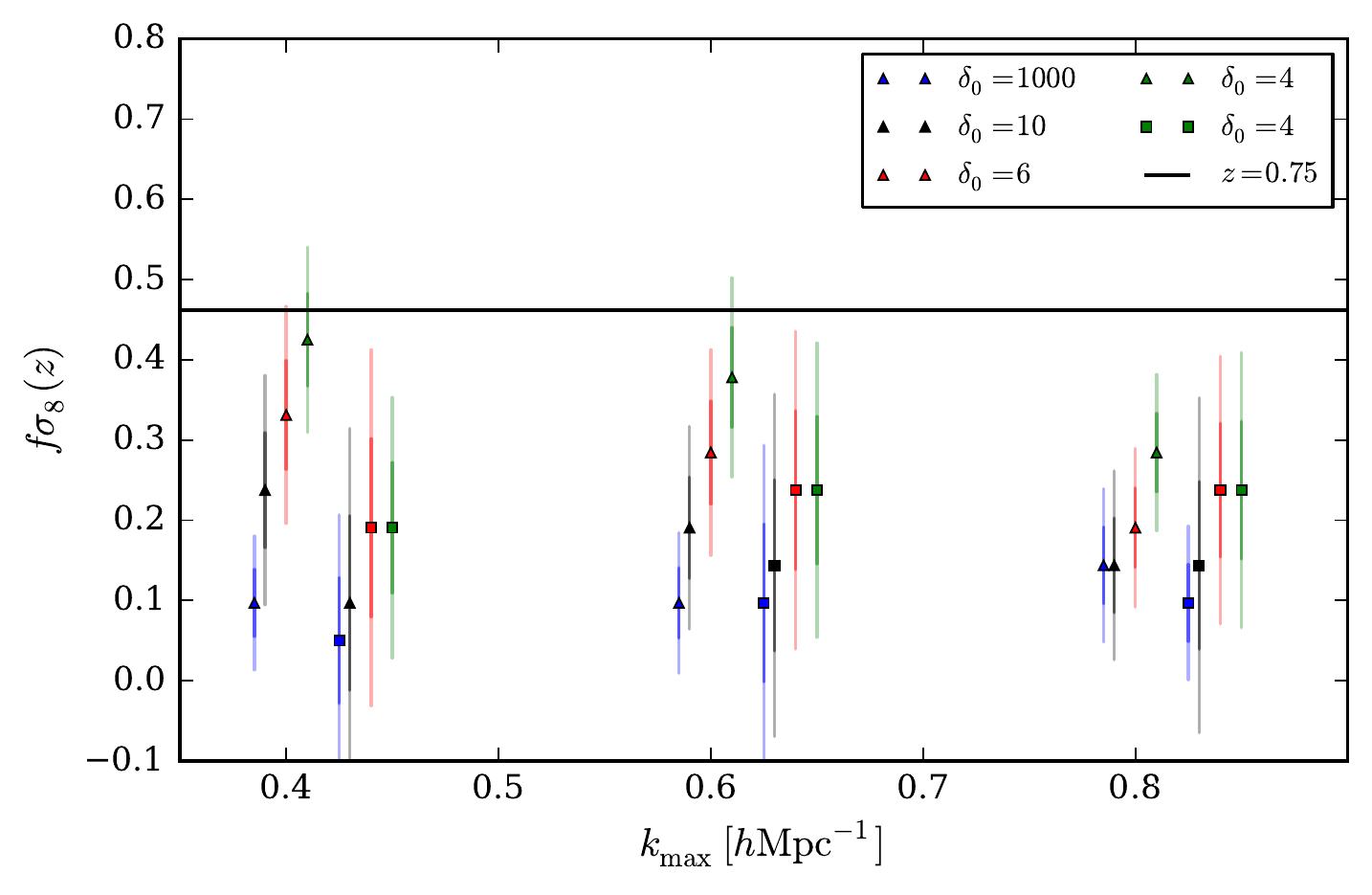}
\caption[Constraints on $f \sigma_8$ for the clipped VIPERS v7 data.]{Same format as Fig. \ref{fig:Clipped_fsig8} but for the VIPERS v7 low-$z$ dataset.  Marked consistency between the two is visible for the W1 field but the trend appears to be for a decreasing $f \sigma_8$ estimate with increasing $k_{\rm{max}}$ (as opposed to the flat trend of the mocks).  The $\chi^2$ of the red and green squares is largely determined by a bump at $k \simeq 0.1 \hompc$, which affects each point in this cumulative plot.}
\label{fig:data_Clipped_fsig8}
\end{figure}
\begin{table}
\centering
\begin{tabular}{|p{0.2\textwidth} | p{0.2\textwidth} | p{0.2\textwidth}| p{0.2\textwidth}|}
\hline
\hline
$\delta_0$ & $f \sigma_8$ & $(f \sigma_8)_s$ & $d(f \sigma_8)_s$ \\
\hline
4 & 0.43 & 0.49 & 0.07\\
\hline
6 & 0.33 & 0.53 & 0.1 \\
\hline
10 & 0.24 & 0.59 & 0.13\\
\hline
1000 & 0.10 & 0.39 & 0.17\\
\hline
\end{tabular}
\caption[VIPERS v7 data: constraints on $f \sigma_8$ for each $\delta_0$ value]{Estimated $f \sigma_8$ values for the W1 low-$z$ dataset for each of $\delta_0 = \{ 1000, 10, 6, 4\}$ and $k_{\rm{max}} = 0.4 \hompc$.  The calibrated values, $(f \sigma_8)_s$, and calibrated errors are also shown.   This calibration is required due to the significant systematic error observed in the mocks for large $\delta_0$, which could be reduced by the use of a more realistic model.}
\label{table:fig8_clip}
\end{table}
\begin{table}
\centering
\begin{tabular}{|p{0.2\textwidth} | p{0.2\textwidth} | p{0.2\textwidth}|}
\hline
\hline
$k_{\rm{max}} [ \hompc ]$ & $f \sigma_8$ & $d(f \sigma_8)$ \\
\hline
0.4 & 0.47 & 0.06 \\
\hline
0.6 & 0.38 & 0.06 \\
\hline
0.8 & 0.32 & 0.05 \\
\hline
\end{tabular}
\caption[VIPERS v7 data: constraints on $f \sigma_8$ from a combination of $\delta_0$ values]{The maximum likelihood values for $f \sigma_8$ from the W1 low-$z$ dataset for the combined constraint provided by all of $\delta_0= \{ 1000, 10, 6, 4\}$.  The systematic bias of individual estimates has been removed by calibration with the mocks, which also allows for the covariance to be obtained.  The $k_{\rm{max}} = 0.4 \hompc$ error represents a $\simeq 16\%$ decrease on the error of the most precise single-threshold ($\delta_0=4$) estimate in this case.  This is robust to the most biased estimate, $\delta_0$, as the errors scale with the calibration factor.}
\label{table:fig8_combclip}
\end{table}

\section{Conclusions and future work}
This chapter details an investigation of the constraints on $f \sigma_8$ obtained from the VIPERS v7 low-$z$ volume when subjected to clipping -- a local transform of the overdensity defined by eqn. (\ref{eqn:clipTransform}).  The results from the mocks are both promising and consistent with previous studies \citep{clippedBispectrum, ClippingCovariance, GAMA_Clipping}. There is ample evidence to suggest that clipping achieves an effective linearisation of the density field: a relative decrease of small-scale power, the migration of the quadrupole zero-crossing to larger $k$ (which is consistent with a suppression of the fingers-of-God) and reduced off-diagonal elements of the correlation matrix.  This is shown by Figures \ref{fig:Clipped_pk} and  \ref{fig:Clipped_MultipoleCovariance} respectively.  

This linearisation simplifies the modelling of the observed signal and potentially increases the validity of perturbation theory models, e.g. \cite{Taruya}, \cite{Okumura} and references therein. Further motivation is provided by `shielded' modified gravity models, which revert to GR in the large (or rapidly changing) curvature regime.  Signatures of modified gravity theories have been shown to be amplified by density-weighted two-point clustering statistics \citep{LombriserClipping};  see \cite{White2016} for further discussion on this point.  Despite these benefits, applying stricter clipping to a mock with a realistic survey selection introduces large-scale effects that are not predicted by simple Gaussian and lognormal models -- there is an upturn of $P_0(k)$ and corresponding downturn in $P_2(k)$ on large scales for $\delta_0 =4 $.  This perhaps results from the difficulty in making an accurate estimate of the overdensity at $z \simeq 0.9$ due to the much reduced $\bar n$ at this redshift.    

With this motivation, the mocks may be used to calibrate the method, estimate the statistical error and ensure there is no significant systematic bias.  Clipping has two degrees-of-freedom: the threshold, $\delta_0$, and the smoothing scale on which the density field is estimated prior to the transform.  \S \ref{sec:clipping_mocks} investigates the dependence of the $f \sigma_8$ estimates on $\delta_0$, for a fiducial smoothing scale of $2 \mpcoh$.  Fig. \ref{fig:Clipped_fsig8} shows that when fitting a Kaiser-Lorentzian model with a linear model for $P_g(k)$ there is a significant bias without clipping -- $f \sigma_8 \simeq 0.2$ is favoured irrespective of $k_{\rm{max}}$;  although the small-scale $P_0(k)$ can be made to fit by a large $b \sigma_8$ and an appropriate $\sigma_p$, the quadrupole shape (the zero-crossing in particular) cannot be simultaneously well modelled.  This figure suggests that the bias can be reduced and eventually removed with sufficient clipping, with $\delta_0=4$ providing the optimal threshold.  The clipped $f \sigma_8$ estimates for the VIPERS v7 dataset are shown in Fig. \ref{fig:data_Clipped_fsig8}, which show reasonable agreement with that seen in the mocks.  This is perhaps not surprising given the similarity of the clipped power spectra for the mocks and data; this is shown in Fig. \ref{fig:data_clipped_pk}.

An $f \sigma_8$ estimate can be obtained by clipping at a range of thresholds, but these are highly correlated due to the overlapping volume.  Moreover, Fig. \ref{fig:Clipped_fsig8} has shown the estimates to be significantly biased for large $\delta_0$.  This bias can be corrected by defining calibration factors using the mocks, which also allow for the covariance to be estimated.  The maximum likelihood values for $f \sigma_8$ from the W1 low-$z$ slice are shown in Table \ref{table:fig8_combclip} for the combined constraint provided by all of $\delta_0= \{ 1000, 10, 6, 4\}$.  The $k_{\rm{max}} = 0.4 \hompc$ error represents a $\simeq 16\%$ decrease on the error of the most precise single-threshold ($\delta_0=4$) estimate in this case, which is shown in Table \ref{table:fig8_clip}.  This is robust to the most biased $\delta_0$ estimate as the errors for each threshold scale with the calibration factor.  These results are highly encouraging to date but the analysis is a work in progress.  The final analysis will be submitted to Astronomy \& Astrophysics as Wilson et al. (2016).  

An original derivation of a `clipped lognormal' model is presented in \S \ref{sec:lognormal}.  This extends the work of \cite{KaiserClusters} and \cite{GAMA_Clipping} to a lognormal field.  I have validated this result by creating and subsequently clipping lognormal mocks.  It is uncertain how this model may be extended to the anisotropic redshift-space $\xi_s(\mathbf{s})$ and therefore I assume a simple Gaussian model when applying clipping to the VIPERS v7 dataset.  

An obvious avenue for future research is to apply this method to current and forthcoming surveys such as eBOSS, DESI and Euclid; but improvements to the method are warranted given the greater statistical power available.  One possibility is to include `rescaled' simulations \citep{Angulo, Mead}; in an ideal world N-body simulations spanning the range of models under test would be available -- including variations in the assumed cosmology, gravitational theory and galaxy formation model.  This is in order to predict the affect of clipping on the small-scale clustering estimator, which is unlikely to be analytically tractable (especially when including realistic selection effects).  But a sufficient number is unlikely to ever become available.  Rescaled simulations represent a compromise in which a N-body simulation is transformed to a new cosmology or gravity model according to simplified models of structure formation.  These may be used to span the regions between a set of points in the parameter space for which N-body simulations are available.  The effect of clipping may then be predicted accurately, rather than requiring simple assumptions to be made \citep{Fergus}.  Evidence that rescaling is sufficiently accurate has been provided by \cite{MeadZ} for example; that showed that the redshift-space power spectrum can be predicted to 5\% accuracy for $k<0.2 h$Mpc$^{-1}$.  This could be bettered when clipping as the volume retained is more amenable to the simplifications assumed. 

Such a method would build upon \cite{SimhaCole} and is similar in spirit to the current BOSS analysis of \cite{Metin}, but it is likely to be superior for a number of reasons; this work assumes a lognormal density field whereas the more accurate halo model may be assumed when rescaling; haloes may be populated according to a realistic halo occupation distribution model or sub-halo abundance matching method (as opposed to assuming a simple Poissonian scheme as in \citejap{Metin}); the likelihood can be calculated for a marked clustering statistic rather than the entire density field, which is likely to become prohibitive when further extensions are included -- given the difficultly in fully sampling a higher-dimensional likelihood space. 

In addition to this, \cite{ClippingCombined} has shown that the $(\Omega_m, \sigma_8)$ degeneracy present in cosmic shear analyses, e.g. \cite{Heymans}, may be broken by jointly fitting the clipped power spectra for a range of $\delta_0$ values with their appropriate covariance.  It is plausible that the $(f \sigma_8, \sigma_p)$ degeneracy present in RSD analyses (see Chapter \ref{chap:RSD}) may be similarly broken with such an approach. The results given above represent a `half-way house' in this respect as the covariance of the $f \sigma_8$ measurements is estimated rather than that of the power spectra directly.  Although this is a higher-order density-dependent clustering analyses it will not break the degeneracy in the manner suggested by \cite{ClippingCombined}.  Therefore a complete joint-likelihood approach with the VIPERS v7 data (or similar) would be a worthwhile avenue to explore. 
\end{chapter}

\backmatter

\singlespace

\phantomsection
\addcontentsline{toc}{chapter}{\bibname}
% Choose a bibliography style to suit your taste here
% This one was downloaded from http://web.reed.edu/cis/help/latex/bibtexstyles.html (June 2012)
%\bibliographystyle{ChicagoReedweb}
\bibliographystyle{mnras}
%\bibliographystyle{aa}
% \bibliography{chapter1/chapter1bib,chapter2/chapter2bib}

\bibliography{biblio}

\end{document}